\documentclass[rmp,aps,twocolumn,groupedaddress,floatfix,letterpaper,nofootinbib]{revtex4-1}
\usepackage{epsf,color,colordvi,amsmath,epsfig}
\usepackage{amsfonts,amsmath,amssymb,amsthm}
\usepackage{epsfig}
\usepackage{feynmf}		
\usepackage{graphics}

\def\numu{\nu_\mu}
\def\nue{\nu_e}
\def\nubar{\overline{\nu}}
\def\numubar{\overline{\nu}_\mu}
\def\nuebar{\overline{\nu}_e}
\def\inbar{\,\vrule height1.5ex width.4pt depth0pt}
\def\IR{\relax{\rm I\kern-.18em R}}
\def\IC{\relax\hbox{$\inbar\kern-.3em{\rm C}$}}

\def\weakangle{\sin^2{\theta}_W}
\def\weakanglesq{\sin^4{\theta}_W}
\def\ket#1{\left\vert #1 \right\rangle}  
\def\bra#1{\left\langle #1 \right\vert}  

\newcommand{\isotope}[2]{$^{#2}{\rm #1}$}

\begin{document}

\title{From eV to EeV: Neutrino Cross-Sections Across Energy Scales}

\author{Joseph A. Formaggio}
\email{josephf@mit.edu}
\affiliation{Laboratory for Nuclear Science\\
Massachusetts Institute of Technology,\\
Cambridge, MA 02139}

\author{G.~P.~Zeller}
\email{gzeller@fnal.gov}
\affiliation{Fermi National Accelerator Laboratory\\
Batavia, IL 60510}

\begin{abstract}  
Since its original postulation by Wolfgang Pauli in 1930, the neutrino has played a prominent role 
in our understanding of nuclear and particle physics. In the intervening 80 years, scientists have 
detected and measured neutrinos from a variety of sources, both man-made and natural. Underlying 
all of these observations, and any inferences we may have made from them, is an understanding 
of how neutrinos interact with matter. Knowledge of neutrino interaction cross-sections is an 
important and necessary ingredient in any neutrino measurement. With the advent of new precision 
experiments, the demands on our understanding of neutrino interactions is becoming even greater. 
The purpose of this article is to survey our current knowledge of neutrino cross-sections across all 
known energy scales: from the very lowest energies to the highest that we hope to observe.  The article covers a wide range of neutrino interactions including coherent scattering, neutrino capture, inverse beta decay, low energy nuclear interactions, quasi-elastic scattering, resonant pion production, kaon production, deep inelastic scattering and ultra-high energy interactions.  Strong emphasis is placed on experimental data whenever such measurements are available.
\end{abstract}                                                                 

\date{\today}
\maketitle
\tableofcontents

\section{Introduction}
\label{sec:intro}

The investigation into the basic properties of the particle known as the neutrino has been a particularly strong and active area of research within nuclear and particle physics. Research conducted over the latter half of the 20th century has revealed, for example, that neutrinos can no longer be considered as massless particles in the Standard Model, representing perhaps the first significant alteration to the theory. Moving into the 21st century, neutrino research continues to expand in new directions. Researchers further investigate the nature of the neutrino mass or explore whether neutrinos can help explain the matter-antimatter asymmetry of the universe. At the heart of many of these experiments is the need for neutrinos to interact with other Standard Model particles.  An understanding of these basic interaction cross sections is often an understated but truly essential element of any experimental neutrino program.

\begin{figure*}[htbp]
\begin{center}
\includegraphics[width=2.15\columnwidth,keepaspectratio=true]{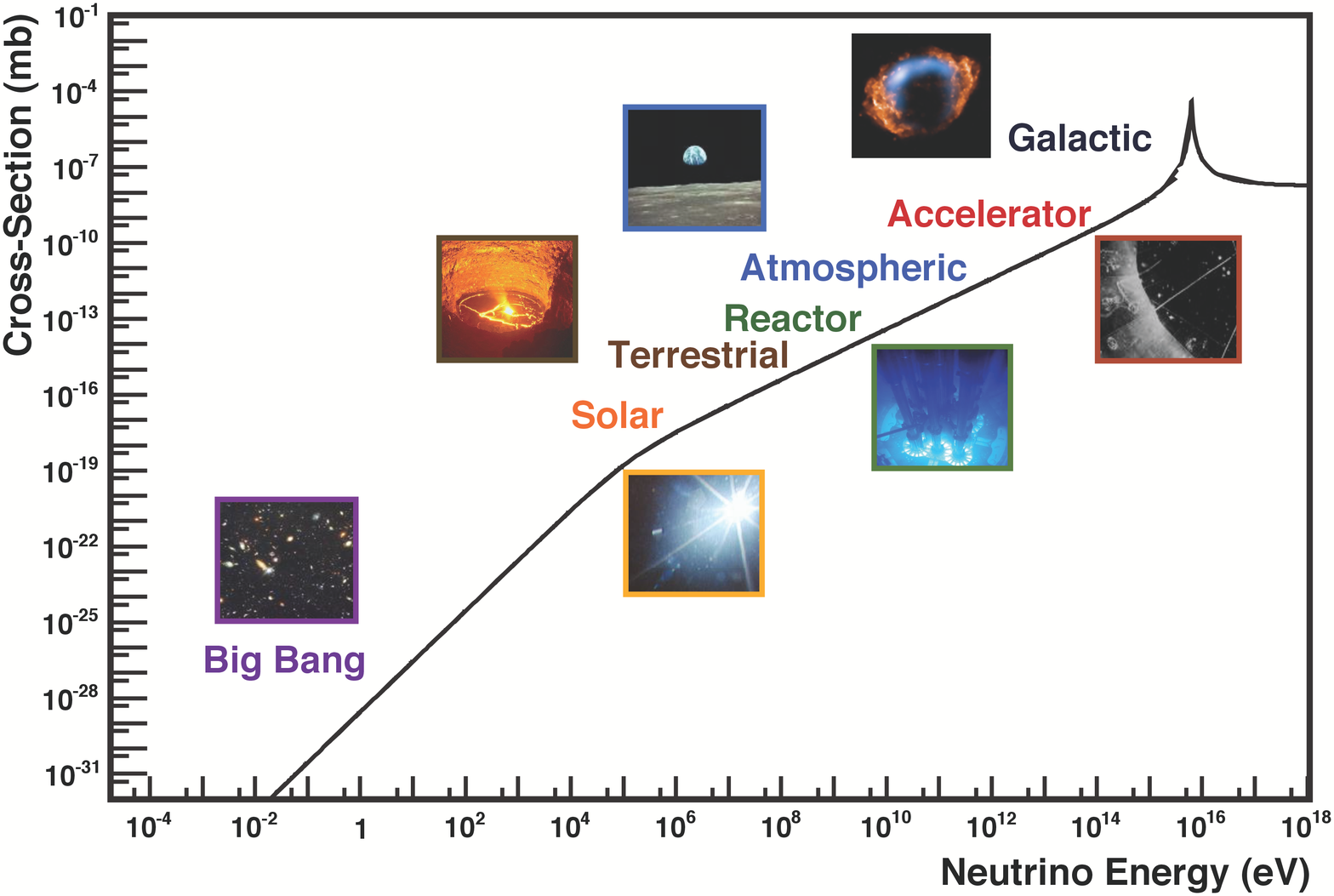} 
\caption{Representative example of various neutrino sources across decades of energy.  The electroweak cross-section for $\bar{\nu}_e e^- \rightarrow \bar{\nu}_e e^-$ scattering on free electrons as a function of neutrino energy (for a massless neutrino) is shown for comparison.  The peak at $10^{16}$ eV is due to the $W^-$ resonance, which we will discuss in greater detail in Section~\ref{sec:UHE}.}
\label{fig:everything}
\end{center}
\end{figure*}

The known reactions of neutrinos with matter fall completely within the purview of  the Standard Model of particle physics. The model of electroweak interactions govern  what those reactions should be, with radiative corrections that can be accurately calculated to many orders. As such, our goal in this review is essentially already complete: we would simply write down the electroweak Lagrangian and we would be finished.  Of course, in practice this is very far from the truth. As with many other disciplines, many factors compound our simple description, including unclear initial state conditions, subtle-but-important nuclear corrections, final state interactions, and other effects. One quickly finds that theoretical approximations which work well in one particular energy regime completely break down elsewhere. Even the language used in describing certain processes in  one context may seem completely foreign in another. Previous neutrino experiments could avoid this issue by virtue of the energy range in which they operated; now, however,  more experiments find themselves ``crossing boundaries" between different  energy regimes. Thus, the need for understanding neutrino cross-sections across many  decades of energy is becoming more imperative. To summarize our current collective understanding, this work provides a review of neutrino cross sections across all explored energy scales.  The range of energies covered, as well as their relevance to various neutrino sources, is highlighted in Figure~\ref{fig:everything}.  We will first establish the formalism of neutrino interactions by considering the simplest case of neutrino-electron scattering. Our focus will then shift to neutrino interaction cross sections at low (0-1 and 1-100 MeV), intermediate (0.1-20 GeV), high  (20-500 GeV) and ultra high (0.5 TeV-1 EeV) energies, emphasizing our current theoretical and experimental understanding of the processes involved.  Though it may be tempting to interpret these delineations as hard and absolute, they are only approximate in nature, meant as a guide for the reader.


\section{A Simple Case:  Neutrino-Lepton Scattering}
\label{sec:NeutrinoLeptonScat}

\subsection{Formalism: Kinematics}

\begin{figure}[!htb]
\begin{center}
\includegraphics[width=0.9\columnwidth,keepaspectratio=true]{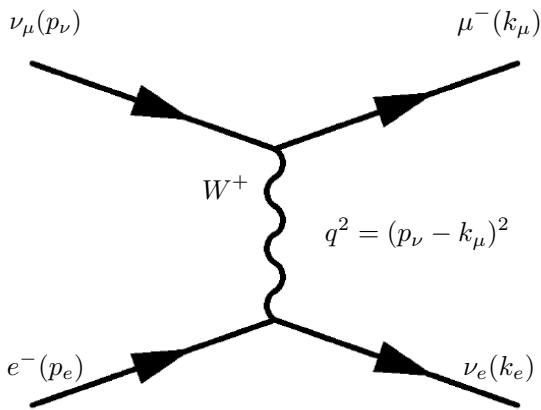} 
\caption{Diagram of 2-body scattering between an incoming muon neutrino with 4-momentum $p_\nu$ and an electron at rest with 4-momentum $p_e$.  See text for details.\label{fig:mikeplot}}
\end{center}
\end{figure}

Let us begin with the simplest of neutrino interactions, neutrino-lepton scattering. As a purely leptonic interaction, neutrino-lepton scattering allows us to establish the formalism and terminology used through the paper, without introducing some of the complexity that often accompanies neutrino-nuclear scattering.  The general form of the two-body scattering process is governed by the dynamics of the process encoded in the matrix elements and the  phase space available in the interaction.  Figure~\ref{fig:mikeplot} shows the tree-level diagram of a neutrino-lepton charged current interaction, known as inverse muon decay.  A muon neutrino with 4-momentum $p_\nu$ (aligned along the $z$-direction) scatters in this example with an electron with 4-momentum $p_e$, which is at rest in the lab frame. This produces an outgoing muon with 4-momentum $k_\mu$ and a scattered electron neutrino with 4-momentum $k_e$.  In the lab frame, the components of these quantities can be written as:

\begin{flalign*}
p_\nu = (E_\nu, \vec{p}_\nu)\\
k_\mu =(E_\mu, \vec{k}_\mu)\\
p_e = (m_e,~0)\\
k_e =  (E_e, \vec{k}_e).
\end{flalign*}

Here we use the convention of the $0^{\rm th}$ component corresponding to the energy portion of the energy-momentum vector, with the usual energy momentum relation $E_i^2 = |\vec{k}|_i^2 + m_i^2$.  From these 4-vector quantities, it is often useful to construct new variables which are invariant under Lorentz transformations:

\begin{flalign*}
s = (p_\nu + p_e)^2 {\rm~~~~(center~of~mass~energy)},\\
Q^2= - q^2 = ( p_\nu - k_\mu)^2 {\rm~~~~(4-momentum~transfer)},\\
y = \frac{p_e \cdot q}{p_e \cdot p_\nu} {\rm~~~~(inelasticity)}.
\end{flalign*}

In the case of two-body collisions between an incoming neutrino and a (stationary) target lepton, the cross-section is given in general by the formula ($\hbar = c = 1$)~\cite{Lifshitz:1974}:

\begin{equation}
\frac{d\sigma}{dq^2} = \frac{1}{16\pi}\frac{|{\cal M}^2|}{(s-(m_e+m_\nu)^2)(s-(m_e-m_\nu)^2)}
\end{equation}

\noindent which, in the context of very small neutrino masses, simplifies to:

\begin{equation}
\frac{d\sigma}{dq^2} = \frac{1}{16\pi}\frac{|{\cal M}^2|}{(s-m_e^2)^2}.
\end{equation}

\noindent Here, $\cal M$ is the matrix element associated with our particular interaction (Figure~\ref{fig:mikeplot}).  In the laboratory frame, it is always possible to express the cross-section in alternative ways by making use of the appropriate Jacobian.  For example, to determine the cross-section as a function of the muon's scattering angle, $\theta_\mu$, the Jacobian is given by:

\begin{equation}
\frac{dq^2}{d\cos{\theta}_\mu} = 2 |\vec{p}_\nu| |\vec{k}_\mu|, 
\end{equation}

\noindent while the Jacobian written in terms of the fraction of the neutrino energy imparted to the outgoing lepton energy ($y$) is given by: 

\begin{equation}
\frac{dq^2}{dy} = 2 m_e E_\nu.
\end{equation}

Pending on what one is interested in studying, the differential cross-sections can be recast to highlight a particular dependence or behavior.

%
%

\subsection{Formalism: Matrix Elements}

The full description of the interaction is encoded within the matrix element.  The Standard Model readily provides a prescription to describe neutrino interactions via the leptonic charged current and neutral current in the weak interaction Lagrangian.  Within the framework of the Standard Model, a variety of neutrino interactions are readily described~\cite{Weinberg:1967tq}.  These interactions all fall within the context of the general gauge theory of $SU(2)_L \times U(1)_Y$.  This readily divides the types of possible interactions for neutrinos into three broad categories.  The first is mediated by the exchange of a charged $W$ boson, otherwise known as a charged current (CC) exchange. The leptonic charged weak current, $j^\mu_W$, is given by the form:

\begin{equation}
j^\mu_W = 2 \sum_{\alpha = {\rm e,\mu,\tau}} \bar{\nu}_{L,\alpha} \gamma^\mu l_{\alpha L}.
\end{equation}

The second type of interaction, known as the neutral current (NC) exchange, is similar in character to the charged current case.  The leptonic neutral current term, $j^\mu_Z$, describes the exchange of the neutral boson, $Z^0$:

\begin{equation}
j^\mu_Z =2 \sum_{\alpha = {\rm e,\mu,\tau}} g_L^\nu \bar{\nu}_{\alpha L} \gamma^\mu \nu_{\alpha L} + g_L^f \bar{l}_{\alpha L} \gamma^\mu l_{\alpha L} + g_R^f \bar{l}_{\alpha R} \gamma^\mu l_{\alpha R}
\end{equation}

Here, $\nu_{\alpha L(R)}$ and $l_{\alpha L (R)}$ correspond to the left (right) neutral and charged leptonic fields, while $g^\nu_L$, $g^f_L$ and $g^f_R$ represent the fermion left and right- handed couplings (for a list of these values, see Table~\ref{tab:couplings}).  Though the charged leptonic fields are of a definite mass eigenstate, this is not necessarily so for the neutrino fields, giving rise to the well-known phenomena of neutrino oscillations.

Historically, the neutrino-lepton charged current and neutral current interactions have been used to study the nature of the weak force in great detail.   Let us return to the case of calculating the charged and neutral current reactions.  These previously defined components enter directly into the Lagrangian via their coupling to the heavy gauge bosons, $W^\pm$ and $Z^0$:

\begin{equation}
{\cal L_{CC}} = -\frac{g}{2\sqrt{2}} (j^\mu_W W_\mu + j^{\mu,\dagger}_W W^{\dagger}_\mu)
\end{equation}
 
\begin{equation}
{\cal L_{NC}} = -\frac{g}{2\cos{\theta_W}} j^\mu_Z Z_\mu 
\end{equation}

\begin{figure}[!htb]
\begin{center}
\vspace{-0.4cm}
\includegraphics[width=0.95\columnwidth,keepaspectratio=true]{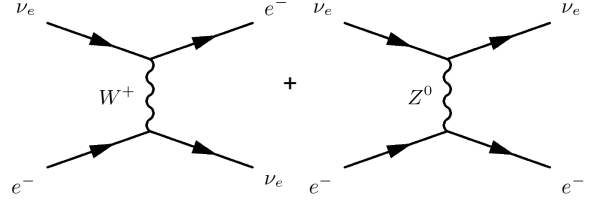} 
\vspace{-1cm}
\caption{Feynman tree-level diagram for charged and neutral current components of $\nu_e + e^- \rightarrow \nu_e + e^-$ scattering. \label{fig:Feynman_CC_NC}}
\end{center}
\end{figure}

Here, $W_\mu$ and $Z_\mu$ represent the heavy gauge boson field, $g$ is the coupling constant while $\theta_W$ is the weak mixing angle.  It is possible to represent these exchanges with the use of Feynman diagrams, as is shown in Figure~\ref{fig:Feynman_CC_NC}. Using this formalism, it is possible to articulate all neutrino interactions~\cite{'tHooft:1971ht} within this simple framework.  

Let us begin by looking at one of the simplest manifestations of the above formalism, where the reaction is a pure charged current interaction:

\begin{equation}
\nu_l + e^- \rightarrow l^- + \nu_e ~~~~{\rm(l=\mu~or~\tau)}
\end{equation}

The corresponding tree-level amplitude can be calculated from the above expressions.  In the case of $\nu_l + e$ (sometimes known as inverse muon or inverse tau decays) on finds:

\begin{equation}
{\cal M}_{CC} = -\frac{G_F}{\sqrt{2}} \{ [\bar{l} \gamma^\mu (1-\gamma^5) \nu_l] [\bar{\nu}_e \gamma_\mu (1-\gamma^5) e] \}
\end{equation}

Here, and in all future cases unless specified, we assume that the 4-momentum of the intermediate boson is much smaller than its mass (i.e. $|q^2| \ll M_{W,Z}^2$) such that propagator effects can be ignored.  In this approximation, the coupling strength is then dictated primarily by the Fermi constant, $G_F$:

\begin{equation}
G_F = \frac{g^2}{4\sqrt{2} M_W^2} = 1.1663788(7) \times 10^{-5} {\rm~GeV}^{-2}.
\end{equation}

By summing over all polarization and spin states, and integrating over all unobserved momenta, one attains the differential cross-section with respect to the fractional energy imparted to the outgoing lepton:

\begin{equation}
\frac{d\sigma(\nu_l e\rightarrow \nu_e l)}{d y} = \frac{2 m_e G_F^2 E_\nu}{\pi} \left( 1-\frac{(m_l^2-m_e^2)}{2m_e E_\nu} \right),
\end{equation}

\noindent where $E_\nu$ is the energy of the incident neutrino and $m_e$ and $m_l$ are the masses of the electron and outgoing lepton, respectively.  The dimensionless inelasticity parameter, $y$, reflects the kinetic energy of the outgoing lepton, which in this particular example is $y=~\frac{E_l~-~\frac{(m_l^2+m_e^2)}{2m_e}}{E_\nu}$.  The limits of $y$ are such that:

\begin{equation}
0 \le y \le y_{\rm max} = 1 - \frac{m_l^2}{2m_e E_\nu + m_e^2}
\end{equation}

Note that in this derivation, we have neglected the contribution from neutrino masses, which in this context is too small to be observed kinematically.  The above cross-section has a threshold energy imposed by the kinematics of the system, $E_{\nu} \ge \frac{(m_l^2-m_e^2)}{2m_e}$.  

In the case where $E_\nu \gg E_{\rm~thresh}$, integration of the above expression yields a simple expression for the total neutrino cross-section as a function of neutrino energy.

\begin{equation}
\sigma \simeq \frac{2 m_e G_F^2 E_\nu}{\pi} = \frac{G_F^2 s}{\pi}
\end{equation}

\noindent where $s$ is the center-of-mass energy of the collision. Note that the neutrino cross-section grows linearly with energy.

Because of the different available spin states, the equivalent expression for the inverse lepton decay of anti-neutrinos:

\begin{equation}
\bar{\nu}_e+ e \rightarrow \bar{\nu}_l + l~~~~{\rm(l=\mu~or~\tau)},
\end{equation}

\noindent has a different dependence on $y$ than its neutrino counterpart, although the matrix elements are equivalent.

\begin{equation}
\frac{d\sigma(\bar{\nu}_e e\rightarrow \bar{\nu}_l l)}{d y} = \frac{2 m_e G_F^2 E_\nu}{\pi} \left((1-y)^2-\frac{(m_l^2-m_e^2)(1-y)}{2m_e E_\nu} \right).
\end{equation}

Upon integration, the total cross-section is approximately a factor of 3 lower than the neutrino-cross-section.  The suppression comes entirely from helicity considerations.

Having just completed a charged current example, let us now turn our attention to a pure neutral current exchange, such as witnessed in the reaction:

\begin{equation}
\bar{\nu}_l+ e \rightarrow \bar{\nu}_l + e~~~~{\rm(l=\mu~or~\tau)}
\end{equation}

In the instance of a pure neutral current interaction, we are no longer at liberty to ignore the left-handed and right-handed leptonic couplings.  As a result, one obtains a more complex expression for the relevant matrix element (for a useful review, see~\cite{Adams:2008cm}).

\begin{equation}
{\cal M}_{NC} = -\sqrt{2}G_F \{ [\bar{\nu_l} \gamma^\mu (g_V^\nu-g_A^\nu\gamma^5) \nu_l] [\bar{e} \gamma_\mu (g_V^f-g_A^f\gamma^5) e] \}
\end{equation}

We have expressed the strength of the coupling in terms of the vector and axial-vector coupling constants ($g_V$ and $g_A$, respectively).  An equivalent formulation can be constructed using left- and right- handed couplings:

\begin{widetext}
\begin{eqnarray}
{\cal M}_{NC} = -\sqrt{2}G_F \{[g_L^\nu  \bar{\nu_l} \gamma^\mu (1-\gamma^5) \nu_l +  g_R^\nu \bar{\nu_l} \gamma^\mu (1+\gamma^5) \nu_l] \times \{ [g_L^f \bar{e} \gamma^\mu (1-\gamma^5) e +  g_R^f \bar{e} \gamma^\mu (1+\gamma^5) e] \}
\end{eqnarray}
\end{widetext}

The relation between the coupling constants are dictated by the Standard Model:

\begin{eqnarray*}
g_L^\nu = \sqrt{\rho} (+\frac{1}{2}),\\
g_R^\nu = 0,\\
g_L^f = \sqrt{\rho} (I_3^f-Q^f \weakangle),\\
g_R^f = \sqrt{\rho} (-Q^f \weakangle),\\
\end{eqnarray*}

\noindent or, equivalently,

\begin{eqnarray*}
g_V^\nu = g_L^\nu + g_R^\nu = \sqrt{\rho} (+\frac{1}{2}),\\
g_A^\nu = g_L^\nu - g_R^\nu = \sqrt{\rho} (+\frac{1}{2}),\\
g_V^f = g_L^f + g_R^f = \sqrt{\rho} (I_3^f-2Q^f \weakangle),\\
g_A^f = g_L^f - g_R^f = \sqrt{\rho} (I^f_3).\\
\end{eqnarray*}

Here, $I_3^f$ and $Q^f$ are the weak isospin and electromagnetic charge of the target lepton, $\rho$ is the relative coupling strength between charged and neutral current interaction (at tree level, $\rho \equiv 1$), while $\theta_W$ is the Weinberg mixing angle.  The Standard Model defines the relation between the electroweak couplings and gauge boson masses $M_W$ and $M_Z$:

\begin{equation}
\weakangle \equiv 1-\frac{M_W^2}{M_Z^2}
\end{equation}

In the observable cross-section for the neutral current reactions highlighted above, we find that they are directly sensitive to the left and right handed couplings.  In the literature, the cross-section is often expressed in terms of their vector and axial-vector currents:

\begin{eqnarray*}
g_V \equiv (2 g_L^\nu g_V^f)\\
g_A \equiv (2 g_L^\nu g_A^f)
\end{eqnarray*}

\begin{widetext}
\begin{eqnarray*}
\frac{d\sigma (\nu_l e \rightarrow \nu_l e)}{d y} = \frac{m_e G_F^2 E_\nu}{2\pi} \left((g_V+g_A)^2+(g_V-g_A)^2(1-y)^2 - (g_V^2-g_A^2)\frac{m_e y}{E_\nu} \right),\\
\frac{d\sigma (\bar{\nu}_l e \rightarrow \bar{\nu}_l e)}{d y} = \frac{m_e G_F^2 E_\nu}{2\pi} \left((g_V-g_A)^2+(g_V+g_A)^2(1-y)^2 - (g_V^2-g_A^2)\frac{m_e y}{E_\nu} \right).
\end{eqnarray*}
\end{widetext}

Though we have limited ourselves to discussing neutrino lepton scattering, the rules governing the coupling strengths are pre-determined by the Standard Model and can be used to describe neutrino-quark interactions as well.  A full list of the different possible coupling strengths for the known fermion fields is shown in Table~\ref{tab:couplings}.  A more in-depth discussion of these topics can be found in a variety of introductory textbooks.  We highlight~\cite{bib:Giunti} as an excellent in-depth resource for the reader.

As such, neutrino-electron scattering is a powerful probe of the nature of the weak interaction, both in terms of the total cross-section as well as its energy dependence~\cite{bib:Marciano03}.  We will briefly examine the experimental tests of these reactions in the next section.

\begin{table}[htdp]
\caption{Values for the $g_V$ (vector), $g_A$ (axial), $g_L$ (left), and $g_R$ (right) coupling constants for the known fermion fields.}
\begin{center}
\begin{tabular}{|c|c|c|c|c|}
\hline
Fermion & $g_L^f$  & $g_R^f$ & $g_V^f$ & $g_A^f$ \\
\hline
& & & & \\
$\nu_e, \nu_\mu, \nu_\tau$ & +$\frac{1}{2}$ &  0 & +$\frac{1}{2}$ & $+\frac{1}{2}$ \\
& & & & \\
$e, \mu, \tau$ & $-\frac{1}{2}+\weakangle$ &  $+\weakangle$ & $-\frac{1}{2} +2\weakangle$ & $-\frac{1}{2}$ \\
& & & & \\
$u,c,t$ & $+\frac{1}{2}-\frac{2}{3}\weakangle$ &  $-\frac{2}{3}\weakangle$ & $+\frac{1}{2} -\frac{4}{3}\weakangle$ & $+\frac{1}{2}$ \\
& & & & \\
$d,s,b$ & $-\frac{1}{2}+\frac{1}{3}\weakangle$ &  +$\frac{1}{3}\weakangle$ & $-\frac{1}{2} +\frac{2}{3}\weakangle$ & $-\frac{1}{2}$ \\
& & & & \\
\hline
\end{tabular}
\end{center}
\label{tab:couplings}
\end{table}%

Before leaving neutrino-lepton interactions completely, we turn our attention to the last possible reaction archetype, where the charged current and neutral current amplitudes interfere with one another.  Such a combined exchange is realized in $\nu_e + e \rightarrow \nu_e + e$ scattering (see Fig.~\ref{fig:Feynman_CC_NC}).  The interference term comes into play by shifting $g_V^f \rightarrow g_V^f + \frac{1}{2}$ and $g_A^f \rightarrow g_A^f +1$.

One remarkable feature of neutrino-electron scattering is that it is highly directional in nature. The outgoing electron is emitted at very small angles with respect to the incoming neutrino direction.  A simple kinematic argument shows that indeed:

\begin{equation}
E_e \theta_e ^2 \leq 2 m_e.
\end{equation}

This remarkable feature has been exploited extensively in various neutrino experiments, particularly for solar neutrino detection.  The Kamiokande neutrino experiment was the first to use this reaction to reconstruct \isotope{B}{8} neutrino events from the sun and point back to the source. The Super-Kamiokande experiment later expanded the technique, creating a photograph of the sun using neutrinos~\cite{bib:SuperKsun}\footnote{The fact that such a picture was taken underground during both day and night is also quite remarkable!}.  The technique was later been used by other solar experiments, such as SNO~\cite{bib:sno1,bib:sno2,bib:sno3} and BOREXINO~\cite{bib:BOREXINO,bib:BorexinoResults}.

\subsection{Experimental Tests of Electroweak Theory}

Neutrino-lepton interactions have played a pivotal role in our understanding of the working of the electroweak force and the Standard Model as a whole.  Consider as an example the first observation of the reaction $\bar{\nu}_\mu + e^- \rightarrow \bar{\nu}_\mu + e^-$ made the CERN bubble chamber neutrino experiment, Gargamelle~\cite{Hasert:1973cr}.  This observation, in conjunction with the observation of neutral current deep inelastic scattering~\cite{Hasert:1973ff,Benvenuti74}, confirmed the existence of weak neutral currents and helped solidify the $SU_L(2) \times U(1)_Y$ structure of the Standard Model~\cite{Weinberg:1967tq,'tHooft:1971ht}.  The very observation of the phenomena made a profound impact on the field of particle physics.  

Subsequent experiments further utilized the information from the observed rates of neutral current reactions as a gauge for measuring $\weakangle$ directly.  Neutrino-lepton scattering is a particularly sensitive probe in this regard because to first order (and even to further orders of $\alpha$, see Section~\ref{sec:radiative}), the cross-sections depend only on one parameter, $\weakangle$.  

Various experimental methods have been employed to measure neutrino-lepton scattering.  Among the first included the observation of $\bar{\nu}_e + e^- \rightarrow e^- + \bar{\nu}_e$ scattering by Reines, Gurr, and Sobel~\cite{Reines:1976pv} at the Savannah River Plant reactor complex.  Making use of the intense $\bar{\nu}_{e}$ flux produced in reactors, a $\pm 20\%$ measurement on the weak mixing angle was extracted.  A more recent result from the TEXANO experiment~\cite{Deniz:2008rw,Deniz:2009mu} also utilizes reactor anti-neutrinos as their source.  There exists an inherent difficulty in extracting these events, as they are often masked by large low-energy backgrounds, particularly those derived from uranium and thorium decays.

\begin{figure}[htbp]
\begin{center}
\includegraphics[width=0.95\columnwidth]{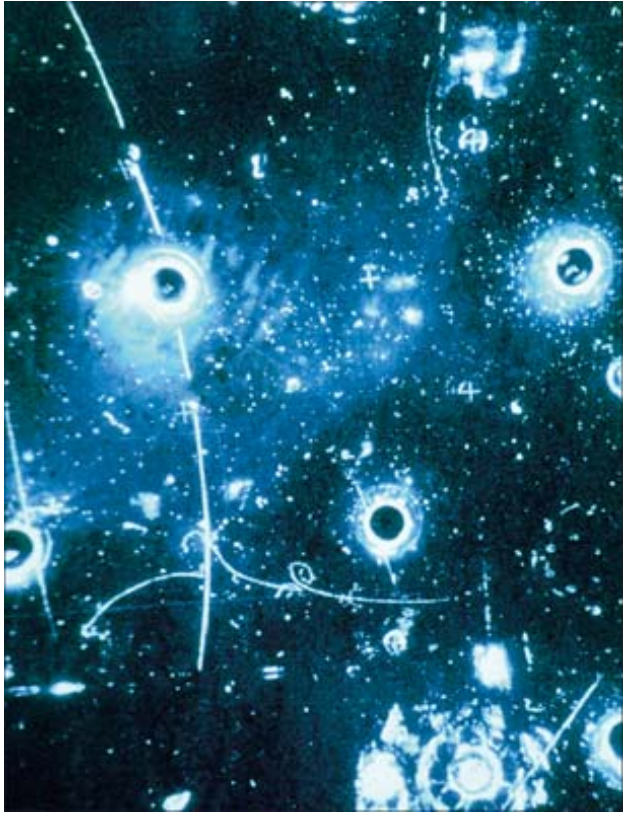}
\caption{The first candidate leptonic neutral current event from the Gargamelle CERN experiment. An incoming muon-antineutrino knocks an electron forwards (towards the left), creating a characteristic electronic shower with electronÐpositron pairs.  Photograph from CERN.}
\label{fig:Gargamelle}
\end{center}
\end{figure}

\begin{table*}[htdp]
\caption{The integrated cross-section for neutrino-lepton scattering interactions.  Corrections due to leptonic masses and radiative correlations are ignored.  Cross-sections are compared to the asymptotic cross-section $\sigma_0 = \frac{G_F^2 s}{\pi}$. Listed are also the experiments which have measured the given reaction, including Gargamelle~\cite{Hasert:1973cr}, the Savannah River Plant~\cite{Reines:1976pv}, Brookhaven National Laboratory (BNL)~\cite{Ahrens:1983py,Abe:1989qk,Ahrens:1990fp}, LAMPF~\cite{Allen:1992qe}, LSND~\cite{PhysRevD.63.112001}, CCFR~\cite{Mishra:1990yf}, CHARM~\cite{Vilain:1996yf,Vilain:1994hm}, NuTeV~\cite{Formaggio:2001jz}, and TEXONO~\cite{Deniz:2008rw}.}
\begin{center}
\begin{tabular}{|c|c|c|c|}
\hline
Reaction & Type & $\sigma (E_\nu \gg E_{\rm thresh})/\sigma_0$ & Experimental Probes \\
\hline
$\nu_e e^- \rightarrow \nu_e e^-$ &CC and NC & $ (\frac{1}{4}+\weakangle+\frac{4}{3}\weakanglesq)$ &CHARM, LAMPF, LSND\\
$\bar{\nu}_e e^- \rightarrow \bar{\nu}_e e^-$ & CC and NC & $ (\frac{1}{12}+\frac{1}{3}\weakangle+\frac{4}{3}\weakanglesq)$ & CHARM, TEXONO, Savannah River\\
$\bar{\nu}_e e^- \rightarrow \bar{\nu}_{\mu} \mu^-$ &CC& $\frac{1}{3}$ & \\
$\nu_{\mu} e^- \rightarrow \nu_e \mu^-$ & CC & $1$ & CHARM, CCFR, NuTeV\\
$\nu_{\mu} e^- \rightarrow \nu_{\mu} e^-$ & NC &  $ (\frac{1}{4}-\weakangle+\frac{4}{3}\weakanglesq)$ & CHARM, LAMPF, LSND, BNL\\
$\bar{\nu}_{\mu} e^- \rightarrow \bar{\nu}_{\mu} e^-$ & NC &  $(\frac{1}{12}-\frac{1}{3}\weakangle+\frac{4}{3}\weakanglesq)$ & Gargamelle, BNL\\
\hline
\end{tabular}
\end{center}
\label{tab:nu-lepton-xsec}
\end{table*}%

The majority of the precision tests of recent have been carried out using high energy neutrino beams.  Experiments such as Gargamelle~\cite{Hasert:1973cr}, Brookhaven's AGS source~~\cite{Ahrens:1983py,Abe:1989qk,Ahrens:1990fp}, CHARM II~\cite{Vilain:1996yf,Vilain:1994hm}, CCFR~\cite{Mishra:1990yf}, and NuTeV~\cite{Formaggio:2001jz}  fall within this category.  Often these experiments exploit the rise in cross-section with energy to increase the sample size collected for analysis.   Stopped pion beams have also been used for these electroweak tests at the Los Alamos National Laboratory in the LAMPF~\cite{Allen:1992qe} and LSND~\cite{PhysRevD.63.112001} experiments.  Table~\ref{tab:nu-lepton-xsec} provides a summary of the types of measurements made using pure neutrino-lepton scattering.  

\subsection{Radiative Corrections and $G_F$}
\label{sec:radiative}

Upon inspection of the cross-section formalisms discussed above, it is clear that, with the exception of ratios, one is critically dependent on certain fundamental constants, such as the strength of the weak coupling constant $G_F$.  Ideally one would like to separate the dependence on the weak mixing angle from the Fermi constant strength.  Fortunately, measurements of the muon lifetime provides such a possibility, as it is inversely proportional to the coupling strength $G_F$ and the muon mass $m_\mu$:

\begin{equation}
(\tau_\mu)^{-1} = \frac{G_F^2 m_\mu^5}{192\pi^3}  f(\rho) (1+\frac{3}{5}\frac{m_\mu^2}{M_W^2})  (1+\Delta(\alpha)) 
\end{equation}

In the above expression, $f(\rho)$ is a phase factor, the $\frac{m_\mu^2}{M_W^2}$ factor encapsulates the W-boson propagator, and $\Delta(\alpha)$ encodes the QED radiative field corrections.  For completeness, we  list these correction factors:

\begin{widetext}
\begin{eqnarray}
f(\rho) = 1 - 8\rho + 8\rho^3 -\rho^4 - 12\rho^2\ln{\rho} \simeq 0.999813\\
\Delta =\frac{\alpha}{\pi} \left( \frac{25}{8}-\frac{\pi^2}{2} - (9 + 4\pi^2 + 12\ln{\rho})\rho + 16\pi^2\rho^{3/2}+{\cal O}(\rho^2)\right) + {\cal O}(\frac{\alpha^2}{\pi^2})+...,
\end{eqnarray}
\end{widetext}


\noindent where $\rho = (\frac{m_e}{m_\mu})^2$ and $\alpha$ is the fine structure constant.  Radiative QED corrections have been calculated to second order and higher in electroweak theory paving the way to precision electroweak tests of the Standard Model.  The best measurement of the muon lifetime to date has been made by the MuLan experiment~\cite{Webber:2010zf}, yielding a value for $G_F$ of $1.1663788(7) \times 10^{-5}$ GeV$^{-2}$, a precision of 0.6 ppm.

At tree level, knowing $G_F$ (and $\alpha$ and $M_Z$) it is possible to exactly predict the value of $\weakangle$ and test this prediction against the relevant cross-section measurements.  However, once one introduces 1-loop radiative contributions, dependencies on the top and Higgs masses are also introduced.  The size of these corrections depend partially on the choice of the normalization scheme.  The two commonly used renormalization schemes include the Sirlin on-shell model~\cite{bib:Sirlin80} or the modified minimal subtraction scheme~\cite{bib:MS}.  In the latter method, the Weinberg angle is defined by $M_W$ and $M_Z$ at some arbitrary renormalized mass scale $\mu$, which is typically set to the electroweak scale $M_Z$. 

\begin{equation}
\weakangle^{\bar{MS}} = 1-\frac{M_W(\mu)^2}{M_Z(\mu)^2}
\end{equation}

Such radiative corrections, although small, often need to be accounted for in order to properly predict the $\weakangle$ value.  Theoretical compilation of such radiative effects can be found in a variety of papers (see, for example,~\cite{bib:Marciano03}).


\section{Threshold-less Processes:  $E_\nu \sim~0 - 1$ MeV}
\label{sec:Thresholdless}

Having established the formalism of basic neutrino interactions, we turn our attention toward describing  neutrino interactions across the various energy scales.  The first step in our journey involves {\em threshold-less} interactions, which can be initiated when the neutrino has essential zero momentum.  Such processes include coherent scattering and neutrino capture~\footnote{Technically, neutrino elastic scattering off of free electrons also falls within this definition, as discussed earlier in this paper.}.

\subsection{Coherent Scattering}

Coherent scattering involves the neutral current exchange where a neutrino interacts {\em coherently} with the nucleus:

\begin{equation}
\nu + A^Z_N \rightarrow~ \nu + A^{*Z}_{N} 
\end{equation}

Shortly after the discovery of neutral-current neutrino reactions, Freedman, Schramm and Tubbs pointed out that neutrino-nucleus interactions should also exist~\cite{bib:Freedman}.  Furthermore, one could take advantage of the fact that at low energies the cross-section should be coherent across all the nucleons present in the nucleus.  As a result, the cross-section grows as the square of the atomic number, $A^2$.  Such an enhancement is possible if the momentum transfer of the reaction is much smaller than the inverse of the target size.  Letting $Q$ represent the momentum transfer and $R$ the nuclear radius, the coherence condition is satisfied when $QR \ll 1$. Under these conditions, the relevant phases have little effect, allowing the scaling to grow as $A^2$.  

Given a recoil kinetic energy $T$ and an incoming neutrino energy $E_\nu$, the differential cross-section can be written compactly as the following expression:

\begin{equation}
\frac{d\sigma}{dT} = \frac{G_F^2}{4\pi} Q^2_W M_A (1-\frac{M_A T}{2 E_\nu^2}) F(Q^2)^2,
\end{equation}

\noindent where $M_A$ is the target mass ($M_A = A M_{\rm nucleon}$), $F(Q^2)$ is the nucleon form factor, and $Q_W$ is the weak current term:

\begin{equation}
Q_W = N - Z(1-4\weakangle)
\end{equation}

The cross-section essentially scales quadratically with neutron ($N$) and proton ($Z$) number; the latter highly suppressed due to the $(1-4\weakangle\simeq 0)$ term.  The form-factor $F(Q^2)$ encodes the coherence across the nucleus and drops quickly to zero as $QR$ becomes large.

Despite the strong coherent enhancement enjoyed by this particular process, this particular interaction has yet to be detected experimentally.  Part of the obstacle stems from the extremely small energies of the emitted recoil.  The maximum recoil energy from such an interaction is limited by the kinematics of the elastic collision:
\begin{equation}
T_{\rm max} = \frac{E_\nu}{1 + \frac{M_A}{2E_\nu}},
\end{equation}
\noindent similar to that of any elastic scatter where the mass of the incoming particle is negligible.  Several experiments have been proposed to detect this interaction; often taking advantage of advances in recoil detection typically utilized by dark matter experiments~\cite{bib:CLEAR,bib:RICOCHET}.  The interaction has also been proposed as a possible mechanism for cosmic relic neutrinos, due to its non-zero cross-section at zero momentum.  However, the $G_F^2$ suppression makes detection beyond the reach of any realizable experiment.

\subsection{Neutrino Capture on Radioactive Nuclei}

Neutrino capture on radioactive nuclei, sometimes referred to as {\em enhanced} or {\it stimulated} beta decay emission constitutes another threshold-less mechanism in our library of possible neutrino interactions.  The process is similar to that of ordinary beta decay:

\begin{equation}
A^Z_N \rightarrow A^{Z+1}_{N-1} + e^- + \bar{\nu}_e,
\end{equation}

\noindent except the neutrino is interacting with the target nucleus.

\begin{equation}
\nu_e + A^Z_N \rightarrow e^- + A^{Z+1}_{N-1}.
\end{equation}

This reaction has the same observable final states as its beta decay counterpart.  What sets this reaction apart from other neutrino interactions is that the process is exothermic and hence no energy is required to initiate the reaction~\footnote{In principle, any elastic interaction on a free target has a finite cross-section at zero momentum, but such interactions would be impossible to discern due to the extremely small transfer of momentum.}.  The cross-section amplitude is directly related to that of beta decay.  Using the formalism of ~\cite{Beacom:1999}, the cross-section can be written as:

\begin{widetext}
\begin{equation}
\frac{d \sigma}{d\cos{\theta}} = \frac{G_F^2 |V_{ud}|^2 F(Z_{\rm f},E_e)}{2\pi\beta_\nu} E_e p_e f^2_V(0) \left((1+\beta_e \beta_\nu \cos{\theta}) +3 \lambda^2 (1-\frac{1}{3}\beta_e \beta_\nu \cos{\theta})\right
)\end{equation}
\end{widetext}

\noindent where $\beta_{e}$ and $\beta_{\nu}$ are the electron and neutrino velocities, respectively, $E_e$, $p_e$, and $\cos{\theta}$ are the electron energy, momentum, and scattering angle, $\lambda^2$ is the axial-to-vector coupling ratio, and $|V_{ud}|^2$ is the Cabbibo angle.  The Fermi function, $F(Z_f,E)$ encapsulates the effects of the Coulomb interaction for a given lepton energy $E_e$ and final state proton number $Z_{\rm f}$.  We will discuss the coupling strengths $f_V(0)$ and $\lambda^2$ later.  

In the above expression, we no longer assume that $\beta_\nu \rightarrow c$. If the neutrino flux is proportional to the neutrino velocity, then the product of the cross-section and the flux results in a finite number of observable events.  If the neutrino and the nucleus each possess negligible energy and momentum, the final-state electron is ejected as a mono-energetic particle whose energy is above the endpoint energy of the reaction.

The interaction cross section of very low energy neutrinos was first suggested by Weinberg~\cite{weinberg1962und}. Recently, this process has attracted particular interest thanks to the work by~\cite{bib:Cocco}, where the authors have considered the process as a means to detect cosmological neutrinos. The reaction has received attention partially due to the advancement of beta decay experiments in extending the reach on neutrino mass scales.  The mechanism, like its coherent counterpart, remains to be observed.


\section{Low Energy Nuclear Processes: $E_\nu \sim$ 1-100 MeV}
\label{sec:LowEnergyNuclear}

As the energy of the neutrino increases, it is possible to probe the target nucleus at smaller and smaller length scales.  Whereas coherent scattering only allows one to ``see" the nucleus as a single coherent structure, higher energies allow one to access nucleons individually.  These low energy interactions have the same fundamental characteristics as those of lepton scattering, though the manner in which they are gauged and calibrated is very different.  And, unlike the thresholdless scattering mechanisms discussed previously, these low energy nuclear processes have been studied extensively in neutrino experiments. 

\subsection{Inverse Beta Decay}

The simplest nuclear interaction that we can study is antineutrino-proton scattering, otherwise known as inverse beta decay:  

\begin{equation}
\bar{\nu}_e + p \rightarrow e^+ + n 
\end{equation}

Inverse beta decay represents one of the earliest reactions to be studied, both theoretically~\cite{bib:Bethe} and experimentally~\cite{Reines:1976pv}. This reaction is typically measured using neutrinos produced from fission in nuclear reactors. The typical neutrino energies used to probe this process range from threshold ($E_\nu \ge 1.806$ MeV) to about 10 MeV~\footnote{The neutrino energy threshold $E_\nu^{\rm thresh}$ in the lab frame is defined by $\frac{(m_n+m_e)^2-m_p^2}{2m_p}$.}.  As this reaction plays an important role in understanding supernova explosion mechanisms, its relevance at slightly higher energies (10-20 MeV) is also of importance.  In this paper, we follow the formalism of~\cite{Beacom:1999}, who expand the cross-section on the proton to first order in nucleon mass in order to study the cross-section's angular dependence.  In this approximation, all relevant form factors approach their zero-momentum values.  The relevant matrix element is given by the expression:

\begin{widetext}
\begin{equation}
{\cal M} = \frac{G_F V_{ud}}{\sqrt{2}} \left[ \bra{\bar{n}} (\gamma_\mu f_V(0) - \gamma_\mu \gamma^5 f_A(0) - \frac{i f_P(0)}{2M_n} \sigma_{\mu\nu}q^\nu)\ket{p} \bra{\bar{\nu}_e} \gamma^\mu (1-\gamma^5) \ket{e} \right].
\end{equation}
\end{widetext}

In the above equation, $f_V, f_A,$ and $f_P$ are nuclear vector, axial-vector, and Pauli (weak magnetism) form factors evaluated at zero momentum transfer 
(for greater detail on the form factor behavior, see Section~\ref{sec:ONT}).  To first order, the differential cross-section can therefore be written down as:

\begin{widetext}
\begin{equation}
\frac{d\sigma(\bar{\nu}_e p \rightarrow e^+ n)}{d\cos{\theta}} = \frac{G_F^2 |V_{ud}|^2 E_e p_e}{2\pi} \left[ f^2_V(0) (1+\beta_e \cos{\theta}) + 3 f^2_A(0) (1-\frac{\beta_e}{3} \cos{\theta}) \right],
\label{eq:IBD}
\end{equation}
\end{widetext}

\noindent where $E_e, p_e, \beta_e,$ and $\cos{\theta}$ refer to the electron's energy, momentum, velocity and scattering angle; respectively.

A few properties in the above formula immediately attract our attention.  First and foremost is that the cross-section neatly divides into two distinct ``components"; a vector-like component, called the Fermi transition, and an axial-vector like component, referred to as Gamow-Teller.  We will talk more about Fermi and Gamow-Teller transitions later.  

A second striking feature is its angular dependence.  The vector portion has a clear $(1+\beta_e \cos{\theta})$ dependence, while the axial portion has a $(1-\frac{\beta_e}{3} \cos{\theta})$ behavior, at least to first order in the nucleon mass.  The overall angular effect is weakly backward scattered for anti-neutrino-proton interactions, showing that the vector and axial-vector terms both contribute at equivalent amplitudes.  This is less so for cases where the interaction is almost purely Gamow-Teller in nature, such as $\nu d$ reactions.  In such reactions, the backwards direction is more prominent.  Such angular distributions have been posited as an experimental tag for supernova detection~\cite{Beacom:2002ix}. 

The final aspect of the cross-section that is worthy to note is that it has a near one-to-one correspondence with the beta decay of the neutron.  We will explore this property in greater detail in the next section.

\subsection{Beta Decay and Its Role in Cross-Section Calibration}

The weak interaction governs both the processes of decay as well as scattering amplitudes.  It goes to show that, especially for simple systems, the two are intimately intertwined, often allowing one process to provide robust predictions for the other.  The most obvious nuclear target where this takes place is in the beta decay of the neutron.  In much the same way as muon decay provided a calibration of the Fermi coupling constant for purely leptonic interactions, neutron beta decay allows one to make a prediction of the inverse beta decay cross-section from experimental considerations alone.

\begin{table*}[htdpb]
\caption{Neutron decay parameters contributing to Equation~\ref{eq:neutrondecay}.  Values extracted from~\cite{bib:SnowARNPS} and~\cite{ga}. }
\begin{center}
\begin{tabular}{|c|c|c|c|}
\hline
Constant & Expression & Numerical Value & Comment \\
\hline
$\lambda$ & $|\frac{g_A}{g_V}|e^{i\phi}$ & $-1.2694 \pm 0.0028$ & axial/vector coupling ratio \\
$a$ & $\frac{1-|\lambda|^2}{1+3|\lambda|^2}$ & $-0.103 \pm 0.004$ & electron-anti-neutrino asymmetry \\
$b$ & 0 & 0 & Fietz interference \\
$A$ & $-2\frac{|\lambda|^2+|\lambda|\cos\phi}{1+3|\lambda|^2}$& $-0.1173\pm 0.0013$ & spin-electron asymmetry \\
$B$ & $+2\frac{|\lambda|^2-|\lambda|\cos\phi}{1+3|\lambda|^2}$& $0.9807\pm 0.0030$ & spin-antineutrino asymmetry \\
$D$ & $2\frac{|\lambda|\sin\phi}{1+3|\lambda|^2}$& $(-4 \pm 6) \times 10^{-4}$ & T-odd triple product \\
$f~(1+\delta_R)$ & & $1.71480 \pm 0.000002$ & theoretical phase space factor \\
$\tau_n$ & $(\frac{m_e^5}{2\pi^3}f_R G_F^2 |V_{ud}|^2 (1+ 3\lambda^2))^{-1}$ &$(885.7\pm 0.8)$ s & neutron lifetime \\
\hline
\end{tabular}
\end{center}
\label{tab:NeutronDecay}
\end{table*}

For the case of neutron beta decay, the double differential decay width at tree-level is given by the expression below~\cite{bib:SnowARNPS}:

\begin{widetext}
\begin{eqnarray*}
\frac{d^3\Gamma}{dE_e d\Omega_e d\Omega_\nu} = G_F^2 |V_{ud}|^2 (1+3\lambda^2) |\vec{p_e}| (T_e + m_e)  (E_0-T_e)^2 \left[ 1+a \frac{\vec{p}_e\cdot \vec{p}_\nu}{T_e E_\nu} + b \frac{m_e}{T_e} + \vec{\sigma}_n \cdot (A  \frac{\vec{p}_e}{T_e} + B \frac{\vec{p}_\nu}{E_\nu}+ D \frac{\vec{p}_e \times \vec{p}_\nu}{T_e E_\nu})\right].
\label{eq:neutrondecay}
\end{eqnarray*}
\end{widetext}

Here, $\vec{p}_e$ and $\vec{p}_\nu$ are the electron and neutrino momenta, $T_e$ is the electron's kinetic energy, $E_\nu$ is the outgoing anti-neutrino energy, $E_0$ is the endpoint energy for beta decay, and $\sigma_n$ is the neutron spin.  The definitions of the other various constants are listed in Table~\ref{tab:NeutronDecay}.

Integrating over the allowed phase space provides a direct measure of the energy-independent portion of the inverse beta decay cross-section, including internal radiative corrections.  That is, Equation~\ref{eq:IBD} can also be written as:

\begin{widetext}
\begin{equation}
\frac{d\sigma(\bar{\nu}_e p \rightarrow e^+ n)}{d\cos{\theta}} = \frac{2\pi^2}{2m_e^5 f(1+\delta_R) \tau_n} E_e p_e \left[(1+\beta_e \cos{\theta}) + 3 \lambda^2 (1-\frac{\beta_e}{3} \cos{\theta}) \right]
\end{equation}
\end{widetext}

The term $f(1+\delta_R)$ is a phase space factor that includes several inner radiative corrections.  Additional radiative corrections and effects due to finite momentum transfer have been evaluated.  From a theoretical standpoint, therefore, the inverse beta decay cross-section is well predicted, with uncertainties around $\pm 0.5\%$\footnote{Some caution should be taken, as currently the most accurate value for the neutron lifetime is $6.5\sigma$ away from the PDG average value~\cite{SEREBROV:2005je}.}. 

The ability for measured beta decay rates to assist in the evaluation of neutrino cross-sections is not limited solely to inverse beta decay.  Beta decay transitions also play a pivotal role in the evaluation of neutrino cross-sections for a variety of other target nuclei.  Nuclei which relate back to super-allowed nuclear transitions stand as one excellent example.  Isotopes that undergo super-allowed Fermi transitions ($0^+ \rightarrow 0^+$) provide the best test of the Conserved Vector Current (CVC) hypothesis~\cite{bib:Gershtein1956,bib:Feynmann1958} and, if one includes measurements of the muon lifetime, the most accurate measurements of the quark mixing matrix element of the CKM matrix, $V_{ud}$~\cite{hardy:129}.  Typically, the value for $V_{ud}$ can be extracted by looking at the combination of the statistical rate function (${\cal F}$) and the partial half-life ($t$) of a given super-allowed transition.  Because the axial current cannot contribute in lowest order to transitions between spin-0 states, the experimental ${\cal F}t$-value is related directly to the vector coupling constant.  For an isospin-1 multiplet, one obtains:

\begin{equation}
|V_{ud}|^2 = \frac{K}{2G_F^2(1+\Delta_R) {\cal F}t}
\end{equation}

\noindent where $\Delta_R$ are the nucleus-independent radiative corrections in $0^+ \rightarrow 0^+$ transitions and $K$ is defined as $K \equiv 2\pi^3 \ln{2}/m_e^5 = (8120.271\pm0.012) \times 10^{-10}$ GeV$^{-4}$ s.  The ${\cal F}t$ value from various transitions are very precisely measured (down to the 0.1\% level) to be $3072.3 \pm 2.0$, while the radiative corrections enter at the 2.4\% level~\cite{bib:Towner1998}.  The process is not directly relatable to that of inverse beta decay because of the lack of the axial form-factor, but it provides a strong constraint on the validity of the CVC hypothesis.

Even excluding neutron decay and super-allowed transitions, beta decay measurements also play an important role in the calculation of  low energy cross-sections simply because they represent a readily measurable analog to their neutrino interaction counterpart.  For example, the $\beta^+$ decay from \isotope{N}{12} to the ground state of \isotope{C}{12} is often used to calibrate calculations of the exclusive cross-section of \isotope{C}{12}$(\nu_e,e^-)$\isotope{N}{12}~\cite{bib:Fukugita1988}.   In the case of deuterium targets, the decay width of tritium beta decay provides an extremely strong constraint on the $\nu d$ cross-section~\cite{Nakamura:2001fr}.  Finally, though not least, both allowed and forbidden $\beta^\pm$ decays often allow a direct measure of the Gamow-Teller contribution to the total cross-section.  Comparisons of neutrino reactions on \isotope{Cl}{37} and the decay process \isotope{Ca}{37}$(\beta^+)$\isotope{K}{37} are prime example of this last constraint technique~\cite{bib:Aufderheide1994}.

\subsection{Theoretical Calculations of Neutrino-Deuterium Cross-Sections}

Next to hydrogen, no nuclear target is better understood than deuterium.  Neutrino-deuterium scattering plays an important role in experimental physics, as heavy water (D$_2$O) was the primary target of the Sudbury Neutrino Observatory (SNO)~\cite{bib:sno1,bib:sno2,bib:sno3,bib:sno4,bib:sno5,bib:sno6,bib:sno7}.  The SNO experiment is able to simultaneously measure the electron and non-electron component of the solar neutrino spectrum by comparing the charged current and neutral current neutrino reactions on deuterium:

\begin{eqnarray}
\nu_e + d \rightarrow e^- + p + p~~~~~{\rm (charged~current)}\\
\nu_x + d \rightarrow \nu_x + n + p~~~~~{\rm (neutral~current)}
\end{eqnarray}

Results from the experiment allowed confirmation of the flavor-changing signature of neutrino oscillations and verification of the MSW mechanism~\cite{bib:MSW1,bib:MSW2}.

Deuterium, with its extremely small binding energy ($E_{\rm bind} \simeq 2.2$ MeV) has no bound final state after scattering.  There exist two prominent methods for calculating such cross-sections. The first method, sometimes referred to in the literature as the elementary-particle treatment (EPT) or at times the standard nuclear physics approach (SNPA) was first introduced by~\cite{Kim:1965zzc} and~\cite{bib:Yamaguchi}.  The technique treats the relevant nuclei as fundamental particles with assigned quantum numbers. A transition matrix element for a given process is parametrized in terms of the nuclear form factors solely based on the transformation properties of the nuclear states, which in turn are constrained from complementary experimental data. Such a technique provides a robust method for calculating $\nu d$ scattering. Typically one divides the problem into two parts; the one-body impulse approximation terms and two-body exchange currents acting on the appropriate nuclear wave functions.  In general, the calculation of these two-body currents presents the most difficulty in terms of verification.  However, data gathered from $n+p \rightarrow d + \gamma$ scattering provides one means of constraining any terms which may arise in $\nu d$ scattering.  An additional means of verification, as discussed previously, involves the reproduction of the experimental tritium beta decay width, which is very precisely measured.

An alternative approach to such calculations has recently emerged on the theoretical scene based on effective field theory (EFT) which has proven to be particularly powerful in the calculations of $\nu d$ scattering~\cite{bib:Chen2000,bib:Chen2001}.  EFT techniques make use of the gap between the long-wavelength and short-range properties of nuclear interactions.  Calculations separate the long-wavelength behavior of the interaction, which can be readily calculated, while absorbing the omitted degrees of freedom into effective operators which are expanded in powers of some cut-off momentum.  Such effective operators can then be related directly to some observable or constraint that fixes the expansion.  In the case of $\nu d$ scattering,  the expansion is often carried out as an expansion of the pion mass, $q/m_\pi$.  EFT separates the two-body current process such that it is dependent on one single parameter, referred to in the literature as $L_{1,A}$.  This low energy constant can be experimentally constrained, and in doing so provides an overall regularization for the entire cross-section.  Comparisons between these two different methods agree to within 1-2\% for energies relevant for solar neutrinos ($< 20$ MeV)~\cite{bib:Mosconi2007ku, MOSCONI:2006jn, Nakamura:2001fr}.  In general, the EFT approach has been extremely successful in providing a solid prediction of the deuterium cross-section, and central to the reduction in the theoretical uncertainties associated with the reaction~\cite{Adelberger:2010qa}.  Given the precision of such cross-sections, one must often include radiative corrections~\cite{bib:Beacom2001,bib:Towner1998,bib:Kurylov2002}.

\subsection{Other Nuclear Targets~\label{sec:ONT}}

So far we have only discussed the most simple of reactions; that is, scattering of anti-neutrinos off of free protons and scattering of neutrinos off of deuterium, both of which do not readily involve any bound states.  In such circumstances, the uncertainties involved are small and well-understood.  But what happens when we wish to expand our arsenal and attempt to evaluate more complex nuclei or nuclei at higher momenta transfer?  The specific technique used depends in part on the type of problem that one is attempting to solve, but it usually falls in one of three main categories:

\begin{enumerate}
\item For the very lowest energies, one must consider the exclusive scattering to particular nuclear bound states and provide an appropriate description of the nuclear response and correlations among nucleons.  The shell model is often invoked here, given its success in describing Fermi and Gamow-Teller amplitudes~\cite{bib:ShellModel}.

\item At higher energies, enumeration of all states becomes difficult and cumbersome.  However, at this stage one can begin to look at the {\it collective} excitation of the nucleus.  Several theoretical tools, such as the random phase approximation (RPA)~\cite{bib:RPA} and extensions of the theory, including continuous random phase approximation (CRPA)~\cite{bib:Kolbe99}, and quasi-particle random phase approximation (QRPA)~\cite{bib:QRPA} have been developed along this strategy.

\item Beyond a certain energy scale, it is possible to begin describing the nucleus in terms of individual, quasi-free nucleons.  Techniques in this regime are discussed later in the text.

\end{enumerate}

Let us first turn our attention to the nature of the matrix elements which describe the cross-section amplitudes of the reaction under study.  In almost all cases, we wish to determine the amplitude of the matrix element that allows us to transition from some initial state $i$ (with initial spin $J_i$) to some final state $f$ (with final spin $J_f$).  For a charged current interaction of the type $\nu_e + A^Z_N \rightarrow~e^- + A^{*Z+1}_{N-1}$, the cross-section can be written in terms of a very general expression:

\begin{equation}
\frac{d \sigma}{d\cos{\theta}} = \frac{E_e p_e}{2\pi}  \sum_{i} \frac{1}{(2J_i+1)} [\sum_{M_i,M_f} |\bra{f} \hat{H_W}\ket{i}|^2]
\end{equation}

\noindent where $E_e$,$p_e$, and $\cos{\theta}$ are the outgoing electron energy, momentum and scattering angle, respectively, and $J_i$ is the total spin of the target nucleus.  The sum is carried over all the accessible spins of the initial and final states.  

The term in the brackets encapsulates the elements due to the hadronic-lepton interaction.  A Fourier transform of the above expression allows one to express the matrix elements of the Hamiltonian in terms of the 4-momenta of the initial and final states of the reaction. The Hamiltonian which governs the strength of the interaction is given by the product of the hadronic current $H(\vec{x})$ and the leptonic current $J(\vec{x})$:  

\begin{eqnarray*}
{\cal H}_W^{\rm CC} = \frac{G_F V_{ud}}{\sqrt{2}} \int  [J^{\rm CC,\mu}(\vec{x}) H_\mu^{\rm CC}(\vec{x}) +h.c.]d\vec{x}, \\
{\cal H}_W^{\rm NC} = \frac{G_F}{\sqrt{2}} \int  [J^{\rm NC,\mu}(\vec{x}) H^{\rm NC}_\mu(\vec{x}) + h.c.] d\vec{x},
\end{eqnarray*}

\noindent  where

\begin{eqnarray*}
H_\mu^{\rm CC}(\vec{x}) = V_\mu^\pm(\vec{x}) + A_\mu^\pm(\vec{x})\\
H_\mu^{\rm NC}(\vec{x}) = (1-2\weakangle)V_\mu^0(\vec{x}) + A^0_\mu(\vec{x})-2\weakangle V_\mu^s
\end{eqnarray*} \\

We will concentrate on the charged current reaction first.  In the above expression, the $V^\pm$ and $A^\pm$ components denote the vector and axial-vector currents, respectively.  The $\pm$ and $0$ index notation denotes the three components of the isospin raising (-lowering) currents for the neutrino (or anti-neutrino) reaction.  The final ingredient, $V^s$ denotes the isoscalar current.   For the case of the impulse approximation, it is possible to write down a general representation of the hadronic weak current in terms of the relevant spin contributions:

\begin{widetext}
\begin{eqnarray*}
\bra{f}V_\mu^{a}(q^2)\ket{i} = \bar{u}(p') \frac{\tau^a}{2}\left[F_1(q^2) \gamma_\mu + i \frac{F_2(q^2)}{2m_n} \sigma_{\mu\nu} q^\nu + i \frac{q_\mu}{M_N} F_S(q^2) \right]u(p)\\
\bra{f}A_\mu^{a}(q^2)\ket{i} = \bar{u}(p') \frac{\tau^a}{2}\left[F_A(q^2) \gamma_\mu\gamma_5 + \frac{F_P}{M_N}(q^2)q_\mu \gamma_5 +  \frac{F_T}{M_N}(q^2) \sigma_{\mu\nu} q^\nu \gamma_5 \right]u(p)
\end{eqnarray*}
\end{widetext}

Here, $\tau^a$ is indexed as $a = \pm, 0$, $\sigma_{\mu\nu}$ are the spin matrices, $ \bar{u}(p')$ and $u(p)$ are the Dirac spinors for the target and final state nucleon, $M_N$ is the (averaged) nucleon mass, and $F_{[S,1,A,2,P,T]}(q^2)$ correspond to the scalar, Dirac, axial-vector, Pauli, pseudoscalar and tensor weak form factors, respectively.  The invariance of the strong interaction under isospin simplifies the picture a bit for the charged current interaction, as both the scalar and tensor components are zero:

\begin{equation}
F_S(q^2) = F_T(q^2) \equiv 0
\end{equation}

In order to proceed further, one needs to make some link between the form factors probed by weak interactions and those from pure electromagnetic interactions.  Fortunately, the Conserved Vector Current (CVC) hypothesis allows us to do just that.

\begin{eqnarray*}
F_1(q^2) = F_1^p(q^2) - F_1^n(q^2)\\
F_2(q^2) = F_2^p(q^2) - F_2^n(q^2)\\
\end{eqnarray*}

\noindent here $F_1^{n,p}$ and $F_2^{n,p}$ are known in the literature as the electromagnetic Dirac and Pauli form factors of the proton and neutron, respectively.  In the limit of zero momentum transfer, the Dirac form factors reduce to the charge of the nucleon, while the Pauli form factors reduce to the nucleon's magnetic moments:

\[F_1^N(0) = q_N = \left\{ 
\begin{array}{l l}
  1 & \quad \mbox{if proton}\\
  0 & \quad \mbox{if neutron}\\ \end{array} \right\}, \]

\[F_2^N(0) = \left\{ 
\begin{array}{l l}
  \frac{\mu_p}{\mu_N}-1 & \quad \mbox{if proton}\\
  \frac{\mu_n}{\mu_N} & \quad \mbox{if neutron}\\ \end{array} \right\}. \]

Here, $q_N$ is the nucleon charge, $\mu_N$ is the nuclear magneton, and $\mu_{p,n}$ are the proton and neutron magnetic form factors.  

To ascertain the $q^2$ dependence of these form factors, it is common to use the Sachs electric and magnetic form factors and relate them back to $F_1^N$ and $F_2^N$:

\begin{eqnarray*}
G_E^N(q^2) = F_1^N(q^2) - \eta F_2^N(q^2)\\
G_M^N(q^2) = F_1^N(q^2) +  F_2^N(q^2)\\
\end{eqnarray*}

\noindent with $\eta \equiv -q^2/4M_N$ and,

\begin{eqnarray*}
G_E^{p}(q^2) = G_D(q^2),~~~~  & G_E^{n}(q^2) = 0,\\
G_M^{p}(q^2) = \frac{\mu_p}{\mu_N} G_D(q^2),~~~~  & G_M^{n}(q^2) = \frac{\mu_n}{\mu_N} G_D(q^2).
\end{eqnarray*}

Here, $G_D(q^2)$ is a dipole function determined by the charge radius of the nucleon.  Empirically, the dipole term can be written as 

\begin{equation}
G_D(q^2) = (1-\frac{q^2}{m_V^2})^{-1}
\end{equation}

\noindent where $m_V \simeq 0.84$ MeV.  

We now turn our attention to the axial portion of the current, where the terms $F_A(q^2)$ and $F_P(q^2)$ play a role.  For $F_A(q^2)$, one also often assumes a dipole-like behavior, but with a different coupling and axial mass term $(m_A)$:

\begin{eqnarray*}
F_A(q^2) = -g_A G_A(q^2),\\
G_A(q^2) = \frac{1}{(1-\frac{q^2}{m_A^2})^2}.
\end{eqnarray*}

The Goldberger-Treiman relation allows one to also relate the pseudoscalar contribution in terms of the axial term as well; typically:

\begin{eqnarray*}
F_P(q^2) = \frac{2 M_N^2}{m_\pi^2 - q^2} F_A(q^2),\\
\end{eqnarray*}

\noindent where $m_\pi$ is the pion mass.  In the limit that the momentum exchange is small (such as in neutron decay or inverse beta decay), the form factors reduce to the constants we had defined previously in this section:

\begin{eqnarray*}
f_V(0) \equiv F_1(0) = 1,\\
f_P(0)  \equiv F_2(0) = \frac{\mu_p - \mu_n}{\mu_N} -1 \simeq 3.706,\\
f_A(0)  \equiv F_A(0) = -g_A,
\end{eqnarray*}

\noindent with $\lambda \equiv {f_A(0)\over f_V(0)} \equiv -1.2694 \pm 0.0028$, as before~\cite{ga}.

The above represents an approach that works quite well when the final states are simple, for example, when one is dealing with a few-nucleon system with no strong bound states or when the momentum exchange is very high (see the next section on quasi-elastic interactions).  

Seminal articles on neutrino (and electron) scattering can be found in earlier review articles by~\cite{bib:Walecka1975,Donnelly19748} and~\cite{bib:Peccei,bib:Peccei1979}.  Peccei and Donnelly equate the relevant form factors to those measured in $(e,e')$ scattering~\cite{bib:deForest1966,Drell196418}, removing some of the model dependence and $q^2$ restrictions prevalent in certain techniques.  This approach is not entirely model-independent, as certain axial form factors are not completely accessible via electron scattering.  This technique has been expanded in describing neutrino scattering at much higher energy scales~\cite{bib:Donnelly,bib:DonnellySuperScaling} with the recent realization that added nuclear effects come into play~\cite{Amaro:2010}.

\subsection{Estimating Fermi and Gamow-Teller Strengths}

For very small momentum transfers, the relevant impact of these various form factors take a back seat to the individual final states accessible to the system.  Under this scheme, it is customary to divide into two general groupings:  the {\it Fermi} transitions (associated with $f_V(0)$) and the Gamow-Teller transitions (associated with $f_A(0)$).  In doing so, the cross-section can be re-written as:

\begin{widetext}
\begin{eqnarray}
\frac{d \sigma}{d\cos{\theta}} \simeq \frac{G_F^2 |V_{ud}|^2 F(Z_{\rm f},E_e) E_e p_e}{2\pi} (f_F(q^2) |M_F|^2 + f_{GT}(q^2)\frac{1}{3}|M_{GT}|^2 +{\rm~interference~terms})
\end{eqnarray}

\noindent where,

\begin{eqnarray}
|M_F|^2 = \frac{1}{2J_i+1} \sum_{M_f,M_i} |\bra{J_f,M_f} \sum_{k=1}^A \tau_\pm(k) e^{iq\cdot r_k} \ket{J_i,M_i}|^2 \label{eq:nucl_eq1}\\
|M_{GT}|^2 =\frac{1}{2J_i+1} \sum_{M_f,M_i} |\bra{J_f,M_f} \sum_{k=1}^A \tau_\pm(k) {\bf \sigma}(k) e^{iq\cdot r_k} \ket{J_i,M_i}|^2 \label{eq:nucl_eq2}
\end{eqnarray}
\end{widetext}

We note to the reader that we have altered our notation slightly to denote explicit summation over individual accessible nuclear states.  Equations~\ref{eq:nucl_eq1} and ~\ref{eq:nucl_eq2} show explicitly the summation across both initial ($\ket{J_i,M_i}$) and final ($\ket{J_f,M_f}$) spin states. In general, the terms associated with the Fermi transitions, $f_F(q^2))$, and the Gamow-Teller transitions, $f_{GT}(q^2)$, are non-trivial combinations of the various form factors described previously (see also~\cite{Kuramoto1990711}).  However, as one approaches zero momentum, we can immediately connect the relevant Fermi and Gamow-Teller amplitudes directly to $\beta$ decay:

\begin{widetext}
\begin{eqnarray}
M_\beta =  f_V(0)^2 |M_F|^2 + f_A(0)^2\frac{1}{3}|M_{GT}|^2\\
|M_F|^2 = \frac{1}{2J_i+1} \sum_{M_f,M_i} |\bra{J_f,M_f} \sum_{k=1}^A \tau_\pm(k) \ket{J_i,M_i}|^2\\
|M_{GT}|^2 = \frac{1}{2J_i+1} \sum_{M_f,M_i} |\bra{J_f,M_f} \sum_{k=1}^A \tau_\pm(k) {\bf \sigma}(k) \ket{J_i,M_i}|^2
\end{eqnarray}
\end{widetext}

and

\begin{equation}
{\cal F}t = \frac{2\pi^3 \ln{2}}{G_F^2 |V_{ud}|^2 m_e^5 M_\beta}.
\end{equation}

Hence, in the most simplistic model, the total charged current cross-section can be calculated directly from evaluating the appropriate the $\beta$-decay reaction and correcting for the spin of the system:

\begin{equation}
\sigma = \frac{2\pi^2 \ln{2}}{m_e^5 {\cal F}t} p_e E_e F(E_e,Z_{\rm f}) \frac{2J_f + 1}{2J_i+1}
\end{equation}

Further information on the relevant coefficients can also be obtained by studying muon capture on the nucleus of interest~\cite{Luyten1963236, Ricci:2010sk,NguyenTienNguyen1975485}, or by imposing sum rules on the total strength of the interaction~\footnote{Examples of known sum rules to this effect include the Ikeda sum rule for the Gamow-Teller strength~\cite{bib:IkedaSumRule}:
\begin{equation*}
\sum_i M_{GT}^2(Z \rightarrow Z+1)_i -  M_{GT}^2(Z \rightarrow Z-1)_i = 3(N-Z).
\end{equation*}.}.  

Another extremely powerful technique in helping discern the contributions to the neutrino cross-section, particularly for Gamow-Teller transitions, has been through $(p,n)$ scattering.  Unlike its $\beta$-decay counterpart, $(p,n)$ scattering does not suffer from being limited to a particular momentum band;  in principle a wider band is accessible via this channel.  Since the processes involved for $(p,n)$ scattering are essentially the same as those for the weak interaction in general, one can obtain an empirical evaluation of the Fermi and Gamow-Teller strengths for a given nucleus.  This is particularly relevant for $(p,n)$ reactions at high incident energies and forward angles, where the direct reaction mechanism dominates.  The use of $(p,n)$ reactions is particularly favorable for studying weak interaction matrix elements for a number of reasons.  The reaction is naturally spin selective and spin sensitive over a wide range of beam energies.  Furthermore, small angle scattering is relatively easy to prove experimentally.  This approach was first explored empirically by Goodman and others ~\cite{bib:Goodman1980,bib:Watson1985} and later expanded in a seminal paper by Taddeucci and collaborators~\cite{Taddeucci1987125}.  Provided that $(p,n)$ forward scattering data on a particular nucleus is available, one can reduce the uncertainties on the corresponding neutrino cross-section considerably.  Data on $(p,n)$ scattering has been taken for a variety of nuclear targets, with particular focus on isotopes relevant for solar neutrino physics and stellar astrophysics.  An example of the latter would be the treatment of the neutrino cross-section at low energies for \isotope{Ga}{71} ~\cite{ Haxton1998110}. 

%
%
%
%
%
%

\subsection{Experimental Tests of Low Energy Cross-Sections on Nuclei}

Low energy neutrino cross-sections feature prominently in a variety of model-building scenarios.  Precise knowledge of the inclusive and differential cross-section feeds into reactor neutrino analysis, supernova modeling, neutrino oscillation tests, and countless others.  Yet, the number of direct experimental tests of these cross-sections is remarkably few.  We describe some examples in the next few sections.

\subsubsection{Hydrogen}

Inverse beta decay holds a special place for experimental neutrino physics, as it is via this channel that neutrinos were first detected~\cite{Cowan20071956,Navarro:2006ke}.  Even to this day, the technique of tagging inverse beta decay is prevalently used in the field for the identification and study of neutrino interactions.  Inverse beta decay and neutrino absorption are still, after 60 years, the main reaction channels used for detecting reactor and solar neutrinos.  Within the context of studying neutrino cross-sections, however, the experimental data is somewhat limited.  Most studies of neutrino interactions on protons (hydrogen) come from reactor experiments, whereby neutrinos are produced from the fission of \isotope{U}{235}, \isotope{Pu}{239}, \isotope{Pu}{241}, and \isotope{U}{238}.  These experiments include ILL-Grenoble~\cite{bib:ILL81,bib:ILL95}\footnote{The ILL experiment revised their original 1986 measurement due to better estimates of power consumption and neutron lifetime.}, G\"osgen~\cite{bib:Gosgen86}, ROVNO~\cite{bib:ROVNO91}, Krasnoyarsk~\cite{bib:Krasnoyarsk}, and Bugey~\cite{bib:Bugey94,bib:Bugey95}, the latter of which had the most precise determination of the cross-section.  In almost all cases, the knowledge of the neutrino flux contributes the largest uncertainty.  A tabulation of extracted cross-sections compared to theoretical predictions is shown in Table~\ref{tab:IBD}.  We currently omit measurements taken from Palo Verde~\cite{bib:Palo}, CHOOZ~\cite{bib:CHOOZ}, and KamLAND~\cite{bib:KamLAND}, as such measurements were performed at distance greater than 100 meters from the reactor core.  Such distances are much more sensitive to oscillation phenomena.  Also, the level of statistical precision from this latter set of experiments is lower than that from the Bugey reactor.

\begin{table*}[htdp]
\caption{Measured inverse beta decay cross-sections from short-baseline ($< 100$ m) reactor experiments.  Data are taken from  ILL-Grenoble~\cite{bib:ILL81,bib:ILL95}, G\"osgen~\cite{bib:Gosgen86}, ROVNO~\cite{bib:ROVNO91}, Krasnoyarsk~\cite{bib:Krasnoyarsk}, and Bugey~\cite{bib:Bugey94,bib:Bugey95}.  Theoretical predictions include original estimates and (in parenthesis) the recalculated predictions from~\cite{Mention:2011rk}.}
\begin{center}
\begin{tabular}{|c|c|c|c|c|c|c|}
\hline
Experiment & \multicolumn{4}{|c|}{Fuel Composition} & Distance & $\sigma_{\rm exp}/\sigma_{\rm theo.}$\\
\hline
& \isotope{U}{235} & \isotope{Pu}{239} & \isotope{U}{239} &  \isotope{Pu}{241} & & \\
\hline
ILL~\cite{bib:ILL81,bib:ILL95} &  93\% & - & - & - & 9 m & $0.800 (0.832) \pm 0.028\pm 0.071$ \\
Bugey~\cite{bib:Bugey94} 94 & 53.8\% & 32.8\%& 7.8\%& 5.6\% & 15 m & $0.987 (0.943) \pm 0.014 \pm 0.027$ \\
Bugey~\cite{bib:Bugey95} 95 & 53.8\% & 32.8\%& 7.8\%& 5.6\% & 15 m & $0.988 (0.943) \pm 0.037 \pm 0.044$ \\
Bugey~\cite{bib:Bugey95} 95 & 53.8\% & 32.8\%& 7.8\%& 5.6\% & 40 m & $0.994 (0.948) \pm 0.010 \pm 0.045$ \\
Bugey~\cite{bib:Bugey95} 95 & 53.8\% & 32.8\%& 7.8\%& 5.6\% & 95 m & $0.915 (0.873) \pm 0.10 \pm 0.041$ \\
G\"osgen~\cite{bib:Gosgen86} I & 61.9\% & 27.2\% & 6.7\% & 4.2\%&  37.9 m & $1.018 (0.971) \pm 0.017 \pm 0.06$ \\ 
G\"osgen~\cite{bib:Gosgen86} II &  58.4\% & 29.8\% & 6.8\% & 5.0\% & 45.9 m & $1.045 (0.997) \pm 0.019 \pm 0.06$ \\ 
G\"osgen~\cite{bib:Gosgen86} III & 54.3\% & 32.9\% & 7.0\% & 5.8\% & 64.7 m & $0.975 (0.930) \pm 0.033 \pm 0.06$ \\ 
ROVNO~\cite{bib:ROVNO91} & 61.4\% & 27.5\% & 3.1\% & 7.4\% & 18 m &  $0.985 (0.940) \pm 0.028 \pm 0.027$ \\ 
Krasnoyarsk~\cite{bib:Krasnoyarsk} I & 99\% & - & - & - & 33 m & $1.013 (0.944) \pm 0.051$ \\
Krasnoyarsk~\cite{bib:Krasnoyarsk} II & 99\% & - & - & - & 57 m & $0.989 (0.954) \pm 0.041$ \\
Krasnoyarsk~\cite{bib:Krasnoyarsk} III & 99\% & - & - & - & 33 m & $1.031 (0.960) \pm 0.20$ \\
\hline
\end{tabular}
\end{center}
\label{tab:IBD}
\end{table*}

\begin{table*}[htdp]
\caption{Measured charged current ($\bar{\nu}_e$CC) and neutral current ($\bar{\nu}_e$NC) neutrino cross-sections on deuterium from short-baseline ($< 100$ m) reactor experiments.  Data are taken from Savannah River~\cite{bib:SavannahRiver}, ROVNO~\cite{bib:RovnoDeut}, Krasnoyarsk~\cite{bib:KrasnoyarskDeut,bib:KrasnoyarskDeut2}, and Bugey~\cite{bib:BugeyDeut}. The comparison with theory is taken from~\cite{bib:KrasnoyarskDeut}.}
\begin{center}
\begin{tabular}{|c|c|c|c|}
\hline
Experiment & Measurement & $\sigma_{\rm fission}~(10^{-44}$ cm$^2$/fission) & $\sigma_{\rm exp}/\sigma_{\rm theory}$ \\
\hline
Savannah River~\cite{bib:SavannahRiver} & $\bar{\nu}_e$CC& $1.5\pm0.4$ & $ 0.7 \pm 0.2$\\
ROVNO~\cite{bib:RovnoDeut} & $\bar{\nu}_e$CC& $1.17\pm0.16$ & $ 1.08 \pm 0.19$\\
Krasnoyarsk ~\cite{bib:KrasnoyarskDeut}& $\bar{\nu}_e$CC& $1.05\pm0.12$ & $ 0.98 \pm 0.18$\\
Bugey ~\cite{bib:BugeyDeut}& $\bar{\nu}_e$CC& $0.95\pm0.20$  & $ 0.97 \pm 0.20$\\
\hline
Savannah River~\cite{bib:SavannahRiver} & $\bar{\nu}_e$NC& $3.8\pm0.9$ & $ 0.8 \pm 0.2$\\
ROVNO~\cite{bib:RovnoDeut} & $\bar{\nu}_e$NC& $2.71\pm0.47$ & $ 0.92 \pm 0.18$\\
Krasnoyarsk ~\cite{bib:KrasnoyarskDeut}& $\bar{\nu}_e$NC& $3.09\pm0.30$ & $ 0.95 \pm 0.33$\\
Bugey ~\cite{bib:BugeyDeut}& $\bar{\nu}_e$NC& $3.15\pm0.40$   & $ 1.01 \pm 0.13$\\
\hline
\end{tabular}
\end{center}
\label{tab:vd}
\end{table*}

\begin{figure*}[htbp]
\begin{center}
\begin{tabular}{ccc}
\includegraphics[width=0.85\columnwidth,keepaspectratio=true]{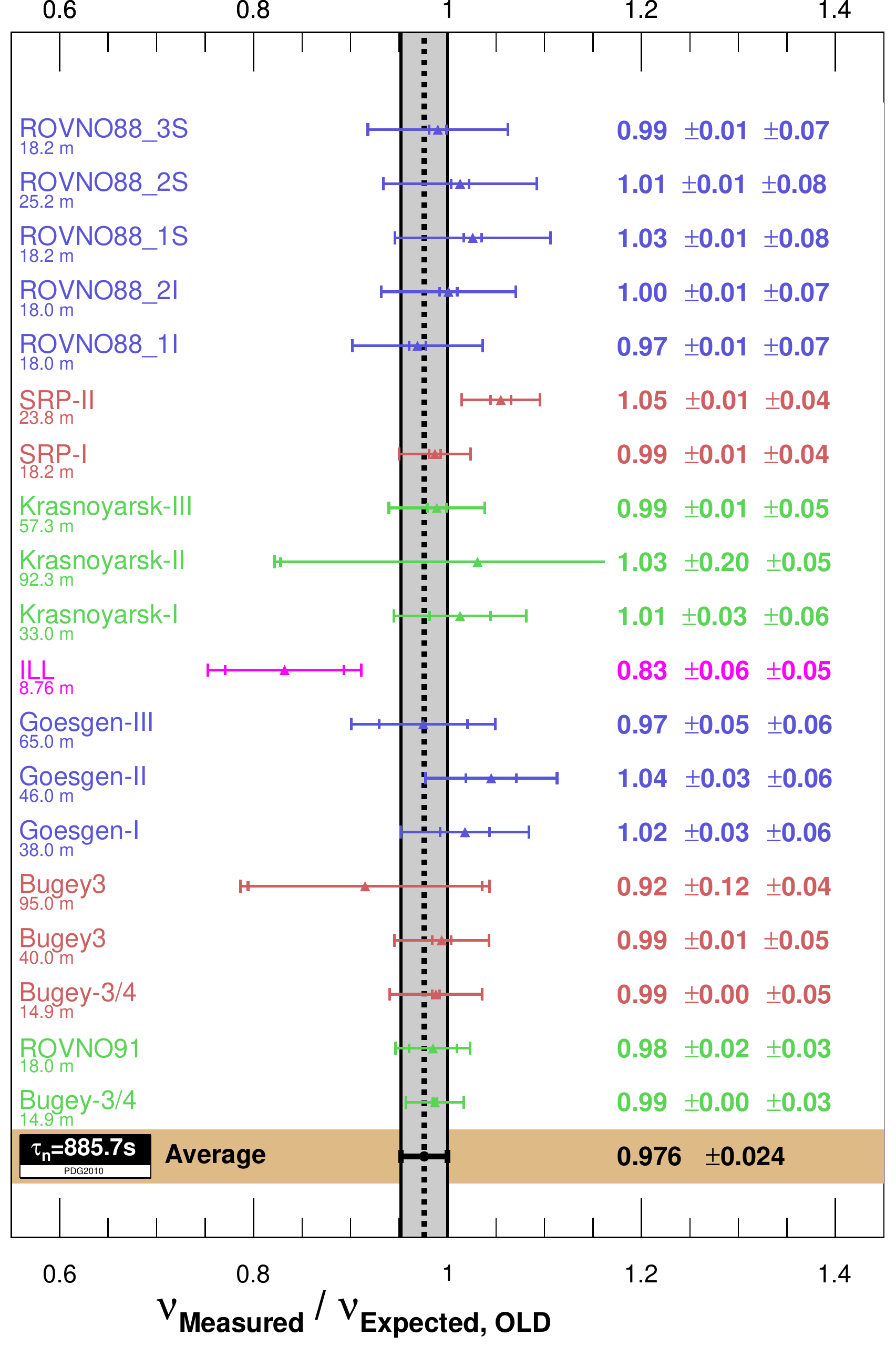} & ~~~~~~~~~~~&
\includegraphics[width=0.85\columnwidth,keepaspectratio=true]{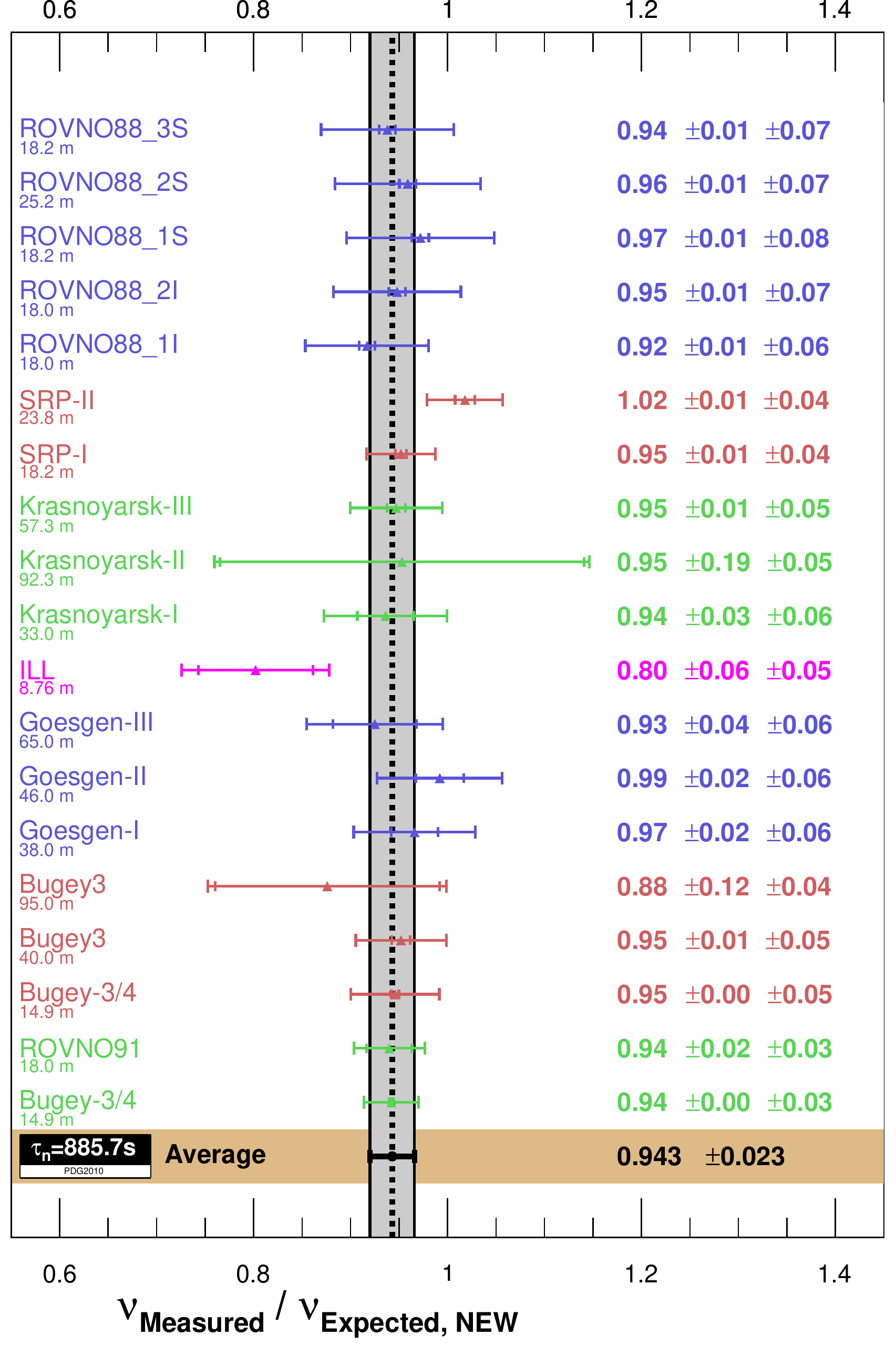} \\
\end{tabular}
\caption{Compilation of world reactor data for neutrino inverse beta decay processes for distances $\le 100$ m based on former theoretical flux predictions (left) and new theoretical prediction from~\cite{Mention:2011rk} (right). The error on the neutron lifetime is shown for comparison.}
\label{fig:ReactorData}
\end{center}
\end{figure*}

Because most experimental tests of inverse beta decay involve neutrinos produced from reactor sources, the conversion from the fission decays of \isotope{U}{235}, \isotope{Pu}{239}, \isotope{U}{239}, and \isotope{Pu}{241} to neutrino fluxes is extremely important.  Most calculations rely on the calculations made by Schreckenbach {\em et al.}~\cite{Schreckenbach:1985ep}.  Recently, a new calculation of the anti-neutrino spectrum has emerged which incorporates a more comprehensive model of fission production~\cite{Mueller:2011nm}.  The new method, which is well constrained by the accompanying electron spectrum measured from fission, has the effect that it systematically raises the expected anti-neutrino flux from reactors~\cite{Mention:2011rk}, providing some tension between the data and theoretical predictions. The new calculation is still under evaluation. In our review, we list the both the shifted and unshifted cross-section ratios (see Table~\ref{tab:IBD} and Figure~\ref{fig:ReactorData}).

\subsubsection{Deuterium}

Direct tests of low energy neutrino interactions on deuterium are of particular importance for both solar processes and solar oscillation probes.  The Sudbury Neutrino Observatory stands as the main example, as it uses heavy water as its main target to study charged current and neutral current interactions from the production of neutrinos from \isotope{B}{8} in the solar core.  The Clinton P. Anderson Meson Physics Facility (LAMPF) at Los Alamos made the only direct measurement of the reaction $\nu_e d \rightarrow e^- p p$ using neutrinos produced from a source of stopped $\mu^+$ decays from stopped pions created at their 720 MeV proton beam stop~\cite{bib:LAMPFvd}.  The cross-section is averaged over the Michel muon decay spectrum.  Their reported measurement of $\left\langle \sigma_\nu \right\rangle = (0.52 \pm 0.18) \times 10^{-40}$ cm$^2$ is in good agreement with theoretical predictions.   

Direct cross-section measurements on deuterium targets have also been carried out using anti-neutrinos produced in nuclear reactors.  Reactor experiments, including Savannah River~\cite{Reines:1976pv}, ROVNO\cite{bib:RovnoDeut}, Krasnoyarsk~\cite{bib:KrasnoyarskDeut}, and Bugey~\cite{bib:BugeyDeut}, have reported cross-sections per fission  for both charged current ($\bar{\nu}_e d \rightarrow e^+ n n$) and neutral current ($\bar{\nu}_e d \rightarrow \bar{\nu}_e p n$) reactions (see Table~\ref{tab:vd}).   

Given the ever-increasing precision gained by large scale solar experiments, however, there has been greater urgency to improve upon the $\pm20\%$ accuracy on the cross-section amplitude achieved by direct beam measurements.  Indirect constraints on the $\nu_e d$ cross-section have therefore emerged, particularly within the context of effective field theory.  As discussed in the previous section, the main uncertainty in the neutrino-deuterium cross-section can be encapsulated in the a single common isovector axial two-body current parameter $L_{1,A}$.  Constraints on $L_{1,A}$ come from a variety of experimental probes.  There are direct extractions, such as from solar neutrino experiments and reactor measurements, as highlighted above.  Constraints can also be extracted from the lifetime of tritium beta decay, muon capture on deuterium, and helio-seismology.  These methods were recently summarized in~\cite{Butler200226} and are reproduced in Table~\ref{tab:L1A}.

\begin{table}[htdp]
\caption{Extraction of the isovector axial two-body current parameter $L_{1,A}$ from various experimental constraints.}
\begin{center}
\begin{tabular}{|c|c|c|}
\hline
Method & Extracted $L_{1,A}$  \\
\hline
Reactor & $3.6 \pm 5.5$ fm$^{3}$ \\
Solar & $4.0 \pm 6.3$ fm$^{3}$\\
Helioseismology & $4.8 \pm 6.7$ fm$^{3}$ \\
\isotope{H}{3}$\rightarrow$ \isotope{He}{3+} $e^- \bar{\nu}_e$ & $6.5 \pm 2.4$ fm$^{3}$ \\
\hline
\end{tabular}
\end{center}
\label{tab:L1A}
\end{table}%

Deuterium represents one of those rare instances where the theoretical predictions are on a more solid footing than even the experimental constraints.  This robustness has translated into direct improvement on the interpretation of collected neutrino data, particularly for solar oscillation phenomena.  As we proceed to other nuclear targets, one immediately appreciates the rarity of this state.


\subsubsection{Additional Nuclear Targets}

The other main nuclear isotope studied in detail is \isotope{C}{12}.  There are a number of neutrino interactions on \isotope{C}{12} that have been investigated experimentally:

\begin{widetext}
\begin{eqnarray}
\nu_{e,\mu} +~{\rm C}^{12}_{\rm g.s.} \rightarrow~(e^-,\mu^-)+{\rm N}^{12}_{\rm g.s.}~~~~~{\rm (Exclusive~Charged~Current)}\label{eq:C12CCX}\\
\nu_{e,\mu} +~{\rm C}^{12}_{\rm g.s.} \rightarrow~(e^-,\mu^-)+{\rm N}^{*12}~~~~~{\rm (Inclusive~Charged~Current)}\label{eq:C12CCI}\\
\nu +~{\rm C}^{12}_{\rm g.s.} \rightarrow~\nu+{\rm C}^{*12}~~~~~{\rm (Neutral~Current)}\label{eq:C12NC}
\end{eqnarray}
\end{widetext}

Reaction~\ref{eq:C12CCX} is a uniquely clean test case for both theory and experiment.  The spin-parity of the ground state of \isotope{C}{12} is $J^\pi = 0^+, T = 0$, while for the final state it is $J^\pi = 1^+, T = 1$.  As such, there exists both an isospin and spin flip in the interaction, the former involving the isovector components of the reaction, while the latter invoking the axial-vector components.  Therefore, both vector and axial-vector components contribute strongly to the interaction.  The isovector components are well-constrained by electron scattering data.  Since the final state of the nucleus is also well-defined, the axial form factors can be equally constrained by looking at the $\beta$ decay of \isotope{N}{12}, as well as the muon capture on \isotope{C}{12}.  Although these constraints occur at a specific momentum transfer, they provide almost all necessary information to calculate the cross-section.  The exclusive reaction is also optimal from an experimental perspective.  The ground state of \isotope{N}{12} beta decays to the ground state of \isotope{C}{12} with a half-life of 11 ms; the emitted secondary electron providing a well-defined tag for event identification.  The neutral current channel has an equally favorable channel, with the emission of a mono-energetic 15.11 MeV photon.  

Studies of the above neutrino cross-sections have been carried out at the LAMPF facility in the United States~\cite{bib:LAMPFvd} and the KARMEN detector at ISIS at the Rutherford Laboratory in the United Kingdom.  The neutrino beam in both experimental facilities is provided from proton beam stops.  High energy proton collisions on a fixed target produce a large $\pi^+$ flux which is subsequently stopped and allowed to decay.  The majority of low energy neutrinos are produced from the decay at rest from stopped $\mu^+$ and $\pi^+$, providing a well-characterized neutrino beam with energies below 50 MeV\footnote{Neutrinos from decay-in-flight muons also allowed for cross-section measurements for energies below 300 MeV.}.  The KARMEN experiment at the ISIS facility additionally benefited from a well-defined proton beam structure, which allowed efficient tagging of neutrino events against cosmic ray backgrounds.  The main uncertainty affecting these cross-section measurements stems primarily from the knowledge of the pion flux produced in the proton-target interactions.   
 
\begin{figure}[htbp]
\begin{center}
\includegraphics[width=0.95\columnwidth,keepaspectratio=true]{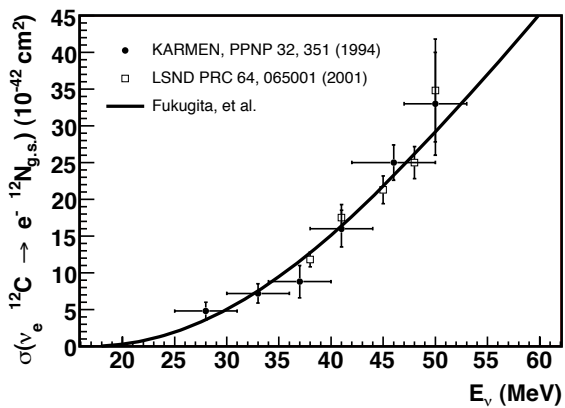} 
\caption{Cross-section as a function of neutrino energy for the exclusive reaction \isotope{C}{12}$(\nu_e,e^-)$\isotope{N}{12} from $\mu^-$ decay-at-rest neutrinos.  Experimental data measured by the KARMEN~\cite{bib:KARMENC12N12} and LSND~\cite{bib:LSNDC12N12_a,bib:LSNDC12N12_b} experiments.  Theoretical prediction taken from Fukugita {\it et al.}~\cite{bib:Fukugita1988}.}
\label{fig:lowe_C12_exclusive}
\end{center}
\end{figure}

\begin{figure}[htbp]
\begin{center}
\includegraphics[width=0.95\columnwidth,keepaspectratio=true]{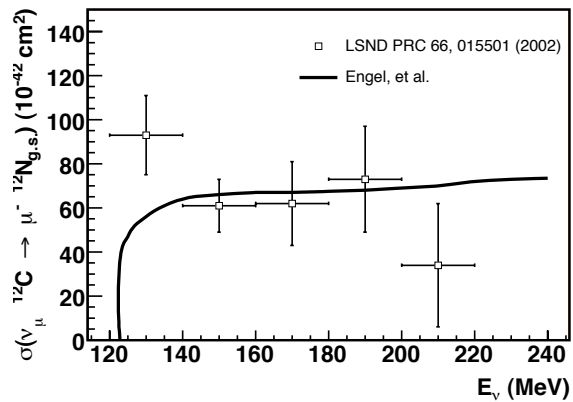} 
\caption{Cross-section as a function of neutrino energy for the exclusive reaction \isotope{C}{12}$(\nu_\mu,\mu^-)$\isotope{N}{12} measured by the LSND~\cite{PhysRevC.66.015501} experiment.  Theoretical prediction taken from~\cite{bib:Engel}.}
\label{fig:lowmu_C12_exclusive}
\end{center}
\end{figure}

Table~\ref{tab:12C} summarizes the measurements to date on the inclusive and exclusive reactions on \isotope{C}{12} at low energies.  Estimates of the cross-sections using a variety of different techniques (shell model, RPA, QRPA, effective particle theory) demonstrate the robustness of the calculations.  Some disagreement can be seen in the inclusive channels; this disagreement is to be expected since the final state is not as well-defined as in the exclusive channels.  More recent predictions employing extensive shell model calculations appear to show better agreement with the experimental data.  A plot showing the collected data from the exclusive reaction \isotope{C}{12}$(\nu_e,e^-)$\isotope{N}{12} and \isotope{C}{12}$(\nu_\mu,\mu^-)$\isotope{N}{12}  are shown in Figures~\ref{fig:lowe_C12_exclusive} and~\ref{fig:lowmu_C12_exclusive}, respectively.

\begin{table*}[htdp]
\begin{center}
\caption{Experimentally measured (flux-averaged) cross-sections on various nuclei at low energies (1-300 MeV).  Experimental data gathered from the LAMPF~\cite{bib:LAMPFvd}, KARMEN~\cite{bib:KARMENC12N12, Maschuw1998183, Armbruster199815,Bodmann1991321,bib:Ruf}, E225~\cite{Krakauer1992}, LSND~\cite{bib:LSNDC12N12_a, bib:LSNDC12N12_b,PhysRevC.66.015501,bib:Distel2003}, GALLEX~\cite{Hampel1998114}, and SAGE~\cite{PhysRevC.59.2246, PhysRevC.73.045805} experiments.  Stopped $\pi/\mu$ beams can access neutrino energies below 53 MeV, while decay-in-flight measurements can extend up to 300 MeV.  The \isotope{Cr}{51} sources have several mono-energetic lines around 430 keV and 750 keV, while the \isotope{Ar}{37} source has its main mono-energetic emission at $E_\nu = 811$ keV.  Selected comparisons to theoretical predictions, using different approaches are also listed.  The theoretical predictions are not meant to be exhaustive.}
\begin{tabular}{|l|l|l|l|l|l|}
\hline
Isotope & Reaction Channel & Source & Experiment & Measurement ($10^{-42}$ cm$^2$) &  Theory ($10^{-42}$ cm$^2$)\\
\hline \hline
\isotope{H}{2}& \isotope{H}{2}($\nu_e,e^-$)pp & Stopped $\pi/\mu$ & LAMPF&  $52 \pm 18$(tot) &  54 (IA) \cite{bib:Tatara90} \\
\hline
\isotope{C}{12}& \isotope{C}{12}($\nu_e,e^-$)\isotope{N}{12}$_{\rm g.s.}$ & Stopped $\pi/\mu$ & KARMEN & $9.1 \pm 0.5 {\rm(stat)} \pm 0.8{\rm(sys)}$ & $9.4$ [Multipole]\cite{bib:Peccei}\\
 & & Stopped $\pi/\mu$ &  E225 & $10.5 \pm 1.0 {\rm(stat)} \pm 1.0{\rm(sys)}$& 9.2 [EPT]~\cite{bib:Fukugita1988}.\\
 & & Stopped $\pi/\mu$ & LSND & $8.9 \pm 0.3 {\rm(stat)} \pm 0.9{\rm(sys)}$ &  8.9 [CRPA]~\cite{bib:Kolbe99}\\
 & & & & & \\
 & \isotope{C}{12}($\nu_e,e^-$)\isotope{N}{12}$^*$ & Stopped $\pi/\mu$ &  KARMEN & $5.1 \pm 0.6 {\rm(stat)} \pm 0.5{\rm(sys)}$ & 5.4-5.6 [CRPA]~\cite{bib:Kolbe99}\\
 & & Stopped $\pi/\mu$ &  E225  & $3.6 \pm 2.0 {\rm(tot)}$& 4.1 [Shell]~\cite{bib:Towner}\\
 & & Stopped $\pi/\mu$ &  LSND & $4.3 \pm 0.4 {\rm(stat)} \pm 0.6{\rm(sys)}$& \\
 & & & & & \\
 & \isotope{C}{12}($\nu_\mu,\nu_\mu$)\isotope{C}{12}$^*$ & Stopped $\pi/\mu$  & KARMEN& $3.2 \pm 0.5 {\rm(stat)} \pm 0.4{\rm(sys)}$ &  2.8 [CRPA]~\cite{bib:Kolbe99}\\
 & \isotope{C}{12}($\nu,\nu$)\isotope{C}{12}$^*$ & Stopped $\pi/\mu$  & KARMEN & $10.5 \pm 1.0 {\rm(stat)} \pm 0.9{\rm(sys)}$ &  10.5 [CRPA]~\cite{bib:Kolbe99}\\
 & & & & & \\
 & \isotope{C}{12}($\nu_\mu,\mu^-$)X& Decay in Flight   & LSND & $1060 \pm 30 {\rm(stat)} \pm 180{\rm(sys)}$ & 1750-1780 [CRPA]~\cite{bib:Kolbe99}\\
 & & & & & 1380 [Shell]~\cite{bib:Towner}\\
 & & & & & 1115 [Green's Function]~\cite{Meucci2004277}\\
 & & & & & \\
 & \isotope{C}{12}($\nu_\mu,\mu^-$)\isotope{N}{12}$_{\rm g.s.}$ & Decay in Flight  & LSND & $56 \pm 8 {\rm(stat)} \pm 10 {\rm(sys)}$ & 68-73 [CRPA]~\cite{bib:Kolbe99}\\
 & & & & & 56 [Shell]~\cite{bib:Towner} \\
\hline
\isotope{Fe}{56}&  \isotope{Fe}{56}($\nu_e,e^-$)\isotope{Co}{56} & Stopped $\pi/\mu$& KARMEN & $256 \pm 108 {\rm(stat)} \pm 43{\rm(sys)}$  & 264  [Shell]~\cite{bib:KolbeFe}\\
\hline
\isotope{Ga}{71}& \isotope{Ga}{71}($\nu_e,e^-$)\isotope{Ge}{71}& \isotope{Cr}{51} source &  GALLEX, ave. & $0.0054 \pm 0.0009$(tot)  & 0.0058 [Shell]~\cite{Haxton1998110}\\
 & &  \isotope{Cr}{51} & SAGE & $0.0055 \pm 0.0007$(tot)  & \\
 & & \isotope{Ar}{37} source &  SAGE & $0.0055 \pm 0.0006$(tot) & 0.0070 [Shell]~\cite{Bahcall:1997eg}  \\
\hline
\isotope{I}{127} &  \isotope{I}{127}($\nu_e,e^-$)\isotope{Xe}{127} & Stopped $\pi/\mu$ & LSND & $284 \pm 91 {\rm(stat)} \pm 25{\rm(sys)}$  & 210-310 [Quasi-particle]~\cite{bib:Engel1994} \\
\hline
\end{tabular}
\label{tab:12C}
\end{center}
\end{table*}

Table~\ref{tab:12C} also lists other nuclei that have been under experimental study.  Proton beam stops at the Los Alamos Meson Physics Facility have also been utilized to study low energy neutrino cross-sections on \isotope{I}{127}.  Cross-sections on iron targets have also been explored with low energy beams at the KARMEN experiment~\cite{bib:Ruf}.

Perhaps the most remarkable of such measurements was the use of MCi radiological sources for low energy electron cross-section measurements.  Both the SAGE~\cite{PhysRevC.59.2246} and GALLEX~\cite{Anselmann1995440} solar neutrino experiments have made use of a MCi $^{51}$Cr source to study the reaction \isotope{Ga}{71}($\nu_e,e^-$)\isotope{Ge}{71} to both the ground and excited states of \isotope{Ge}{71}.  The source strength of $^{51}$Cr is typically determined using calorimetric techniques and the uncertainty on the final activity is constrained to about 1-2\%.  The SAGE collaboration subsequently have also made use of a gaseous \isotope{Ar}{37} MCi source. Its activity, using a variety of techniques, is constrained to better than 0.5\%~\cite{Haxton1998110,Barsanov:2007fp}.  Since \isotope{Ar}{37} provides a mono-energetic neutrino at slightly higher energies that its \isotope{Cr}{51} counterpart, it provides a much cleaner check on the knowledge of such low energy cross-sections~\cite{Barsanov:2007fp}.  Experimental measurements are in general in agreement with the theory, although the experimental values are typically lower than the corresponding theoretical predictions.

Finally, although the cross-section was not measured explicitly using a terrestrial source, neutrino capture on chlorine constitutes an important channel used in experimental neutrino physics.  The reaction \isotope{Cl}{37}$(\nu_e,e^-)$\isotope{Ar}{37} was the first reaction used to detect solar neutrinos~\cite{bib:Davis}.

In summary, the level at which low energy cross-sections are probed using nuclear targets is relatively few, making the ability to test the robustness of theoretical models and techniques  somewhat limited.  The importance of such low energy cross-sections is continually stressed by advances in astrophysics, particularly in the calculation of elemental abundances and supernova physics~\cite{Heger:2005dj, Langanke:2004ek}. Measurements of neutrino cross-sections on nuclear targets is currently being revisited now that new high intensity stopped pion/muon sources are once again becoming available~\cite{bib:ORLAND}. 


\subsection{Transitioning to Higher Energy Scales...}

As we transition from low energy neutrino interactions to higher energies, the reader may notice that our approach is primarily focused on the scattering off a particular target, whether that target be a nucleus, a nucleon, or a parton.  This approach is not accidental, as it is theoretically a much more well-defined problem when the target constituents are treated individually.  With that said, we acknowledge that the approach is also limited, as it fails to incorporate the nucleus as a whole.  Such departmentalization is part of the reason why the spheres of low energy and high energy physics appear so disjointed in both approach and terminology.  Until a full, comprehensive model of the entire neutrino-target interaction is formulated, we are constrained to also follow this approach.

\section{Intermediate Energy Cross Sections: $E_\nu \sim 0.1-20$ GeV}
\label{chapter:medium-energy}

\begin{figure}[h]
\begin{center}
\includegraphics[scale=0.45]{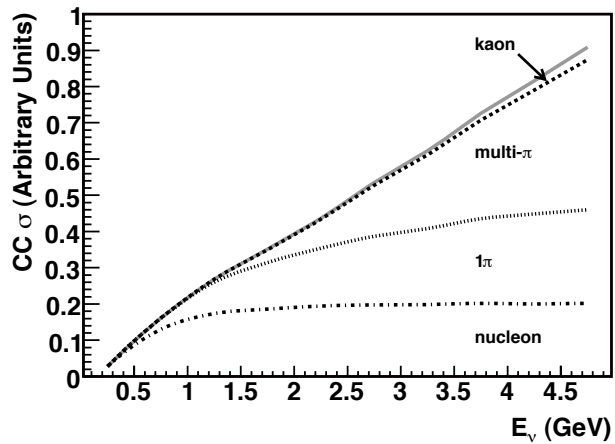}
\end{center}
\vspace{-0.2in}
\caption{ Predicted processes to the total CC inclusive scattering cross 
          section at intermediate
         energies. The underlying quasi-elastic, resonance, and deep inelastic
         scattering contributions can produce a variety of possible
         final states including the emission of nucleons, single pions, 
         multi-pions, kaons, as well as other mesons (not shown). Combined, 
         the inclusive cross section exhibits a linear dependence on neutrino 
         energy as the neutrino energy increases.}
\label{fig:all-xsecs}
\end{figure}

As we move up farther still in energy, the description of neutrino 
scattering becomes increasingly more diverse and complicated. At these 
intermediate energies, several distinct neutrino scattering mechanisms 
start to play a role. The possibilities fall into three main categories:

\begin{itemize}
 \item {\em elastic and quasi-elastic scattering:} Neutrinos can elastically scatter 
       off an entire nucleon liberating a nucleon (or multiple nucleons) from 
       the target. In the case of charged current neutrino scattering, this 
       process is referred to as ``quasi-elastic scattering'' and is a 
       mechanism we first alluded to in Section~\ref{sec:ONT}, whereas for neutral current scattering this is traditionally referred to as ``elastic scattering". 
  \item {\em resonance production:} Neutrinos can excite the target nucleon to 
        a resonance state. The resultant baryonic resonance ($\Delta$, $N^*$) 
        decays to a variety of possible mesonic final states producing
        combinations of nucleons and mesons.
  \item {\em deep inelastic scattering:} Given enough energy, the neutrino can resolve
        the individual quark constituents of the nucleon. This is called deep 
        inelastic scattering and manifests in the creation of a hadronic 
        shower.
\end{itemize}
\noindent
As a result of these competing processes, the products of neutrino 
interactions include a variety of final states ranging from 
the emission of nucleons to more complex final states including pions, 
kaons, and collections of mesons (Figure~\ref{fig:all-xsecs}). This energy regime is often referred to as the ``transition region'' because it 
corresponds to the boundary between quasi-elastic scattering (in which the 
target is a nucleon) on the one end and deep inelastic scattering (in which 
the target is the constituent parton inside the nucleon) on the other. 
Historically, adequate theoretical descriptions of quasi-elastic, 
resonance-mediated, and deep inelastic scattering have been formulated, 
however there is no uniform description which globally describes the 
transition between these processes or how they should be combined. Moreover, the full extend to which 
nuclear effects impact this region is a topic that has only recently been 
appreciated. Therefore, in this section, we will focus on what is currently known, both 
experimentally and theoretically, about each of the exclusive final state 
processes that participate in this region.

\begin{figure*}[h]
\begin{center}
\includegraphics[width=1.5\columnwidth]{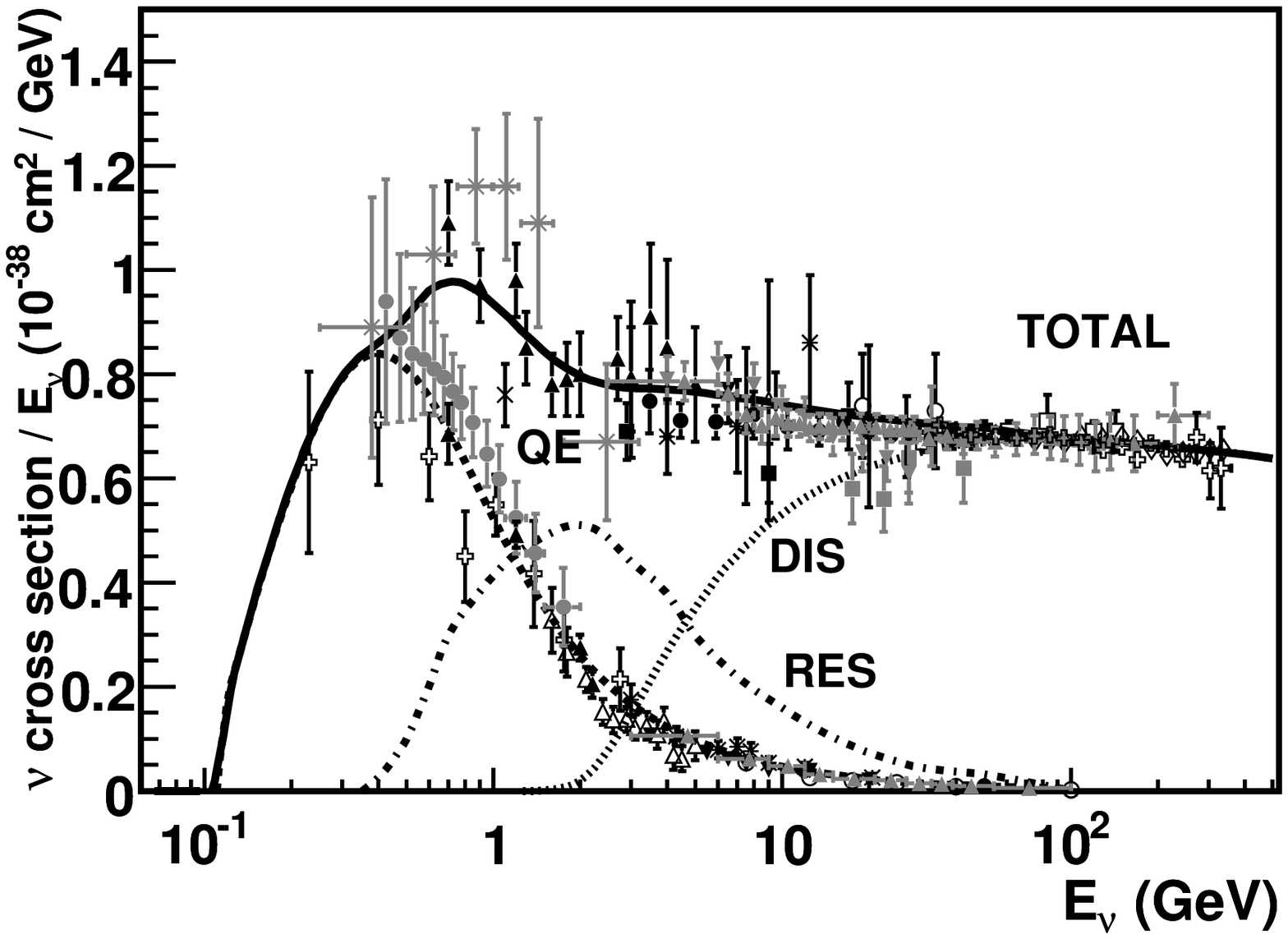}
\includegraphics[width=1.5\columnwidth]{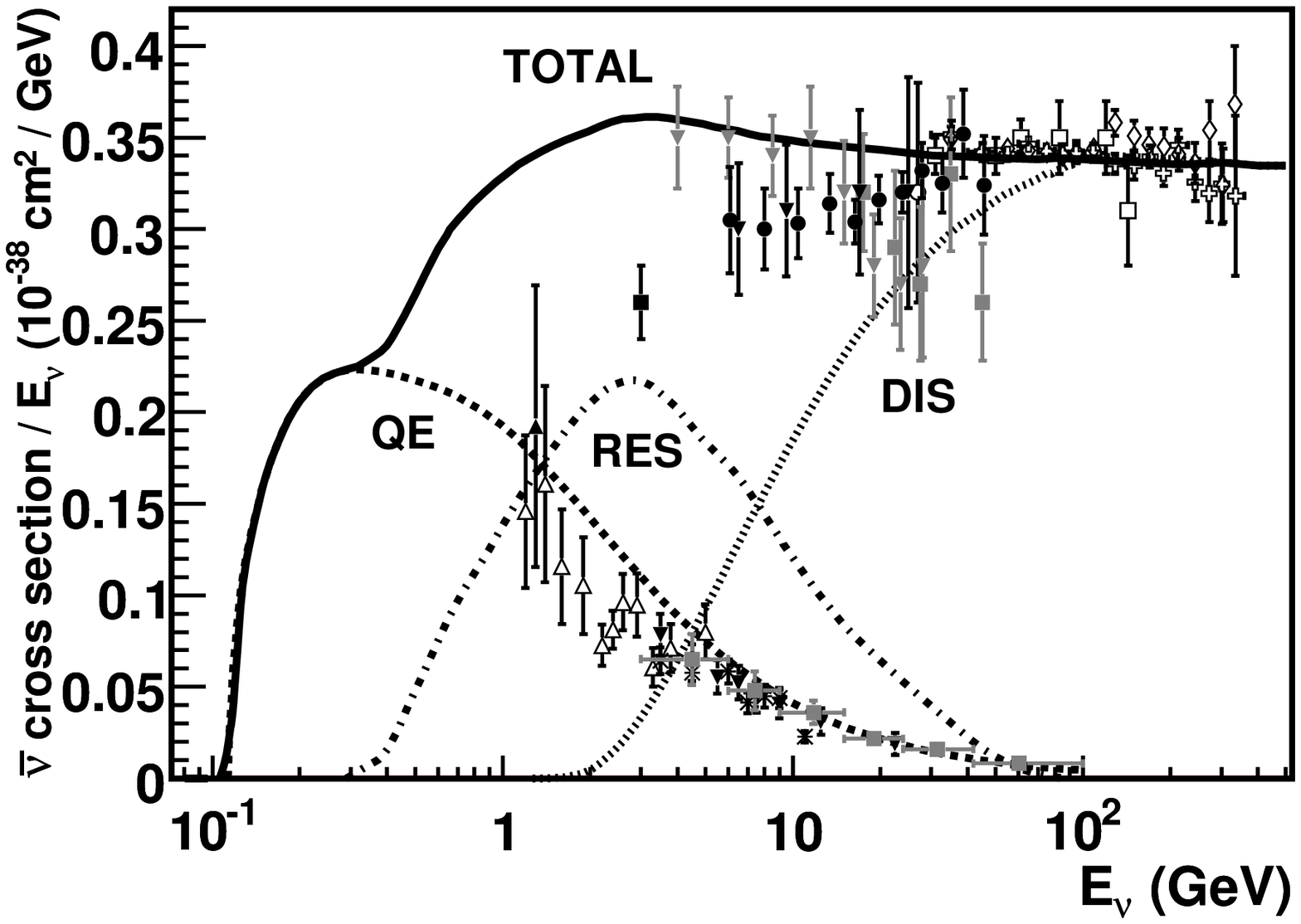}
\end{center}
\vspace{-0.2in}
\caption{Total neutrino and antineutrino per nucleon CC cross sections  
       (for an isoscalar target) divided by neutrino energy and 
       plotted as a function of energy. Data are the same as in 
       Figures~\ref{fig:dis}, \ref{fig:qe-nu}, and \ref{fig:qe-nubar}
       with the inclusion of additional lower energy CC inclusive data 
       from $\blacktriangle$~\cite{Baker:1982ty}, 
       $\ast$~\cite{Baranov:1978sx},
       $\blacksquare$~\cite{Ciampolillo:1979wp},
       and $\star$~\cite{Nakajima:2010fp}.
       Also shown are the various contributing processes that will 
       be investigated in the remaining sections of this review. These 
       contributions include quasi-elastic scattering (dashed), resonance 
       production (dot-dash), and deep inelastic scattering (dotted). 
       Example predictions for each are provided by the NUANCE 
       generator~\cite{nuance}. Note that the quasi-elastic scattering
       data and predictions have been averaged over neutron and proton
       targets and hence have been divided by a factor of two for the
       purposes of this plot.}
\label{fig:cc-inclusive}
\end{figure*}

To start, Figure~\ref{fig:cc-inclusive} summarizes the existing measurements 
of CC neutrino and antineutrino cross sections across this intermediate
energy range:

\begin{eqnarray}
   \numu \, N \rightarrow \mu^- \, X \\
   \numubar \, N \rightarrow \mu^+ \, X
\end{eqnarray}

\noindent
These results have been accumulated over many decades using a variety of
neutrino targets and detector technologies. We can immediately 
notice three things from this figure. First, the total cross sections 
approaches a linear dependence on neutrino energy. This scaling behavior 
is a prediction of the quark parton model~\cite{qpm}, a topic we will return 
to later, and is expected if point-like scattering off quarks dominates 
the scattering mechanism, for example in the case of deep inelastic 
scattering. Such assumptions break down, of course,  at lower neutrino 
energies (i.e., lower momentum transfers). Second, the neutrino cross sections 
at the lower energy end of this region are not typically are typically not as well-measured as their high 
energy counterparts. This is generally due to the lack of high statistics data 
historically available in this energy range and the challenges that arise 
when trying to describe all of the various underlying physical processes that can participate in this region. Third, antineutrino cross sections are typically less 
well-measured than their neutrino counterparts. This is generally due to 
lower statistics and larger background contamination present in that case. \\

Most of our knowledge of neutrino cross sections in this intermediate
energy range comes from early experiments that collected relatively small 
data samples (tens-to-a-few-thousand events). These measurements were 
conducted in the 1970's and 1980's using either bubble chamber or spark
chamber detectors and represent a large fraction of the data presented 
in the summary plots we will show. Over the years, interest in this energy 
region waned as efforts migrated to higher energies to yield larger event 
samples and the focus centered on measurement of electroweak parameters 
($\sin^2\theta_W$) and structure functions in the deep inelastic scattering 
region. With the discovery of neutrino oscillations and the advent of higher 
intensity neutrino beams, however, this situation has been rapidly changing. 
The processes we will discuss here are important because they form some of the dominant signal and 
background channels for experiments searching for neutrino oscillations. This is especially true for experiments that use atmospheric or accelerator-based sources of neutrinos. With a view 
to better understanding these neutrino cross sections, new experiments such 
as ArgoNeuT, K2K, MiniBooNE, MINER$\nu$A, MINOS, NOMAD, SciBooNE, 
and T2K have started to study this intermediate energy region in greater 
detail. New theoretical approaches have also recently emerged. 

\begin{figure}[h]
\begin{center}
\includegraphics[scale=0.45]{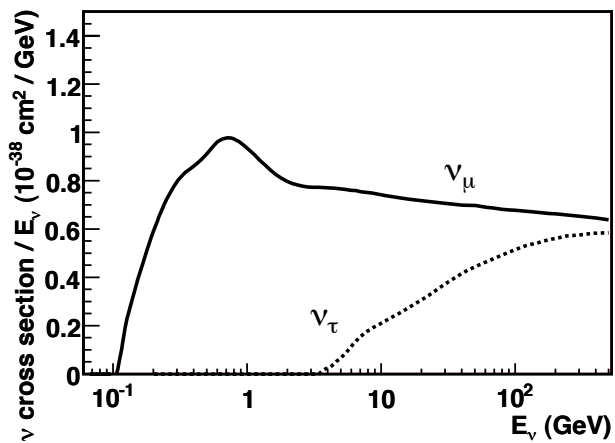}
\end{center}
\vspace{-0.2in}
\caption{Plot comparing the total charged current $\nu_\mu$ (solid) and $\nu_\tau$ (dashed) per nucleon cross sections divided by neutrino energy and plotted as a function of neutrino energy.}
\label{fig:tau}
\end{figure}

We start by describing the key processes which can contribute to the total cross section at these intermediate neutrino energies. 
Here, we will focus on several key processes: quasi-elastic, NC elastic 
scattering, resonant single pion production, coherent pion production, 
multi-pion production, and kaon production before turning our discussion 
to deep inelastic scattering in the following chapter on high energy 
neutrino interactions. For comparison purposes, we will also include
predictions from the NUANCE event generator~\cite{nuance}, chosen as a 
representative of the type of models used in modern neutrino experiments
to describe this energy region. The bulk of our discussions center around measurements of $\nu_\mu$-nucleon scattering.  Many of these arguments also carry over to $\nu_\tau$ scattering, except for one key difference; the energy threshold for the reaction.  Unlike for the muon case, the charged current $\nu_\tau$ interaction cross section is severely altered because of the large $\tau$ lepton mass.  Figure~\ref{fig:tau} reflects some of the large differences in the cross section that come about due to this threshold energy.\\

\subsection{Quasi-Elastic Scattering}
\label{sec:qe}

For neutrino energies less than $\sim2$ GeV, neutrino-hadron interactions 
are predominantly quasi-elastic (QE), hence they provide a large source
of signal events in many neutrino oscillation experiments operating
in this energy range. In a QE interaction, the neutrino scatters off an 
entire nucleon rather than its constituent partons. In a charged current 
neutrino QE interaction, the target neutron is converted to a proton.
In the case of an antineutrino scattering, the target proton 
is converted into a neutron:

\begin{eqnarray}
  \numu \, n \rightarrow \mu^- \,  p, \hspace{0.2in}
  \numubar \, p \rightarrow \mu^+ \, n
\end{eqnarray}

\noindent
Such simple interactions were extensively studied in the 1970-1990's
primarily using deuterium-filled bubbble chambers. The main interest at the
time was in testing the V-A nature of the weak interaction and in measuring 
the axial-vector form factor of the nucleon, topics that were considered 
particularly important in providing an anchor for the study of NC interactions
(Section~\ref{sec:ncel}). As examples, references~\cite{singh-qe,nomad-qe} 
provide valuable summaries of some of these early QE investigations. \\

In predicting the QE scattering cross section, early experiments 
relied heavily on the formalism first written down by Llewellyn-Smith in 
1972~\cite{ll-smith}. In the case of QE scattering off free nucleons,
the QE differential cross section can be expressed as:

\begin{eqnarray}
\label{eqn:xsectqe}
  \frac{d\sigma}{dQ^2} = \frac{G_F^2 M^2 |V_{ud}|^2}{8\pi E_\nu^2} 
                        \left[ A \pm \frac{(s-u)}{M^2}B + 
                               \frac{(s-u)^2}{M^4} C \right] 
\end{eqnarray}

\noindent
where $(-)+$ refers to (anti)neutrino scattering, $G_F$ is the Fermi 
coupling constant, $Q^2$ is the squared four-momentum transfer ($Q^2=-q^2>0$), 
$M$ is the nucleon mass, $m$ is the lepton mass, $E_\nu$ is the incident 
neutrino energy, and $(s - u) = 4ME_\nu - Q^2 - m^2$. The factors $A$, 
$B$, and $C$ are functions of the familiar  vector ($F_1$ and $F_2$),
axial-vector ($F_A$), and pseudoscalar ($F_P$) form factors of the nucleon:

\begin{eqnarray}
  A &=& {(m^2+Q^2)\over M^2}\left[ \left(1 + \eta \right)F_A^2 
        - \left( 1 - \eta \right)F_1^2 \right. \nonumber \\ 
    && \left. + \eta 
        \left( 1 - \eta \right) F_2^2 
        + 4 \eta F_1 F_2 \right. \nonumber \\
    && \left.\mbox{} - {m^2\over 4M^2} \left( (F_1 + F_2)^2 + 
        (F_A + 2F_P)^2 \right. \right. \nonumber \\
    && \left. \left. \hspace{0.9in} 
       - \left( {Q^2\over M^2} +4 \right) F_P^2 \right)\right] \\
 B &=& \frac{Q^2}{M^2} F_A (F_1 + F_2)  \\
 C &=& {1\over 4}\left(F_A^2 + F_1^2 + \eta F_2^2 \right) 
\label{eqn:xsectabc}
\end{eqnarray}

\noindent
where $\eta=Q^2/4M^2$. Much of these equations should be familiar from Section IV. Historically, this formalism was used to analyze 
neutrino QE scattering data on deuterium, subject to minor modifications 
for nuclear effects. In this way, experiments studying neutrino QE scattering 
could in principle measure the vector, axial-vector, and pseudoscalar form 
factors given that the weak hadronic current contains all three of these 
components. In practice, the pseudoscalar contribution was typically neglected
in the analysis of $\numu$ QE scattering as it enters the cross section 
multiplied by $m^2/M^2$. Using CVC, the vector form factors could be obtained 
from electron scattering, thus leaving the neutrino experiments to measure 
the axial-vector form factor of the nucleon. For the axial-vector form factor,
it was (and still is) customary to assume a dipole form:

\begin{eqnarray}
    F_A(Q^2) &=&  {{g_{A}} \over {\left(\displaystyle{1 + {{Q^2}\over{M_A^2} }}
             \right)^2}}
  \label{eqn:qe-fa}
\end{eqnarray}

\noindent
which depends on two empirical parameters: the value of the axial-vector
form factor at $Q^2=0$, $g_A=F_A(0)=1.2694 \pm 0.0028$~\cite{ga}, and an 
``axial mass'', $M_A$. With the vector form factors under control from 
electron scattering and $g_A$ determined with high precision from nuclear 
beta decay, measurement of the axial-vector form factor (and hence $M_A$) 
became the focus of the earliest measurements of neutrino QE scattering. 
Values of $M_A$ ranging from 0.65 to 1.09 GeV were obtained in the period 
from the late 1960's to early 1990's resulting from fits both to the total 
rate of observed events and the shape of their measured $Q^2$ dependence (for a recent review, see~\cite{nomad-qe}). 
In addition to providing the first measurements of $M_A$ and the QE cross 
section, many of these experiments also performed checks of CVC, fit for 
the presence of second-class currents, and experimented with different forms 
for the axial-vector form factor. By the end of this period, the neutrino QE 
cross section could be accurately and consistently described by V-A theory 
assuming a dipole axial-vector form factor with $M_A=1.026 \pm 0.021$ 
GeV~\cite{bernard}. These conclusions were largely driven by experimental 
measurements on deuterium, but less-precise data on other heavier targets 
also contributed. More recently, some attention has been given to re-analyzing
this same data using modern vector form factors as input. The use of updated 
vector form factors slightly shifts the best-fit axial mass values 
obtained from this data; however the conclusion is still that $M_A \sim 1.0$ 
GeV~\footnote{A value of $M_A=1.014 \pm 0.014$ GeV is obtained from a recent 
global fit to the deuterium data in Reference~\cite{bodek-ma}, while a 
consistent value of $M_A=0.999 \pm 0.011$ GeV is obtained in 
Reference~\cite{kuzmin-ma} from a fit that additionally includes some 
of the early heavy target data.}. \\

Modern day neutrino experiments no longer include deuterium but use complex 
nuclei as their neutrino targets. As a result, nuclear effects become much 
more important and produce sizable modifications to the QE differential cross 
section from Equation~\ref{eqn:xsectqe}. With QE events forming the 
largest contribution to signal samples in many neutrino oscillation 
experiments, there has been renewed interest in the measurement and 
modeling of QE scattering on nuclear targets. In such situations, the nucleus 
is typically described in terms of individual quasi-free nucleons that 
participate in the scattering process (the so-called ``impulse approximation''
approach~\cite{modern-qe-theory}). Most neutrino experiments use a relativistic Fermi Gas 
model~\cite{smith-moniz} when simulating their QE scattering events, although
many other independent particle approaches have been developed in recent years 
that incorporate more sophisticated treatments. These 
include spectral function~\cite{Nakamura02,Benhar05,Benhar07,Ankowski06,Juszczak10}, superscaling~\cite{bib:DonnellySuperScaling}, RPA~\cite{Nieves06, Nieves04,Alvarez-Ruso07,Athar10}, and PWIA-based calculations~\cite{Butkevich10}. 
In concert, the added nuclear effects from these improved calculations tend 
to reduce the predicted neutrino QE cross section beyond the Fermi-Gas model 
based predictions. These reductions are typically on the order of 
$10-20\%$~\cite{reduced-qe-cross-section}. \\

Using Fermi-Gas model based simulations and analyzing higher statistics QE 
data on a variety of nuclear targets, new experiments have begun to repeat 
the axial-vector measurements that fueled much of the early investigations 
of QE scattering. Axial mass values ranging from 1.05 to 1.35 GeV have been 
recently obtained~\cite{nomad-qe, modern-ma, modern-ma-1,modern-ma-2,modern-ma-3,mb-qe}, with most of the experiments systematically measuring higher $M_A$ values than those found in the deuterium fits. This has recently sparked 
some debate, especially given that higher $M_A$ values naturally imply higher 
cross sections and hence larger event yields for neutrino 
experiments~\footnote{Note that modern determinations of $M_A$ have largely 
been obtained from fits to the shape of the observed $Q^2$ distribution 
of QE events and not their normalization.}. We will come back to this point 
later.\\

\begin{figure}[h]
\begin{center}
\includegraphics[scale=0.45]{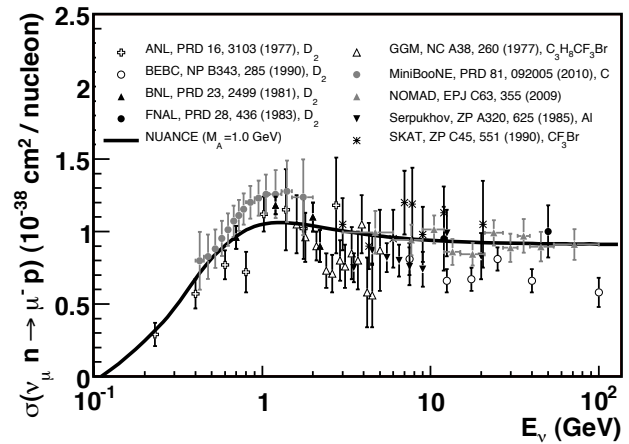}
\end{center}
\caption{ Existing measurements of the $\numu$ quasi-elastic scattering cross 
         section, $\numu \, n \rightarrow \mu^- \, p$,  as a function of 
         neutrino energy on a variety of nuclear targets. The free nucleon 
         scattering prediction assuming $M_A=1.0$ GeV is shown for 
         comparison~\cite{nuance}.}
\label{fig:qe-nu}
\end{figure}

Neutrino experiments have also begun to remeasure the absolute QE 
scattering cross section making use of more reliable incoming neutrino 
fluxes made available in modern experimental setups. 
Figure~\ref{fig:qe-nu} summarizes the existing measurements 
of $\numu$ QE scattering cross sections as a function of neutrino energy 
from both historical and recent measurements. As expected, we observe a 
linearly rising cross section that is damped by the form factors at higher 
neutrino energies. What is not expected is the disparity observed between 
recent measurements. High statistics measurements of the QE scattering cross 
section by the MiniBooNE~\cite{mb-qe} and NOMAD~\cite{nomad-qe} experiments, 
both on carbon, appear to differ in normalization by about $30\%$. The low 
energy MiniBooNE results are higher than expected from the Fermi Gas 
model~\cite{smith-moniz} and more sophisticated impulse approximation
calculations~\cite{modern-qe-theory,modern-qe-theory-1,Nieves04,modern-qe-theory-3,modern-qe-theory-4,modern-qe-theory-5,Benhar07,modern-qe-theory-7,modern-qe-theory-8,Alvarez-Ruso07,modern-qe-theory-10} assuming an axial mass, $M_A=1.0$ GeV, 
from deuterium-based measurements as input. \\

How can it be that new, high statistics measurements of this simple process
are not coming out as expected? The fact that modern measurements of QE 
scattering have seemingly raised more questions than they have answered has 
been recently noted in the literature~\cite{qe-annual-review, Sobczyk11}. It is 
currently believed that nuclear effects beyond the impulse approximation approach 
are responsible for the discrepancies noted in the experimental data. In 
particular, it is now being recognized that nucleon-nucleon  
correlations and two-body exchange currents must be included in order to 
provide a more accurate description of neutrino-nucleus QE scattering. 
These effects yield significantly enhanced cross sections (larger than the free scattering case) 
which, in some cases, appear to better match the experimental data~\cite{mb-qe} at low 
neutrino energies~\cite{Amaro:2010,Barbaro:2011st,Giusti:2011dj,Nieves2011sc,Martini:2011wp,Sobczyk:2012ah,Bodek:2011ps}. They also produce final
states that include multiple nucleons, especially when it comes to scattering off of nuclei. The final state need not just include a single nucleon, hence why one needs to be careful in defining a ``quasi-elastic'' event especially when it comes to scattering 
off nuclei. 

In hindsight, the increased neutrino QE cross sections and 
harder $Q^2$ distributions (high $M_A$) observed in the much of the 
experimental data should probably have not come as a surprise. 
Such effects were also measured in transverse electron-nucleus quasi-elastic 
scattering many years prior~\cite{carlson}. The possible connection between electron and neutrino QE scattering observations has only been recently appreciated. Today, the role that additional nuclear effects may play in neutrino-nucleus QE scattering remains the subject
of much theoretical and experimental scrutiny. Improved
theoretical calculations and experimental measurements are already underway. 
As an example, the first double differential cross section distributions 
for $\numu$ QE scattering were recently reported by the MiniBooNE 
experiment~\cite{mb-qe}. It is generally recognized that such model
model-independent measurements are more useful than comparing $M_A$ values.
Such differential cross section data are also providing an important new testing 
ground for improved nuclear model calculations~\cite{Amaro:2011aa,Giusti:2011dj,Martini:2011ui,Nieves:2011yp,Sobczyk:2012ah}. Moving forward, additional
differential cross section measurements, detailed measurements of nucleon 
emission, and studies of antineutrino QE scattering are needed before 
a solid description can be secured.

\begin{figure}[h]
\begin{center}
\includegraphics[scale=0.45]{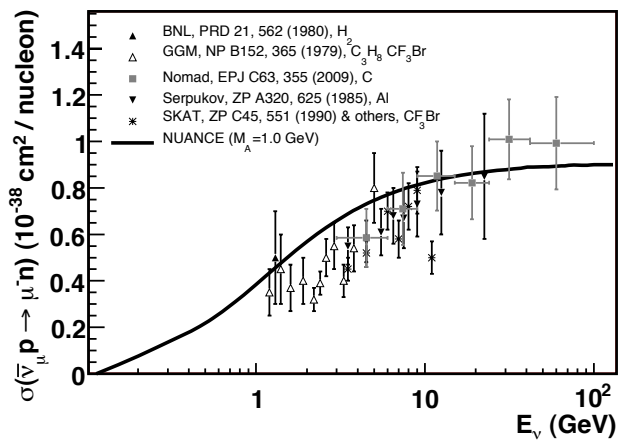}
\end{center}
\vspace{-0.2in}
\caption{ Same as Figure~\ref{fig:qe-nu} except for antineutrino QE
         scattering, $\numubar \, p \rightarrow \mu^+ \, n$.}
\label{fig:qe-nubar}
\end{figure}

So far, we have focused on neutrino QE scattering. Figure~\ref{fig:qe-nubar} 
shows the status of measurements of the anti-neutrino QE scattering cross 
section. Recent results from the NOMAD experiment have 
expanded the reach out to higher neutrino energies, however, there are
currently no existing measurements of the antineutrino QE scattering cross 
section below 1 GeV. Given that the newly appreciated effects of 
nucleon-nucleon correlations are expected to be different for neutrinos 
and antineutrinos, a high priority has been recently given to the study of 
antineutrino QE scattering at these energies. A precise handle
on neutrino and antineutrino QE interaction cross sections will be 
particularly important in the quest for the detection of CP violation in 
the leptonic sector going into the future. \\

So far we have discussed the case where nucleons can be ejected in
the elastic scattering of neutrinos from a given target. The final state 
is traditionally a single nucleon, but can also include multiple nucleons, 
especially in the case of neutrino-nucleus scattering. For antineutrino QE scattering, it should be noted that the Cabibbo-suppressed production of hyperons is also possible, for example:

\begin{eqnarray}
  \numubar \, p &\rightarrow& \mu^+ \, \Lambda^0 \\
  \numubar \, n &\rightarrow& \mu^+ \, \Sigma^- \\
  \numubar \, p &\rightarrow& \mu^+ \, \Sigma^0 
\end{eqnarray}

\noindent
Cross sections for QE hyperon production by neutrinos were calculated very
early on~\cite{ll-smith,cabibbo-hyperon} and verified in low statistics
measurements by a variety of bubble chamber 
experiments~\cite{eichten,erriques,ammosov,Brunner:1989kw}. New calculations have also 
recently surfaced in the literature~\cite{new-hyperon-models, Kuzmin:2008zz, Mintz:2007zz}.  We will say more about strange particle production later when we discuss kaon production (Section~\ref{sec:kaon}). \\

Combined, all experimental measurements of QE scattering cross sections
have been conducted using beams of muon neutrinos and antineutrinos. No 
direct measurements of $\nue$ or $\nuebar$ QE scattering cross sections 
have yet been performed at these energies. 

\subsection{NC Elastic Scattering}
\label{sec:ncel}

Neutrinos can also elastically scatter from nucleons via neutral current (NC) interactions:

\begin{eqnarray}
  \nu \, p &\rightarrow& \nu \, p, \hspace{0.2in}
  \nubar \, p \rightarrow \nubar \, p \\
  \nu \, n &\rightarrow& \nu \, n, \hspace{0.2in}
  \nubar \, n \rightarrow \nubar \, n 
\end{eqnarray}


\noindent
Equations~\ref{eqn:xsectqe}-\ref{eqn:xsectabc} still apply in describing
NC elastic scattering from free nucleons with the exception that, in this 
case, the form factors include additional coupling factors and a contribution 
from strange quarks:

\begin{eqnarray}
 F_1(Q^2) &=& \left(\frac{1}{2}-\sin^2\theta_W\right)
          \left[\frac{\tau_3(1+\eta\,(1+\mu_p-\mu_n))}
                     {(1+\eta)\left(1+Q^2/M_V^2\right)^2}\right]
\nonumber \\ 
&& \hspace{-0.4in} -\sin^2\theta_W \left[\frac{1+\eta\,(1+\mu_p+\mu_n)}
                       {(1+\eta)\left(1+Q^2/M_V^2\right)^2}\right] 
  - \frac{F_1^s(Q^2)}{2} \nonumber \\
 F_2(Q^2) &=& \left(\frac{1}{2}-\sin^2\theta_W\right){\tau_3\,{(\mu_p-\mu_n)} 
      \over {(1+\eta) \left(\displaystyle{1 + {{Q^2}\over{M_V^2} }}\right)^2}} 
\nonumber \\
      && \hspace{-0.4in} -\sin^2\theta_W {{\mu_p+ \mu_n} 
      \over {(1+\eta) \left(\displaystyle{1 + {{Q^2}\over{M_V^2} }}\right)^2}} 
      - \frac{F_2^s(Q^2)}{2} \nonumber \\
 F_A(Q^2) &=&   {{g_{A} \: \tau_3} 
             \over {2 \left(\displaystyle{1 + {{Q^2}\over{M_A^2} }}\right)^2}} 
             - \frac{F_A^s(Q^2)}{2} \nonumber
\end{eqnarray}

\noindent
Here, $\tau_3=+1 (-1)$ for proton (neutron) scattering, $\sin^2\theta_W$ is
the weak mixing angle, and $F_{1,2}^s(Q^2)$ are the strange vector form 
factors, here assuming a dipole form. The strange axial vector form factor 
is commonly denoted as:

\begin{eqnarray}
  F_A^s(Q^2)= \frac{\Delta s}{\left(\displaystyle{1 + {{Q^2}\over{M_A^2} }}
  \right)^2}
\end{eqnarray}

\noindent
where $\Delta s$ is the strange quark contribution to the nucleon spin
and $M_A$ is the same axial mass appearing in the expression for CC
QE scattering (Equation~\ref{eqn:qe-fa}). \\  

Over the years, experiments have typically measured NC elastic cross section
ratios with respect to QE scattering to help minimize systematics. 
Table~\ref{table:nc-elastic} lists a collection of historical measurements 
of the NC elastic/QE cross section ratio, 
$(\numu \, p \rightarrow \numu \, p)/(\numu \, n \rightarrow \mu^- \, p)$.
These ratios have been integrated over the kinematic range of the experiment. More recently, the MiniBooNE experiment has measured the NC elastic/QE 
ratio on carbon in bins of $Q^2$~\cite{mb-ncel}.  \\

\begin{table}[h]
\begin{tabular}{|l|c|c|c|}
\hline
Experiment   & Target  & Ratio  & $Q^2$(GeV$^2$) \\
\hline\hline
BNL E734  & $CH_2$ & 0.153 $\pm$ 0.018  
                               & $0.5-1.0$ \\
BNL CIB  & $Al$     & 0.11 $\pm$ 0.03   
                               & $0.3-0.9$ \\ 
Aachen   & $Al$     & 0.10 $\pm$ 0.03   
                               & $0.2-1.0$ \\
BNL E613  & $CH_2$   & 0.11 $\pm$ 0.02   
                               & $0.4-0.9$ \\
Gargamelle & $CF_3Br$ & 0.12 $\pm$ 0.06   
                               & $0.3-1.0$ \\
\hline
\end{tabular}
\caption{ Measurements of the ratio, 
   $(\numu \, p \rightarrow \numu \, p)/(\numu \, n \rightarrow \mu^- \, p)$ taken from BNL E734~\cite{nc-elastic-1,nc-elastic-2,nc-elastic-3},BNL E613~\cite{nc-elastic-4}, and Gargamelle~\cite{nc-elastic-5}. 
   Also indicated is the $Q^2$ interval over which the ratio was measured.}
\label{table:nc-elastic}
\end{table}

Experiments such as BNL E734 and MiniBooNE have additionally reported 
measurements of flux-averaged absolute differential cross sections, $d\sigma/dQ^2$, 
for NC elastic scattering on carbon. From these distributions, measurements 
of parameters appearing in the cross section for this process, $M_A$ and 
$\Delta s$, can be directly obtained. Table~\ref{table:ncel-ma-deltas} 
summarizes those findings. As with QE scattering, a new appreciation for the presence of nuclear effects in such neutral current interactions has also recently arisen with many new calculations of this cross section on nuclear targets~\cite{Meucci11,Butkevich11,Benhar11,Amaro06}. Just like in the charged current case, nuclear corrections can be on the order of 20\% or more

\begin{table}[h]
\begin{tabular}{|l|c|c|}
\hline
Experiment   & $M_A$ (GeV)     & $\Delta s$ \\
\hline\hline
BNL E734     & $1.06 \pm 0.05$ & $-0.15 \pm 0.09$ \\
MiniBooNE    & $1.39 \pm 0.11$ & $0.08 \pm 0.26$ \\
\hline
\end{tabular}
\caption{Measurements of the axial mass and strange quark content to the 
         nucleon spin from neutrino NC elastic scattering data taken from 
         BNL E734~\cite{nc-elastic-1} and MiniBooNE~\cite{mb-ncel}. BNL-E734
         reported a measurement of $\eta=0.12 \pm 0.07$ which implies
         $\Delta s = -g_A \eta = -0.15 \pm 0.09$. Note that updated
         fits to the BNL-E734 data were also later performed 
         by several groups~\cite{bnl-e734-update,bnl-e734-update-2}.}
\label{table:ncel-ma-deltas}
\end{table}

\subsection{Resonant Single Pion Production}
\label{sec:single-pi}

Now that we have discussed quasi-elastic and elastic scattering mechanisms, 
let us consider another interaction possibility: this time an inelastic
interaction. Given enough energy, neutrinos can excite the struck 
nucleon to an excited state. In this case, the neutrino interaction produces 
a baryon resonance ($N^*$). The baryon resonance quickly decays, most
often to to a nucleon and single pion final state:

\begin{eqnarray}
   \numu \, N \rightarrow \mu^- &N^*& \\
                                &N^*& \rightarrow \pi \, N^\prime 
\end{eqnarray}

\noindent
where $N,N^\prime=n,p$. Other higher multiplicity decay modes are also
possible and will be discussed later. \\

The most common means of single pion production in intermediate energy 
neutrino scattering arises through this mechanism. In scattering off of free nucleons, there are seven possible 
resonant single pion reaction channels  (seven each for neutrino and 
antineutrino scattering), three charged current:

\begin{eqnarray}
   \numu \, p &\rightarrow& \mu^- \, p \, \pi^+, \hspace{0.2in}
   \numubar \, p \rightarrow \mu^+ \, p \, \pi^- \\ 
   \numu \, n &\rightarrow& \mu^- \, p \, \pi^0,  \hspace{0.2in}
   \numubar \, p \rightarrow \mu^+ \, n \, \pi^0 \\
   \numu \, n &\rightarrow& \mu^- \, n \, \pi^+, \hspace{0.18in}
   \numubar \, n \rightarrow \mu^+ \, n \, \pi^- 
\end{eqnarray}

\noindent
and four neutral current:

\begin{eqnarray}
   \numu \, p &\rightarrow& \numu \, p \, \pi^0, \hspace{0.22in} 
   \numubar \, p \rightarrow \numubar \, p \, \pi^0 \\
   \numu \, p &\rightarrow& \numu \, n \, \pi^+, \hspace{0.2in}
   \numubar \, n \rightarrow \numubar \, n \, \pi^0 \\
   \numu \, n &\rightarrow& \numu \, n \, \pi^0, \hspace{0.2in}
   \numubar \, n \rightarrow \numubar \, n \, \pi^0  \\
   \numu \, n &\rightarrow& \numu \, p \, \pi^-, \hspace{0.2in}
   \numubar \, n \rightarrow \numubar \, p \, \pi^- 
\end{eqnarray}


\noindent
To describe such resonance production processes, neutrino experiments
most commonly use calculations from the Rein and Sehgal 
model~\cite{rein-sehgal, rein-sehgal-1, rein-sehgal-2} with the additional 
inclusion of lepton mass terms. This model gives predictions for both CC 
and NC resonance production and a prescription for handling interferences 
between overlapping resonances. The cross sections for the production of numerous 
different resonances are typically evaluated, though at the lowest energies 
the process is dominated by production of the $\Delta(1232)$. \\

\begin{table*}[h]
\newcommand{\m}{\hphantom{$-$}}
\newcommand{\cc}[1]{\multicolumn{1}{c}{#1}}
\begin{tabular}{@{}llll}
\hline
Channel  & Experiment  & Target  & \# Events \\
\hline\hline
$\numubar \, p \rightarrow \mu^+ \, p \, \pi^-$:  &  & \\
   & BEBC      & $D_2$    & 300 \\
   & BEBC      & $H_2$    & 609 \\
   & GGM       & $CF_3Br$ & 282 \\
   & FNAL      & $H_2$    & 175 \\
   & SKAT      & $CF_3Br$ & 145 \\
$\numubar \, p \rightarrow \mu^+ \, n \, \pi^0$:  &  & \\
   & GGM       & $CF_3Br$ & 179 \\
   & SKAT      & $CF_3Br$ & 83 \\
$\numubar \, n \rightarrow \mu^+ \, n \, \pi^-$:  &  & \\
   & BEBC     & $D_2$    & 545 \\
   & GGM      & $CF_3Br$ & 266 \\
   & SKAT     & $CF_3Br$ & 178 \\
\hline
\end{tabular}
\caption{Measurements of antineutrino CC single pion production from BEBC~\cite{Allen:1985ti,Allasia:1983qh,Jones:1989vt}, FNAL~\cite{Barish:1979ny}, Gargamelle~\cite{Bolognese:1979gf}, and SKAT~\cite{Grabosch:1988gw}.} 
\label{table:ccpion-nubar}
\end{table*}

\begin{figure}[h]
\begin{center}
\includegraphics[scale=0.45]{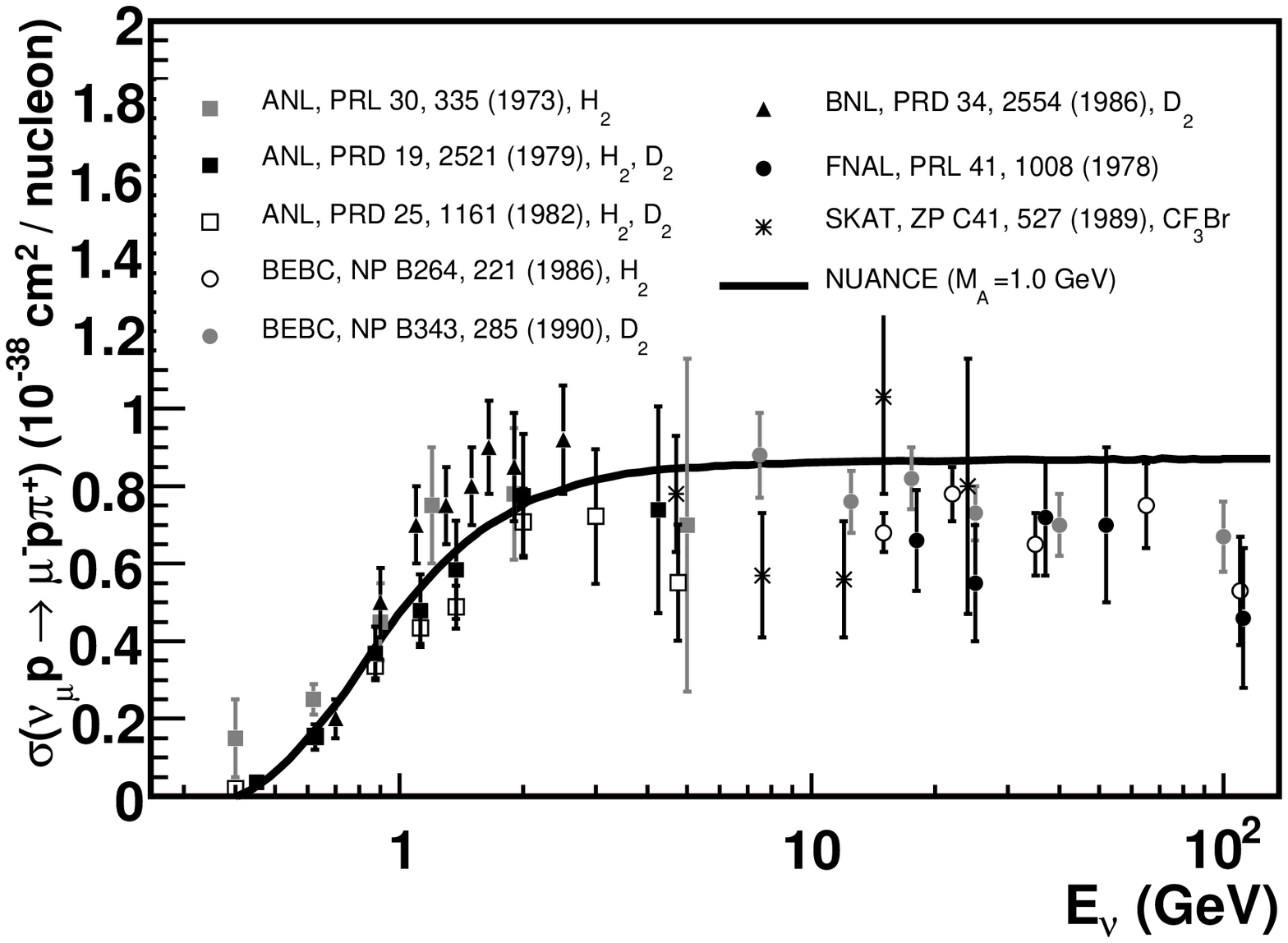}
\end{center}
\vspace{-0.2in}
\caption{ Existing measurements of the cross section for the CC process,
         $\numu \, p \rightarrow \mu^- \, p \, \pi^+$, as a function of 
         neutrino energy. Also shown is the prediction from ~\cite{nuance} assuming $M_A=1.1$ GeV.}
\label{fig:ccpi-ch3}
\end{figure}

\begin{figure}[h]
\begin{center}
\includegraphics[scale=0.45]{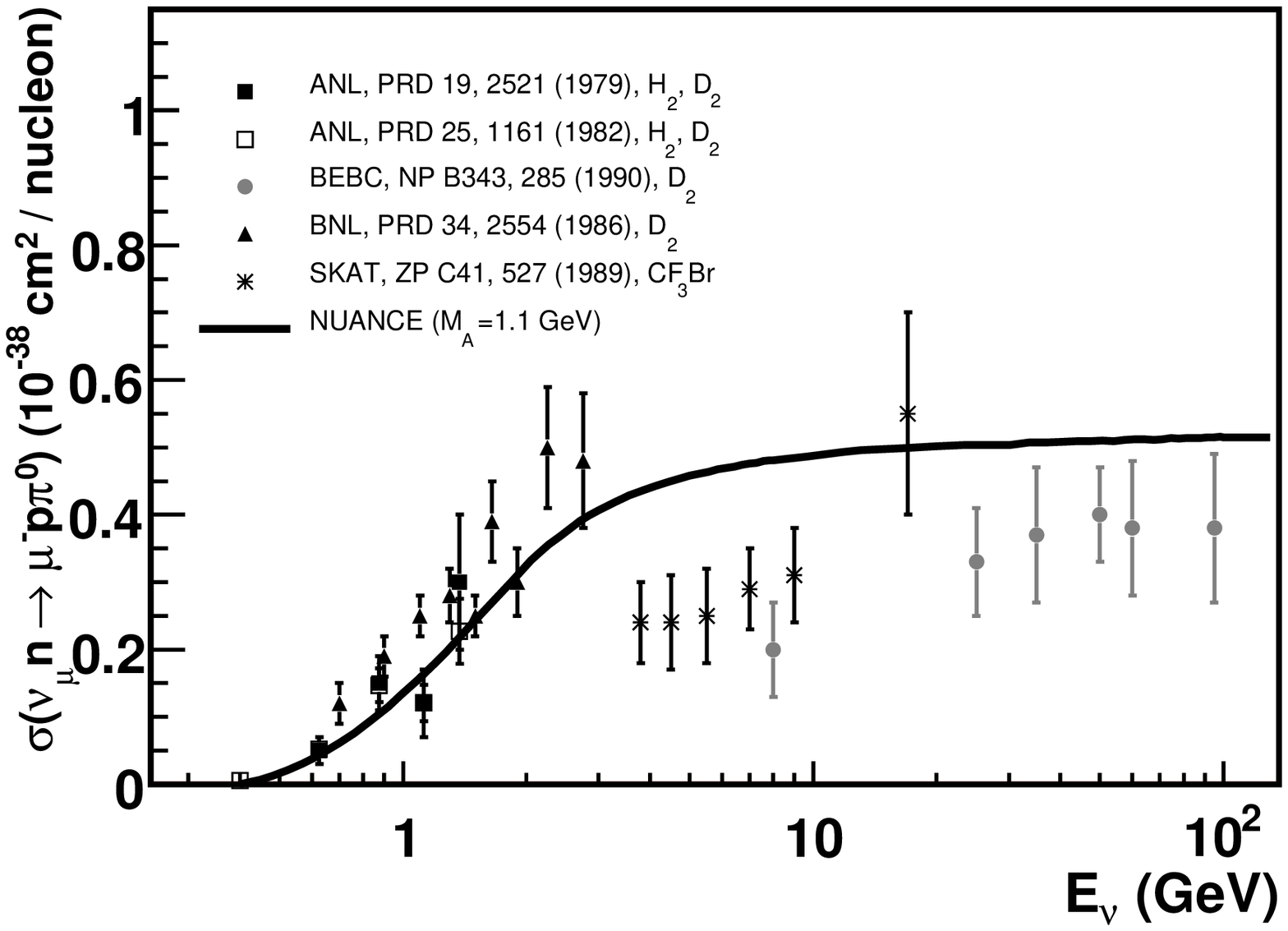}
\end{center}
\vspace{-0.2in}
\caption{ Existing measurements of the cross section for the CC process,
         $\numu \, n \rightarrow \mu^- \, p \, \pi^0$, as a function of 
         neutrino energy. Also shown is the prediction from ~\cite{nuance} assuming $M_A=1.1$ GeV.}
\label{fig:ccpi-ch4}
\end{figure}

\begin{figure}[h]
\begin{center}
\includegraphics[scale=0.45]{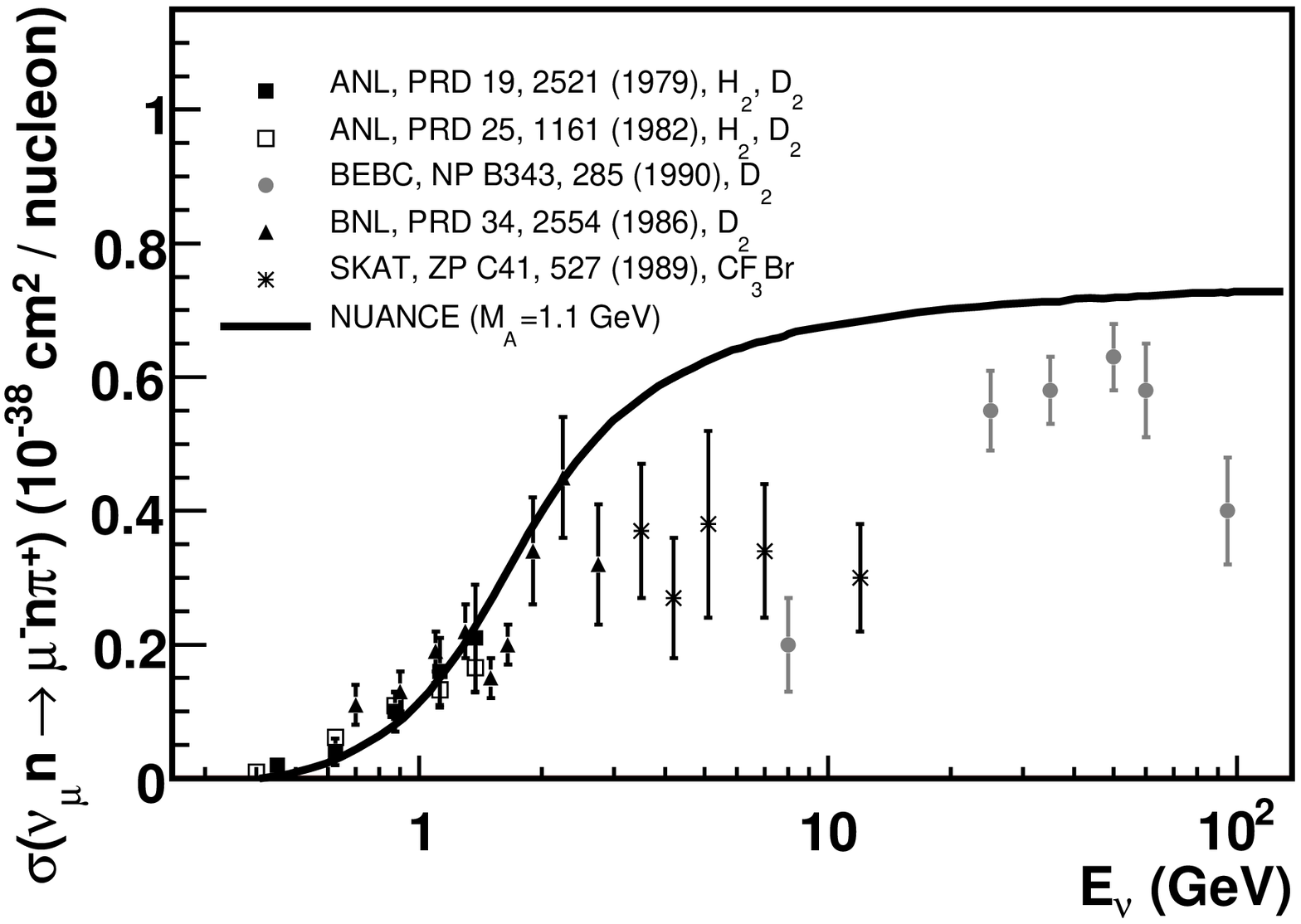}
\end{center}
\vspace{-0.2in}
\caption{ Existing measurements of the cross section for the CC process,
         $\numu \, n \rightarrow \mu^- \, n \, \pi^+$ as a function of 
         neutrino energy. Also shown is the prediction from~\cite{nuance} assuming $M_A=1.1$ GeV.}
\label{fig:ccpi-ch5}
\end{figure}

Figures~\ref{fig:ccpi-ch3}-\ref{fig:ccpi-ch5} summarize the historical
measurements of CC neutrino single pion production cross sections as a 
function of neutrino energy. Table~\ref{table:ccpion-nubar} lists corresponding measurements in antineutrino scattering.  Many of these measurements were conducted 
on light (hydrogen or deuterium) targets and served as a crucial 
verification of cross section predictions at the time. Measurements of 
the axial mass were often repeated using these samples. Experiments 
also performed tests of resonance production models by measuring 
invariant mass and angular distributions. However, many of these tests 
were often limited in statistics. \\

\begin{figure}[h]
\begin{center}
\includegraphics[scale=0.45]{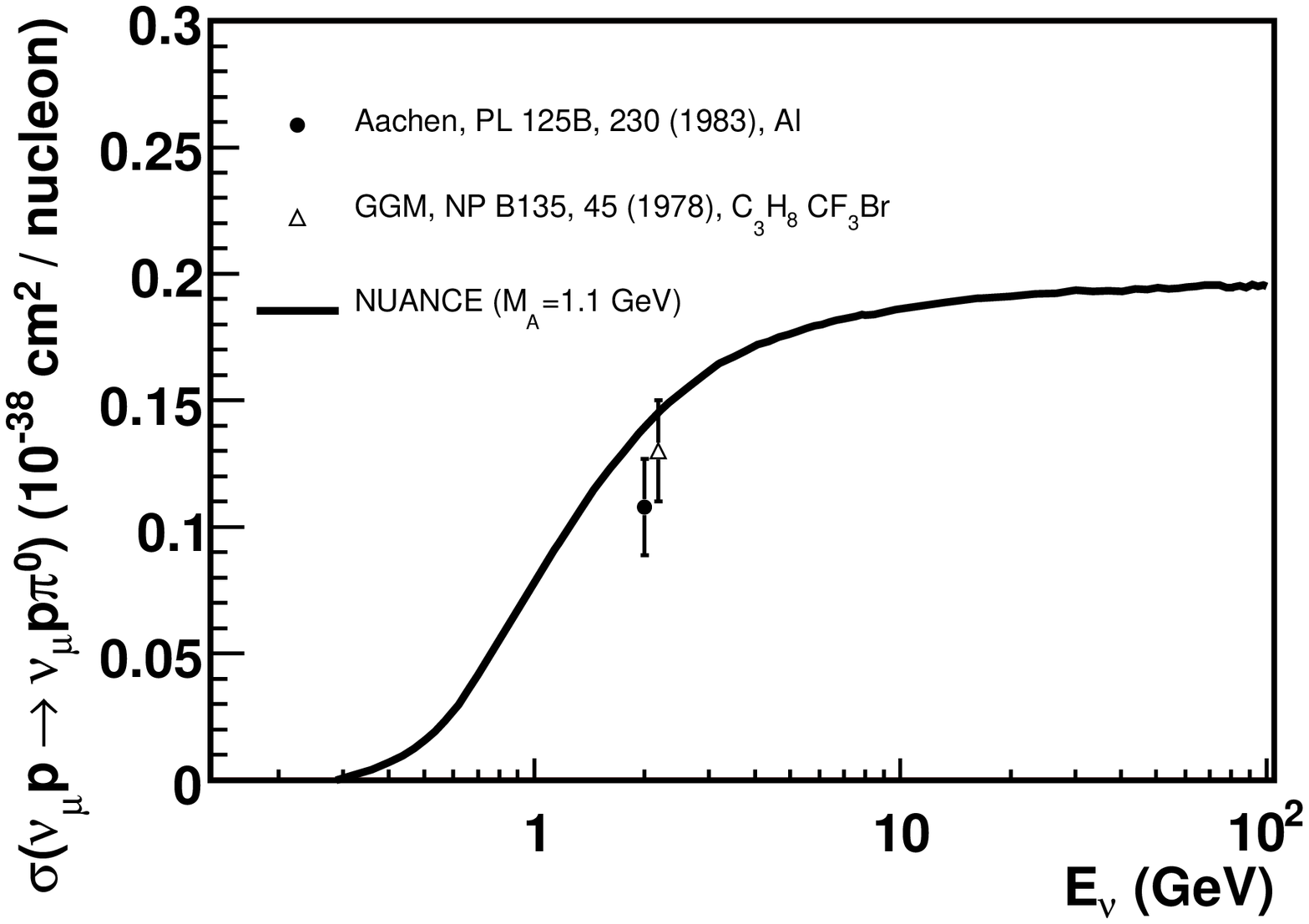}
\end{center}
\vspace{-0.2in}
\caption{ Existing measurements of the cross section for the NC process,
         $\numu \, p \rightarrow \numu \, p \, \pi^0$, as a function of 
         neutrino energy. Also shown is the prediction from 
      Reference~\cite{nuance} assuming $M_A=1.1$ GeV. 
      The Gargamelle measurement comes from a more recent re-analysis 
      of this data~\cite{hawker}.}
\label{fig:ncpi-ch6}
\end{figure}

\begin{figure}[h]
\begin{center}
\includegraphics[scale=0.45]{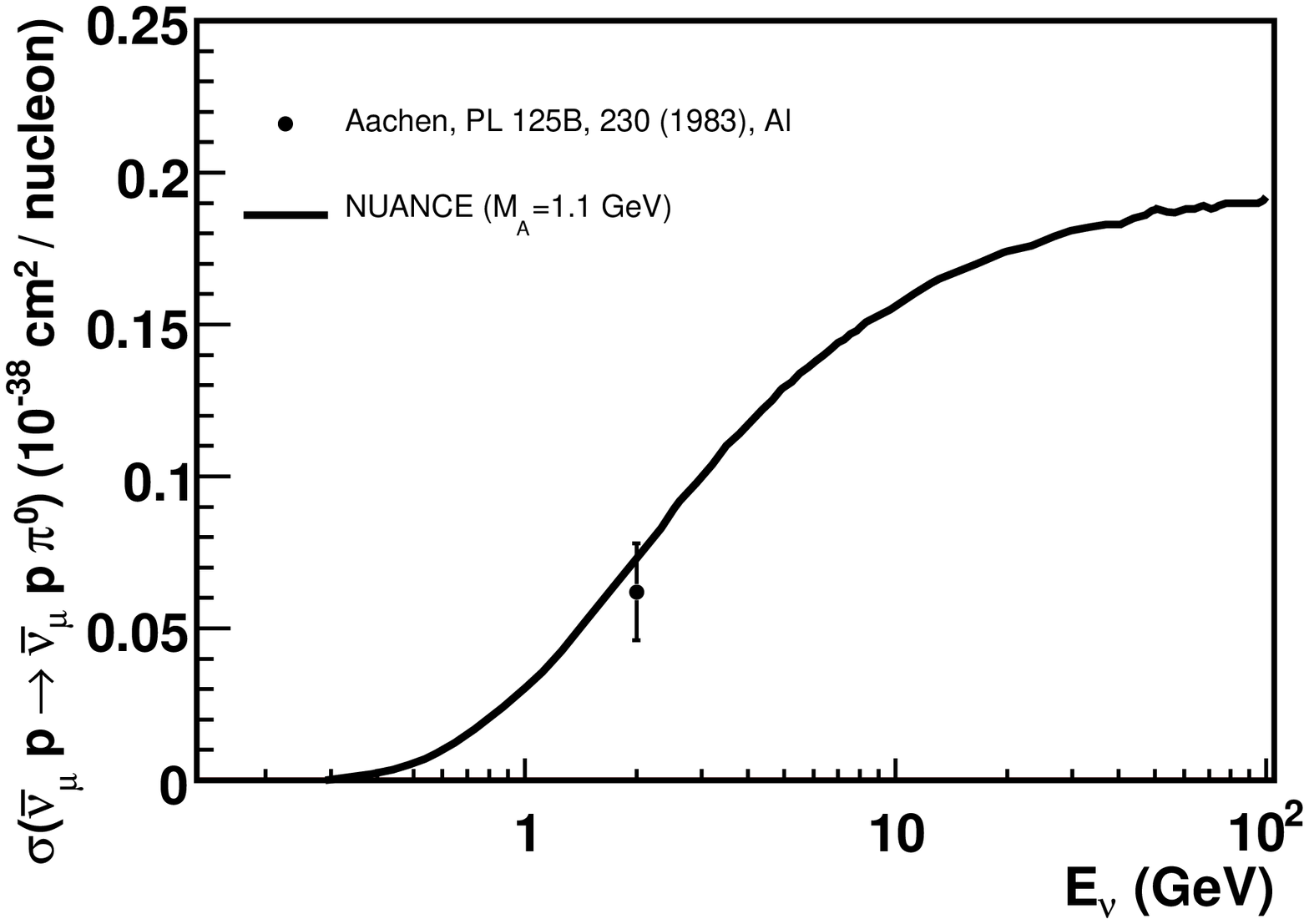}
\end{center}
\vspace{-0.2in}
\caption{ Existing measurements of the cross section for the NC process,
         $\numubar \, p \rightarrow \numubar \, p \, \pi^0$, as a function of 
         neutrino energy. Also shown is the prediction from 
         Reference~\cite{nuance} assuming $M_A=1.1$ GeV.}
\label{fig:ncpi-ch6-nubar}
\end{figure}

\begin{figure}[h]
\begin{center}
\includegraphics[scale=0.45]{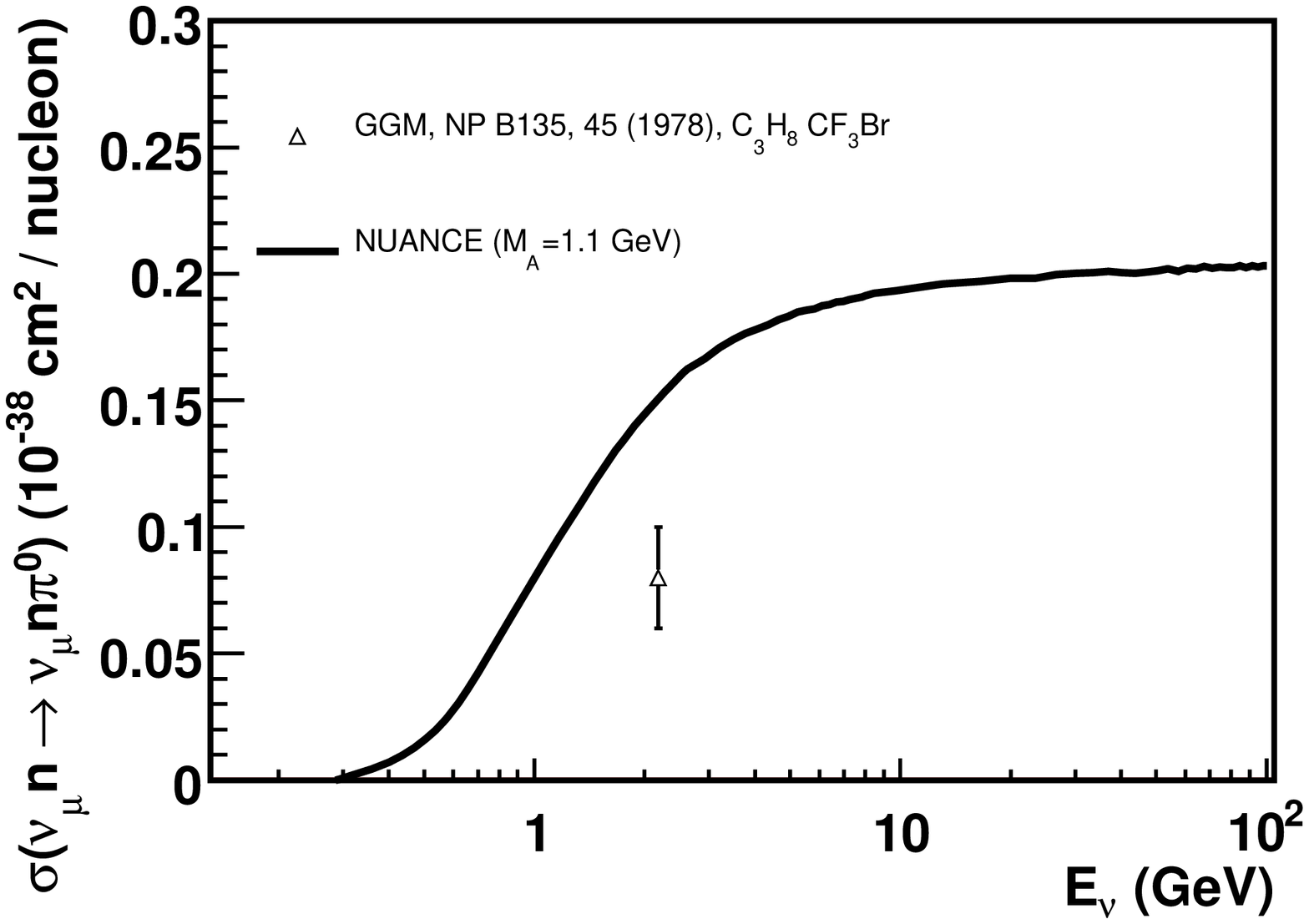}
\end{center}
\vspace{-0.2in}
\caption{ Existing measurements of the cross section for the NC process,
         $\numu \, n \rightarrow \numu \, n \, \pi^0$, as a function of 
         neutrino energy. Also shown is the prediction from 
         Reference~\cite{nuance} assuming $M_A=1.1$ GeV.
      The Gargamelle measurement comes from a more recent re-analysis 
      of this data~\cite{hawker}.}
\label{fig:ncpi-ch8}
\end{figure}

\begin{figure}[h]
\begin{center}
\includegraphics[scale=0.45]{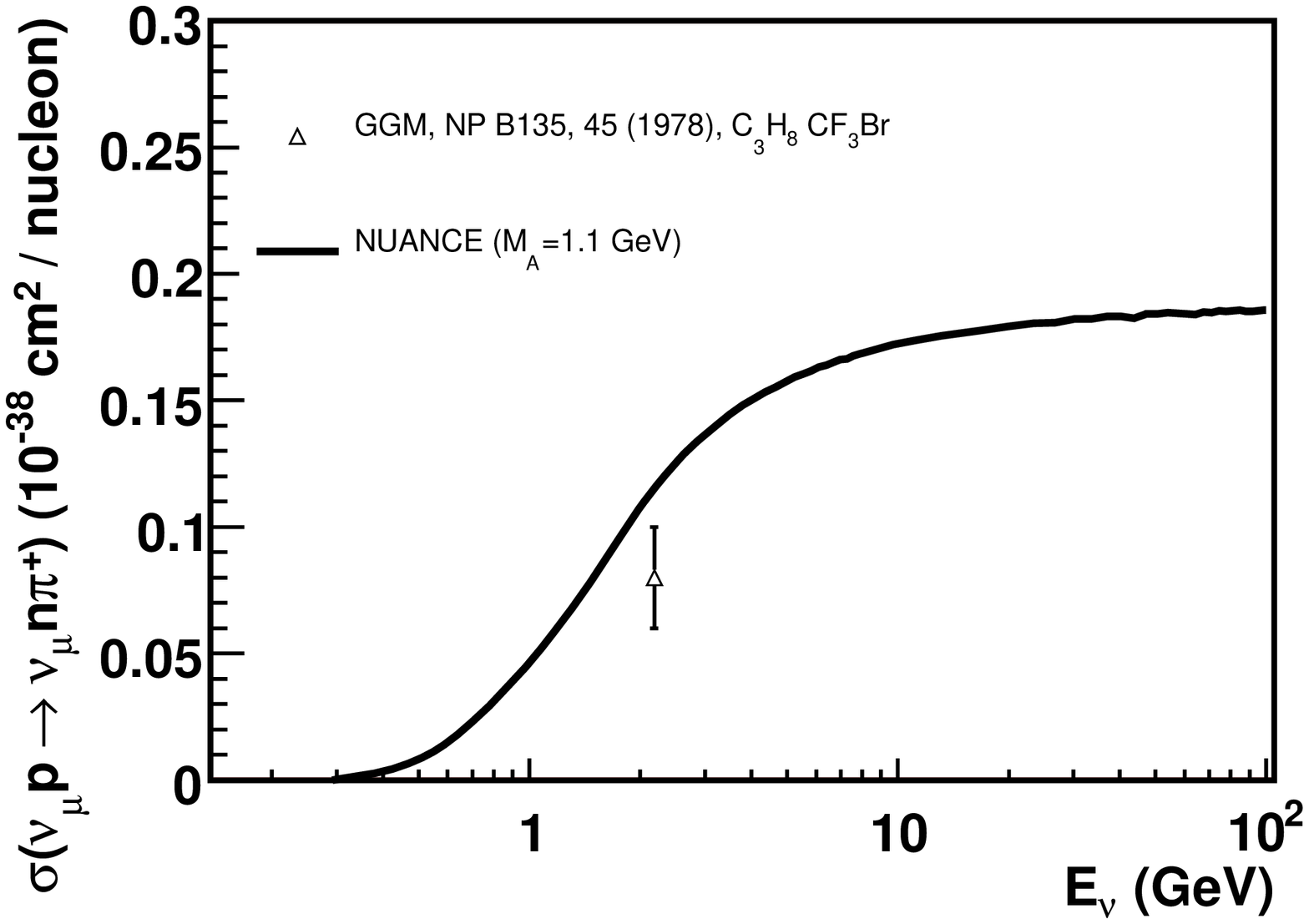}
\end{center}
\vspace{-0.2in}
\caption{ Existing measurements of the cross section for the NC process,
         $\numu \, p \rightarrow \numu \, n \, \pi^+$, as a function of 
         neutrino energy. Also shown is the prediction from 
         Reference~\cite{nuance} assuming $M_A=1.1$ GeV.}
\label{fig:ncpi-ch7}
\end{figure}

\begin{figure}[h]
\begin{center}
\includegraphics[scale=0.45]{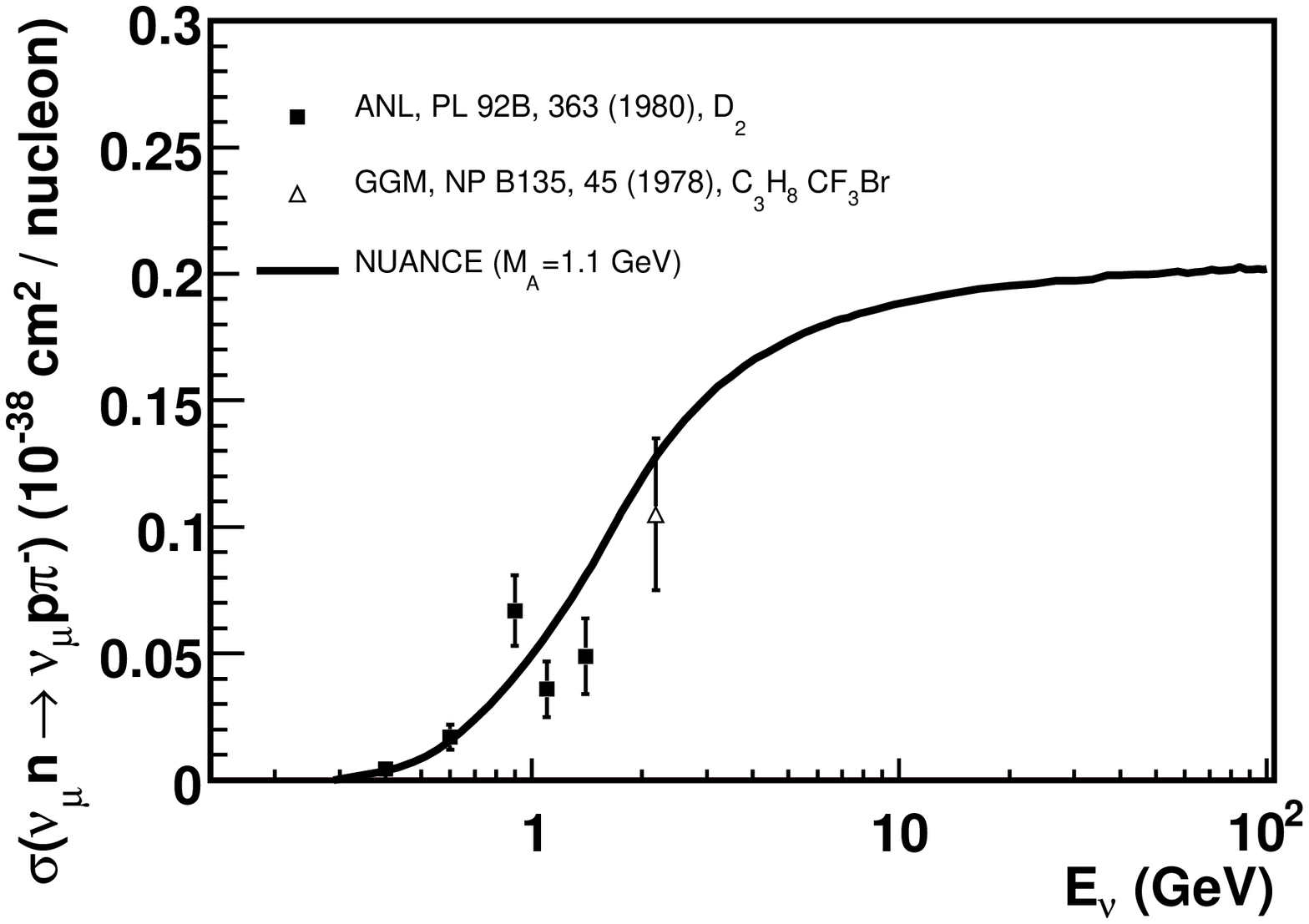}
\end{center}
\vspace{-0.2in}
\caption{ Existing measurements of the cross section for the NC process,
         $\numu \, n \rightarrow \numu \, p \, \pi^-$, as a function of 
         neutrino energy. Also shown is the prediction from 
         Reference~\cite{nuance} assuming $M_A=1.1$ GeV.}
\label{fig:ncpi-ch9}
\end{figure}

\begin{figure}[h]
\begin{center}
\includegraphics[scale=0.45]{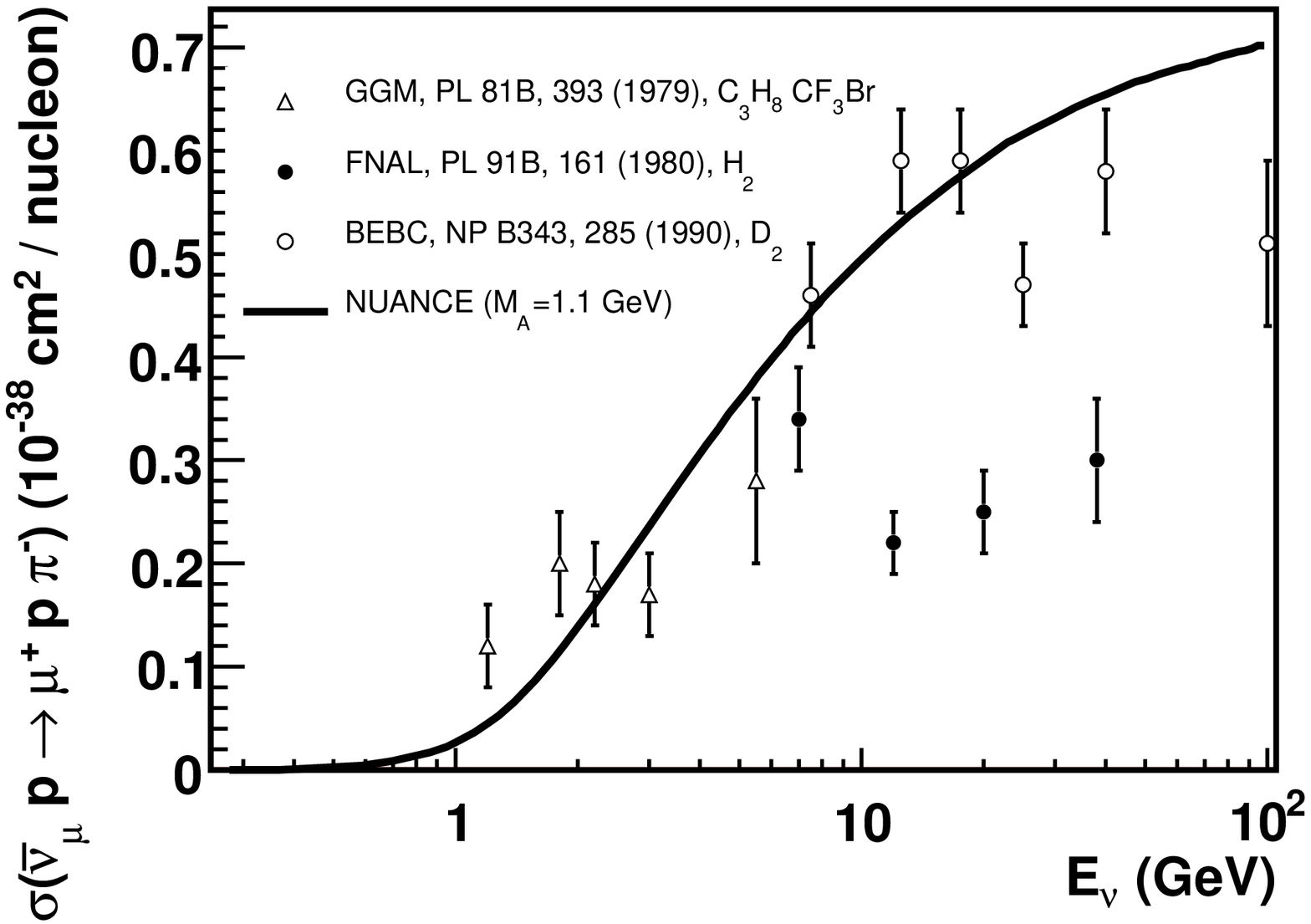}
\end{center}
\vspace{-0.2in}
\caption{ Existing measurements of the cross section for the NC process,
         $\numubar \, n \rightarrow \numubar \, p \, \pi^-$, as a function of 
         neutrino energy. Also shown is the prediction from 
        Reference~\cite{nuance} assuming $M_A=1.1$ GeV.}
\label{fig:ncpi-ch9-nubar}
\end{figure}

Compared to their charged current counterparts, measurements of neutral
current single pion cross sections tend to be much more sparse. Most of 
this data exists in the form of NC/CC cross section ratios (Table~\ref{table:ncpion}); 
however a limited number of absolute cross section measurements were also 
performed over the years 
(Figures~\ref{fig:ncpi-ch6}-\ref{fig:ncpi-ch9-nubar}). 

\begin{table*}[h]
\newcommand{\m}{\hphantom{$-$}}
\newcommand{\cc}[1]{\multicolumn{1}{c}{#1}}
\begin{tabular}{@{}lllll}
\hline
Experiment  & Target  & NC/CC Ratio & Value  & Reference \\
\hline\hline
ANL         & $H_2$ 
            & $\sigma(\numu \, p \rightarrow \numu \, p \, \pi^0)/
               \sigma(\numu p \, \rightarrow \mu^- \, p \, \pi^+)$  
            & $0.51 \pm 0.25$  & \cite{nc-cc-single-pi-1} \\
ANL         & $H_2$ 
            & $\sigma(\numu \, p \rightarrow \numu \, p \, \pi^0)/
               \sigma(\numu p \, \rightarrow \mu^- \, p \, \pi^+)$  
            & $0.09 \pm 0.05^*$
              & \cite{nc-cc-single-pi-2} \\
\hline
ANL         & $H_2$ 
            & $\sigma(\numu \, p \rightarrow \numu \, n \, \pi^+)/
               \sigma(\numu p \, \rightarrow \mu^- \, p \, \pi^+)$  
            & $0.17 \pm 0.08$  & \cite{nc-cc-single-pi-1} \\
ANL         & $H_2$ 
            & $\sigma(\numu \, p \rightarrow \numu \, n \, \pi^+)/
               \sigma(\numu p \, \rightarrow \mu^- \, p \, \pi^+)$  
            & $0.12 \pm 0.04$  & \cite{nc-cc-single-pi-2} \\
\hline
ANL         & $D_2$ 
            & $\sigma(\numu \, n \rightarrow \numu \, p \, \pi^-)/
               \sigma(\numu n \, \rightarrow \mu^- \, n \, \pi^+)$  
            & $0.38 \pm 0.11$  & 
              \cite{nc-cc-single-pi-3} \\
\hline
GGM         & $C_3H_8$ $CF_3Br$ 
            & $\sigma(\numu \, N \rightarrow \numu \, N \, \pi^0)/
               2\, \sigma(\numu n \, \rightarrow \mu^- \, p \, \pi^0)$  
            & $0.45 \pm 0.08$  & \cite{nc-cc-single-pi-4} \\
CERN PS     & $Al$ 
            & $\sigma(\numu \, N \rightarrow \numu \, N \, \pi^0)/
               2\, \sigma(\numu n \, \rightarrow \mu^- \, p \, \pi^0)$  
            & $0.40 \pm 0.06$  & \cite{nc-cc-single-pi-3} \\
BNL         & $Al$ 
            &  $\sigma(\numu \, N \rightarrow \numu \, N \, \pi^0)/
               2\, \sigma(\numu n \, \rightarrow \mu^- \, p \, \pi^0)$  
            & $0.17 \pm 0.04$  & \cite{nc-cc-single-pi-5} \\
BNL         & $Al$ 
            & $\sigma(\numu \, N \rightarrow \numu \, N \, \pi^0)/
               2\, \sigma(\numu n \, \rightarrow \mu^- \, p \, \pi^0)$  
            & $0.25 \pm 0.09^{**}$ & \cite{nienaber} \\
\hline
ANL         & $D_2$ 
            & $\sigma(\numu \, n \rightarrow \numu \, p \, \pi^-)/
               \sigma(\numu p \, \rightarrow \mu^- \, p \, \pi^+)$  
            & $0.11 \pm 0.022$  & \cite{nc-cc-single-pi-2} \\
\hline
\end{tabular}
\caption{ Measurements of NC/CC single pion cross section ratios ($N=n,p$).
         The Gargamelle data has been corrected to a free nucleon   
         ratio~\cite{nc-cc-single-pi-4}. $^*$In their later
         paper~\cite{nc-cc-single-pi-2}, Derrick {\em et al.}
         remark that while this result is $1.6\sigma$ smaller than their
         previous result~\cite{nc-cc-single-pi-1}, the neutron 
         background was later better understood. $^{**}$The BNL NC 
         $\pi^0$ data~\cite{nc-cc-single-pi-5} was later reanalyzed after 
         properly taking into account
         multi-pion backgrounds and found to have a larger fractional cross 
         section~\cite{nienaber}.} 
\label{table:ncpion}
\end{table*}

Today, improved measurements and predictions of neutrino-induced single pion 
production has become increasingly important because of the role such
processes play in the interpretation of neutrino oscillation 
data~\cite{ccpi-walter}. In this case, both NC and
CC processes contribute. NC $\pi^0$ production is often the largest 
$\numu$-induced background in experiments searching for 
$\numu \rightarrow \nue$ oscillations. In addition, CC $\pi$ production processes can present a non-negligible complication in the determination of neutrino energy in 
experiments measuring parameters associated with $\numu$ and $\numubar$ 
disappearance. Since such neutrino oscillation experiments use heavy materials
as their neutrino targets, measuring and modeling nuclear effects in pion 
production processes has become paramount. Such effects are sizable, not
well-known, and ultimately complicate the description of neutrino 
interactions. Once created in the initial neutrino interaction, the pion 
must escape the nucleus before it can be detected. Along its journey, the 
pion can rescatter, get absorbed, or charge-exchange thus altering its 
identity and kinematics. Improved calculations of such ``final state 
interactions'' (FSI) have been undertaken by a number of 
groups~\cite{1pi-fsi-models,1pi-fsi-models-1,1pi-fsi-models-2,1pi-fsi-models-3,1pi-fsi-models-4,1pi-fsi-models-5}. The impact of in-medium effects on the 
$\Delta$ width and the possibility for intranuclear $\Delta$ re-interactions 
($\Delta N \rightarrow N \, N$) also play a role. The combined result 
are sizable distortions to the interaction cross section and kinematics 
of final state hadrons that are produced in a nuclear environment. \\

While new calculations of pion production have proliferated, new approaches
to the experimental measurement of these processes have also surfaced in
recent years. Modern experiments have realized the importance of
final state effects, often directly reporting the distributions 
of final state particles they observe. Such ``observable'' cross sections 
are more useful in that they measure the combined effects of nuclear 
processes and are much less model dependent. Table~\ref{tab:new-pion} 
lists the collection of some of these most recent pion production cross 
section reportings. Measurements have been produced both in the form 
of ratios and absolute cross sections, all on carbon-based targets. Similar measurements on additional nuclear targets are clearly needed to help round out our understanding of nuclear effects in pion production interactions.\\

\begin{table*}[h]
\begin{tabular}{|l|c|c|r|}
\hline
Experiment   & Target  & Process & Cross Section Measurements \\
\hline\hline
K2K       & $C_8H_8$ & $\numu$ CC $\pi^+$/QE & $\sigma$, $\sigma(E_\nu)$ \\
K2K       & $C_8H_8$ & $\numu$ CC $\pi^0$/QE & $\sigma$ \\
K2K       &          & $\numu$ NC $\pi^0$/CC & $\sigma$ \\
\hline
MiniBooNE & $CH_2$   & $\numu$ CC $\pi^+$/QE & $\sigma(E_\nu)$\\
MiniBooNE & $CH_2$   & $\numu$ CC $\pi^+$    & $\sigma$, $\sigma(E_\nu)$, 
            $\frac{d\sigma}{dQ^2}$, $\frac{d\sigma}{dT_\mu}$, 
            $\frac{d\sigma}{dT_\pi}$, 
            $\frac{d^2\sigma}{dT_\mu\,d\cos\theta_\mu}$,
            $\frac{d^2\sigma}{dT_\pi\,d\cos\theta_\pi}$\\
MiniBooNE & $CH_2$   & $\numu$ CC $\pi^0$    & $\sigma$, $\sigma(E_\nu)$, 
            $\frac{d\sigma}{dQ^2}$, $\frac{d\sigma}{dE_\mu}$, 
            $\frac{d\sigma}{d\cos\theta_\mu}$, $\frac{d\sigma}{dp_\pi}$, 
            $\frac{d\sigma}{d\cos\theta_\pi}$ \\
MiniBooNE & $CH_2$   & $\numu$ NC $\pi^0$    & $\sigma$, 
            $\frac{d\sigma}{dp_\pi}$, 
            $\frac{d\sigma}{d\cos\theta_\pi}$ \\
MiniBooNE & $CH_2$   & $\numubar$ NC $\pi^0$ & $\sigma$,  
            $\frac{d\sigma}{dp_\pi}$, 
            $\frac{d\sigma}{d\cos\theta_\pi}$\\
\hline
SciBooNE  & $C_8H_8$ & $\numu$ NC $\pi^0$/CC & $\sigma$ \\
\hline
\end{tabular}
\caption{ Modern measurements of single pion production by neutrinos, at the time of this writing.
        In the last column, $\sigma$ refers to a measurement of the
        total flux-integrated cross section.  Measurements are listed from K2K~\cite{k2k-ccpip,k2k-ccpi0,k2k-ncpi0}, MiniBooNE~\cite{mb-ccpip-qe,mb-ccpip,mb-ccpi0,mb-ncpi0}, and SciBooNE~\cite{sb-ncpi0}.}
\label{tab:new-pion}
\end{table*}

Before we move on, it should be noted that many of the same baryon resonances that decay to 
single pion final states can also decay to photons 
(e.g., $\Delta \rightarrow N \, \gamma$ and $N^* \rightarrow N \, \gamma$). 
Such radiative decay processes have small branching fractions ($<1\%$) yet, 
like NC $\pi^0$ production, they still pose non-negligible sources of 
background to $\numu \rightarrow \nue$ oscillation searches. There have been 
no direct experimental measurements of neutrino-induced resonance 
radiative decay to-date; however, studies of photon production in deep 
inelastic neutrino interactions have been performed at higher 
energies~\cite{nu-photon}. New experimental investigations are clearly needed. As an example, such a photon search was recently reported by the NOMAD collaboration~\cite{Kullenberg12}. New calculations have already begun to explore the possibility for 
Standard Model-based sources of photon production in neutrino scattering~\cite{Hill11,Harvey07,Jenkins09,Efronsinin09,Ankowski12}. \\

In addition to photon decay, baryon resonances can also decay in a variety 
of other modes. This includes multi-pion and kaon final states which we will return to later in this chapter.

\subsection{Coherent Pion Production}
\label{sec:coherent-pi}

In addition to resonance production, neutrinos can also coherently produce
single pion final states. In this case, the neutrino coherently scatters from 
the entire nucleus, transferring negligible energy to the target ($A$). These
low-$Q^2$ interactions produce no nuclear recoil and a distinctly 
forward-scattered pion, compared to their resonance-mediated counterparts. 
Both NC and CC coherent pion production processes are possible:

\begin{eqnarray}
   \numu \, A &\rightarrow& \numu \, A \, \pi^0 \hspace{0.2in} 
   \numubar \, A \rightarrow \numubar \, A \, \pi^0 \\
   \numu \, A &\rightarrow& \mu^- \, A \, \pi^+  \hspace{0.2in}
   \numubar \, A \rightarrow \mu^+ \, A \, \pi^-
\end{eqnarray}

\begin{figure}[h]
\begin{center}
\includegraphics[scale=0.45]{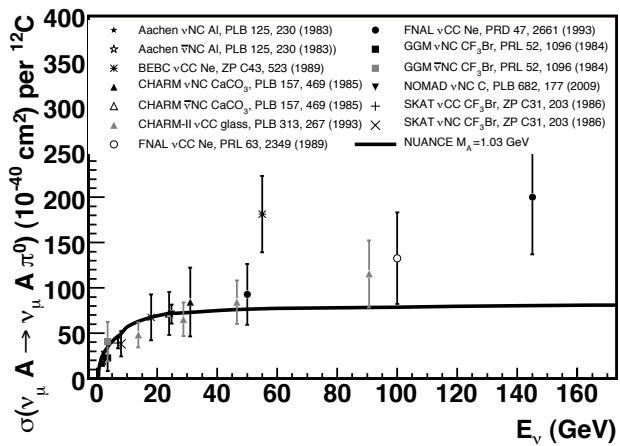}
\end{center}
\vspace{-0.2in}
\caption{ Measurements of absolute coherent pion production cross sections from a variety of nuclear
         targets and samples, as indicated in the legend. Both NC and CC
         data are displayed on the same plot after rescaling the CC data
         using the prediction that 
         $\sigma_{NC}=\frac{1}{2}\sigma_{CC}$~\cite{rein-sehgal-coherent}.
         In addition, data from various targets have been corrected to
         carbon cross sections assuming $A^{1/3}$ 
         scaling~\cite{rein-sehgal-coherent}.  
         Also shown is the prediction from~\cite{nuance}.}
\label{fig:coherent-pi}
\end{figure}

\noindent
While the cross sections for these processes are predicted to be comparatively
small, coherent pion production has been observed
across a broad energy range in both NC and CC interactions of neutrinos and 
antineutrinos. Figure~\ref{fig:coherent-pi} shows a collection of existing 
measurements of coherent pion production cross sections for a variety of 
nuclei. A valuable compilation of the same is also available in~\cite{vilain}. Most of these historical measurements were performed 
at higher energies ($E_\nu>2$ GeV). Table~\ref{tab:new-coherent-pion} provides
a listing of more recent measurements of coherent pion production, most in
the form of cross section ratios that were measured at low energy 
($E_\nu<\sim2$ GeV). Experiments measuring coherent pion production at these 
very low neutrino energies have typically observed less coherent pion 
production than predicted by models which well-describe the high energy data. 
In addition, the production of CC coherent pion events at low energy has 
been seemingly absent from much of the experimental 
data~\cite{k2k-coh,sb-coh-ccpip}, although refined searches have indicated 
some evidence for their existence~\cite{sb-new-coh}. \\

To date, it has been a challenge to develop a single description that can 
successfully describe existing coherent pion production measurements across 
all energies. The most common theoretical approach for describing coherent 
pion production is typically based on Adler's PCAC theorem~\cite{adler-pcac} 
which relates neutrino-induced coherent pion production to pion-nucleus 
elastic scattering in the limit $Q^2=0$. A nuclear form factor is then 
invoked to extrapolate to non-zero values of $Q^2$. Such PCAC-based 
models~\cite{rein-sehgal-coherent} have existed for many years and have been 
rather successful in describing coherent pion production at high energy
(the prediction shown in Figure~\ref{fig:coherent-pi} is such a model). 
With the accumulation of increasingly large amounts of low energy neutrino 
data, revised approaches have been applied to describe the reduced
level of coherent pion production observed by low energy experiments. 
Two such approaches have been developed. The first class of models are again
based on PCAC~\cite{rein-sehgal-coherent,pcac-coh-models,pcac-coh-models-1,pcac-coh-models-2,pcac-coh-models-3,pcac-coh-models-4,pcac-coh-models-5,pcac-coh-models-6}. The other class 
are microscopic models involving $\Delta$ resonance 
production~\cite{delta-coh-models,delta-coh-models-1,delta-coh-models-2,delta-coh-models-3,delta-coh-models-4,Martini:2011wp,delta-coh-models-6,delta-coh-models-7}. Because this latter class involves 
$\Delta$ formation, their validity is limited to the low energy region. An excellent review of the current experimental and theoretical situation is available in~\cite{Alvarez-Ruso11}. The study and prediction of coherent pion production is important as it
provides another source of potential background for neutrino oscillation
experiments.

\begin{table*}[h]
\begin{tabular}{|l|c|c|r|}
\hline
Experiment   & Target  & $E_\nu$ & Measurement \\
\hline\hline
K2K       & $C_8H_8$  & 1.3 GeV  & $\sigma$(CC coherent $\pi^+$/CC) \\
SciBooNE  & $C_8H_8$  & 1.1 GeV  & $\sigma$(CC coherent $\pi^+$/CC) \\
SciBooNE  & $C_8H_8$  & 1.1, 2.2 GeV & $\sigma$(CC coherent $\pi^+$/
                     NC coherent $\pi^0$) \\
\hline
MiniBooNE & $CH_2$    & 1.1 GeV  & $\sigma$(NC coherent $\pi^0$/NC $\pi^0$) \\
NOMAD     & $C$-based & 24.8 GeV & $\sigma$(NC coherent $\pi^0$) \\
SciBooNE  & $C_8H_8$  & 1.1 GeV  & $\sigma$(NC coherent $\pi^0$/CC) \\
\hline
\end{tabular}
\caption{ Modern measurements of CC (top) and NC (bottom) coherent pion production by neutrinos, 
        at the time of this writing. Measurements are listed from K2K~\cite{k2k-coh}, MiniBooNE~\cite{mb-coh}, NOMAD~\cite{nomad-coh}, and SciBooNE~\cite{sb-coh-ccpip,sb-ncpi0,sb-coh-ncpi0}.  All are ratio measurements performed
        at low energy, with the exception of the absolute coherent pion 
        production cross section measurement recently reported by 
        NOMAD. Higher energy coherent pion production results have also been recently reported by the MINOS experiment~\cite{Cherdack11}.}
\label{tab:new-coherent-pion}
\end{table*}

\subsection{Multi-Pion Production}
\label{sec:multi-pi}

\begin{figure}[h]
\begin{center}
\includegraphics[scale=0.45]{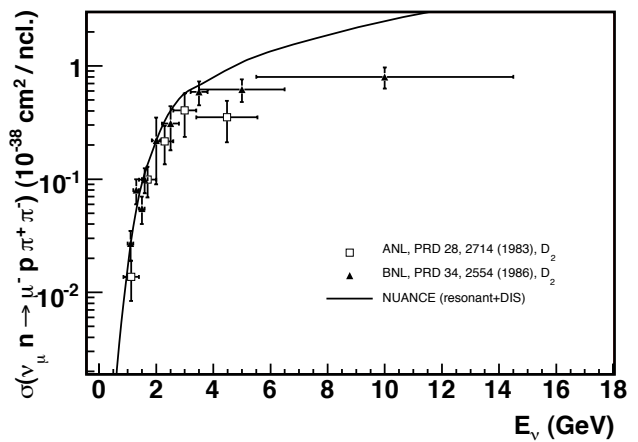}
\end{center}
\vspace{-0.2in}
\caption{ Existing measurements of the 
         $\numu \, n \rightarrow \mu^- \, p \, \pi^+ \, \pi^-$
         scattering cross section as a function of neutrino energy.
         Also shown is the prediction from 
         Reference~\cite{nuance} including both resonant and DIS contributions
         to this reaction channel.}
\label{fig:two-pi-1}
\end{figure}

\begin{figure}[h]
\begin{center}
\includegraphics[scale=0.45]{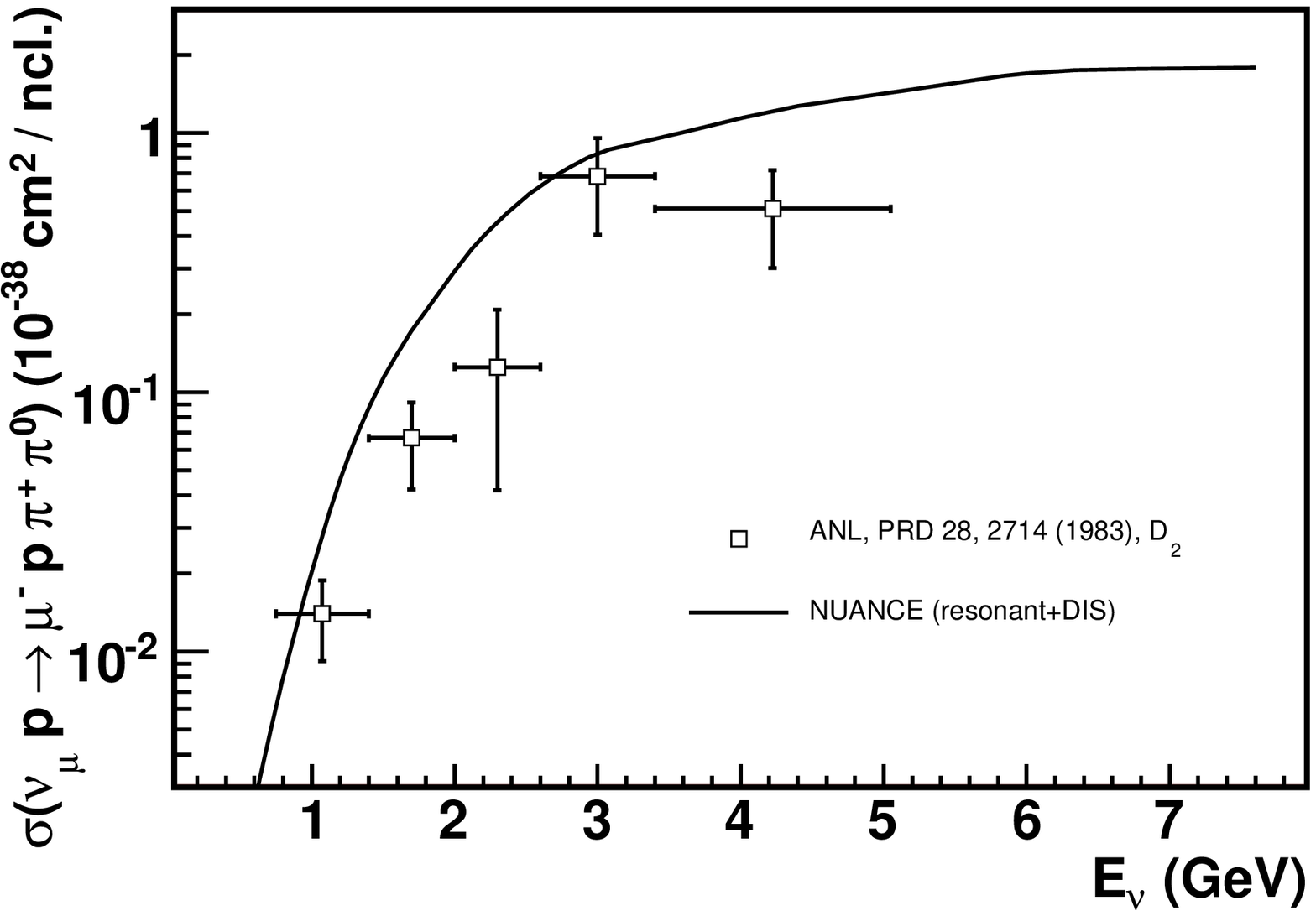}
\end{center}
\vspace{-0.2in}
\caption{ Existing measurements of the 
         $\numu \, p \rightarrow \mu^- \, p \, \pi^+ \, \pi^0$
         scattering cross section as a function of neutrino energy.
         Also shown is the prediction from 
         Reference~\cite{nuance} including both resonant and DIS contributions
         to this reaction channel.}
\label{fig:two-pi-2}
\end{figure}

\begin{figure}[h]
\begin{center}
\includegraphics[scale=0.45]{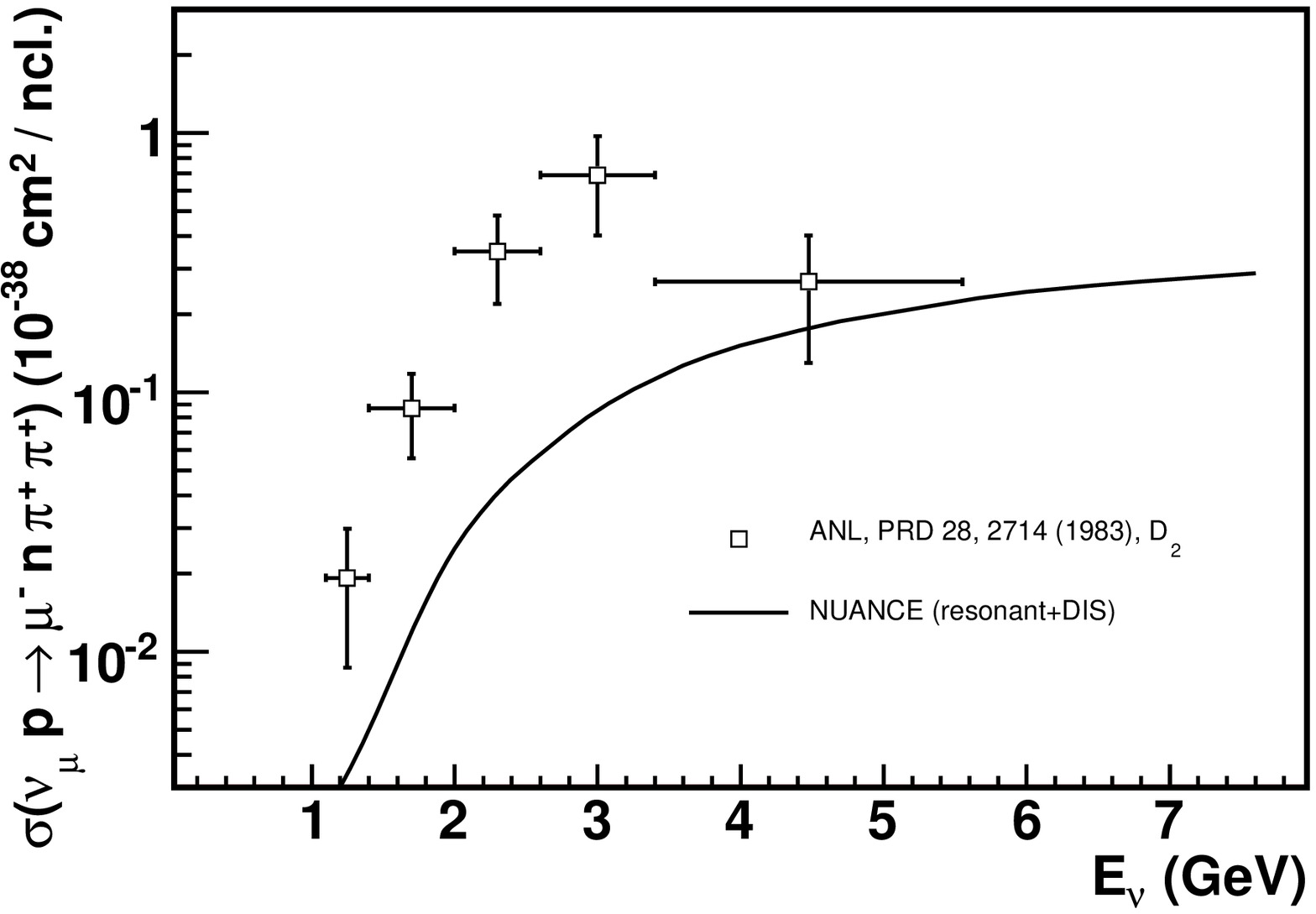}
\end{center}
\vspace{-0.2in}
\caption{ Existing measurements of the 
         $\numu \, p \rightarrow \mu^- \, n \, \pi^+ \, \pi^+$
         scattering cross section as a function of neutrino energy.
         Also shown is the prediction from 
         Reference~\cite{nuance} including both resonant and DIS contributions
         to this reaction channel.}
\label{fig:two-pi-3}
\end{figure}

As mentioned earlier, the baryonic resonances created in neutrino-nucleon 
interactions can potentially decay to multi-pion final states. Other inelastic 
processes, such as deep inelastic scattering, can also contribute a 
copious source of multi-pion final states depending on the neutrino energy. 
Figures~\ref{fig:two-pi-1}-~\ref{fig:two-pi-3} show existing measurements 
of neutrino-induced di-pion production cross sections compared to an example 
prediction including contributions from both resonant and deep inelastic 
scattering mechanisms. Due to the inherent complexity of reconstructing 
multiple pions final states, there are not many existing experimental 
measurements of this process. All of the existing measurements have been 
performed strictly using deuterium-filled bubble chambers. Improved 
measurements will be important because they test our understanding of 
the transition region and will provide a constraint on potential backgrounds 
for neutrino oscillation experiments operating in higher energy beams. 

\subsection{Kaon Production}
\label{sec:kaon}

Neutrino interactions in this energy range can also produce final states 
involving strange quarks. Some of the contributing strange production 
channels at intermediate energies include the following processes:

\begin{eqnarray}
  && \mathrm{CC:} \hspace{1.0in}  \mathrm{NC:} \nonumber \\
  \numu \, n &\rightarrow& \mu^- \, K^+ \, \Lambda^0 \hspace{0.2in}
  \numu \, p \rightarrow \numu \, K^+ \Lambda^0 
\nonumber \\
  \numu \, p &\rightarrow& \mu^- \, K^+ \, p \hspace{0.27in} 
  \numu \, n \rightarrow \numu \, K^0 \, \Lambda^0
\nonumber \\
  \numu \, n &\rightarrow& \mu^- \, K^0 \, p \hspace{0.29in}
  \numu \, p \rightarrow \numu \, K^+ \, \Sigma^0
\nonumber \\\
  \numu \, n &\rightarrow& \mu^- \, K^+ \, n \hspace{0.27in}
  \numu \, p \rightarrow \numu \, K^0 \, \Sigma^+
\nonumber \\
  \numu \, p &\rightarrow& \mu^- \, K^+ \, \Sigma^+ \hspace{0.17in}
  \numu \, n \rightarrow \numu \, K^0 \, \Sigma^0 
\nonumber \\
  \numu \, n &\rightarrow& \mu^- \, K^+ \, \Sigma^0 \hspace{0.18in}
  \numu \, n \rightarrow \numu \, K^+ \, \Sigma^-
\nonumber \\
  \numu \, n &\rightarrow& \mu^- \, K^0 \, \Sigma^+ \hspace{0.17in}
  \numu \, n \rightarrow \numu \, K^- \, \Sigma^+
\end{eqnarray}

\noindent
These reactions typically have small cross sections due in part to the kaon 
mass and because the kaon channels are not enhanced by any dominant resonance.
Measuring neutrino-induced kaon production is of interest primarily
as a source of potential background for proton decay searches. Proton decay 
modes containing a final state kaon, $p \rightarrow K^+ \nu$, have large 
branching ratios in many SUSY GUT models. Because there is a non-zero 
probability that an atmospheric neutrino interaction can mimic such a proton 
decay signature, estimating these background rates has become an increasingly
important component to such searches. \\

\begin{figure}[h]
\begin{center}
\includegraphics[scale=0.45]{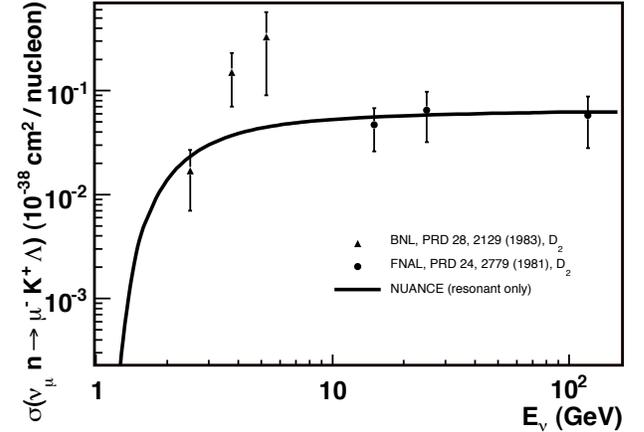}
\end{center}
\vspace{-0.2in}
\caption{ Measurements of the associated production cross section, 
        $\numu \, n \rightarrow \mu^- \, K^+ \Lambda^0$, as a function
        of neutrino energy. Also shown is the prediction from 
        Reference~\cite{nuance} which includes both resonant and DIS
        contributions.}
\label{fig:kaon}
\end{figure}

Figure~\ref{fig:kaon} shows the only two experiments which have published
cross sections on the dominant associated production channel, 
$\numu \, n \rightarrow \mu^- \, K^+ \Lambda^0$. Both bubble chamber 
measurements were performed on a deuterium target and are based on less than 
30 events combined. Many other measurements of strange particle production 
yields have been performed throughout the years, most using bubble 
chambers~\cite{kaon-bc,kaon-bc-1,kaon-bc-2,kaon-bc-3,kaon-bc-4,kaon-bc-5,kaon-bc-6,kaon-bc-7,kaon-bc-8,kaon-bc-9,kaon-bc-10,kaon-bc-11,kaon-bc-12,kaon-bc-13,kaon-bc-14,kaon-bc-15,kaon-bc-16,Bell:1978qu,Barish:1974ye}.
More recently, NOMAD has reported NC and CC strange particle production 
yields and multiplicities for a variety of reaction 
kinematics~\cite{nomad-strange,nomad-strange-1,nomad-strange-2}. \\

As far as theoretical calculations go, predictions for these neutrino-induced kaon production processes have existed 
for several decades~\cite{single-k-models,single-k-models-1,single-k-models-2,single-k-models-3,single-k-models-4}, although there have been several 
revised calculations in recent years~\cite{new-k-models,new-k-models-1, Alam12}.

\subsection{Outlook}

In summary, neutrino scattering at intermediate energies is notoriously 
complex and the level to which these contributing processes have been 
studied remains incomplete~\cite{Ruso12,Benhar12}. Improved experimental measurements and 
theoretical calculations will be especially important for reducing systematics
in future precision neutrino oscillation experiments. Luckily, such studies 
are already underway making use of new intense accelerator-based sources 
of neutrinos. However, for such updated cross section measurements to be 
robust, they must be accompanied by an equally precise knowledge of the 
incoming neutrino flux. Improved hadro-production measurements are key to 
providing the level of precision necessary. In addition, further scrutiny of 
nuclear effects in intermediate energy neutrino and antineutrino interactions 
is absolutely essential. Analysis of data from the MINER$\nu$A experiment will
soon enable the first detailed look at nuclear effects in neutrino 
interactions. Together, theoretical advances and new data taken on a variety 
of nuclear targets from the ArgoNeuT, K2K, MicroBooNE, MINER$\nu$A, MiniBooNE,
MINOS, NOMAD, NOvA, and SciBooNE experiments should provide both a necessary
and broad foundation going into the future.  In order to make the most progress in our understanding in this energy regime, experiments should strive towards model-independent measurements of differential cross sections.


\section{High Energy Cross Sections: $E_\nu \sim 20-500$ GeV}
\label{chapter:high-energy}

Up to now, we have largely discussed neutrino scattering from composite
entities such as nucleons or nuclei. Given enough energy, the neutrino can 
actually begin to resolve the internal structure of the target. In the most
common high energy interaction, the neutrino can scatter off an individual 
quark inside the nucleon, a process called deep inelastic scattering (DIS). An 
excellent review of this subject has been previously published in this
journal~\cite{dis-rmp}, therefore we will provide only a brief summary of the 
DIS cross section, relevant kinematics, and most recent experimental 
measurements here.

\subsection{Deep Inelastic Scattering}
\label{sec:dis}

\begin{figure}[h]
\begin{center}
\includegraphics[scale=0.35]{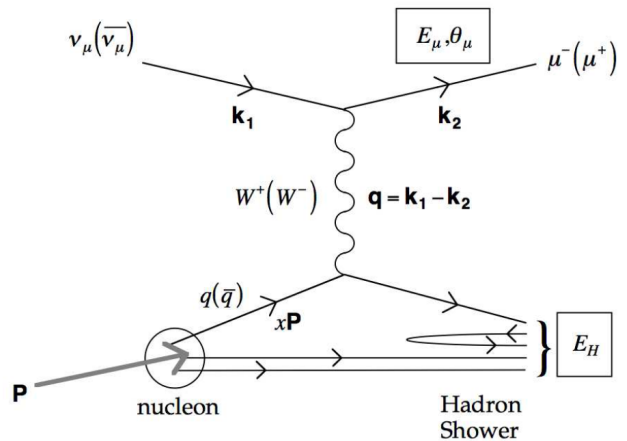}
\end{center}
\vspace{-0.2in}
\caption{ Feynman diagram for a CC neutrino DIS process. In the case of NC DIS, the outgoing
         lepton is instead a neutrino and the exchange particle is a Z 
         boson. Diagram is reproduced from~\cite{dis-rmp}.}
\label{fig:dis-picture}
\end{figure}

Neutrino deep inelastic scattering has long been used to validate the Standard
Model and probe nucleon structure. Over the years, experiments have measured 
cross sections, electroweak parameters, coupling constants, nucleon structure 
functions, and scaling variables using such processes. In deep inelastic 
scattering (Figure~\ref{fig:dis-picture}), the neutrino scatters off a quark 
in the nucleon via the exchange of a virtual $W$ or $Z$ boson producing a 
lepton and a hadronic system in the final state~\footnote{Quarks cannot be 
individually detected; they quickly recombine and thus appear as a hadronic 
shower.}. Both CC and NC processes are possible:

\begin{eqnarray}
  \numu \, N &\rightarrow& \mu^- \, X \hspace{0.3in} 
  \numubar \, N \rightarrow \mu^+ \, X \\
  \numu \, N &\rightarrow& \numu \, X \hspace{0.3in} 
  \numubar \, N \rightarrow \numubar \, X 
\end{eqnarray}

\noindent
Here, we restrict ourselves to the case of $\numu$ scattering, as an example,
though $\nu_e$ and $\nu_\tau$ DIS interactions are also possible. 

Following the formalism introduced in Section~\ref{sec:NeutrinoLeptonScat}, 
DIS processes can be completely described in terms of three dimensionless 
kinematic invariants.  The first two, the inelasticity $(y)$ and the 
4-momentum transfer ($Q^2 = -q^2$) have already been defined.  We now define 
the Bjorken scaling variable, $x$:
\begin{equation}
x = \frac{Q^2}{2 p_e\cdot q}  {\rm~~~~(Bjorken~scaling~variable)}
\end{equation}

The Bjorken scaling variable plays a prominent role in deep 
inelastic neutrino scattering, where the target can carry a portion of the 
incoming energy-momentum of the struck nucleus. \\

On a practical level, these Lorentz-invariant parameters cannot be readily 
determined from 4-vectors, but they can be reconstructed using readily measured observables in a given experiment:

\begin{eqnarray}
    x & =\frac{Q^2}{2M\nu} = \frac{Q^2}{2ME{_\nu}y} \\
    y & = E_{had}/E_\nu \\
    Q^2 & =  -m_\mu^2 + 2E_\nu(E_\mu-p_\mu \cos\theta_\mu)
\end{eqnarray}

\noindent where $E_\nu$ is the incident neutrino energy, $M_N$ is the nucleon mass,
$\nu=E_{had}$ is the energy of the hadronic system, and $E_\mu$, $p_\mu$, 
and $\cos\theta_\mu$ are the energy, momentum, and scattering angle 
of the outgoing muon in the laboratory frame. In the case of NC scattering, 
the outgoing neutrino is not reconstructed. Thus, experimentally, all of 
the event information must be inferred from the hadronic shower in that 
case.  \\

Using these variables, the inclusive cross section for DIS scattering 
of neutrinos and antineutrinos can then be written as:

\begin{widetext}
\begin{eqnarray}
 \label{eqn:xsectcc}
  \frac{d^2\sigma^{\nu,\,\nubar}}{dx\:dy} = 
        \frac{G_F^2ME_\nu}{\pi\,(1+Q^2/M_{W,Z}^2)^2} \, 
        \left[\begin{array}{c} \frac{y^2}{2} 2x F_1(x,Q^2) + 
             \left(1-y-\frac{Mxy}{2E}\right) F_2(x,Q^2)  \\
             \pm y\left(1-\frac{y}{2}\right) xF_3(x,Q^2) \end{array}\right] 
\end{eqnarray}
\end{widetext}

\noindent
where $G_F$ is again the Fermi weak coupling constant, $M_{W,Z}$ is the mass of
the $W^{\pm}$ ($Z^0$ boson) in the case of CC (NC) scattering, 
and the +(--) sign in the last term refers to neutrino (antineutrino) 
interactions. In the above expression, $F_i(x,Q^2)$ are the dimensionless
nucleon structure functions that encompass the underlying structure of 
the target. For electron scattering, there are two structure functions 
while for neutrino scattering there is additionally a third structure 
function, $xF_3(x,Q^2)$, which represents the V,A interference term. \\

Assuming the quark parton model, in which the nucleon consists of partons
(quarks and gluons), $F_i(x,Q^2)$ can be expressed in terms of the 
quark composition of the target. They depend on the target and 
type of scattering interaction and are functions of $x$ and $Q^2$. In 
the simplest case, the nucleon structure functions can then be expressed 
as the sum of the probabilities:

\begin{eqnarray}
   F_2(x,Q^2) = 2 \sum_{i=u,d,...} (xq(x,Q^2)+x\overline{q}(x,Q^2)) \\
   xF_3(x,Q^2) = 2 \sum_{i=u,d,...} (xq(x,Q^2)-x\overline{q}(x,Q^2))
\end{eqnarray}

\noindent
where the sum is over all quark species. The struck quark carries a fraction, 
$x$, of the nucleon's momentum, such that $xq$ ($x\overline{q}$) is 
the probability of finding the quark (antiquark) with a given momentum 
fraction. These probabilities are known as parton distribution functions 
or PDFs, for short. In this way, $F_2(x,Q^2)$ measures the sum of the quark
and antiquark PDFs in the nucleon, while $xF_3(x,Q^2)$ measures their 
difference and is therefore sensitive to the valence quark PDFs.  The 
third structure function, $2xF_1(x,Q^2)$, is commonly related to $F_2(x,Q^2)$
via a longitudinal structure function, $R_L(x,Q^2)$:

\begin{eqnarray}
   F_2(x,Q^2) = \frac{1+R_L(x,Q^2)}{1+4M^2x^2/Q^2} \, 2xF_1(x,Q^2)
\end{eqnarray}

\noindent
where $R_L(x,Q^2)$ is the ratio of cross sections for scattering off
longitudinally and transversely polarized exchange bosons. \\

Measurement of these structure functions has been the focus of many
charged lepton and neutrino DIS experiments, which together have 
probed $F_2(x,Q^2)$, $R_L(x,Q^2)$, and $xF_3$ (in the case of neutrino
scattering) over a wide range of $x$ and $Q^2$ values~\cite{ga}. Neutrino 
scattering is unique, however, in that it measures the valence quark 
distributions through measurement of $xF_3$ and the strange quark 
distribution through detection of neutrino-induced dimuon production. 
These provide important constraints that cannot be obtained from either
electron or muon scattering experiments. \\

While Equation~\ref{eqn:xsectcc} provides a tidy picture of neutrino
DIS, additional effects must be included in any realistic description 
of these processes. The inclusion of 
lepton masses~\cite{Albright:1974ts,Kretzer:2002fr}, 
higher order QCD processes~\cite{Moch:1999eb,Dobrescu:2003ta,McFarland:2003jw},
nuclear effects, radiative corrections~\cite{DeRujula:1979jj,Sirlin:1981yz,radcorr-1,radcorr-2,radcorr-3}, target mass effects~\cite{tmass}, heavy quark production~\cite{charm,charm-1,Gottschalk:1980rv}, and non-perturbative higher twist effects~\cite{Buras:1979yt} further modify the scattering kinematics and cross sections. In general, these contributions are typically well-known and do not add large uncertainties to the predicted cross sections.\\

\begin{figure*}[h]
\begin{center}
\includegraphics[scale=0.7]{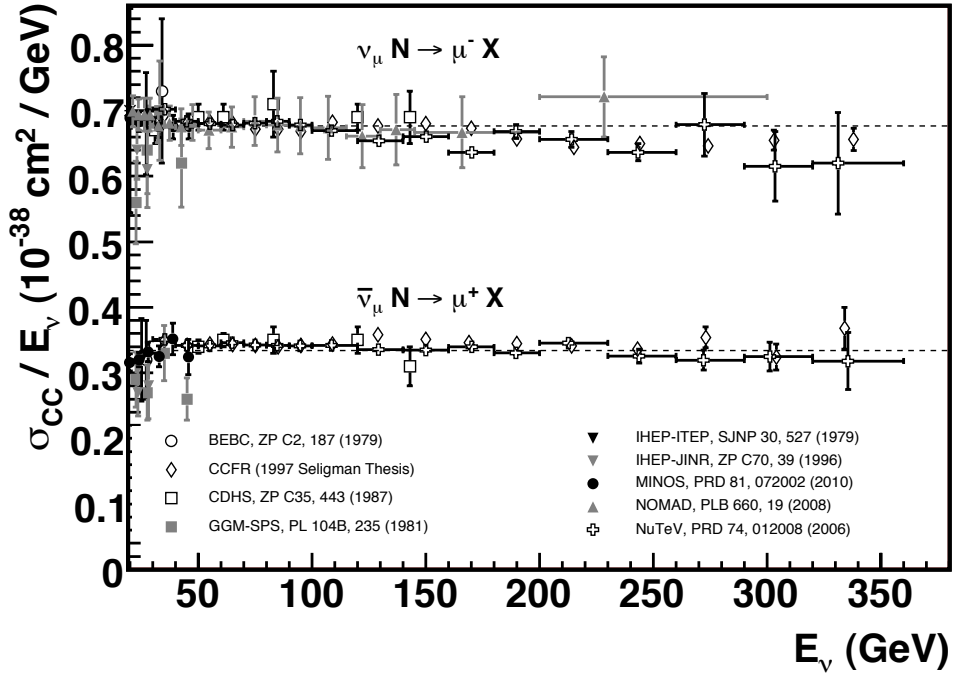}
\end{center}
\vspace{-0.2in}
\caption{ Measurements of the inclusive neutrino and antineutrino CC cross sections 
        ($\numu \, N \rightarrow \mu^- \, X$ and $\numubar \, N \rightarrow 
        \mu^+ \, X$) divided by neutrino energy plotted as a function of 
        neutrino energy. Here, $N$ refers to an isoscalar nucleon within
        the target. The dotted lines indicate the world-averaged cross 
        sections, $\sigma^{\nu}/E_\nu=(0.677\pm 0.014)\times 10^{-38}$ 
        cm$^2$/GeV and $\sigma^{\nubar}/E_\nu=(0.334\pm 0.008)\times 10^{-38}$
        cm$^2$/GeV, for neutrinos and antineutrinos, respectively~\cite{ga}.
        For an extension to lower neutrino energies, see the complete 
        compilation in Figure~\ref{fig:cc-inclusive}.}
\label{fig:dis}
\end{figure*}

Having completed a very brief description of DIS, let us next turn to
some of the experimental measurements. Table~\ref{table:dis-modern} lists 
the most recent experiments that have probed such high energy neutrino
scattering. To isolate DIS events, neutrino experiments typically apply 
kinematic cuts to remove quasi-elastic scattering (Section~\ref{sec:qe}) 
and resonance-mediated (Section~\ref{sec:single-pi}) contributions from 
their data. Using high statistics samples of DIS events, these experiments 
have provided measurements of the weak mixing angle, $\sin^2\theta_W$, 
from NC DIS samples as well as measurements of structure functions, inclusive
cross sections, and double differential cross sections for CC single 
muon and dimuon production. Figure~\ref{fig:dis} specifically shows 
measurements of the inclusive CC cross section from the NOMAD, NuTeV, 
and MINOS experiments compared to historical data. As can be seen, 
the CC cross section is measured to a few percent in this region. 
A linear dependence of the cross section on neutrino energy is also 
exhibited at these energies, a confirmation of the quark parton 
model predictions.\\

In addition to such inclusive measurements as a function of neutrino energy, experiments have reported differential cross sections, for example, most recently~\cite{Tzanov:2005kr}. Also, over the years, exclusive processes such as opposite-sign dimuon production have been measured~\cite{Dore:2011qe}. Such dimuon investigations have been performed in counter experiments like CCFR~\cite{Bazarko:1994tt,Rabinowitz:1993xx,Foudas:1989rk}, CDHS~\cite{Abramowicz:1982zr}, CHARM II~\cite{Vilain:1998uw}, E616~\cite{Lang:1986at}, HPWF~\cite{Benvenuti:1978gy,Aubert:1974zz}, NOMAD~\cite{Astier:2000us} and NuTeV~\cite{Mason:2007zz,Goncharov:2001qe}, in bubble chambers like BEBC~cite{Gerbier:1985rj}, FNAL~\cite{Baker:1985rx,Ballagh:1981yh} and Gargamelle~\cite{Haatuft:1983it} as well as in nuclear emulsion detectors such as E531~\cite{Ushida:1982ty} and CHORUS~\cite{KayisTopaksu:2011mx,Topaksu:2008xp,KayisTopaksu:2005je,Onengut:2005sy,Onengut:2004vd}. This latter class of measurements is particularly important for constraining the strange and anti-strange quark content of the nucleon and their momentum dependence.\\

In the near future, high statistics measurements of neutrino and antineutrino
DIS are expected from the MINER$\nu$A experiment~\cite{minerva}. With 
multiple nuclear targets, MINER$\nu$A will also be able to complete the 
first detailed examination of nuclear effects in neutrino DIS.

\begin{table*}[h]
\begin{tabular}{|l|c|c|c|c|l|}
\hline
Experiment   & Target  & $E_\nu$ (GeV) &  Statistics &  Year & Results \\
\hline\hline
CHORUS & Pb & 10-200 & $8.7\times10^5 \: \nu$, 
  $1.5\times10^5 \: \nubar$ & 1995-1998 & $F_2(x,Q^2)$, $xF_3(x,Q^2)$ \\
MINOS & Fe & 3-50 & $19.4\times10^5 \: \nu$, 
  $1.6\times10^5 \: \nubar$ & 2005--present & $\sigma(E_\nu)$ \\
NOMAD  & C & 3-300 & $10.4\times10^5 \: \nu$ & 1995-1998 & 
  $\sigma(E_\nu)$ \\
NuTeV & Fe & 30-360 & 
  $8.6\times10^5 \: \nu$, $2.3\times10^5 \: \nubar$ & 1996-1997 & 
  $F_2(x,Q^2)$, $xF_3(x,Q^2)$, $\sigma(E_\nu)$, 
  $\frac{d^2\sigma}{dxdy}$, $\sin^2\theta_W$  \\
\hline
\end{tabular}
\caption{ Attributes of neutrino experiments that have recently studied DIS, including CHORUS~\cite{chorus-dis,chorus-dimuon}, MINOS~\cite{minos-dis}, NOMAD~\cite{nomad-dis}, and NuTeV~\cite{nutev-dis,nutev-stw,nutev-dimuon}.}
\label{table:dis-modern}
\end{table*}

\section{Ultra High Energy Neutrinos: 0.5 TeV - 1 EeV}
\label{sec:UHE}

In reaching the ultra high energy scale, we find ourselves, remarkably, back to the beginning of our journey at extremely low energies.  Neutrinos at this energy scale have yet to manifest themselves as confirmed observations, though our present technology is remarkably close to dispelling that fact.  To date, the highest energy neutrino recorded is several hundred TeV~\cite{bib:Tyce_pc}.  However, experimentalists and theorists have their aspirations set much higher, to energies above $10^{15}$ eV.  On the theoretical side, this opens the door for what could be called ``neutrino astrophysics".  A variety of astrophysical objects and mechanisms become accessible at these energies, providing information that is complementary to that already obtained from electromagnetic or hadronic observations.  

\begin{figure}[htbp]
\begin{center}
\includegraphics[width=1.00\columnwidth,keepaspectratio=true]{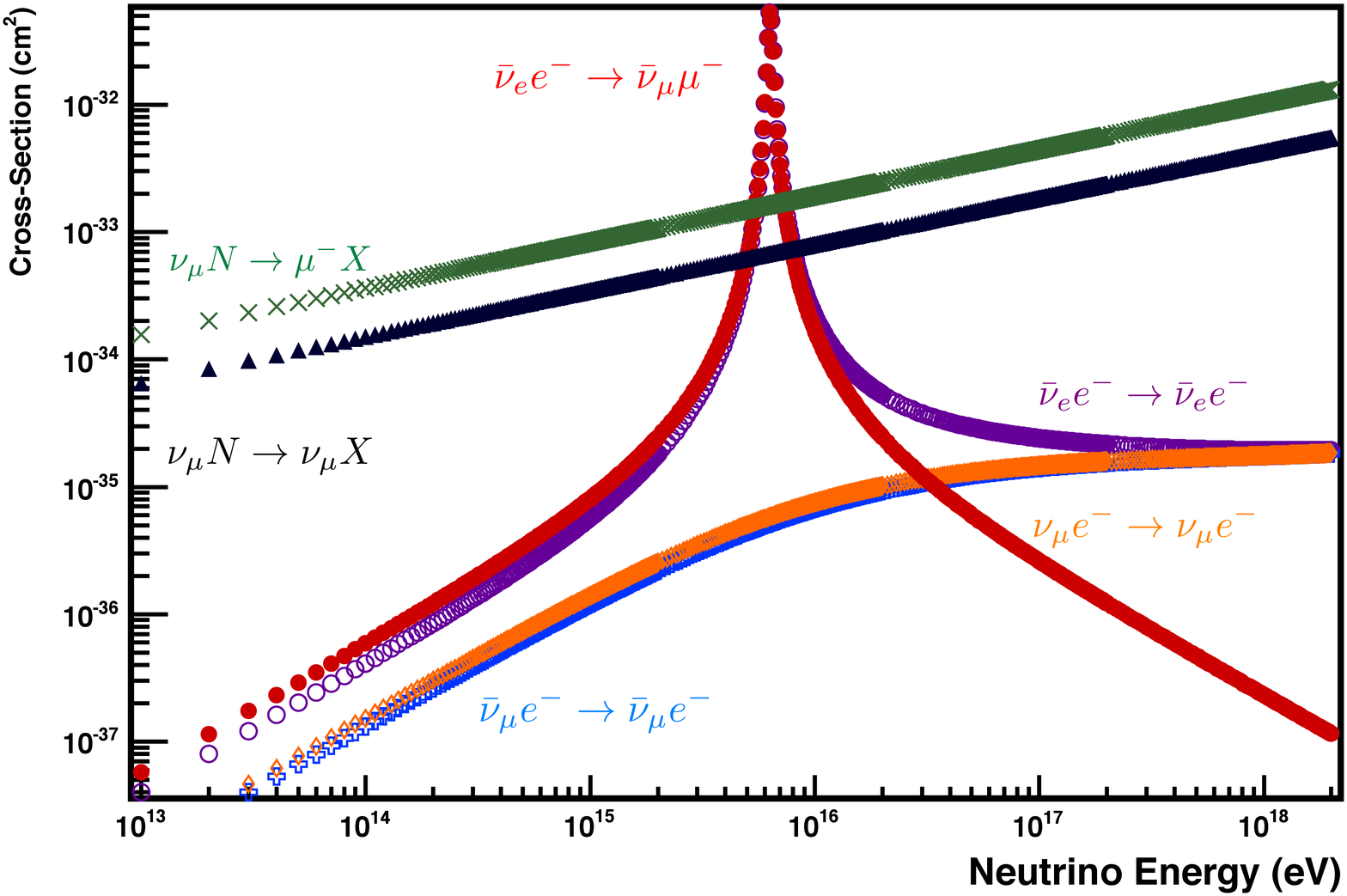} 
\caption{Neutrino electron and nucleon scattering in the ultra high energy regime ($E_\nu \ge 10^4$ GeV).  Shown are the electron interactions $\bar{\nu}_\mu e^- \rightarrow \bar{\nu}_\mu e^-$ (crosses, blue), $\nu_\mu e^- \rightarrow \nu_\mu e^-$ (diamonds, orange), $\bar{\nu}_e e^- \rightarrow \bar{\nu}_e e^-$ (hollow circles, violet), $\bar{\nu}_e e^- \rightarrow \bar{\nu}_\mu e^-$ (filled circles, red), and the nucleon charged current (cross markers, green) and neutral current (filled triangles, black) interactions.  The leptonic $W$ resonance channel is clearly evident~\cite{Gandhi199681,Butkevich88}.}
\label{fig:uhe}
\end{center}
\end{figure}

In response to the call, the experimental community has forged ahead with a number of observational programs and techniques geared toward the observation of ultra high energy neutrinos from astrophysical sources.  The range of these techniques include detectors scanning for ultra high energy cosmic neutrino induced events in large volumes of water (Baikal~\cite{bib:Baikal-1,bib:Baikal-2}, Antares~\cite{bib:Antares}), ice (AMANDA~\cite{bib:AMANDA}, IceCube~\cite{bib:IceCube}, RICE~\cite{bib:RICE}, FORTE~\cite{bib:FORTE}, ANITA~\cite{bib:ANITA}), the Earth's atmosphere (Pierre Auger~\cite{bib:PierreAuger}, HiRes~\cite{bib:HiRes}) and the lunar regolith (GLUE~\cite{bib:GLUE}). Even more future programs are in the planning stages.  As such, the knowledge of neutrino cross-section in this high energy region is becoming ever-increasing in importance. Once first detection is firmly established, the emphasis is likely to shift toward obtaining more detailed information about the observed astrophysical objects, and thus the neutrino fluxes will need to be examined in much greater detail.

The neutrino cross-sections in this energy range~\footnote{Typically, the high energy region is demarcated by $E_\nu \ge 10^6$ GeV.} are essentially extensions of the high energy parton model that was discussed in Section~\ref{chapter:high-energy}.  However, at these energies, the propagation term from interaction vertex is no longer dominated by the W-Z boson mass.  As a result, the cross-section no longer grows linearly with neutrino energy. The propagator term in fact suppresses the cross-sections for energies above 10 TeV.  Likewise, the $(1-y)^2$ suppression that typically allows distinction between neutrino and anti-neutrino interactions is much less pronounced, making the two cross-section ($\nu N$ and $\bar{\nu}N$) nearly identical.

For a rough estimate of the neutrino cross-section at these high energies ($10^{16} {\rm~eV} \le E_\nu \le 10^{21}$ eV),  the following power law dependence provides a reasonable approximation~\cite{Gandhi199681}:

\begin{widetext}
\begin{eqnarray}
\sigma_{\nu N}^{CC} = 5.53 \times 10^{-36} {\rm~cm}^2 (\frac{E_\nu}{\rm 1~GeV})^\alpha,\\
\sigma_{\nu N}^{NC} = 2.31 \times 10^{-36} {\rm~cm}^2 (\frac{E_\nu}{\rm 1~GeV})^\alpha,
\end{eqnarray}
\end{widetext}

\noindent where $\alpha \simeq 0.363$.   

There is one peculiar oddity that is worth highlighting for neutrino cross-sections at such high energies.  Neutrino-electron scattering is usually sub-dominant to any neutrino-nucleus interaction because of its small target mass.  There is one notable exception, however; when the neutrino undergoes a resonant enhancement from the formation of an intermediate $W$-boson in $\bar{\nu}_e e^-$ interactions.  This resonance formation takes place at $E_{\rm res} = M_W^2/2m_e = 6.3$ PeV and is by far more prominent than any $\nu N$ interaction up to $10^{21}$ eV (see Figure~\ref{fig:uhe}).  The mechanism was first suggested by Glashow in 1960 as a means to directly detect the $W$ boson~\cite{bib:Glashow1960}.  The cross-section was later generalized by Berezinsky and Gazizov~\cite{bib:Berezinsky} to other possible channels.

\begin{widetext}
\begin{equation}
\frac{d\sigma(\bar{\nu}_e e^- \rightarrow \bar{\nu}_e e^-)}{dy} = \frac{2G_F^2 m_e E_\nu}{\pi} \left[ \frac{g_R^2}{(1+2 m_e E_\nu y / M_Z^2)^2} + |\frac{g_L}{1+2 m_e E_\nu y / M_Z^2} + \frac{1}{1-2m_e E_\nu / M_W^2 + i \Gamma_W/M_W}|^2 \right]
\end{equation}
\end{widetext}

\noindent where $g_{L,R}$ are the left and right handed fermion couplings, $M_W$ is the W-boson mass and $\Gamma_W$ is the W-decay width ($\sim 2.08$ GeV).  This resonance occurs only for s-channel processes mediated by W-exchange:

\begin{widetext}
\begin{eqnarray*}
\frac{d\sigma (\nu_l e \rightarrow \nu_l e)}{d y} = \frac{2m_e G_F^2 E_\nu}{\pi} \frac{1}{(1+2m_eE_\nu y / M_Z^2)^2} \left(g_L^2+g_R^2(1-y)^2 \right),\\
\frac{d\sigma (\bar{\nu}_l e \rightarrow \bar{\nu}_l e)}{d y} = \frac{2m_e G_F^2 E_\nu}{\pi} \frac{1}{(1+2m_eE_\nu y / M_Z^2)^2} \left(g_R^2+g_L^2(1-y)^2 \right).
\end{eqnarray*}
\end{widetext}

When compared to that of neutrino-nucleon scattering or even the non-resonant neutrino-lepton scattering, the $\bar{\nu}_e$ scattering dominates.  Such high cross-sections can often cause the Earth to be opaque to neutrinos in certain energy regimes and depart substantially from Standard Model predictions if new physics is present.~\cite{Gandhi199681}.

\subsection{Uncertainties and Projections}

For a more accurate prediction of the cross-section, a well-formulated model of the relevant quark structure functions is needed.  This predictive power is especially important in the search for new physics.  At such high energies, the neutrino cross-section can depart substantially from the Standard Model prediction if new physics is at play.  Study of such high energy neutrinos can be a possible probe into new physics.  

Direct neutrino scattering measurements at such extreme energies are, of course, unavailable.  Therefore, predictions rely heavily on the existing knowledge of parton distribution functions and, as the reader can imagine, extrapolation can introduce substantial uncertainties to these predictions. The best constraints on the relevant parton distribution functions stem from data collected from high energy $ep$ scattering experiments such as HERA~\cite{bib:HERA}.  The challenge rests on the ability to fit existing data to as low values of $x$ as possible.  At high energies, the propagator term limits the maximum $Q^2$ to the $M_{W,Z}$ mass.  The relevant range for $x$ then falls inversely with neutrino energy:

\begin{equation}
x \sim \frac{M_W}{E_\nu}
\end{equation}

\noindent which, for EeV scales, implies $x$ down to $10^{-8}$ or lower.  The ZEUS collaboration has recently extended their analysis of parton distribution function data down to $x \simeq 10^{-5}$, allowing a more robust extrapolation of the neutrino cross-section to higher energies~\cite{bib:Sarkar2008}.  Uncertainties in their parton distribution function translate into $\pm 4\%$ uncertainties for the neutrino cross-section for center-of-mass energy of $10^{4}$ GeV and $\pm 14\%$ uncertainties at $\sqrt{s} = 10^{6}$ GeV.

An equal factor in the precise evaluation of these cross-sections is the selection of an adequate parton distribution function (PDF) itself.  The conventional PDF makes use of the Dokshitzer-Gribov-Lipatov-Altarelli-Parisi (DGLAP) formalism~\cite{Altarelli:1977zs,Dokshitzer:1977sg}, which is a next-to-leading order QCD calculation.  As one pushes further down in $x$, the PDFs introduce greater uncertainties, whereby other approaches can be used, such as the formalism adopted by the Balitsky-Fadin-Kuraev-Lipatov (BFKL) group~\cite{Kuraev:1977fs,Ciafaloni:2006yk}.  In reality, the approaches of both DGLAP/BFKL need to be combined in order to properly account for the $Q^2$ and $x$ evolution of these PDFs.  

One of the more difficult effects to account for in these parametrization schemes is that of gluon recombination ($gg\rightarrow g$).  Such a saturation must take place at the very highest energies in order to preserve unitarity.  Groups have made use of non-linear color glass condensate models as a way to model these effects~\cite{Iancu:2003xm}. Such techniques have been successfully applied to RHIC data~\cite{JalilianMarian:2005jf}.  


\section{Summary}
\label{sec:Summary}

In this work, we have presented a comprehensive review of neutrino 
interaction cross sections. Our discussion has ranged all the way from 
eV to EeV energy scales and therefore spanned a broad range of underlying 
physics processes, theoretical calculations, and experimental measurements.

While our knowledge of neutrino scattering may not be equally precise
at all energies, one cannot help but marvel at how far our theoretical 
frameworks extends. From literally zero-point energy to unfathomable reaches,
it appears that our models can shed some light in the darkness. Equally 
remarkable is the effort by which we seek to ground our theories.  Where 
data does not exist, we seek other anchors by which we can assess their 
validity.  When even that approach fails, we pile model against model 
in the hopes of finding weaknesses that ultimately will strengthen our 
foundations.

As the journey continues into the current millennium, we find that more and more 
direct data is being collected to guide our theoretical understanding.  
Even as this article is being written, new experiments are coming online  
to shed more light on neutrino interactions.  Therefore, these authors 
believe that, as comprehensive as we have tried to make this review, it 
is certainly an incomplete story whose chapters continue to be written.

\section{Acknowledgments}
\label{sec:Acknowledgments}

The authors would like to thank S. Brice, S. Dytman, D. Naples, J.P. Krane, G. Mention and R. Tayloe for help in gathering experimental data used in this review.  The authors would also wish to thank W. Haxton, W. Donnelly, and R. G. H. Robertson for their comments and suggestions pertaining to this work.  J. A. Formaggio is supported by the United States Department of Energy under Grant No. DE-FG02-06ER-41420. G. P. Zeller is supported via the Fermi Research Alliance, LLC under Contract  No. DE-AC02-07CH11359 with the United States Department of Energy.

\bibliographystyle{apsrmp}

\bibliography{nuXSec_resubmit}

\begin{thebibliography}{367}
\expandafter\ifx\csname natexlab\endcsname\relax\def\natexlab#1{#1}\fi
\expandafter\ifx\csname bibnamefont\endcsname\relax
  \def\bibnamefont#1{#1}\fi
\expandafter\ifx\csname bibfnamefont\endcsname\relax
  \def\bibfnamefont#1{#1}\fi
\expandafter\ifx\csname citenamefont\endcsname\relax
  \def\citenamefont#1{#1}\fi
\expandafter\ifx\csname url\endcsname\relax
  \def\url#1{\texttt{#1}}\fi
\expandafter\ifx\csname urlprefix\endcsname\relax\def\urlprefix{URL }\fi
\providecommand{\bibinfo}[2]{#2}
\providecommand{\eprint}[2][]{\url{#2}}

\bibitem[{Abbasi \emph{et~al.}(2004)\citenamefont{Abbasi}
  \emph{et~al.}}]{bib:HiRes}
\bibinfo{author}{\bibnamefont{Abbasi}, \bibfnamefont{R.~U.}}, \emph{et~al.}
  (\bibinfo{collaboration}{High Resolution FlyÕs Eye Collaboration}),
  \bibinfo{year}{2004}, \bibinfo{journal}{Phys. Rev. Lett.}
  \textbf{\bibinfo{volume}{92}}(\bibinfo{number}{15}), \bibinfo{pages}{151101}.

\bibitem[{\citenamefont{Abdurashitov}
  \emph{et~al.}(2006)\citenamefont{Abdurashitov, Gavrin, Girin, Gorbachev,
  Gurkina, Ibragimova, Kalikhov, Khairnasov, Knodel, Matveev, Mirmov, Shikhin}
  \emph{et~al.}}]{PhysRevC.73.045805}
\bibinfo{author}{\bibnamefont{Abdurashitov}, \bibfnamefont{J.~N.}},
  \bibinfo{author}{\bibfnamefont{V.~N.} \bibnamefont{Gavrin}},
  \bibinfo{author}{\bibfnamefont{S.~V.} \bibnamefont{Girin}},
  \bibinfo{author}{\bibfnamefont{V.~V.} \bibnamefont{Gorbachev}},
  \bibinfo{author}{\bibfnamefont{P.~P.} \bibnamefont{Gurkina}},
  \bibinfo{author}{\bibfnamefont{T.~V.} \bibnamefont{Ibragimova}},
  \bibinfo{author}{\bibfnamefont{A.~V.} \bibnamefont{Kalikhov}},
  \bibinfo{author}{\bibfnamefont{N.~G.} \bibnamefont{Khairnasov}},
  \bibinfo{author}{\bibfnamefont{T.~V.} \bibnamefont{Knodel}},
  \bibinfo{author}{\bibfnamefont{V.~A.} \bibnamefont{Matveev}},
  \bibinfo{author}{\bibfnamefont{I.~N.} \bibnamefont{Mirmov}},
  \bibinfo{author}{\bibfnamefont{A.~A.} \bibnamefont{Shikhin}}, \emph{et~al.},
  \bibinfo{year}{2006}, \bibinfo{journal}{Phys. Rev. C}
  \textbf{\bibinfo{volume}{73}}(\bibinfo{number}{4}), \bibinfo{pages}{045805}.

\bibitem[{\citenamefont{Abdurashitov}
  \emph{et~al.}(1999)\citenamefont{Abdurashitov, Gavrin, Girin, Gorbachev,
  Ibragimova, Kalikhov, Khairnasov, Knodel, Kornoukhov, Mirmov, Shikhin,
  Veretenkin} \emph{et~al.}}]{PhysRevC.59.2246}
\bibinfo{author}{\bibnamefont{Abdurashitov}, \bibfnamefont{J.~N.}},
  \bibinfo{author}{\bibfnamefont{V.~N.} \bibnamefont{Gavrin}},
  \bibinfo{author}{\bibfnamefont{S.~V.} \bibnamefont{Girin}},
  \bibinfo{author}{\bibfnamefont{V.~V.} \bibnamefont{Gorbachev}},
  \bibinfo{author}{\bibfnamefont{T.~V.} \bibnamefont{Ibragimova}},
  \bibinfo{author}{\bibfnamefont{A.~V.} \bibnamefont{Kalikhov}},
  \bibinfo{author}{\bibfnamefont{N.~G.} \bibnamefont{Khairnasov}},
  \bibinfo{author}{\bibfnamefont{T.~V.} \bibnamefont{Knodel}},
  \bibinfo{author}{\bibfnamefont{V.~N.} \bibnamefont{Kornoukhov}},
  \bibinfo{author}{\bibfnamefont{I.~N.} \bibnamefont{Mirmov}},
  \bibinfo{author}{\bibfnamefont{A.~A.} \bibnamefont{Shikhin}},
  \bibinfo{author}{\bibfnamefont{E.~P.} \bibnamefont{Veretenkin}},
  \emph{et~al.} (\bibinfo{collaboration}{(The SAGE Collaboration)}),
  \bibinfo{year}{1999}, \bibinfo{journal}{Phys. Rev. C}
  \textbf{\bibinfo{volume}{59}}(\bibinfo{number}{4}), \bibinfo{pages}{2246}.

\bibitem[{\citenamefont{Abe} \emph{et~al.}(1989)\citenamefont{Abe, Ahrens,
  Amako, Aronson, Beier} \emph{et~al.}}]{Abe:1989qk}
\bibinfo{author}{\bibnamefont{Abe}, \bibfnamefont{K.}},
  \bibinfo{author}{\bibfnamefont{L.}~\bibnamefont{Ahrens}},
  \bibinfo{author}{\bibfnamefont{K.}~\bibnamefont{Amako}},
  \bibinfo{author}{\bibfnamefont{S.}~\bibnamefont{Aronson}},
  \bibinfo{author}{\bibfnamefont{E.}~\bibnamefont{Beier}}, \emph{et~al.},
  \bibinfo{year}{1989}, \bibinfo{journal}{Phys.Rev.Lett.}
  \textbf{\bibinfo{volume}{62}}, \bibinfo{pages}{1709}.

\bibitem[{Abraham \emph{et~al.}(2008)\citenamefont{Abraham}
  \emph{et~al.}}]{bib:PierreAuger}
\bibinfo{author}{\bibnamefont{Abraham}, \bibfnamefont{J.}}, \emph{et~al.}
  (\bibinfo{collaboration}{Pierre Auger Collaboration}), \bibinfo{year}{2008},
  \bibinfo{journal}{Phys. Rev. Lett.}
  \textbf{\bibinfo{volume}{100}}(\bibinfo{number}{21}),
  \bibinfo{pages}{211101}.

\bibitem[{\citenamefont{Abramowicz}
  \emph{et~al.}(1982)\citenamefont{Abramowicz, de~Groot, Knobloch, May,
  Palazzi} \emph{et~al.}}]{Abramowicz:1982zr}
\bibinfo{author}{\bibnamefont{Abramowicz}, \bibfnamefont{H.}},
  \bibinfo{author}{\bibfnamefont{J.}~\bibnamefont{de~Groot}},
  \bibinfo{author}{\bibfnamefont{J.}~\bibnamefont{Knobloch}},
  \bibinfo{author}{\bibfnamefont{J.}~\bibnamefont{May}},
  \bibinfo{author}{\bibfnamefont{P.}~\bibnamefont{Palazzi}}, \emph{et~al.},
  \bibinfo{year}{1982}, \bibinfo{journal}{Z.Phys.}
  \textbf{\bibinfo{volume}{C15}}, \bibinfo{pages}{19}.

\bibitem[{\citenamefont{Achkar} \emph{et~al.}(1995)\citenamefont{Achkar,
  Aleksan, Avenier, Bagieu, Bouchez, Brissot, Cavaignac, Collot, Cousinou,
  Cussonneau, Declais, Dufour} \emph{et~al.}}]{bib:Bugey95}
\bibinfo{author}{\bibnamefont{Achkar}, \bibfnamefont{B.}},
  \bibinfo{author}{\bibfnamefont{R.}~\bibnamefont{Aleksan}},
  \bibinfo{author}{\bibfnamefont{M.}~\bibnamefont{Avenier}},
  \bibinfo{author}{\bibfnamefont{G.}~\bibnamefont{Bagieu}},
  \bibinfo{author}{\bibfnamefont{J.}~\bibnamefont{Bouchez}},
  \bibinfo{author}{\bibfnamefont{R.}~\bibnamefont{Brissot}},
  \bibinfo{author}{\bibfnamefont{J.~F.} \bibnamefont{Cavaignac}},
  \bibinfo{author}{\bibfnamefont{J.}~\bibnamefont{Collot}},
  \bibinfo{author}{\bibfnamefont{M.-C.} \bibnamefont{Cousinou}},
  \bibinfo{author}{\bibfnamefont{J.~P.} \bibnamefont{Cussonneau}},
  \bibinfo{author}{\bibfnamefont{Y.}~\bibnamefont{Declais}},
  \bibinfo{author}{\bibfnamefont{Y.}~\bibnamefont{Dufour}}, \emph{et~al.},
  \bibinfo{year}{1995}, \bibinfo{journal}{Nuclear Physics B}
  \textbf{\bibinfo{volume}{434}}(\bibinfo{number}{3}), \bibinfo{pages}{503 },
  ISSN \bibinfo{issn}{0550-3213},
  \urlprefix\url{http://www.sciencedirect.com/science/article/B6TVC-3YS905D-F/%
2/e5b54776228f070b30ab6095afa02cdc}.

\bibitem[{Achterberg \emph{et~al.}(2007)\citenamefont{Achterberg}
  \emph{et~al.}}]{bib:AMANDA}
\bibinfo{author}{\bibnamefont{Achterberg}, \bibfnamefont{A.}}, \emph{et~al.}
  (\bibinfo{collaboration}{IceCube}), \bibinfo{year}{2007},
  \bibinfo{journal}{Phys. Rev.} \textbf{\bibinfo{volume}{D75}},
  \bibinfo{pages}{102001}.

\bibitem[{Adams \emph{et~al.}(2009)\citenamefont{Adams}
  \emph{et~al.}}]{Adams:2008cm}
\bibinfo{author}{\bibnamefont{Adams}, \bibfnamefont{T.}}, \emph{et~al.}
  (\bibinfo{collaboration}{NuSOnG}), \bibinfo{year}{2009},
  \bibinfo{journal}{Int. J. Mod. Phys.} \textbf{\bibinfo{volume}{A24}},
  \bibinfo{pages}{671}.

\bibitem[{Adamson \emph{et~al.}(2010)\citenamefont{Adamson}
  \emph{et~al.}}]{minos-dis}
\bibinfo{author}{\bibnamefont{Adamson}, \bibfnamefont{P.}}, \emph{et~al.},
  \bibinfo{year}{2010}, \bibinfo{journal}{Phys. Rev.}
  \textbf{\bibinfo{volume}{D81}}, \bibinfo{pages}{072002}.

\bibitem[{\citenamefont{Adelberger}
  \emph{et~al.}(2010)\citenamefont{Adelberger, Balantekin, Bemmerer, Bertulani,
  Chen} \emph{et~al.}}]{Adelberger:2010qa}
\bibinfo{author}{\bibnamefont{Adelberger}, \bibfnamefont{E.}},
  \bibinfo{author}{\bibfnamefont{A.}~\bibnamefont{Balantekin}},
  \bibinfo{author}{\bibfnamefont{D.}~\bibnamefont{Bemmerer}},
  \bibinfo{author}{\bibfnamefont{C.}~\bibnamefont{Bertulani}},
  \bibinfo{author}{\bibfnamefont{J.-W.} \bibnamefont{Chen}}, \emph{et~al.},
  \bibinfo{year}{2010}, \bibinfo{note}{* Temporary entry *},
  \eprint{1004.2318}.

\bibitem[{Adera \emph{et~al.}(2010)\citenamefont{Adera}
  \emph{et~al.}}]{new-k-models-1}
\bibinfo{author}{\bibnamefont{Adera}, \bibfnamefont{G.~B.}}, \emph{et~al.},
  \bibinfo{year}{2010}, \bibinfo{journal}{Phys. Rev.}
  \textbf{\bibinfo{volume}{C82}}, \bibinfo{pages}{025501}.

\bibitem[{Aderholz \emph{et~al.}(1992)\citenamefont{Aderholz}
  \emph{et~al.}}]{kaon-bc-12}
\bibinfo{author}{\bibnamefont{Aderholz}, \bibfnamefont{M.}}, \emph{et~al.},
  \bibinfo{year}{1992}, \bibinfo{journal}{Phys. Rev.}
  \textbf{\bibinfo{volume}{D45}}, \bibinfo{pages}{2232}.

\bibitem[{\citenamefont{Adler}(1964)}]{adler-pcac}
\bibinfo{author}{\bibnamefont{Adler}, \bibfnamefont{S.~L.}},
  \bibinfo{year}{1964}, \bibinfo{journal}{Phys. Rev. B}
  \textbf{\bibinfo{volume}{135}}, \bibinfo{pages}{963}.

\bibitem[{\citenamefont{A.Fujii and Y.Yamaguchi}(1964)}]{bib:Yamaguchi}
\bibinfo{author}{\bibnamefont{A.Fujii}}, and
  \bibinfo{author}{\bibnamefont{Y.Yamaguchi}}, \bibinfo{year}{1964},
  \bibinfo{journal}{Prog. Theor. Phys.} \textbf{\bibinfo{volume}{31}},
  \bibinfo{pages}{107}.

\bibitem[{Agababyan \emph{et~al.}(2006)\citenamefont{Agababyan}
  \emph{et~al.}}]{kaon-bc-16}
\bibinfo{author}{\bibnamefont{Agababyan}, \bibfnamefont{N.~M.}}, \emph{et~al.},
  \bibinfo{year}{2006}, \bibinfo{journal}{Phys. Atom. Nucl.}
  \textbf{\bibinfo{volume}{69}}, \bibinfo{pages}{35}.

\bibitem[{Aguilar-Arevalo
  \emph{et~al.}(2010{\natexlab{a}})\citenamefont{Aguilar-Arevalo}
  \emph{et~al.}}]{mb-qe}
\bibinfo{author}{\bibnamefont{Aguilar-Arevalo}, \bibfnamefont{A.}},
  \emph{et~al.}, \bibinfo{year}{2010}{\natexlab{a}}, \bibinfo{journal}{Phys.
  Rev.} \textbf{\bibinfo{volume}{D81}}, \bibinfo{pages}{092005}.

\bibitem[{Aguilar-Arevalo
  \emph{et~al.}(2010{\natexlab{b}})\citenamefont{Aguilar-Arevalo}
  \emph{et~al.}}]{mb-ncel}
\bibinfo{author}{\bibnamefont{Aguilar-Arevalo}, \bibfnamefont{A.}},
  \emph{et~al.}, \bibinfo{year}{2010}{\natexlab{b}}, \bibinfo{journal}{Phys.
  Rev.} \textbf{\bibinfo{volume}{D82}}, \bibinfo{pages}{092005}.

\bibitem[{\citenamefont{Aguilar-Arevalo}(2008)}]{modern-ma-2}
\bibinfo{author}{\bibnamefont{Aguilar-Arevalo}, \bibfnamefont{A.~A.}},
  \bibinfo{year}{2008}, \bibinfo{journal}{Phys. Rev. Lett.}
  \textbf{\bibinfo{volume}{100}}, \bibinfo{pages}{032301}.

\bibitem[{Aguilar-Arevalo \emph{et~al.}(2008)\citenamefont{Aguilar-Arevalo}
  \emph{et~al.}}]{mb-coh}
\bibinfo{author}{\bibnamefont{Aguilar-Arevalo}, \bibfnamefont{A.~A.}},
  \emph{et~al.}, \bibinfo{year}{2008}, \bibinfo{journal}{Phys. Lett.}
  \textbf{\bibinfo{volume}{B664}}, \bibinfo{pages}{41}.

\bibitem[{Aguilar-Arevalo \emph{et~al.}(2009)\citenamefont{Aguilar-Arevalo}
  \emph{et~al.}}]{mb-ccpip-qe}
\bibinfo{author}{\bibnamefont{Aguilar-Arevalo}, \bibfnamefont{A.~A.}},
  \emph{et~al.}, \bibinfo{year}{2009}, \bibinfo{journal}{Phys. Rev. Lett.}
  \textbf{\bibinfo{volume}{103}}, \bibinfo{pages}{081801}.

\bibitem[{Aguilar-Arevalo
  \emph{et~al.}(2010{\natexlab{c}})\citenamefont{Aguilar-Arevalo}
  \emph{et~al.}}]{mb-ccpip}
\bibinfo{author}{\bibnamefont{Aguilar-Arevalo}, \bibfnamefont{A.~A.}},
  \emph{et~al.}, \bibinfo{year}{2010}{\natexlab{c}}, \bibinfo{journal}{Phys.
  Rev.} \textbf{\bibinfo{volume}{D83}}, \bibinfo{pages}{052007}.

\bibitem[{Aguilar-Arevalo
  \emph{et~al.}(2010{\natexlab{d}})\citenamefont{Aguilar-Arevalo}
  \emph{et~al.}}]{mb-ccpi0}
\bibinfo{author}{\bibnamefont{Aguilar-Arevalo}, \bibfnamefont{A.~A.}},
  \emph{et~al.}, \bibinfo{year}{2010}{\natexlab{d}}, \bibinfo{journal}{Phys.
  Rev.} \textbf{\bibinfo{volume}{D83}}, \bibinfo{pages}{052009}.

\bibitem[{Aguilar-Arevalo
  \emph{et~al.}(2010{\natexlab{e}})\citenamefont{Aguilar-Arevalo}
  \emph{et~al.}}]{mb-ncpi0}
\bibinfo{author}{\bibnamefont{Aguilar-Arevalo}, \bibfnamefont{A.~A.}},
  \emph{et~al.}, \bibinfo{year}{2010}{\natexlab{e}}, \bibinfo{journal}{Phys.
  Rev.} \textbf{\bibinfo{volume}{D81}}, \bibinfo{pages}{013005}.

\bibitem[{Aharmin \emph{et~al.}(2005)\citenamefont{Aharmin}
  \emph{et~al.}}]{bib:sno6}
\bibinfo{author}{\bibnamefont{Aharmin}, \bibfnamefont{B.}}, \emph{et~al.},
  \bibinfo{year}{2005}, \bibinfo{journal}{Phys. Rev. C.}
  \textbf{\bibinfo{volume}{72}}, \bibinfo{pages}{055502}.

\bibitem[{Aharmin \emph{et~al.}(2007)\citenamefont{Aharmin}
  \emph{et~al.}}]{bib:sno4}
\bibinfo{author}{\bibnamefont{Aharmin}, \bibfnamefont{B.}}, \emph{et~al.},
  \bibinfo{year}{2007}, \bibinfo{journal}{Phys. Rev. C.}
  \textbf{\bibinfo{volume}{75}}, \bibinfo{pages}{045502}.

\bibitem[{Aharmin \emph{et~al.}(2008)\citenamefont{Aharmin}
  \emph{et~al.}}]{bib:sno7}
\bibinfo{author}{\bibnamefont{Aharmin}, \bibfnamefont{B.}}, \emph{et~al.},
  \bibinfo{year}{2008}, \bibinfo{journal}{Phys. Rev. Lett.}
  \textbf{\bibinfo{volume}{101}}, \bibinfo{pages}{11130}.

\bibitem[{Ahmad \emph{et~al.}(2001)\citenamefont{Ahmad}
  \emph{et~al.}}]{bib:sno1}
\bibinfo{author}{\bibnamefont{Ahmad}, \bibfnamefont{Q.}}, \emph{et~al.},
  \bibinfo{year}{2001}, \bibinfo{journal}{Phys. Rev. Lett.}
  \textbf{\bibinfo{volume}{87}}, \bibinfo{pages}{071301}.

\bibitem[{Ahmad \emph{et~al.}(2002{\natexlab{a}})\citenamefont{Ahmad}
  \emph{et~al.}}]{bib:sno2}
\bibinfo{author}{\bibnamefont{Ahmad}, \bibfnamefont{Q.}}, \emph{et~al.},
  \bibinfo{year}{2002}{\natexlab{a}}, \bibinfo{journal}{Phys. Rev. Lett.}
  \textbf{\bibinfo{volume}{89}}, \bibinfo{pages}{011301}.

\bibitem[{Ahmad \emph{et~al.}(2002{\natexlab{b}})\citenamefont{Ahmad}
  \emph{et~al.}}]{bib:sno3}
\bibinfo{author}{\bibnamefont{Ahmad}, \bibfnamefont{Q.}}, \emph{et~al.},
  \bibinfo{year}{2002}{\natexlab{b}}, \bibinfo{journal}{Phys. Rev. Lett.}
  \textbf{\bibinfo{volume}{89}}, \bibinfo{pages}{011302}.

\bibitem[{Ahmed \emph{et~al.}(2004)\citenamefont{Ahmed}
  \emph{et~al.}}]{bib:sno5}
\bibinfo{author}{\bibnamefont{Ahmed}, \bibfnamefont{S.}}, \emph{et~al.},
  \bibinfo{year}{2004}, \bibinfo{journal}{Phys. Rev. Lett.}
  \textbf{\bibinfo{volume}{92}}, \bibinfo{pages}{181301}.

\bibitem[{\citenamefont{Ahrens} \emph{et~al.}(1983)\citenamefont{Ahrens,
  Aronson, Connolly, Erickson, Gibbard} \emph{et~al.}}]{Ahrens:1983py}
\bibinfo{author}{\bibnamefont{Ahrens}, \bibfnamefont{L.}},
  \bibinfo{author}{\bibfnamefont{S.}~\bibnamefont{Aronson}},
  \bibinfo{author}{\bibfnamefont{P.}~\bibnamefont{Connolly}},
  \bibinfo{author}{\bibfnamefont{T.}~\bibnamefont{Erickson}},
  \bibinfo{author}{\bibfnamefont{B.}~\bibnamefont{Gibbard}}, \emph{et~al.},
  \bibinfo{year}{1983}, \bibinfo{journal}{Phys.Rev.Lett.}
  \textbf{\bibinfo{volume}{51}}, \bibinfo{pages}{1514}.

\bibitem[{\citenamefont{Ahrens} \emph{et~al.}(1990)\citenamefont{Ahrens,
  Aronson, Connolly, Gibbard, Murtagh} \emph{et~al.}}]{Ahrens:1990fp}
\bibinfo{author}{\bibnamefont{Ahrens}, \bibfnamefont{L.}},
  \bibinfo{author}{\bibfnamefont{S.}~\bibnamefont{Aronson}},
  \bibinfo{author}{\bibfnamefont{P.}~\bibnamefont{Connolly}},
  \bibinfo{author}{\bibfnamefont{B.}~\bibnamefont{Gibbard}},
  \bibinfo{author}{\bibfnamefont{M.}~\bibnamefont{Murtagh}}, \emph{et~al.},
  \bibinfo{year}{1990}, \bibinfo{journal}{Phys.Rev.}
  \textbf{\bibinfo{volume}{D41}}, \bibinfo{pages}{3297}.

\bibitem[{Ahrens \emph{et~al.}(1988)\citenamefont{Ahrens}
  \emph{et~al.}}]{nc-elastic-1}
\bibinfo{author}{\bibnamefont{Ahrens}, \bibfnamefont{L.~A.}}, \emph{et~al.},
  \bibinfo{year}{1988}, \bibinfo{journal}{Phys. Lett.}
  \textbf{\bibinfo{volume}{B202}}, \bibinfo{pages}{284}.

\bibitem[{\citenamefont{Alam} \emph{et~al.}(2012)\citenamefont{Alam, Simo,
  Athar, and Vicente~Vacas}}]{Alam12}
\bibinfo{author}{\bibnamefont{Alam}, \bibfnamefont{M.}},
  \bibinfo{author}{\bibfnamefont{I.}~\bibnamefont{Simo}},
  \bibinfo{author}{\bibfnamefont{M.}~\bibnamefont{Athar}}, and
  \bibinfo{author}{\bibfnamefont{M.}~\bibnamefont{Vicente~Vacas}},
  \bibinfo{year}{2012}, \bibinfo{journal}{Phys.Rev.}
  \textbf{\bibinfo{volume}{D85}}, \bibinfo{pages}{013014}, \bibinfo{note}{15
  pages and 6 figures. This version matches accepted version for publication in
  Physical Review D}.

\bibitem[{Alam \emph{et~al.}(2010)\citenamefont{Alam}
  \emph{et~al.}}]{new-k-models}
\bibinfo{author}{\bibnamefont{Alam}, \bibfnamefont{R.}}, \emph{et~al.},
  \bibinfo{year}{2010}, \bibinfo{journal}{Phys. Rev.}
  \textbf{\bibinfo{volume}{D82}}, \bibinfo{pages}{033001}.

\bibitem[{Alberico \emph{et~al.}(1999)\citenamefont{Alberico}
  \emph{et~al.}}]{bnl-e734-update-2}
\bibinfo{author}{\bibnamefont{Alberico}, \bibfnamefont{W.~M.}}, \emph{et~al.},
  \bibinfo{year}{1999}, \bibinfo{journal}{Nucl. Phys.}
  \textbf{\bibinfo{volume}{A651}}, \bibinfo{pages}{277}.

\bibitem[{\citenamefont{Albright}(1975)}]{single-k-models-1}
\bibinfo{author}{\bibnamefont{Albright}, \bibfnamefont{C.}},
  \bibinfo{year}{1975}, \bibinfo{journal}{Phys. Rev.}
  \textbf{\bibinfo{volume}{D12}}, \bibinfo{pages}{1329}.

\bibitem[{\citenamefont{Albright and Jarlskog}(1975)}]{Albright:1974ts}
\bibinfo{author}{\bibnamefont{Albright}, \bibfnamefont{C.~H.}}, and
  \bibinfo{author}{\bibfnamefont{C.}~\bibnamefont{Jarlskog}},
  \bibinfo{year}{1975}, \bibinfo{journal}{Nucl.Phys.}
  \textbf{\bibinfo{volume}{B84}}, \bibinfo{pages}{467}.

\bibitem[{Alimonti \emph{et~al.}(2002)\citenamefont{Alimonti}
  \emph{et~al.}}]{bib:BOREXINO}
\bibinfo{author}{\bibnamefont{Alimonti}, \bibfnamefont{G.}}, \emph{et~al.}
  (\bibinfo{collaboration}{Borexino}), \bibinfo{year}{2002},
  \bibinfo{journal}{Astropart. Phys.} \textbf{\bibinfo{volume}{16}},
  \bibinfo{pages}{205}.

\bibitem[{Allasia \emph{et~al.}(1983)\citenamefont{Allasia}
  \emph{et~al.}}]{Allasia:1983qh}
\bibinfo{author}{\bibnamefont{Allasia}, \bibfnamefont{D.}}, \emph{et~al.}
  (\bibinfo{collaboration}{AMSTERDAM-BOLOGNA-PADUA-PISA-SACLAY-TURIN
  COLLABORATION}), \bibinfo{year}{1983}, \bibinfo{journal}{Z.Phys.}
  \textbf{\bibinfo{volume}{C20}}, \bibinfo{pages}{95}.

\bibitem[{Allen \emph{et~al.}(1986)\citenamefont{Allen}
  \emph{et~al.}}]{Allen:1985ti}
\bibinfo{author}{\bibnamefont{Allen}, \bibfnamefont{P.}}, \emph{et~al.}
  (\bibinfo{collaboration}{Aachen-Birmingham-Bonn-CERN-London-Munich-Oxford
  Collaboration}), \bibinfo{year}{1986}, \bibinfo{journal}{Nucl.Phys.}
  \textbf{\bibinfo{volume}{B264}}, \bibinfo{pages}{221}.

\bibitem[{\citenamefont{Allen} \emph{et~al.}(1993)\citenamefont{Allen, Chen,
  Doe, Hausammann, Lee} \emph{et~al.}}]{Allen:1992qe}
\bibinfo{author}{\bibnamefont{Allen}, \bibfnamefont{R.}},
  \bibinfo{author}{\bibfnamefont{H.}~\bibnamefont{Chen}},
  \bibinfo{author}{\bibfnamefont{P.}~\bibnamefont{Doe}},
  \bibinfo{author}{\bibfnamefont{R.}~\bibnamefont{Hausammann}},
  \bibinfo{author}{\bibfnamefont{W.}~\bibnamefont{Lee}}, \emph{et~al.},
  \bibinfo{year}{1993}, \bibinfo{journal}{Phys.Rev.}
  \textbf{\bibinfo{volume}{D47}}, \bibinfo{pages}{11}.

\bibitem[{\citenamefont{Altarelli and Parisi}(1977)}]{Altarelli:1977zs}
\bibinfo{author}{\bibnamefont{Altarelli}, \bibfnamefont{G.}}, and
  \bibinfo{author}{\bibfnamefont{G.}~\bibnamefont{Parisi}},
  \bibinfo{year}{1977}, \bibinfo{journal}{Nucl. Phys.}
  \textbf{\bibinfo{volume}{B126}}, \bibinfo{pages}{298}.

\bibitem[{\citenamefont{Alvarez-Ruso}(2010)}]{reduced-qe-cross-section}
\bibinfo{author}{\bibnamefont{Alvarez-Ruso}, \bibfnamefont{L.}},
  \bibinfo{year}{2010}, \eprint{nucl-th/1012.3871}.

\bibitem[{\citenamefont{Alvarez-Ruso}(2011{\natexlab{a}})}]{Alvarez-Ruso11}
\bibinfo{author}{\bibnamefont{Alvarez-Ruso}, \bibfnamefont{L.}},
  \bibinfo{year}{2011}{\natexlab{a}}, \bibinfo{journal}{AIP Conf.Proc.}
  \textbf{\bibinfo{volume}{1405}}, \bibinfo{pages}{140}.

\bibitem[{\citenamefont{Alvarez-Ruso}(2011{\natexlab{b}})}]{Ruso12}
\bibinfo{author}{\bibnamefont{Alvarez-Ruso}, \bibfnamefont{L.}},
  \bibinfo{year}{2011}{\natexlab{b}}, \bibinfo{journal}{AIP Conf.Proc.}
  \textbf{\bibinfo{volume}{1382}}, \bibinfo{pages}{161}.

\bibitem[{Alvarez-Ruso \emph{et~al.}(2007)\citenamefont{Alvarez-Ruso}
  \emph{et~al.}}]{delta-coh-models-2}
\bibinfo{author}{\bibnamefont{Alvarez-Ruso}, \bibfnamefont{L.}}, \emph{et~al.},
  \bibinfo{year}{2007}, \bibinfo{journal}{Phys. Rev.}
  \textbf{\bibinfo{volume}{C75}}, \bibinfo{pages}{055501}.

\bibitem[{\citenamefont{Amaro}
  \emph{et~al.}(2011{\natexlab{a}})\citenamefont{Amaro, Barbaro, Caballero, and
  Donnelly}}]{Amaro:2011aa}
\bibinfo{author}{\bibnamefont{Amaro}, \bibfnamefont{J.}},
  \bibinfo{author}{\bibfnamefont{M.}~\bibnamefont{Barbaro}},
  \bibinfo{author}{\bibfnamefont{J.}~\bibnamefont{Caballero}}, and
  \bibinfo{author}{\bibfnamefont{T.}~\bibnamefont{Donnelly}},
  \bibinfo{year}{2011}{\natexlab{a}}, \eprint{1112.2123}.

\bibitem[{\citenamefont{Amaro}
  \emph{et~al.}(2011{\natexlab{b}})\citenamefont{Amaro, Barbaro, Caballero,
  Donnelly, and Udias}}]{Amaro:2010}
\bibinfo{author}{\bibnamefont{Amaro}, \bibfnamefont{J.}},
  \bibinfo{author}{\bibfnamefont{M.}~\bibnamefont{Barbaro}},
  \bibinfo{author}{\bibfnamefont{J.}~\bibnamefont{Caballero}},
  \bibinfo{author}{\bibfnamefont{T.}~\bibnamefont{Donnelly}}, and
  \bibinfo{author}{\bibfnamefont{J.}~\bibnamefont{Udias}},
  \bibinfo{year}{2011}{\natexlab{b}}, \bibinfo{journal}{Phys.Rev.}
  \textbf{\bibinfo{volume}{D84}}, \bibinfo{pages}{033004}.

\bibitem[{\citenamefont{Amaro} \emph{et~al.}(2006)\citenamefont{Amaro, Barbaro,
  Caballero, and Donnelly}}]{Amaro06}
\bibinfo{author}{\bibnamefont{Amaro}, \bibfnamefont{J.~E.}},
  \bibinfo{author}{\bibfnamefont{M.}~\bibnamefont{Barbaro}},
  \bibinfo{author}{\bibfnamefont{J.}~\bibnamefont{Caballero}}, and
  \bibinfo{author}{\bibfnamefont{T.}~\bibnamefont{Donnelly}},
  \bibinfo{year}{2006}, \bibinfo{journal}{Phys.Rev.}
  \textbf{\bibinfo{volume}{C73}}, \bibinfo{pages}{035503}.

\bibitem[{\citenamefont{Amaro} \emph{et~al.}(2007)\citenamefont{Amaro, Barbaro,
  Caballero, Donnelly, and Udias}}]{bib:Donnelly}
\bibinfo{author}{\bibnamefont{Amaro}, \bibfnamefont{J.~E.}},
  \bibinfo{author}{\bibfnamefont{M.}~\bibnamefont{Barbaro}},
  \bibinfo{author}{\bibfnamefont{J.}~\bibnamefont{Caballero}},
  \bibinfo{author}{\bibfnamefont{T.}~\bibnamefont{Donnelly}}, and
  \bibinfo{author}{\bibfnamefont{J.}~\bibnamefont{Udias}},
  \bibinfo{year}{2007}, \bibinfo{journal}{Phys.Rev.}
  \textbf{\bibinfo{volume}{C75}}, \bibinfo{pages}{034613}.

\bibitem[{\citenamefont{Amaro} \emph{et~al.}(2005)\citenamefont{Amaro, Barbaro,
  Caballero, Donnelly, Molinari, and Sick}}]{bib:DonnellySuperScaling}
\bibinfo{author}{\bibnamefont{Amaro}, \bibfnamefont{J.~E.}},
  \bibinfo{author}{\bibfnamefont{M.~B.} \bibnamefont{Barbaro}},
  \bibinfo{author}{\bibfnamefont{J.~A.} \bibnamefont{Caballero}},
  \bibinfo{author}{\bibfnamefont{T.~W.} \bibnamefont{Donnelly}},
  \bibinfo{author}{\bibfnamefont{A.}~\bibnamefont{Molinari}}, and
  \bibinfo{author}{\bibfnamefont{I.}~\bibnamefont{Sick}}, \bibinfo{year}{2005},
  \bibinfo{journal}{Phys. Rev. C} \textbf{\bibinfo{volume}{71}},
  \bibinfo{pages}{015501},
  \urlprefix\url{http://link.aps.org/doi/10.1103/PhysRevC.71.015501}.

\bibitem[{Amaro \emph{et~al.}(2009)\citenamefont{Amaro}
  \emph{et~al.}}]{delta-coh-models-3}
\bibinfo{author}{\bibnamefont{Amaro}, \bibfnamefont{J.~E.}}, \emph{et~al.},
  \bibinfo{year}{2009}, \bibinfo{journal}{Phys. Rev.}
  \textbf{\bibinfo{volume}{D79}}, \bibinfo{pages}{013002}.

\bibitem[{\citenamefont{Amer}(1978)}]{single-k-models-3}
\bibinfo{author}{\bibnamefont{Amer}, \bibfnamefont{A.~A.}},
  \bibinfo{year}{1978}, \bibinfo{journal}{Phys. Rev.}
  \textbf{\bibinfo{volume}{D18}}, \bibinfo{pages}{2290}.

\bibitem[{Ammosov \emph{et~al.}(1987)\citenamefont{Ammosov}
  \emph{et~al.}}]{ammosov}
\bibinfo{author}{\bibnamefont{Ammosov}, \bibfnamefont{V.~V.}}, \emph{et~al.},
  \bibinfo{year}{1987}, \bibinfo{journal}{Z. Phys.}
  \textbf{\bibinfo{volume}{C36}}, \bibinfo{pages}{377}.

\bibitem[{\citenamefont{Ankowski} \emph{et~al.}(2012)\citenamefont{Ankowski,
  Benhar, Mori, Yamaguchi, and Sakuda}}]{Ankowski12}
\bibinfo{author}{\bibnamefont{Ankowski}, \bibfnamefont{A.~M.}},
  \bibinfo{author}{\bibfnamefont{O.}~\bibnamefont{Benhar}},
  \bibinfo{author}{\bibfnamefont{T.}~\bibnamefont{Mori}},
  \bibinfo{author}{\bibfnamefont{R.}~\bibnamefont{Yamaguchi}}, and
  \bibinfo{author}{\bibfnamefont{M.}~\bibnamefont{Sakuda}},
  \bibinfo{year}{2012}, \bibinfo{journal}{Phys.Rev.Lett.}
  \textbf{\bibinfo{volume}{108}}, \bibinfo{pages}{052505}, \bibinfo{note}{5
  pages, 4 figures}.

\bibitem[{\citenamefont{Ankowski and Sobczyk}(2006)}]{Ankowski06}
\bibinfo{author}{\bibnamefont{Ankowski}, \bibfnamefont{A.~M.}}, and
  \bibinfo{author}{\bibfnamefont{J.~T.} \bibnamefont{Sobczyk}},
  \bibinfo{year}{2006}, \bibinfo{journal}{Phys.Rev.}
  \textbf{\bibinfo{volume}{C74}}, \bibinfo{pages}{054316}.

\bibitem[{\citenamefont{Ankowski and Sobczyk}(2008)}]{modern-qe-theory-7}
\bibinfo{author}{\bibnamefont{Ankowski}, \bibfnamefont{A.~M.}}, and
  \bibinfo{author}{\bibfnamefont{J.~T.} \bibnamefont{Sobczyk}},
  \bibinfo{year}{2008}, \bibinfo{journal}{Phys. Rev.}
  \textbf{\bibinfo{volume}{C77}}, \bibinfo{pages}{044311}.

\bibitem[{\citenamefont{Anselmann} \emph{et~al.}(1995)\citenamefont{Anselmann,
  Fockenbrock, Hampel, Heusser, Kiko, Kirsten, Laubenstein, Pernicka, Pezzoni,
  R?nn, Sann, Spielker} \emph{et~al.}}]{Anselmann1995440}
\bibinfo{author}{\bibnamefont{Anselmann}, \bibfnamefont{P.}},
  \bibinfo{author}{\bibfnamefont{R.}~\bibnamefont{Fockenbrock}},
  \bibinfo{author}{\bibfnamefont{W.}~\bibnamefont{Hampel}},
  \bibinfo{author}{\bibfnamefont{G.}~\bibnamefont{Heusser}},
  \bibinfo{author}{\bibfnamefont{J.}~\bibnamefont{Kiko}},
  \bibinfo{author}{\bibfnamefont{T.}~\bibnamefont{Kirsten}},
  \bibinfo{author}{\bibfnamefont{M.}~\bibnamefont{Laubenstein}},
  \bibinfo{author}{\bibfnamefont{E.}~\bibnamefont{Pernicka}},
  \bibinfo{author}{\bibfnamefont{S.}~\bibnamefont{Pezzoni}},
  \bibinfo{author}{\bibfnamefont{U.}~\bibnamefont{R?nn}},
  \bibinfo{author}{\bibfnamefont{M.}~\bibnamefont{Sann}},
  \bibinfo{author}{\bibfnamefont{F.}~\bibnamefont{Spielker}}, \emph{et~al.},
  \bibinfo{year}{1995}, \bibinfo{journal}{Physics Letters B}
  \textbf{\bibinfo{volume}{342}}(\bibinfo{number}{1-4}), \bibinfo{pages}{440 },
  ISSN \bibinfo{issn}{0370-2693},
  \urlprefix\url{http://www.sciencedirect.com/science/article/B6TVN-40329BY-39%
/2/56816933534cdcc200410599d6822b0e}.

\bibitem[{Antipin \emph{et~al.}(2007)\citenamefont{Antipin}
  \emph{et~al.}}]{bib:Baikal-2}
\bibinfo{author}{\bibnamefont{Antipin}, \bibfnamefont{K.}}, \emph{et~al.},
  \bibinfo{year}{2007}, \bibinfo{journal}{Nucl. Phys. Proc. Suppl.}
  \textbf{\bibinfo{volume}{168}}, \bibinfo{pages}{296}.

\bibitem[{Antonello \emph{et~al.}(2009)\citenamefont{Antonello}
  \emph{et~al.}}]{1pi-fsi-models-2}
\bibinfo{author}{\bibnamefont{Antonello}, \bibfnamefont{M.}}, \emph{et~al.},
  \bibinfo{year}{2009}, \bibinfo{journal}{Acta Phys. Polon}
  \textbf{\bibinfo{volume}{B40}}, \bibinfo{pages}{2519}.

\bibitem[{\citenamefont{Apollonio} \emph{et~al.}(2003)\citenamefont{Apollonio,
  Baldini, Bemporad, Caffau, Cei, DŽclais, Kerret, Dieterle, Etenko, Foresti,
  George, Giannini} \emph{et~al.}}]{bib:CHOOZ}
\bibinfo{author}{\bibnamefont{Apollonio}, \bibfnamefont{M.}},
  \bibinfo{author}{\bibfnamefont{A.}~\bibnamefont{Baldini}},
  \bibinfo{author}{\bibfnamefont{C.}~\bibnamefont{Bemporad}},
  \bibinfo{author}{\bibfnamefont{E.}~\bibnamefont{Caffau}},
  \bibinfo{author}{\bibfnamefont{F.}~\bibnamefont{Cei}},
  \bibinfo{author}{\bibfnamefont{Y.}~\bibnamefont{DŽclais}},
  \bibinfo{author}{\bibfnamefont{H.~d.} \bibnamefont{Kerret}},
  \bibinfo{author}{\bibfnamefont{B.}~\bibnamefont{Dieterle}},
  \bibinfo{author}{\bibfnamefont{A.}~\bibnamefont{Etenko}},
  \bibinfo{author}{\bibfnamefont{L.}~\bibnamefont{Foresti}},
  \bibinfo{author}{\bibfnamefont{J.}~\bibnamefont{George}},
  \bibinfo{author}{\bibfnamefont{G.}~\bibnamefont{Giannini}}, \emph{et~al.},
  \bibinfo{year}{2003}, \bibinfo{journal}{The European Physical Journal C -
  Particles and Fields} \textbf{\bibinfo{volume}{27}}, \bibinfo{pages}{331},
  ISSN \bibinfo{issn}{1434-6044}, \bibinfo{note}{10.1140/epjc/s2002-01127-9},
  \urlprefix\url{http://dx.doi.org/10.1140/epjc/s2002-01127-9}.

\bibitem[{Arbuzov \emph{et~al.}(2005)\citenamefont{Arbuzov}
  \emph{et~al.}}]{radcorr-3}
\bibinfo{author}{\bibnamefont{Arbuzov}, \bibfnamefont{A.~B.}}, \emph{et~al.},
  \bibinfo{year}{2005}, \bibinfo{journal}{JHEP 078.}
  \textbf{\bibinfo{volume}{0506}}.

\bibitem[{\citenamefont{Armbruster}
  \emph{et~al.}(1998)\citenamefont{Armbruster, Blair, Bodmann, Booth, Drexlin,
  Eberhard, Edgington, Eichner, Eitel, Finckh, Gemmeke, Hš§l}
  \emph{et~al.}}]{Armbruster199815}
\bibinfo{author}{\bibnamefont{Armbruster}, \bibfnamefont{B.}},
  \bibinfo{author}{\bibfnamefont{I.}~\bibnamefont{Blair}},
  \bibinfo{author}{\bibfnamefont{B.~A.} \bibnamefont{Bodmann}},
  \bibinfo{author}{\bibfnamefont{N.~E.} \bibnamefont{Booth}},
  \bibinfo{author}{\bibfnamefont{G.}~\bibnamefont{Drexlin}},
  \bibinfo{author}{\bibfnamefont{V.}~\bibnamefont{Eberhard}},
  \bibinfo{author}{\bibfnamefont{J.~A.} \bibnamefont{Edgington}},
  \bibinfo{author}{\bibfnamefont{C.}~\bibnamefont{Eichner}},
  \bibinfo{author}{\bibfnamefont{K.}~\bibnamefont{Eitel}},
  \bibinfo{author}{\bibfnamefont{E.}~\bibnamefont{Finckh}},
  \bibinfo{author}{\bibfnamefont{H.}~\bibnamefont{Gemmeke}},
  \bibinfo{author}{\bibfnamefont{J.}~\bibnamefont{Hš§l}}, \emph{et~al.},
  \bibinfo{year}{1998}, \bibinfo{journal}{Physics Letters B}
  \textbf{\bibinfo{volume}{423}}(\bibinfo{number}{1-2}), \bibinfo{pages}{15 },
  ISSN \bibinfo{issn}{0370-2693},
  \urlprefix\url{http://www.sciencedirect.com/science/article/B6TVN-3TTV8P8-3/%
2/dea35f415d9c85513635b161d178a5c5}.

\bibitem[{Arpesella \emph{et~al.}(2008)\citenamefont{Arpesella}
  \emph{et~al.}}]{bib:BorexinoResults}
\bibinfo{author}{\bibnamefont{Arpesella}, \bibfnamefont{C.}}, \emph{et~al.}
  (\bibinfo{collaboration}{The Borexino}), \bibinfo{year}{2008},
  \bibinfo{journal}{Phys. Rev. Lett.} \textbf{\bibinfo{volume}{101}},
  \bibinfo{pages}{091302}.

\bibitem[{Aslanides \emph{et~al.}(1999)\citenamefont{Aslanides}
  \emph{et~al.}}]{bib:Antares}
\bibinfo{author}{\bibnamefont{Aslanides}, \bibfnamefont{E.}}, \emph{et~al.}
  (\bibinfo{collaboration}{ANTARES}), \bibinfo{year}{1999},
  \eprint{astro-ph/9907432}.

\bibitem[{Astier \emph{et~al.}(2000)\citenamefont{Astier}
  \emph{et~al.}}]{Astier:2000us}
\bibinfo{author}{\bibnamefont{Astier}, \bibfnamefont{P.}}, \emph{et~al.}
  (\bibinfo{collaboration}{NOMAD Collaboration}), \bibinfo{year}{2000},
  \bibinfo{journal}{Phys.Lett.} \textbf{\bibinfo{volume}{B486}},
  \bibinfo{pages}{35}.

\bibitem[{Astier \emph{et~al.}(2002)\citenamefont{Astier}
  \emph{et~al.}}]{nomad-strange-2}
\bibinfo{author}{\bibnamefont{Astier}, \bibfnamefont{P.}}, \emph{et~al.},
  \bibinfo{year}{2002}, \bibinfo{journal}{Nucl. Phys.}
  \textbf{\bibinfo{volume}{B621}}, \bibinfo{pages}{3}.

\bibitem[{Athanassopoulos \emph{et~al.}(1997)\citenamefont{Athanassopoulos}
  \emph{et~al.}}]{bib:LSNDC12N12_a}
\bibinfo{author}{\bibnamefont{Athanassopoulos}, \bibfnamefont{C.}},
  \emph{et~al.}, \bibinfo{year}{1997}, \bibinfo{journal}{Phys. Rev. C}
  \textbf{\bibinfo{volume}{55}}, \bibinfo{pages}{2080}.

\bibitem[{Athar \emph{et~al.}(2010)\citenamefont{Athar}
  \emph{et~al.}}]{modern-qe-theory-10}
\bibinfo{author}{\bibnamefont{Athar}, \bibfnamefont{M.~S.}}, \emph{et~al.},
  \bibinfo{year}{2010}, \bibinfo{journal}{Eur. Phys. J.}
  \textbf{\bibinfo{volume}{A43}}, \bibinfo{pages}{209}.

\bibitem[{\citenamefont{Aubert} \emph{et~al.}(1974)\citenamefont{Aubert,
  Benvenuti, Cline, Ford, Imlay} \emph{et~al.}}]{Aubert:1974zz}
\bibinfo{author}{\bibnamefont{Aubert}, \bibfnamefont{B.}},
  \bibinfo{author}{\bibfnamefont{D.}~\bibnamefont{Benvenuti}},
  \bibinfo{author}{\bibfnamefont{D.}~\bibnamefont{Cline}},
  \bibinfo{author}{\bibfnamefont{W.}~\bibnamefont{Ford}},
  \bibinfo{author}{\bibfnamefont{R.}~\bibnamefont{Imlay}}, \emph{et~al.},
  \bibinfo{year}{1974}, \bibinfo{journal}{AIP Conf.Proc.}
  \textbf{\bibinfo{volume}{22}}, \bibinfo{pages}{201}.

\bibitem[{\citenamefont{Auerbach} \emph{et~al.}(2001)\citenamefont{Auerbach,
  Burman, Caldwell, Church, Donahue, Fazely, Garvey, Gunasingha, Imlay, Louis,
  Majkic, Malik} \emph{et~al.}}]{PhysRevD.63.112001}
\bibinfo{author}{\bibnamefont{Auerbach}, \bibfnamefont{L.~B.}},
  \bibinfo{author}{\bibfnamefont{R.~L.} \bibnamefont{Burman}},
  \bibinfo{author}{\bibfnamefont{D.~O.} \bibnamefont{Caldwell}},
  \bibinfo{author}{\bibfnamefont{E.~D.} \bibnamefont{Church}},
  \bibinfo{author}{\bibfnamefont{J.~B.} \bibnamefont{Donahue}},
  \bibinfo{author}{\bibfnamefont{A.}~\bibnamefont{Fazely}},
  \bibinfo{author}{\bibfnamefont{G.~T.} \bibnamefont{Garvey}},
  \bibinfo{author}{\bibfnamefont{R.~M.} \bibnamefont{Gunasingha}},
  \bibinfo{author}{\bibfnamefont{R.}~\bibnamefont{Imlay}},
  \bibinfo{author}{\bibfnamefont{W.~C.} \bibnamefont{Louis}},
  \bibinfo{author}{\bibfnamefont{R.}~\bibnamefont{Majkic}},
  \bibinfo{author}{\bibfnamefont{A.}~\bibnamefont{Malik}}, \emph{et~al.}
  (\bibinfo{collaboration}{(LSND Collaboration)}), \bibinfo{year}{2001},
  \bibinfo{journal}{Phys. Rev. D}
  \textbf{\bibinfo{volume}{63}}(\bibinfo{number}{11}), \bibinfo{pages}{112001}.

\bibitem[{\citenamefont{Auerbach} \emph{et~al.}(2002)\citenamefont{Auerbach,
  Burman, Caldwell, Church, Donahue, Fazely, Garvey, Gunasingha, Imlay, Louis,
  Majkic, Malik} \emph{et~al.}}]{PhysRevC.66.015501}
\bibinfo{author}{\bibnamefont{Auerbach}, \bibfnamefont{L.~B.}},
  \bibinfo{author}{\bibfnamefont{R.~L.} \bibnamefont{Burman}},
  \bibinfo{author}{\bibfnamefont{D.~O.} \bibnamefont{Caldwell}},
  \bibinfo{author}{\bibfnamefont{E.~D.} \bibnamefont{Church}},
  \bibinfo{author}{\bibfnamefont{J.~B.} \bibnamefont{Donahue}},
  \bibinfo{author}{\bibfnamefont{A.}~\bibnamefont{Fazely}},
  \bibinfo{author}{\bibfnamefont{G.~T.} \bibnamefont{Garvey}},
  \bibinfo{author}{\bibfnamefont{R.~M.} \bibnamefont{Gunasingha}},
  \bibinfo{author}{\bibfnamefont{R.}~\bibnamefont{Imlay}},
  \bibinfo{author}{\bibfnamefont{W.~C.} \bibnamefont{Louis}},
  \bibinfo{author}{\bibfnamefont{R.}~\bibnamefont{Majkic}},
  \bibinfo{author}{\bibfnamefont{A.}~\bibnamefont{Malik}}, \emph{et~al.}
  (\bibinfo{collaboration}{(LSND Collaboration)}), \bibinfo{year}{2002},
  \bibinfo{journal}{Phys. Rev. C}
  \textbf{\bibinfo{volume}{66}}(\bibinfo{number}{1}), \bibinfo{pages}{015501}.

\bibitem[{Auerbach \emph{et~al.}(2001)\citenamefont{Auerbach}
  \emph{et~al.}}]{bib:LSNDC12N12_b}
\bibinfo{author}{\bibnamefont{Auerbach}, \bibfnamefont{L.~B.}}, \emph{et~al.},
  \bibinfo{year}{2001}, \bibinfo{journal}{Phys. Rev. C}
  \textbf{\bibinfo{volume}{64}}, \bibinfo{pages}{065501}.

\bibitem[{\citenamefont{Auerbach and Klein}(1983)}]{bib:RPA}
\bibinfo{author}{\bibnamefont{Auerbach}, \bibfnamefont{N.}}, and
  \bibinfo{author}{\bibfnamefont{A.}~\bibnamefont{Klein}},
  \bibinfo{year}{1983}, \bibinfo{journal}{Nuclear Physics A}
  \textbf{\bibinfo{volume}{395}}(\bibinfo{number}{1}), \bibinfo{pages}{77 },
  ISSN \bibinfo{issn}{0375-9474},
  \urlprefix\url{http://www.sciencedirect.com/science/article/B6TVB-472PNRC-2R%
/2/2778d56b433f439560ef24bd7ff76c41}.

\bibitem[{\citenamefont{Aufderheide}
  \emph{et~al.}(1994)\citenamefont{Aufderheide, Bloom, Resler, and
  Goodman}}]{bib:Aufderheide1994}
\bibinfo{author}{\bibnamefont{Aufderheide}, \bibfnamefont{M.~B.}},
  \bibinfo{author}{\bibfnamefont{S.~D.} \bibnamefont{Bloom}},
  \bibinfo{author}{\bibfnamefont{D.~A.} \bibnamefont{Resler}}, and
  \bibinfo{author}{\bibfnamefont{C.~D.} \bibnamefont{Goodman}},
  \bibinfo{year}{1994}, \bibinfo{journal}{Phys. Rev. C}
  \textbf{\bibinfo{volume}{49}}(\bibinfo{number}{2}), \bibinfo{pages}{678}.

\bibitem[{Avignone \emph{et~al.}(2000)\citenamefont{Avignone}
  \emph{et~al.}}]{bib:ORLAND}
\bibinfo{author}{\bibnamefont{Avignone}, \bibfnamefont{F.~T.}}, \emph{et~al.},
  \bibinfo{year}{2000}, \bibinfo{journal}{Phys. Atom. Nucl.}
  \textbf{\bibinfo{volume}{63}}, \bibinfo{pages}{1007}.

\bibitem[{\citenamefont{Aynutdinov}
  \emph{et~al.}(2009)\citenamefont{Aynutdinov, Avrorin, Balkanov, Belolaptikov,
  Bogorodsky, Budnev, Danilchenko, Domogatsky, Doroshenko, Dyachok,
  Dzhilkibaev, Fialkovsky} \emph{et~al.}}]{bib:Baikal-1}
\bibinfo{author}{\bibnamefont{Aynutdinov}, \bibfnamefont{V.}},
  \bibinfo{author}{\bibfnamefont{A.}~\bibnamefont{Avrorin}},
  \bibinfo{author}{\bibfnamefont{V.}~\bibnamefont{Balkanov}},
  \bibinfo{author}{\bibfnamefont{I.}~\bibnamefont{Belolaptikov}},
  \bibinfo{author}{\bibfnamefont{D.}~\bibnamefont{Bogorodsky}},
  \bibinfo{author}{\bibfnamefont{N.}~\bibnamefont{Budnev}},
  \bibinfo{author}{\bibfnamefont{I.}~\bibnamefont{Danilchenko}},
  \bibinfo{author}{\bibfnamefont{G.}~\bibnamefont{Domogatsky}},
  \bibinfo{author}{\bibfnamefont{A.}~\bibnamefont{Doroshenko}},
  \bibinfo{author}{\bibfnamefont{A.}~\bibnamefont{Dyachok}},
  \bibinfo{author}{\bibfnamefont{Z.-A.} \bibnamefont{Dzhilkibaev}},
  \bibinfo{author}{\bibfnamefont{S.}~\bibnamefont{Fialkovsky}}, \emph{et~al.},
  \bibinfo{year}{2009}, \bibinfo{journal}{Nuclear Instruments and Methods in
  Physics Research Section A: Accelerators, Spectrometers, Detectors and
  Associated Equipment} \textbf{\bibinfo{volume}{602}}(\bibinfo{number}{1}),
  \bibinfo{pages}{14 }, ISSN \bibinfo{issn}{0168-9002},
  \bibinfo{note}{proceedings of the 3rd International Workshop on a Very Large
  Volume Neutrino Telescope for the Mediterranean Sea},
  \urlprefix\url{http://www.sciencedirect.com/science/article/B6TJM-4V59VT0-3/%
2/0250f6748281a8ebc1728548be54e89e}.

\bibitem[{\citenamefont{Bahcall}(1997)}]{Bahcall:1997eg}
\bibinfo{author}{\bibnamefont{Bahcall}, \bibfnamefont{J.~N.}},
  \bibinfo{year}{1997}, \bibinfo{journal}{Phys. Rev.}
  \textbf{\bibinfo{volume}{C56}}, \bibinfo{pages}{3391}.

\bibitem[{\citenamefont{Baker} \emph{et~al.}(1985)\citenamefont{Baker, Cnops,
  Connolly, Kahn, Kirk} \emph{et~al.}}]{Baker:1985rx}
\bibinfo{author}{\bibnamefont{Baker}, \bibfnamefont{N.}},
  \bibinfo{author}{\bibfnamefont{A.}~\bibnamefont{Cnops}},
  \bibinfo{author}{\bibfnamefont{P.}~\bibnamefont{Connolly}},
  \bibinfo{author}{\bibfnamefont{S.}~\bibnamefont{Kahn}},
  \bibinfo{author}{\bibfnamefont{H.}~\bibnamefont{Kirk}}, \emph{et~al.},
  \bibinfo{year}{1985}, \bibinfo{journal}{Phys.Rev.}
  \textbf{\bibinfo{volume}{D32}}, \bibinfo{pages}{531}.

\bibitem[{\citenamefont{Baker} \emph{et~al.}(1982)\citenamefont{Baker,
  Connolly, Kahn, Murtagh, Palmer} \emph{et~al.}}]{Baker:1982ty}
\bibinfo{author}{\bibnamefont{Baker}, \bibfnamefont{N.}},
  \bibinfo{author}{\bibfnamefont{P.}~\bibnamefont{Connolly}},
  \bibinfo{author}{\bibfnamefont{S.}~\bibnamefont{Kahn}},
  \bibinfo{author}{\bibfnamefont{M.}~\bibnamefont{Murtagh}},
  \bibinfo{author}{\bibfnamefont{R.}~\bibnamefont{Palmer}}, \emph{et~al.},
  \bibinfo{year}{1982}, \bibinfo{journal}{Phys.Rev.}
  \textbf{\bibinfo{volume}{D25}}, \bibinfo{pages}{617}.

\bibitem[{Baker \emph{et~al.}(1981)\citenamefont{Baker}
  \emph{et~al.}}]{kaon-bc-6}
\bibinfo{author}{\bibnamefont{Baker}, \bibfnamefont{N.~J.}}, \emph{et~al.},
  \bibinfo{year}{1981}, \bibinfo{journal}{Phys. Rev.}
  \textbf{\bibinfo{volume}{D24}}, \bibinfo{pages}{2779}.

\bibitem[{Baker \emph{et~al.}(1986)\citenamefont{Baker}
  \emph{et~al.}}]{kaon-bc-11}
\bibinfo{author}{\bibnamefont{Baker}, \bibfnamefont{N.~J.}}, \emph{et~al.},
  \bibinfo{year}{1986}, \bibinfo{journal}{Phys. Rev.}
  \textbf{\bibinfo{volume}{D34}}, \bibinfo{pages}{1251}.

\bibitem[{\citenamefont{Ballagh} \emph{et~al.}(1981)\citenamefont{Ballagh,
  Bingham, Lawry, Lynch, Lys} \emph{et~al.}}]{Ballagh:1981yh}
\bibinfo{author}{\bibnamefont{Ballagh}, \bibfnamefont{H.}},
  \bibinfo{author}{\bibfnamefont{H.}~\bibnamefont{Bingham}},
  \bibinfo{author}{\bibfnamefont{T.}~\bibnamefont{Lawry}},
  \bibinfo{author}{\bibfnamefont{G.}~\bibnamefont{Lynch}},
  \bibinfo{author}{\bibfnamefont{J.}~\bibnamefont{Lys}}, \emph{et~al.},
  \bibinfo{year}{1981}, \bibinfo{journal}{Phys.Rev.}
  \textbf{\bibinfo{volume}{D24}}, \bibinfo{pages}{7}.

\bibitem[{Ballagh \emph{et~al.}(1983)\citenamefont{Ballagh}
  \emph{et~al.}}]{nu-photon}
\bibinfo{author}{\bibnamefont{Ballagh}, \bibfnamefont{H.~C.}}, \emph{et~al.},
  \bibinfo{year}{1983}, \bibinfo{journal}{Phys. Rev. Lett.}
  \textbf{\bibinfo{volume}{50}}, \bibinfo{pages}{1963}.

\bibitem[{\citenamefont{Baranov} \emph{et~al.}(1979)\citenamefont{Baranov,
  Bugorsky, Ivanilov, Kochetkov, Konyushko} \emph{et~al.}}]{Baranov:1978sx}
\bibinfo{author}{\bibnamefont{Baranov}, \bibfnamefont{D.}},
  \bibinfo{author}{\bibfnamefont{A.}~\bibnamefont{Bugorsky}},
  \bibinfo{author}{\bibfnamefont{A.}~\bibnamefont{Ivanilov}},
  \bibinfo{author}{\bibfnamefont{V.}~\bibnamefont{Kochetkov}},
  \bibinfo{author}{\bibfnamefont{V.}~\bibnamefont{Konyushko}}, \emph{et~al.},
  \bibinfo{year}{1979}, \bibinfo{journal}{Phys.Lett.}
  \textbf{\bibinfo{volume}{B81}}, \bibinfo{pages}{255}.

\bibitem[{\citenamefont{Barbaro} \emph{et~al.}(2011)\citenamefont{Barbaro,
  Amaro, Caballero, Donnelly, Udias} \emph{et~al.}}]{Barbaro:2011st}
\bibinfo{author}{\bibnamefont{Barbaro}, \bibfnamefont{M.}},
  \bibinfo{author}{\bibfnamefont{J.}~\bibnamefont{Amaro}},
  \bibinfo{author}{\bibfnamefont{J.}~\bibnamefont{Caballero}},
  \bibinfo{author}{\bibfnamefont{T.}~\bibnamefont{Donnelly}},
  \bibinfo{author}{\bibfnamefont{J.}~\bibnamefont{Udias}}, \emph{et~al.},
  \bibinfo{year}{2011}, \eprint{1110.4739}.

\bibitem[{\citenamefont{Bardin and Dokuchaeva}(1986)}]{radcorr-1}
\bibinfo{author}{\bibnamefont{Bardin}, \bibfnamefont{D.~Y.}}, and
  \bibinfo{author}{\bibfnamefont{V.~A.} \bibnamefont{Dokuchaeva}},
  \bibinfo{year}{1986}, \bibinfo{journal}{preprint JINR-E-26;} ,
  \bibinfo{pages}{2}.

\bibitem[{\citenamefont{Barish} \emph{et~al.}(1980)\citenamefont{Barish, Brock,
  Engler, Kikuchi, Kraemer} \emph{et~al.}}]{Barish:1979ny}
\bibinfo{author}{\bibnamefont{Barish}, \bibfnamefont{S.}},
  \bibinfo{author}{\bibfnamefont{R.}~\bibnamefont{Brock}},
  \bibinfo{author}{\bibfnamefont{A.}~\bibnamefont{Engler}},
  \bibinfo{author}{\bibfnamefont{T.}~\bibnamefont{Kikuchi}},
  \bibinfo{author}{\bibfnamefont{R.}~\bibnamefont{Kraemer}}, \emph{et~al.},
  \bibinfo{year}{1980}, \bibinfo{journal}{Phys.Lett.}
  \textbf{\bibinfo{volume}{B91}}, \bibinfo{pages}{161}.

\bibitem[{\citenamefont{Barish} \emph{et~al.}(1974)\citenamefont{Barish,
  Derrick, Hyman, Schreiner, Singer} \emph{et~al.}}]{Barish:1974ye}
\bibinfo{author}{\bibnamefont{Barish}, \bibfnamefont{S.}},
  \bibinfo{author}{\bibfnamefont{M.}~\bibnamefont{Derrick}},
  \bibinfo{author}{\bibfnamefont{L.}~\bibnamefont{Hyman}},
  \bibinfo{author}{\bibfnamefont{P.}~\bibnamefont{Schreiner}},
  \bibinfo{author}{\bibfnamefont{R.}~\bibnamefont{Singer}}, \emph{et~al.},
  \bibinfo{year}{1974}, \bibinfo{journal}{Phys.Rev.Lett.}
  \textbf{\bibinfo{volume}{33}}, \bibinfo{pages}{1446}.

\bibitem[{Barish \emph{et~al.}(1974)\citenamefont{Barish}
  \emph{et~al.}}]{nc-cc-single-pi-1}
\bibinfo{author}{\bibnamefont{Barish}, \bibfnamefont{S.~J.}}, \emph{et~al.},
  \bibinfo{year}{1974}, \bibinfo{journal}{Phys. Rev. Lett.}
  \textbf{\bibinfo{volume}{33}}, \bibinfo{pages}{448}.

\bibitem[{\citenamefont{Barnett}(1976)}]{charm}
\bibinfo{author}{\bibnamefont{Barnett}, \bibfnamefont{R.~M.}},
  \bibinfo{year}{1976}, \bibinfo{journal}{Phys. Rev.}
  \textbf{\bibinfo{volume}{D14}}, \bibinfo{pages}{70}.

\bibitem[{\citenamefont{Barsanov} \emph{et~al.}(2007)\citenamefont{Barsanov,
  Dzhanelidze, Zlokazov, Kotelnikov, Markov, Selin, Shakirov, Abdurashitov,
  Veretenkin, Gavrin, Gorbachev, Ibragimova} \emph{et~al.}}]{Barsanov:2007fp}
\bibinfo{author}{\bibnamefont{Barsanov}, \bibfnamefont{V.~I.}},
  \bibinfo{author}{\bibfnamefont{A.~A.} \bibnamefont{Dzhanelidze}},
  \bibinfo{author}{\bibfnamefont{S.~B.} \bibnamefont{Zlokazov}},
  \bibinfo{author}{\bibfnamefont{N.~A.} \bibnamefont{Kotelnikov}},
  \bibinfo{author}{\bibfnamefont{S.~Y.} \bibnamefont{Markov}},
  \bibinfo{author}{\bibfnamefont{V.~V.} \bibnamefont{Selin}},
  \bibinfo{author}{\bibfnamefont{Z.~N.} \bibnamefont{Shakirov}},
  \bibinfo{author}{\bibfnamefont{D.~N.} \bibnamefont{Abdurashitov}},
  \bibinfo{author}{\bibfnamefont{E.~P.} \bibnamefont{Veretenkin}},
  \bibinfo{author}{\bibfnamefont{V.~N.} \bibnamefont{Gavrin}},
  \bibinfo{author}{\bibfnamefont{V.~V.} \bibnamefont{Gorbachev}},
  \bibinfo{author}{\bibfnamefont{T.~V.} \bibnamefont{Ibragimova}},
  \emph{et~al.}, \bibinfo{year}{2007}, \bibinfo{journal}{Physics of Atomic
  Nuclei} \textbf{\bibinfo{volume}{70}}(\bibinfo{number}{2}),
  \bibinfo{pages}{300}.

\bibitem[{Barwick \emph{et~al.}(2006)\citenamefont{Barwick}
  \emph{et~al.}}]{bib:ANITA}
\bibinfo{author}{\bibnamefont{Barwick}, \bibfnamefont{S.~W.}}, \emph{et~al.}
  (\bibinfo{collaboration}{ANITA}), \bibinfo{year}{2006},
  \bibinfo{journal}{Phys. Rev. Lett.} \textbf{\bibinfo{volume}{96}},
  \bibinfo{pages}{171101}.

\bibitem[{Bazarko \emph{et~al.}(1995)\citenamefont{Bazarko}
  \emph{et~al.}}]{Bazarko:1994tt}
\bibinfo{author}{\bibnamefont{Bazarko}, \bibfnamefont{A.}}, \emph{et~al.}
  (\bibinfo{collaboration}{CCFR Collaboration}), \bibinfo{year}{1995},
  \bibinfo{journal}{Z.Phys.} \textbf{\bibinfo{volume}{C65}},
  \bibinfo{pages}{189}.

\bibitem[{\citenamefont{Beacom} \emph{et~al.}(2002)\citenamefont{Beacom, Farr,
  and Vogel}}]{Beacom:2002ix}
\bibinfo{author}{\bibnamefont{Beacom}, \bibfnamefont{J.}},
  \bibinfo{author}{\bibfnamefont{W.}~\bibnamefont{Farr}}, and
  \bibinfo{author}{\bibfnamefont{P.}~\bibnamefont{Vogel}},
  \bibinfo{year}{2002}, \bibinfo{journal}{Physical Review D}
  \textbf{\bibinfo{volume}{66}}(\bibinfo{number}{3}).

\bibitem[{\citenamefont{Beacom and Vogel}(1999)}]{Beacom:1999}
\bibinfo{author}{\bibnamefont{Beacom}, \bibfnamefont{J.}}, and
  \bibinfo{author}{\bibfnamefont{P.}~\bibnamefont{Vogel}},
  \bibinfo{year}{1999}, \bibinfo{journal}{Physical Review D}
  \textbf{\bibinfo{volume}{60}}(\bibinfo{number}{053003}).

\bibitem[{\citenamefont{Beacom and Parke}(2001)}]{bib:Beacom2001}
\bibinfo{author}{\bibnamefont{Beacom}, \bibfnamefont{J.~F.}}, and
  \bibinfo{author}{\bibfnamefont{S.~J.} \bibnamefont{Parke}},
  \bibinfo{year}{2001}, \bibinfo{journal}{Physical Review D}
  \textbf{\bibinfo{volume}{64}}, \bibinfo{pages}{091302}.

\bibitem[{\citenamefont{Belkov and Kopeliovich}(1987)}]{pcac-coh-models}
\bibinfo{author}{\bibnamefont{Belkov}, \bibfnamefont{A.~A.}}, and
  \bibinfo{author}{\bibfnamefont{B.~Z.} \bibnamefont{Kopeliovich}},
  \bibinfo{year}{1987}, \bibinfo{journal}{Sov. J. Nucl. Phys.}
  \textbf{\bibinfo{volume}{46}}, \bibinfo{pages}{499}.

\bibitem[{\citenamefont{Bell} \emph{et~al.}(1978)\citenamefont{Bell, Coffin,
  Diamond, French, Louis} \emph{et~al.}}]{Bell:1978qu}
\bibinfo{author}{\bibnamefont{Bell}, \bibfnamefont{J.}},
  \bibinfo{author}{\bibfnamefont{C.}~\bibnamefont{Coffin}},
  \bibinfo{author}{\bibfnamefont{R.}~\bibnamefont{Diamond}},
  \bibinfo{author}{\bibfnamefont{H.}~\bibnamefont{French}},
  \bibinfo{author}{\bibfnamefont{W.}~\bibnamefont{Louis}}, \emph{et~al.},
  \bibinfo{year}{1978}, \bibinfo{journal}{Phys.Rev.Lett.}
  \textbf{\bibinfo{volume}{41}}, \bibinfo{pages}{1008}.

\bibitem[{\citenamefont{Benhar}(2010)}]{Benhar12}
\bibinfo{author}{\bibnamefont{Benhar}, \bibfnamefont{O.}},
  \bibinfo{year}{2010}, \eprint{1012.2032}.

\bibitem[{\citenamefont{Benhar} \emph{et~al.}(2005)\citenamefont{Benhar,
  Farina, Nakamura, Sakuda, and Seki}}]{Benhar05}
\bibinfo{author}{\bibnamefont{Benhar}, \bibfnamefont{O.}},
  \bibinfo{author}{\bibfnamefont{N.}~\bibnamefont{Farina}},
  \bibinfo{author}{\bibfnamefont{H.}~\bibnamefont{Nakamura}},
  \bibinfo{author}{\bibfnamefont{M.}~\bibnamefont{Sakuda}}, and
  \bibinfo{author}{\bibfnamefont{R.}~\bibnamefont{Seki}}, \bibinfo{year}{2005},
  \bibinfo{journal}{Phys.Rev.} \textbf{\bibinfo{volume}{D72}},
  \bibinfo{pages}{053005}.

\bibitem[{\citenamefont{Benhar and Meloni}(2007)}]{Benhar07}
\bibinfo{author}{\bibnamefont{Benhar}, \bibfnamefont{O.}}, and
  \bibinfo{author}{\bibfnamefont{D.}~\bibnamefont{Meloni}},
  \bibinfo{year}{2007}, \bibinfo{journal}{Nucl.Phys.}
  \textbf{\bibinfo{volume}{A789}}, \bibinfo{pages}{379}.

\bibitem[{\citenamefont{Benhar and Veneziano}(2011)}]{Benhar11}
\bibinfo{author}{\bibnamefont{Benhar}, \bibfnamefont{O.}}, and
  \bibinfo{author}{\bibfnamefont{G.}~\bibnamefont{Veneziano}},
  \bibinfo{year}{2011}, \bibinfo{journal}{Phys.Lett.}
  \textbf{\bibinfo{volume}{B702}}, \bibinfo{pages}{433}.

\bibitem[{\citenamefont{Benvenuti and al.}(1974)}]{Benvenuti74}
\bibinfo{author}{\bibnamefont{Benvenuti}, \bibfnamefont{A.}}, and
  \bibinfo{author}{\bibfnamefont{e.}~\bibnamefont{al.}}, \bibinfo{year}{1974},
  \bibinfo{journal}{Phys.Rev.Lett.} \textbf{\bibinfo{volume}{32}},
  \bibinfo{pages}{800}.

\bibitem[{\citenamefont{Benvenuti} \emph{et~al.}(1978)\citenamefont{Benvenuti,
  Bobisut, Cline, Cooper, Gilchriese} \emph{et~al.}}]{Benvenuti:1978gy}
\bibinfo{author}{\bibnamefont{Benvenuti}, \bibfnamefont{A.}},
  \bibinfo{author}{\bibfnamefont{F.}~\bibnamefont{Bobisut}},
  \bibinfo{author}{\bibfnamefont{D.}~\bibnamefont{Cline}},
  \bibinfo{author}{\bibfnamefont{P.}~\bibnamefont{Cooper}},
  \bibinfo{author}{\bibfnamefont{M.}~\bibnamefont{Gilchriese}}, \emph{et~al.},
  \bibinfo{year}{1978}, \bibinfo{journal}{Phys.Rev.Lett.}
  \textbf{\bibinfo{volume}{41}}, \bibinfo{pages}{1204}.

\bibitem[{\citenamefont{Berestetskii}
  \emph{et~al.}(1974)\citenamefont{Berestetskii, Lifshitz, and
  Pitaevski}}]{Lifshitz:1974}
\bibinfo{author}{\bibnamefont{Berestetskii}, \bibfnamefont{V.~B.}},
  \bibinfo{author}{\bibfnamefont{E.~M.} \bibnamefont{Lifshitz}}, and
  \bibinfo{author}{\bibfnamefont{L.}~\bibnamefont{Pitaevski}},
  \bibinfo{year}{1974}, \emph{\bibinfo{title}{{Relativistic quantum theory,
  Part I}}} (\bibinfo{publisher}{Pergamon Press}).

\bibitem[{\citenamefont{Berezinsky and Gazizov}(1977)}]{bib:Berezinsky}
\bibinfo{author}{\bibnamefont{Berezinsky}, \bibfnamefont{V.~S.}}, and
  \bibinfo{author}{\bibfnamefont{A.~Z.} \bibnamefont{Gazizov}},
  \bibinfo{year}{1977}, \bibinfo{journal}{JETP Lett.}
  \textbf{\bibinfo{volume}{25}}, \bibinfo{pages}{254}.

\bibitem[{Berge \emph{et~al.}(1976)\citenamefont{Berge}
  \emph{et~al.}}]{kaon-bc-3}
\bibinfo{author}{\bibnamefont{Berge}, \bibfnamefont{J.}}, \emph{et~al.},
  \bibinfo{year}{1976}, \bibinfo{journal}{Phys. Rev. Lett.}
  \textbf{\bibinfo{volume}{36}}, \bibinfo{pages}{127}.

\bibitem[{Berge \emph{et~al.}(1978)\citenamefont{Berge}
  \emph{et~al.}}]{kaon-bc-4}
\bibinfo{author}{\bibnamefont{Berge}, \bibfnamefont{J.~P.}}, \emph{et~al.},
  \bibinfo{year}{1978}, \bibinfo{journal}{Phys. Rev.}
  \textbf{\bibinfo{volume}{D18}}, \bibinfo{pages}{1359}.

\bibitem[{\citenamefont{Berger and Sehgal}(2009)}]{pcac-coh-models-4}
\bibinfo{author}{\bibnamefont{Berger}, \bibfnamefont{C.}}, and
  \bibinfo{author}{\bibfnamefont{L.~M.} \bibnamefont{Sehgal}},
  \bibinfo{year}{2009}, \bibinfo{journal}{Phys. Rev.}
  \textbf{\bibinfo{volume}{D79}}, \bibinfo{pages}{053003}.

\bibitem[{Bernard \emph{et~al.}(2002)\citenamefont{Bernard}
  \emph{et~al.}}]{bernard}
\bibinfo{author}{\bibnamefont{Bernard}, \bibfnamefont{V.}}, \emph{et~al.},
  \bibinfo{year}{2002}, \bibinfo{journal}{J. Phys. G.}
  \textbf{\bibinfo{volume}{28}}, \bibinfo{pages}{R1}.

\bibitem[{\citenamefont{Bethe and Peierls}(1934)}]{bib:Bethe}
\bibinfo{author}{\bibnamefont{Bethe}, \bibfnamefont{H.~A.}}, and
  \bibinfo{author}{\bibfnamefont{R.~E.} \bibnamefont{Peierls}},
  \bibinfo{year}{1934}, \bibinfo{journal}{Nature (London)}
  \textbf{\bibinfo{volume}{133}}, \bibinfo{pages}{532}.

\bibitem[{Blietschau \emph{et~al.}(1976)\citenamefont{Blietschau}
  \emph{et~al.}}]{kaon-bc-1}
\bibinfo{author}{\bibnamefont{Blietschau}, \bibfnamefont{J.}}, \emph{et~al.},
  \bibinfo{year}{1976}, \bibinfo{journal}{Phys. Lett.}
  \textbf{\bibinfo{volume}{B60}}, \bibinfo{pages}{207}.

\bibitem[{\citenamefont{Bodek and Budd}(2011)}]{Bodek:2011ps}
\bibinfo{author}{\bibnamefont{Bodek}, \bibfnamefont{A.}}, and
  \bibinfo{author}{\bibfnamefont{H.}~\bibnamefont{Budd}}, \bibinfo{year}{2011},
  \bibinfo{journal}{Eur.Phys.J.} \textbf{\bibinfo{volume}{C71}},
  \bibinfo{pages}{1726}.

\bibitem[{Bodek \emph{et~al.}(2007)\citenamefont{Bodek}
  \emph{et~al.}}]{bodek-ma}
\bibinfo{author}{\bibnamefont{Bodek}, \bibfnamefont{A.}}, \emph{et~al.},
  \bibinfo{year}{2007}, \eprint{0709.3538 [hep-ex]}.

\bibitem[{\citenamefont{Bodmann} \emph{et~al.}(1991)\citenamefont{Bodmann,
  Booth, Burtak, Dodd, Drexlin, Eberhard, Edgington, Finckh, Gemmeke,
  Giorginis, Glombik, Gorringe} \emph{et~al.}}]{Bodmann1991321}
\bibinfo{author}{\bibnamefont{Bodmann}, \bibfnamefont{B.}},
  \bibinfo{author}{\bibfnamefont{N.~E.} \bibnamefont{Booth}},
  \bibinfo{author}{\bibfnamefont{F.}~\bibnamefont{Burtak}},
  \bibinfo{author}{\bibfnamefont{A.}~\bibnamefont{Dodd}},
  \bibinfo{author}{\bibfnamefont{G.}~\bibnamefont{Drexlin}},
  \bibinfo{author}{\bibfnamefont{V.}~\bibnamefont{Eberhard}},
  \bibinfo{author}{\bibfnamefont{J.~A.} \bibnamefont{Edgington}},
  \bibinfo{author}{\bibfnamefont{E.}~\bibnamefont{Finckh}},
  \bibinfo{author}{\bibfnamefont{H.}~\bibnamefont{Gemmeke}},
  \bibinfo{author}{\bibfnamefont{G.}~\bibnamefont{Giorginis}},
  \bibinfo{author}{\bibfnamefont{A.}~\bibnamefont{Glombik}},
  \bibinfo{author}{\bibfnamefont{T.}~\bibnamefont{Gorringe}}, \emph{et~al.},
  \bibinfo{year}{1991}, \bibinfo{journal}{Physics Letters B}
  \textbf{\bibinfo{volume}{267}}(\bibinfo{number}{3}), \bibinfo{pages}{321 },
  ISSN \bibinfo{issn}{0370-2693},
  \urlprefix\url{http://www.sciencedirect.com/science/article/B6TVN-46YKYS6-4J%
D/2/4814a1f8f6dc6c753157ea67bfcd5aa2}.

\bibitem[{\citenamefont{Boehm} \emph{et~al.}(2001)\citenamefont{Boehm,
  Busenitz, Cook, Gratta, Henrikson, Kornis, Lawrence, Lee, McKinny, Miller,
  Novikov, Piepke} \emph{et~al.}}]{bib:Palo}
\bibinfo{author}{\bibnamefont{Boehm}, \bibfnamefont{F.}},
  \bibinfo{author}{\bibfnamefont{J.}~\bibnamefont{Busenitz}},
  \bibinfo{author}{\bibfnamefont{B.}~\bibnamefont{Cook}},
  \bibinfo{author}{\bibfnamefont{G.}~\bibnamefont{Gratta}},
  \bibinfo{author}{\bibfnamefont{H.}~\bibnamefont{Henrikson}},
  \bibinfo{author}{\bibfnamefont{J.}~\bibnamefont{Kornis}},
  \bibinfo{author}{\bibfnamefont{D.}~\bibnamefont{Lawrence}},
  \bibinfo{author}{\bibfnamefont{K.~B.} \bibnamefont{Lee}},
  \bibinfo{author}{\bibfnamefont{K.}~\bibnamefont{McKinny}},
  \bibinfo{author}{\bibfnamefont{L.}~\bibnamefont{Miller}},
  \bibinfo{author}{\bibfnamefont{V.}~\bibnamefont{Novikov}},
  \bibinfo{author}{\bibfnamefont{A.}~\bibnamefont{Piepke}}, \emph{et~al.},
  \bibinfo{year}{2001}, \bibinfo{journal}{Phys. Rev. D}
  \textbf{\bibinfo{volume}{64}}(\bibinfo{number}{11}), \bibinfo{pages}{112001}.

\bibitem[{\citenamefont{Bolognese} \emph{et~al.}(1979)\citenamefont{Bolognese,
  Engel, Guyonnet, and Riester}}]{Bolognese:1979gf}
\bibinfo{author}{\bibnamefont{Bolognese}, \bibfnamefont{T.}},
  \bibinfo{author}{\bibfnamefont{J.}~\bibnamefont{Engel}},
  \bibinfo{author}{\bibfnamefont{J.}~\bibnamefont{Guyonnet}}, and
  \bibinfo{author}{\bibfnamefont{J.}~\bibnamefont{Riester}},
  \bibinfo{year}{1979}, \bibinfo{journal}{Phys.Lett.}
  \textbf{\bibinfo{volume}{B81}}, \bibinfo{pages}{393}.

\bibitem[{Bosetti \emph{et~al.}(1982)\citenamefont{Bosetti}
  \emph{et~al.}}]{kaon-bc-8}
\bibinfo{author}{\bibnamefont{Bosetti}, \bibfnamefont{P.}}, \emph{et~al.},
  \bibinfo{year}{1982}, \bibinfo{journal}{Nucl. Phys.}
  \textbf{\bibinfo{volume}{B209}}, \bibinfo{pages}{29}.

\bibitem[{Brock \emph{et~al.}(1982)\citenamefont{Brock}
  \emph{et~al.}}]{kaon-bc-7}
\bibinfo{author}{\bibnamefont{Brock}, \bibfnamefont{R.}}, \emph{et~al.},
  \bibinfo{year}{1982}, \bibinfo{journal}{Phys. Rev.}
  \textbf{\bibinfo{volume}{D25}}, \bibinfo{pages}{1753}.

\bibitem[{Brunner \emph{et~al.}(1990)\citenamefont{Brunner}
  \emph{et~al.}}]{Brunner:1989kw}
\bibinfo{author}{\bibnamefont{Brunner}, \bibfnamefont{J.}}, \emph{et~al.}
  (\bibinfo{collaboration}{SKAT Collaboration}), \bibinfo{year}{1990},
  \bibinfo{journal}{Z.Phys.} \textbf{\bibinfo{volume}{C45}},
  \bibinfo{pages}{551}.

\bibitem[{\citenamefont{Buras}(1980)}]{Buras:1979yt}
\bibinfo{author}{\bibnamefont{Buras}, \bibfnamefont{A.~J.}},
  \bibinfo{year}{1980}, \bibinfo{journal}{Rev.Mod.Phys.}
  \textbf{\bibinfo{volume}{52}}, \bibinfo{pages}{199}, \bibinfo{note}{academic
  Training Lectures presented at Fermilab, Batavia, Ill., Feb 1979}.

\bibitem[{\citenamefont{Butkevich}(2010)}]{Butkevich10}
\bibinfo{author}{\bibnamefont{Butkevich}, \bibfnamefont{A.}},
  \bibinfo{year}{2010}, \bibinfo{journal}{Phys.Rev.}
  \textbf{\bibinfo{volume}{C82}}, \bibinfo{pages}{055501}.

\bibitem[{\citenamefont{Butkevich} \emph{et~al.}(1988)\citenamefont{Butkevich,
  Krastev, Leonov-Vendrovsky, Zheleznykh, and Kaidalov}}]{Butkevich88}
\bibinfo{author}{\bibnamefont{Butkevich}, \bibfnamefont{A.}},
  \bibinfo{author}{\bibfnamefont{P.}~\bibnamefont{Krastev}},
  \bibinfo{author}{\bibfnamefont{A.}~\bibnamefont{Leonov-Vendrovsky}},
  \bibinfo{author}{\bibfnamefont{I.}~\bibnamefont{Zheleznykh}}, and
  \bibinfo{author}{\bibfnamefont{A.}~\bibnamefont{Kaidalov}},
  \bibinfo{year}{1988}, \bibinfo{journal}{Z.Phys.}
  \textbf{\bibinfo{volume}{C39}}, \bibinfo{pages}{241}.

\bibitem[{\citenamefont{Butkevich and Perevalov}(2011)}]{Butkevich11}
\bibinfo{author}{\bibnamefont{Butkevich}, \bibfnamefont{A.}}, and
  \bibinfo{author}{\bibfnamefont{D.}~\bibnamefont{Perevalov}},
  \bibinfo{year}{2011}, \bibinfo{journal}{Phys.Rev.}
  \textbf{\bibinfo{volume}{C84}}, \bibinfo{pages}{015501}.

\bibitem[{\citenamefont{Butkevich}(2009)}]{modern-qe-theory-8}
\bibinfo{author}{\bibnamefont{Butkevich}, \bibfnamefont{A.~V.}},
  \bibinfo{year}{2009}, \bibinfo{journal}{Phys. Rev.}
  \textbf{\bibinfo{volume}{C80}}, \bibinfo{pages}{014610}.

\bibitem[{\citenamefont{Butler and Chen}(2000)}]{bib:Chen2000}
\bibinfo{author}{\bibnamefont{Butler}, \bibfnamefont{M.}}, and
  \bibinfo{author}{\bibfnamefont{J.-W.} \bibnamefont{Chen}},
  \bibinfo{year}{2000}, \bibinfo{journal}{Nuclear Physics A}
  \textbf{\bibinfo{volume}{675}}(\bibinfo{number}{3-4}), \bibinfo{pages}{575 },
  ISSN \bibinfo{issn}{0375-9474},
  \urlprefix\url{http://www.sciencedirect.com/science/article/B6TVB-43F33YY-5/%
2/e8d180bde3f4ab3cd5345b8982081300}.

\bibitem[{\citenamefont{Butler} \emph{et~al.}(2001)\citenamefont{Butler, Chen,
  and Kong}}]{bib:Chen2001}
\bibinfo{author}{\bibnamefont{Butler}, \bibfnamefont{M.}},
  \bibinfo{author}{\bibfnamefont{J.-W.} \bibnamefont{Chen}}, and
  \bibinfo{author}{\bibfnamefont{X.}~\bibnamefont{Kong}}, \bibinfo{year}{2001},
  \bibinfo{journal}{Phys. Rev. C} \textbf{\bibinfo{volume}{63}},
  \bibinfo{pages}{035501}.

\bibitem[{\citenamefont{Butler} \emph{et~al.}(2002)\citenamefont{Butler, Chen,
  and Vogel}}]{Butler200226}
\bibinfo{author}{\bibnamefont{Butler}, \bibfnamefont{M.}},
  \bibinfo{author}{\bibfnamefont{J.-W.} \bibnamefont{Chen}}, and
  \bibinfo{author}{\bibfnamefont{P.}~\bibnamefont{Vogel}},
  \bibinfo{year}{2002}, \bibinfo{journal}{Physics Letters B}
  \textbf{\bibinfo{volume}{549}}(\bibinfo{number}{1-2}), \bibinfo{pages}{26 },
  ISSN \bibinfo{issn}{0370-2693},
  \urlprefix\url{http://www.sciencedirect.com/science/article/B6TVN-473FPMH-8/%
2/fce079dc1cc97cacbce67dfc22e96af8}.

\bibitem[{\citenamefont{Cabibbo and Chilton}(1965)}]{cabibbo-hyperon}
\bibinfo{author}{\bibnamefont{Cabibbo}, \bibfnamefont{N.}}, and
  \bibinfo{author}{\bibfnamefont{F.}~\bibnamefont{Chilton}},
  \bibinfo{year}{1965}, \bibinfo{journal}{Phys. Rev.}
  \textbf{\bibinfo{volume}{137}}, \bibinfo{pages}{1628}.

\bibitem[{Carlson \emph{et~al.}(2002)\citenamefont{Carlson}
  \emph{et~al.}}]{carlson}
\bibinfo{author}{\bibnamefont{Carlson}, \bibfnamefont{J.}}, \emph{et~al.},
  \bibinfo{year}{2002}, \bibinfo{journal}{Phys. Rev.}
  \textbf{\bibinfo{volume}{C65}}, \bibinfo{pages}{024002}.

\bibitem[{\citenamefont{Casper}(2002)}]{nuance}
\bibinfo{author}{\bibnamefont{Casper}, \bibfnamefont{D.}},
  \bibinfo{year}{2002}, \bibinfo{journal}{Nucl. Phys. Proc. Suppl.}
  \textbf{\bibinfo{volume}{112}}, \bibinfo{pages}{161}.

\bibitem[{\citenamefont{Caurier} \emph{et~al.}(2005)\citenamefont{Caurier,
  Mart\'\i{}nez-Pinedo, Nowacki, Poves, and Zuker}}]{bib:ShellModel}
\bibinfo{author}{\bibnamefont{Caurier}, \bibfnamefont{E.}},
  \bibinfo{author}{\bibfnamefont{G.}~\bibnamefont{Mart\'\i{}nez-Pinedo}},
  \bibinfo{author}{\bibfnamefont{F.}~\bibnamefont{Nowacki}},
  \bibinfo{author}{\bibfnamefont{A.}~\bibnamefont{Poves}}, and
  \bibinfo{author}{\bibfnamefont{A.~P.} \bibnamefont{Zuker}},
  \bibinfo{year}{2005}, \bibinfo{journal}{Rev. Mod. Phys.}
  \textbf{\bibinfo{volume}{77}}(\bibinfo{number}{2}), \bibinfo{pages}{427}.

\bibitem[{Chekanov \emph{et~al.}(2003)\citenamefont{Chekanov}
  \emph{et~al.}}]{bib:HERA}
\bibinfo{author}{\bibnamefont{Chekanov}, \bibfnamefont{S.}}, \emph{et~al.}
  (\bibinfo{collaboration}{ZEUS}), \bibinfo{year}{2003},
  \bibinfo{journal}{Phys. Rev.} \textbf{\bibinfo{volume}{D67}},
  \bibinfo{pages}{012007}.

\bibitem[{\citenamefont{Cherdack}(2011)}]{Cherdack11}
\bibinfo{author}{\bibnamefont{Cherdack}, \bibfnamefont{D.}},
  \bibinfo{year}{2011}, \bibinfo{journal}{AIP Conf.Proc.}
  \textbf{\bibinfo{volume}{1405}}, \bibinfo{pages}{115}.

\bibitem[{Chukanov \emph{et~al.}(2006)\citenamefont{Chukanov}
  \emph{et~al.}}]{nomad-strange}
\bibinfo{author}{\bibnamefont{Chukanov}, \bibfnamefont{A.}}, \emph{et~al.},
  \bibinfo{year}{2006}, \bibinfo{journal}{Eur. Phys. J.}
  \textbf{\bibinfo{volume}{C46}}, \bibinfo{pages}{69}.

\bibitem[{\citenamefont{Ciafaloni} \emph{et~al.}(2006)\citenamefont{Ciafaloni,
  Colferai, Salam, and Stasto}}]{Ciafaloni:2006yk}
\bibinfo{author}{\bibnamefont{Ciafaloni}, \bibfnamefont{M.}},
  \bibinfo{author}{\bibfnamefont{D.}~\bibnamefont{Colferai}},
  \bibinfo{author}{\bibfnamefont{G.~P.} \bibnamefont{Salam}}, and
  \bibinfo{author}{\bibfnamefont{A.~M.} \bibnamefont{Stasto}},
  \bibinfo{year}{2006}, \bibinfo{journal}{Phys. Lett.}
  \textbf{\bibinfo{volume}{B635}}, \bibinfo{pages}{320}.

\bibitem[{Ciampolillo \emph{et~al.}(1979)\citenamefont{Ciampolillo}
  \emph{et~al.}}]{Ciampolillo:1979wp}
\bibinfo{author}{\bibnamefont{Ciampolillo}, \bibfnamefont{S.}}, \emph{et~al.}
  (\bibinfo{collaboration}{Gargamelle Neutrino Propane Collaboration,
  Aachen-Brussels-CERN-Ecole Poly-Orsay-Padua Collaboration}),
  \bibinfo{year}{1979}, \bibinfo{journal}{Phys.Lett.}
  \textbf{\bibinfo{volume}{B84}}, \bibinfo{pages}{281}.

\bibitem[{\citenamefont{Cleveland} \emph{et~al.}(1998)\citenamefont{Cleveland,
  Daily, Raymond~Davis, Distel, Lande, Lee, Wildenhain, and
  Ullman}}]{bib:Davis}
\bibinfo{author}{\bibnamefont{Cleveland}, \bibfnamefont{B.~T.}},
  \bibinfo{author}{\bibfnamefont{T.}~\bibnamefont{Daily}},
  \bibinfo{author}{\bibfnamefont{J.}~\bibnamefont{Raymond~Davis}},
  \bibinfo{author}{\bibfnamefont{J.~R.} \bibnamefont{Distel}},
  \bibinfo{author}{\bibfnamefont{K.}~\bibnamefont{Lande}},
  \bibinfo{author}{\bibfnamefont{C.~K.} \bibnamefont{Lee}},
  \bibinfo{author}{\bibfnamefont{P.~S.} \bibnamefont{Wildenhain}}, and
  \bibinfo{author}{\bibfnamefont{J.}~\bibnamefont{Ullman}},
  \bibinfo{year}{1998}, \bibinfo{journal}{The Astrophysical Journal}
  \textbf{\bibinfo{volume}{496}}(\bibinfo{number}{1}), \bibinfo{pages}{505},
  \urlprefix\url{http://stacks.iop.org/0004-637X/496/i=1/a=505}.

\bibitem[{\citenamefont{Cocco} \emph{et~al.}(2007)\citenamefont{Cocco, Mangano,
  and Messina}}]{bib:Cocco}
\bibinfo{author}{\bibnamefont{Cocco}, \bibfnamefont{A.}},
  \bibinfo{author}{\bibfnamefont{G.}~\bibnamefont{Mangano}}, and
  \bibinfo{author}{\bibfnamefont{M.}~\bibnamefont{Messina}},
  \bibinfo{year}{2007}, \bibinfo{journal}{Arxiv preprint hep-ph/0703075} .

\bibitem[{\citenamefont{Conrad} \emph{et~al.}(1998)\citenamefont{Conrad,
  Shaevitz, and Bolton}}]{dis-rmp}
\bibinfo{author}{\bibnamefont{Conrad}, \bibfnamefont{J.~M.}},
  \bibinfo{author}{\bibfnamefont{M.~S.} \bibnamefont{Shaevitz}}, and
  \bibinfo{author}{\bibfnamefont{T.}~\bibnamefont{Bolton}},
  \bibinfo{year}{1998}, \bibinfo{journal}{Rev. Mod. Phys.}
  \textbf{\bibinfo{volume}{70}}, \bibinfo{pages}{1341}.

\bibitem[{\citenamefont{Cooper-Sarkar and Sarkar}(2008)}]{bib:Sarkar2008}
\bibinfo{author}{\bibnamefont{Cooper-Sarkar}, \bibfnamefont{A.}}, and
  \bibinfo{author}{\bibfnamefont{S.}~\bibnamefont{Sarkar}},
  \bibinfo{year}{2008}, \bibinfo{journal}{Journal of High Energy Physics}
  \textbf{\bibinfo{volume}{2008}}(\bibinfo{number}{01}), \bibinfo{pages}{075},
  \urlprefix\url{http://stacks.iop.org/1126-6708/2008/i=01/a=075}.

\bibitem[{Coteus \emph{et~al.}(1981)\citenamefont{Coteus}
  \emph{et~al.}}]{nc-elastic-2}
\bibinfo{author}{\bibnamefont{Coteus}, \bibfnamefont{P.}}, \emph{et~al.},
  \bibinfo{year}{1981}, \bibinfo{journal}{Phys. Rev.}
  \textbf{\bibinfo{volume}{D24}}, \bibinfo{pages}{1420}.

\bibitem[{\citenamefont{Cowan} \emph{et~al.}(1956)\citenamefont{Cowan, Reines,
  Harrison, Kruse, and McGuire}}]{Cowan20071956}
\bibinfo{author}{\bibnamefont{Cowan}, \bibfnamefont{C.~L.}},
  \bibinfo{author}{\bibfnamefont{F.}~\bibnamefont{Reines}},
  \bibinfo{author}{\bibfnamefont{F.~B.} \bibnamefont{Harrison}},
  \bibinfo{author}{\bibfnamefont{H.~W.} \bibnamefont{Kruse}}, and
  \bibinfo{author}{\bibfnamefont{A.~D.} \bibnamefont{McGuire}},
  \bibinfo{year}{1956}, \bibinfo{journal}{Science}
  \textbf{\bibinfo{volume}{124}}(\bibinfo{number}{3212}), \bibinfo{pages}{103},
  \urlprefix\url{http://www.sciencemag.org/content/124/3212/103.short}.

\bibitem[{\citenamefont{De~Rujula} \emph{et~al.}(1979)\citenamefont{De~Rujula,
  Petronzio, and Savoy-Navarro}}]{DeRujula:1979jj}
\bibinfo{author}{\bibnamefont{De~Rujula}, \bibfnamefont{A.}},
  \bibinfo{author}{\bibfnamefont{R.}~\bibnamefont{Petronzio}}, and
  \bibinfo{author}{\bibfnamefont{A.}~\bibnamefont{Savoy-Navarro}},
  \bibinfo{year}{1979}, \bibinfo{journal}{Nucl.Phys.}
  \textbf{\bibinfo{volume}{B154}}, \bibinfo{pages}{394}.

\bibitem[{\citenamefont{Declais} \emph{et~al.}(1994)\citenamefont{Declais,
  de~Kerret, Lefivre, Obolensky, Etenko, Kozlov, Machulin, Martemianov,
  Mikaelyan, Skorokhvatov, Sukhotin, and Vyrodov}}]{bib:Bugey94}
\bibinfo{author}{\bibnamefont{Declais}, \bibfnamefont{Y.}},
  \bibinfo{author}{\bibfnamefont{H.}~\bibnamefont{de~Kerret}},
  \bibinfo{author}{\bibfnamefont{B.}~\bibnamefont{Lefivre}},
  \bibinfo{author}{\bibfnamefont{M.}~\bibnamefont{Obolensky}},
  \bibinfo{author}{\bibfnamefont{A.}~\bibnamefont{Etenko}},
  \bibinfo{author}{\bibfnamefont{Y.}~\bibnamefont{Kozlov}},
  \bibinfo{author}{\bibfnamefont{I.}~\bibnamefont{Machulin}},
  \bibinfo{author}{\bibfnamefont{V.}~\bibnamefont{Martemianov}},
  \bibinfo{author}{\bibfnamefont{L.}~\bibnamefont{Mikaelyan}},
  \bibinfo{author}{\bibfnamefont{M.}~\bibnamefont{Skorokhvatov}},
  \bibinfo{author}{\bibfnamefont{S.}~\bibnamefont{Sukhotin}}, and
  \bibinfo{author}{\bibfnamefont{V.}~\bibnamefont{Vyrodov}},
  \bibinfo{year}{1994}, \bibinfo{journal}{Physics Letters B}
  \textbf{\bibinfo{volume}{338}}(\bibinfo{number}{2-3}), \bibinfo{pages}{383 },
  ISSN \bibinfo{issn}{0370-2693},
  \urlprefix\url{http://www.sciencedirect.com/science/article/B6TVN-4729D5F-1H%
/2/b7bd28e082d3b1d8d629883fdf168ef2}.

\bibitem[{Deden \emph{et~al.}(1975)\citenamefont{Deden}
  \emph{et~al.}}]{kaon-bc-2}
\bibinfo{author}{\bibnamefont{Deden}, \bibfnamefont{H.}}, \emph{et~al.},
  \bibinfo{year}{1975}, \bibinfo{journal}{Phys. Lett.}
  \textbf{\bibinfo{volume}{B58}}, \bibinfo{pages}{361}.

\bibitem[{\citenamefont{Deniz and Wong}(2008)}]{Deniz:2008rw}
\bibinfo{author}{\bibnamefont{Deniz}, \bibfnamefont{M.}}, and
  \bibinfo{author}{\bibfnamefont{H.~T.} \bibnamefont{Wong}}
  (\bibinfo{collaboration}{TEXONO}), \bibinfo{year}{2008}, \eprint{0810.0809}.

\bibitem[{Deniz \emph{et~al.}(2010)\citenamefont{Deniz}
  \emph{et~al.}}]{Deniz:2009mu}
\bibinfo{author}{\bibnamefont{Deniz}, \bibfnamefont{M.}}, \emph{et~al.}
  (\bibinfo{collaboration}{TEXONO}), \bibinfo{year}{2010},
  \bibinfo{journal}{Phys. Rev.} \textbf{\bibinfo{volume}{D81}},
  \bibinfo{pages}{072001}.

\bibitem[{DeProspo \emph{et~al.}(1994)\citenamefont{DeProspo}
  \emph{et~al.}}]{kaon-bc-15}
\bibinfo{author}{\bibnamefont{DeProspo}, \bibfnamefont{D.}}, \emph{et~al.},
  \bibinfo{year}{1994}, \bibinfo{journal}{Phys. Rev.}
  \textbf{\bibinfo{volume}{D50}}, \bibinfo{pages}{6691}.

\bibitem[{Derrick \emph{et~al.}(1981)\citenamefont{Derrick}
  \emph{et~al.}}]{nc-cc-single-pi-2}
\bibinfo{author}{\bibnamefont{Derrick}, \bibfnamefont{M.}}, \emph{et~al.},
  \bibinfo{year}{1981}, \bibinfo{journal}{Phys. Rev.}
  \textbf{\bibinfo{volume}{D23}}, \bibinfo{pages}{569}.

\bibitem[{\citenamefont{Dewan}(1981)}]{single-k-models-4}
\bibinfo{author}{\bibnamefont{Dewan}, \bibfnamefont{H.~K.}},
  \bibinfo{year}{1981}, \bibinfo{journal}{Phys. Rev.}
  \textbf{\bibinfo{volume}{D24}}, \bibinfo{pages}{2369}.

\bibitem[{\citenamefont{DeYoung}(2011)}]{bib:Tyce_pc}
\bibinfo{author}{\bibnamefont{DeYoung}, \bibfnamefont{T.}},
  \bibinfo{year}{2011}, \bibinfo{note}{private communication}.

\bibitem[{Diener \emph{et~al.}(2004)\citenamefont{Diener}
  \emph{et~al.}}]{radcorr-2}
\bibinfo{author}{\bibnamefont{Diener}, \bibfnamefont{K.-P.~O.}}, \emph{et~al.},
  \bibinfo{year}{2004}, \bibinfo{journal}{Phys. Rev.}
  \textbf{\bibinfo{volume}{D69}}, \bibinfo{pages}{073005}.

\bibitem[{\citenamefont{Distel} \emph{et~al.}(2003)\citenamefont{Distel,
  Cleveland, Lande, Lee, Wildenhain, Allen, and Burman}}]{bib:Distel2003}
\bibinfo{author}{\bibnamefont{Distel}, \bibfnamefont{J.~R.}},
  \bibinfo{author}{\bibfnamefont{B.~T.} \bibnamefont{Cleveland}},
  \bibinfo{author}{\bibfnamefont{K.}~\bibnamefont{Lande}},
  \bibinfo{author}{\bibfnamefont{C.~K.} \bibnamefont{Lee}},
  \bibinfo{author}{\bibfnamefont{P.~S.} \bibnamefont{Wildenhain}},
  \bibinfo{author}{\bibfnamefont{G.~E.} \bibnamefont{Allen}}, and
  \bibinfo{author}{\bibfnamefont{R.~L.} \bibnamefont{Burman}},
  \bibinfo{year}{2003}, \bibinfo{journal}{Phys. Rev. C}
  \textbf{\bibinfo{volume}{68}}(\bibinfo{number}{5}), \bibinfo{pages}{054613}.

\bibitem[{\citenamefont{Dobrescu and Ellis}(2004)}]{Dobrescu:2003ta}
\bibinfo{author}{\bibnamefont{Dobrescu}, \bibfnamefont{B.~A.}}, and
  \bibinfo{author}{\bibfnamefont{R.}~\bibnamefont{Ellis}},
  \bibinfo{year}{2004}, \bibinfo{journal}{Phys.Rev.}
  \textbf{\bibinfo{volume}{D69}}, \bibinfo{pages}{114014}.

\bibitem[{\citenamefont{Dokshitzer}(1977)}]{Dokshitzer:1977sg}
\bibinfo{author}{\bibnamefont{Dokshitzer}, \bibfnamefont{Y.~L.}},
  \bibinfo{year}{1977}, \bibinfo{journal}{Sov. Phys. JETP}
  \textbf{\bibinfo{volume}{46}}, \bibinfo{pages}{641}.

\bibitem[{\citenamefont{Donnelly} \emph{et~al.}(1974)\citenamefont{Donnelly,
  Hitlin, Schwartz, Walecka, and Wiesner}}]{Donnelly19748}
\bibinfo{author}{\bibnamefont{Donnelly}, \bibfnamefont{T.}},
  \bibinfo{author}{\bibfnamefont{D.}~\bibnamefont{Hitlin}},
  \bibinfo{author}{\bibfnamefont{M.}~\bibnamefont{Schwartz}},
  \bibinfo{author}{\bibfnamefont{J.}~\bibnamefont{Walecka}}, and
  \bibinfo{author}{\bibfnamefont{S.}~\bibnamefont{Wiesner}},
  \bibinfo{year}{1974}, \bibinfo{journal}{Physics Letters B}
  \textbf{\bibinfo{volume}{49}}(\bibinfo{number}{1}), \bibinfo{pages}{8 }, ISSN
  \bibinfo{issn}{0370-2693},
  \urlprefix\url{http://www.sciencedirect.com/science/article/pii/037026937490%
567X}.

\bibitem[{\citenamefont{Donnelly and Peccei}(1979)}]{bib:Peccei}
\bibinfo{author}{\bibnamefont{Donnelly}, \bibfnamefont{T.}}, and
  \bibinfo{author}{\bibfnamefont{R.}~\bibnamefont{Peccei}},
  \bibinfo{year}{1979}, \bibinfo{journal}{Phys. Rep.}
  \textbf{\bibinfo{volume}{50}}(\bibinfo{number}{1}), \bibinfo{pages}{1}.

\bibitem[{\citenamefont{Donnelly and Walecka}(1975)}]{bib:Walecka1975}
\bibinfo{author}{\bibnamefont{Donnelly}, \bibfnamefont{T.}}, and
  \bibinfo{author}{\bibfnamefont{J.}~\bibnamefont{Walecka}},
  \bibinfo{year}{1975}, \bibinfo{journal}{Annual Review of Nuclear and Science}
  \textbf{\bibinfo{volume}{25}}, \bibinfo{pages}{329}.

\bibitem[{\citenamefont{Dore}(2011)}]{Dore:2011qe}
\bibinfo{author}{\bibnamefont{Dore}, \bibfnamefont{U.}}, \bibinfo{year}{2011},
  \eprint{1103.4572}.

\bibitem[{\citenamefont{Dorman}(2009)}]{modern-ma-3}
\bibinfo{author}{\bibnamefont{Dorman}, \bibfnamefont{M.}},
  \bibinfo{year}{2009}, \bibinfo{journal}{AIP Conf. Proc.}
  \textbf{\bibinfo{volume}{1189}}, \bibinfo{pages}{133}.

\bibitem[{Drakoulakos \emph{et~al.}(2004)\citenamefont{Drakoulakos}
  \emph{et~al.}}]{minerva}
\bibinfo{author}{\bibnamefont{Drakoulakos}, \bibfnamefont{D.}}, \emph{et~al.},
  \bibinfo{year}{2004}, \eprint{hep-ex/0405002}.

\bibitem[{\citenamefont{Drell and Walecka}(1964)}]{Drell196418}
\bibinfo{author}{\bibnamefont{Drell}, \bibfnamefont{S.~D.}}, and
  \bibinfo{author}{\bibfnamefont{J.~D.} \bibnamefont{Walecka}},
  \bibinfo{year}{1964}, \bibinfo{journal}{Annals of Physics}
  \textbf{\bibinfo{volume}{28}}(\bibinfo{number}{1}), \bibinfo{pages}{18 },
  ISSN \bibinfo{issn}{0003-4916},
  \urlprefix\url{http://www.sciencedirect.com/science/article/B6WB1-4DF4W8Y-5F%
/2/eb1a665e935ebdf37a1afa7a047db7b2}.

\bibitem[{\citenamefont{Dytman}(2009)}]{1pi-fsi-models-1}
\bibinfo{author}{\bibnamefont{Dytman}, \bibfnamefont{S.}},
  \bibinfo{year}{2009}, \bibinfo{journal}{Acta Phys. Polon}
  \textbf{\bibinfo{volume}{B40}}, \bibinfo{pages}{2445}.

\bibitem[{\citenamefont{Efrosinin} \emph{et~al.}(2009)\citenamefont{Efrosinin,
  Kudenko, and Khotjantsev}}]{Efronsinin09}
\bibinfo{author}{\bibnamefont{Efrosinin}, \bibfnamefont{V.}},
  \bibinfo{author}{\bibfnamefont{Y.}~\bibnamefont{Kudenko}}, and
  \bibinfo{author}{\bibfnamefont{A.}~\bibnamefont{Khotjantsev}},
  \bibinfo{year}{2009}, \bibinfo{journal}{Phys.Atom.Nucl.}
  \textbf{\bibinfo{volume}{72}}, \bibinfo{pages}{459}.

\bibitem[{Eichten \emph{et~al.}(1972)\citenamefont{Eichten}
  \emph{et~al.}}]{eichten}
\bibinfo{author}{\bibnamefont{Eichten}, \bibfnamefont{T.}}, \emph{et~al.},
  \bibinfo{year}{1972}, \bibinfo{journal}{Phys. Lett.}
  \textbf{\bibinfo{volume}{40B}}, \bibinfo{pages}{593}.

\bibitem[{\citenamefont{Engel} \emph{et~al.}(1996)\citenamefont{Engel, Kolbe,
  Langanke, and Vogel}}]{bib:Engel}
\bibinfo{author}{\bibnamefont{Engel}, \bibfnamefont{J.}},
  \bibinfo{author}{\bibfnamefont{E.}~\bibnamefont{Kolbe}},
  \bibinfo{author}{\bibfnamefont{K.}~\bibnamefont{Langanke}}, and
  \bibinfo{author}{\bibfnamefont{P.}~\bibnamefont{Vogel}},
  \bibinfo{year}{1996}, \bibinfo{journal}{Phys. Rev. C}
  \textbf{\bibinfo{volume}{54}}(\bibinfo{number}{5}), \bibinfo{pages}{2740}.

\bibitem[{\citenamefont{Engel} \emph{et~al.}(1994)\citenamefont{Engel, Pittel,
  and Vogel}}]{bib:Engel1994}
\bibinfo{author}{\bibnamefont{Engel}, \bibfnamefont{J.}},
  \bibinfo{author}{\bibfnamefont{S.}~\bibnamefont{Pittel}}, and
  \bibinfo{author}{\bibfnamefont{P.}~\bibnamefont{Vogel}},
  \bibinfo{year}{1994}, \bibinfo{journal}{Phys. Rev. C}
  \textbf{\bibinfo{volume}{50}}(\bibinfo{number}{3}), \bibinfo{pages}{1702}.

\bibitem[{Entenberg \emph{et~al.}(1979)\citenamefont{Entenberg}
  \emph{et~al.}}]{nc-elastic-4}
\bibinfo{author}{\bibnamefont{Entenberg}, \bibfnamefont{A.}}, \emph{et~al.},
  \bibinfo{year}{1979}, \bibinfo{journal}{Phys. Rev. Lett.}
  \textbf{\bibinfo{volume}{42}}, \bibinfo{pages}{1198}.

\bibitem[{Erriques \emph{et~al.}(1977)\citenamefont{Erriques}
  \emph{et~al.}}]{erriques}
\bibinfo{author}{\bibnamefont{Erriques}, \bibfnamefont{O.}}, \emph{et~al.},
  \bibinfo{year}{1977}, \bibinfo{journal}{Phys. Lett.}
  \textbf{\bibinfo{volume}{70B}}, \bibinfo{pages}{383}.

\bibitem[{\citenamefont{Espinal and Sanchez}(2007)}]{modern-ma-1}
\bibinfo{author}{\bibnamefont{Espinal}, \bibfnamefont{X.}}, and
  \bibinfo{author}{\bibfnamefont{F.}~\bibnamefont{Sanchez}},
  \bibinfo{year}{2007}, \bibinfo{journal}{AIP Conf. Proc.}
  \textbf{\bibinfo{volume}{967}}, \bibinfo{pages}{117}.

\bibitem[{Ezawa \emph{et~al.}(1975)\citenamefont{Ezawa}
  \emph{et~al.}}]{single-k-models-2}
\bibinfo{author}{\bibnamefont{Ezawa}, \bibfnamefont{Y.}}, \emph{et~al.},
  \bibinfo{year}{1975}, \bibinfo{journal}{Prog. Theor. Phys.}
  \textbf{\bibinfo{volume}{53}}, \bibinfo{pages}{1455}.

\bibitem[{Faissner \emph{et~al.}(1980)\citenamefont{Faissner}
  \emph{et~al.}}]{nc-elastic-3}
\bibinfo{author}{\bibnamefont{Faissner}, \bibfnamefont{H.}}, \emph{et~al.},
  \bibinfo{year}{1980}, \bibinfo{journal}{Phys. Rev.}
  \textbf{\bibinfo{volume}{D21}}, \bibinfo{pages}{555}.

\bibitem[{\citenamefont{Feynman}(1969)}]{qpm}
\bibinfo{author}{\bibnamefont{Feynman}, \bibfnamefont{R.}},
  \bibinfo{year}{1969}, \bibinfo{journal}{Phys. Rev. Lett.}
  \textbf{\bibinfo{volume}{23}}, \bibinfo{pages}{1415}.

\bibitem[{\citenamefont{Feynman and Gell-Mann}(1958)}]{bib:Feynmann1958}
\bibinfo{author}{\bibnamefont{Feynman}, \bibfnamefont{R.~P.}}, and
  \bibinfo{author}{\bibfnamefont{M.}~\bibnamefont{Gell-Mann}},
  \bibinfo{year}{1958}, \bibinfo{journal}{Phys. Rep.}
  \textbf{\bibinfo{volume}{109}}, \bibinfo{pages}{193}.

\bibitem[{\citenamefont{Fogli and Nardulli}(1980)}]{nc-cc-single-pi-3}
\bibinfo{author}{\bibnamefont{Fogli}, \bibfnamefont{G.~L.}}, and
  \bibinfo{author}{\bibfnamefont{G.}~\bibnamefont{Nardulli}},
  \bibinfo{year}{1980}, \bibinfo{journal}{Nucl. Phys.}
  \textbf{\bibinfo{volume}{B165}}, \bibinfo{pages}{162}.

\bibitem[{Formaggio \emph{et~al.}(2001)\citenamefont{Formaggio}
  \emph{et~al.}}]{Formaggio:2001jz}
\bibinfo{author}{\bibnamefont{Formaggio}, \bibfnamefont{J.}}, \emph{et~al.}
  (\bibinfo{collaboration}{NuTeV Collaboration}), \bibinfo{year}{2001},
  \bibinfo{journal}{Phys.Rev.Lett.} \textbf{\bibinfo{volume}{87}},
  \bibinfo{pages}{071803}.

\bibitem[{\citenamefont{Formaggio} \emph{et~al.}(2012)\citenamefont{Formaggio,
  Figueroa-Feliciano, and Anderson}}]{bib:RICOCHET}
\bibinfo{author}{\bibnamefont{Formaggio}, \bibfnamefont{J.~A.}},
  \bibinfo{author}{\bibfnamefont{E.}~\bibnamefont{Figueroa-Feliciano}}, and
  \bibinfo{author}{\bibfnamefont{A.~J.} \bibnamefont{Anderson}},
  \bibinfo{year}{2012}, \bibinfo{journal}{Phys. Rev. D}
  \textbf{\bibinfo{volume}{85}}, \bibinfo{pages}{013009},
  \urlprefix\url{http://link.aps.org/doi/10.1103/PhysRevD.85.013009}.

\bibitem[{\citenamefont{Foudas} \emph{et~al.}(1990)\citenamefont{Foudas,
  Bachmann, Bernstein, Blair, Lefmann} \emph{et~al.}}]{Foudas:1989rk}
\bibinfo{author}{\bibnamefont{Foudas}, \bibfnamefont{C.}},
  \bibinfo{author}{\bibfnamefont{K.}~\bibnamefont{Bachmann}},
  \bibinfo{author}{\bibfnamefont{R.}~\bibnamefont{Bernstein}},
  \bibinfo{author}{\bibfnamefont{R.}~\bibnamefont{Blair}},
  \bibinfo{author}{\bibfnamefont{W.}~\bibnamefont{Lefmann}}, \emph{et~al.},
  \bibinfo{year}{1990}, \bibinfo{journal}{Phys.Rev.Lett.}
  \textbf{\bibinfo{volume}{64}}, \bibinfo{pages}{1207}.

\bibitem[{\citenamefont{Freedman} \emph{et~al.}(1977)\citenamefont{Freedman,
  Schramm, and Tubbs}}]{bib:Freedman}
\bibinfo{author}{\bibnamefont{Freedman}, \bibfnamefont{D.~Z.}},
  \bibinfo{author}{\bibfnamefont{D.~N.} \bibnamefont{Schramm}}, and
  \bibinfo{author}{\bibfnamefont{D.~L.} \bibnamefont{Tubbs}},
  \bibinfo{year}{1977}, \bibinfo{journal}{Annual Review of Nuclear Science}
  \textbf{\bibinfo{volume}{27}}(\bibinfo{number}{1}), \bibinfo{pages}{167},
  \urlprefix\url{http://www.annualreviews.org/doi/abs/10.1146/annurev.ns.27.12%
0177.001123}.

\bibitem[{\citenamefont{Frullani and Mougey}(1984)}]{modern-qe-theory}
\bibinfo{author}{\bibnamefont{Frullani}, \bibfnamefont{S.}}, and
  \bibinfo{author}{\bibfnamefont{J.}~\bibnamefont{Mougey}},
  \bibinfo{year}{1984}, \bibinfo{journal}{Adv. Nucl. Phys.}
  \textbf{\bibinfo{volume}{14}}, \bibinfo{pages}{1}.

\bibitem[{\citenamefont{Fukuda} \emph{et~al.}(1998)\citenamefont{Fukuda,
  Hayakawa, Ichihara, Inoue, Ishihara, Ishino, Itow, Kajita, Kameda, Kasuga,
  Kobayashi, Kobayashi} \emph{et~al.}}]{bib:SuperKsun}
\bibinfo{author}{\bibnamefont{Fukuda}, \bibfnamefont{Y.}},
  \bibinfo{author}{\bibfnamefont{T.}~\bibnamefont{Hayakawa}},
  \bibinfo{author}{\bibfnamefont{E.}~\bibnamefont{Ichihara}},
  \bibinfo{author}{\bibfnamefont{K.}~\bibnamefont{Inoue}},
  \bibinfo{author}{\bibfnamefont{K.}~\bibnamefont{Ishihara}},
  \bibinfo{author}{\bibfnamefont{H.}~\bibnamefont{Ishino}},
  \bibinfo{author}{\bibfnamefont{Y.}~\bibnamefont{Itow}},
  \bibinfo{author}{\bibfnamefont{T.}~\bibnamefont{Kajita}},
  \bibinfo{author}{\bibfnamefont{J.}~\bibnamefont{Kameda}},
  \bibinfo{author}{\bibfnamefont{S.}~\bibnamefont{Kasuga}},
  \bibinfo{author}{\bibfnamefont{K.}~\bibnamefont{Kobayashi}},
  \bibinfo{author}{\bibfnamefont{Y.}~\bibnamefont{Kobayashi}}, \emph{et~al.},
  \bibinfo{year}{1998}, \bibinfo{journal}{Phys. Rev. Lett.}
  \textbf{\bibinfo{volume}{81}}(\bibinfo{number}{6}), \bibinfo{pages}{1158}.

\bibitem[{\citenamefont{Fukugita} \emph{et~al.}(1988)\citenamefont{Fukugita,
  Kohyama, and Kubodera}}]{bib:Fukugita1988}
\bibinfo{author}{\bibnamefont{Fukugita}, \bibfnamefont{M.}},
  \bibinfo{author}{\bibfnamefont{Y.}~\bibnamefont{Kohyama}}, and
  \bibinfo{author}{\bibfnamefont{K.}~\bibnamefont{Kubodera}},
  \bibinfo{year}{1988}, \bibinfo{journal}{Phys. Lett. B}
  \textbf{\bibinfo{volume}{212}}, \bibinfo{pages}{319}.

\bibitem[{\citenamefont{Gallagher} \emph{et~al.}(2011)\citenamefont{Gallagher,
  Garvey, and Zeller}}]{qe-annual-review}
\bibinfo{author}{\bibnamefont{Gallagher}, \bibfnamefont{H.}},
  \bibinfo{author}{\bibfnamefont{G.}~\bibnamefont{Garvey}}, and
  \bibinfo{author}{\bibfnamefont{G.~P.} \bibnamefont{Zeller}},
  \bibinfo{year}{2011}, \bibinfo{journal}{Annual Review}
  \textbf{\bibinfo{volume}{61}}, \bibinfo{pages}{355}.

\bibitem[{\citenamefont{Gandhi} \emph{et~al.}(1996)\citenamefont{Gandhi, Quigg,
  Reno, and Sarcevic}}]{Gandhi199681}
\bibinfo{author}{\bibnamefont{Gandhi}, \bibfnamefont{R.}},
  \bibinfo{author}{\bibfnamefont{C.}~\bibnamefont{Quigg}},
  \bibinfo{author}{\bibfnamefont{M.~H.} \bibnamefont{Reno}}, and
  \bibinfo{author}{\bibfnamefont{I.}~\bibnamefont{Sarcevic}},
  \bibinfo{year}{1996}, \bibinfo{journal}{Astroparticle Physics}
  \textbf{\bibinfo{volume}{5}}(\bibinfo{number}{2}), \bibinfo{pages}{81 }, ISSN
  \bibinfo{issn}{0927-6505},
  \urlprefix\url{http://www.sciencedirect.com/science/article/B6TJ1-3VRVT8H-1/%
2/2b2c37e04d0320d0372846664919ad0e}.

\bibitem[{\citenamefont{Gando} \emph{et~al.}(2011)\citenamefont{Gando, Gando,
  Ichimura, Ikeda, Inoue, Kibe, Kishimoto, Koga, Minekawa, Mitsui, Morikawa,
  Nagai} \emph{et~al.}}]{bib:KamLAND}
\bibinfo{author}{\bibnamefont{Gando}, \bibfnamefont{A.}},
  \bibinfo{author}{\bibfnamefont{Y.}~\bibnamefont{Gando}},
  \bibinfo{author}{\bibfnamefont{K.}~\bibnamefont{Ichimura}},
  \bibinfo{author}{\bibfnamefont{H.}~\bibnamefont{Ikeda}},
  \bibinfo{author}{\bibfnamefont{K.}~\bibnamefont{Inoue}},
  \bibinfo{author}{\bibfnamefont{Y.}~\bibnamefont{Kibe}},
  \bibinfo{author}{\bibfnamefont{Y.}~\bibnamefont{Kishimoto}},
  \bibinfo{author}{\bibfnamefont{M.}~\bibnamefont{Koga}},
  \bibinfo{author}{\bibfnamefont{Y.}~\bibnamefont{Minekawa}},
  \bibinfo{author}{\bibfnamefont{T.}~\bibnamefont{Mitsui}},
  \bibinfo{author}{\bibfnamefont{T.}~\bibnamefont{Morikawa}},
  \bibinfo{author}{\bibfnamefont{N.}~\bibnamefont{Nagai}}, \emph{et~al.}
  (\bibinfo{collaboration}{The KamLAND Collaboration}), \bibinfo{year}{2011},
  \bibinfo{journal}{Phys. Rev. D}
  \textbf{\bibinfo{volume}{83}}(\bibinfo{number}{5}), \bibinfo{pages}{052002}.

\bibitem[{Garvey \emph{et~al.}(1983)\citenamefont{Garvey}
  \emph{et~al.}}]{bnl-e734-update}
\bibinfo{author}{\bibnamefont{Garvey}, \bibfnamefont{G.~T.}}, \emph{et~al.},
  \bibinfo{year}{1983}, \bibinfo{journal}{Phys. Rev.}
  \textbf{\bibinfo{volume}{C48}}, \bibinfo{pages}{761}.

\bibitem[{\citenamefont{Georgi and Politzer}(1976)}]{charm-1}
\bibinfo{author}{\bibnamefont{Georgi}, \bibfnamefont{H.}}, and
  \bibinfo{author}{\bibfnamefont{D.}~\bibnamefont{Politzer}},
  \bibinfo{year}{1976}, \bibinfo{journal}{Phys. Rev.}
  \textbf{\bibinfo{volume}{D14}}, \bibinfo{pages}{1829}.

\bibitem[{\citenamefont{Gerstein and Zeldovich}(1956)}]{bib:Gershtein1956}
\bibinfo{author}{\bibnamefont{Gerstein}, \bibfnamefont{S.}}, and
  \bibinfo{author}{\bibfnamefont{Y.~B.} \bibnamefont{Zeldovich}},
  \bibinfo{year}{1956}, \bibinfo{journal}{Sov. Phys. JETP}
  \textbf{\bibinfo{volume}{2}}, \bibinfo{pages}{576}.

\bibitem[{\citenamefont{Giunti and Kim}(2007)}]{bib:Giunti}
\bibinfo{author}{\bibnamefont{Giunti}, \bibfnamefont{C.}}, and
  \bibinfo{author}{\bibfnamefont{C.~W.} \bibnamefont{Kim}},
  \bibinfo{year}{2007}, \emph{\bibinfo{title}{{Fundamentals of Neutrino Physics
  and Astrophysics}}} (\bibinfo{publisher}{Oxford University Press}),
  \bibinfo{note}{iSBN-9780198508717}.

\bibitem[{\citenamefont{Giusti and Meucci}(2011)}]{Giusti:2011dj}
\bibinfo{author}{\bibnamefont{Giusti}, \bibfnamefont{C.}}, and
  \bibinfo{author}{\bibfnamefont{A.}~\bibnamefont{Meucci}},
  \bibinfo{year}{2011}, \eprint{1110.4005}.

\bibitem[{\citenamefont{Glashow}(1960)}]{bib:Glashow1960}
\bibinfo{author}{\bibnamefont{Glashow}, \bibfnamefont{S.~L.}},
  \bibinfo{year}{1960}, \bibinfo{journal}{Phys. Rev.}
  \textbf{\bibinfo{volume}{118}}, \bibinfo{pages}{316}.

\bibitem[{Goncharov \emph{et~al.}(2001)\citenamefont{Goncharov}
  \emph{et~al.}}]{Goncharov:2001qe}
\bibinfo{author}{\bibnamefont{Goncharov}, \bibfnamefont{M.}}, \emph{et~al.}
  (\bibinfo{collaboration}{NuTeV Collaboration}), \bibinfo{year}{2001},
  \bibinfo{journal}{Phys.Rev.} \textbf{\bibinfo{volume}{D64}},
  \bibinfo{pages}{112006}.

\bibitem[{\citenamefont{Goodman} \emph{et~al.}(1980)\citenamefont{Goodman,
  Goulding, Greenfield, Rapaport, Bainum, Foster, Love, and
  Petrovich}}]{bib:Goodman1980}
\bibinfo{author}{\bibnamefont{Goodman}, \bibfnamefont{C.~D.}},
  \bibinfo{author}{\bibfnamefont{C.~A.} \bibnamefont{Goulding}},
  \bibinfo{author}{\bibfnamefont{M.~B.} \bibnamefont{Greenfield}},
  \bibinfo{author}{\bibfnamefont{J.}~\bibnamefont{Rapaport}},
  \bibinfo{author}{\bibfnamefont{D.~E.} \bibnamefont{Bainum}},
  \bibinfo{author}{\bibfnamefont{C.~C.} \bibnamefont{Foster}},
  \bibinfo{author}{\bibfnamefont{W.~G.} \bibnamefont{Love}}, and
  \bibinfo{author}{\bibfnamefont{F.}~\bibnamefont{Petrovich}},
  \bibinfo{year}{1980}, \bibinfo{journal}{Phys. Rev. Lett.}
  \textbf{\bibinfo{volume}{44}}(\bibinfo{number}{26}), \bibinfo{pages}{1755}.

\bibitem[{\citenamefont{Gorham} \emph{et~al.}(2004)\citenamefont{Gorham,
  Hebert, Liewer, Naudet, Saltzberg, and Williams}}]{bib:GLUE}
\bibinfo{author}{\bibnamefont{Gorham}, \bibfnamefont{P.~W.}},
  \bibinfo{author}{\bibfnamefont{C.~L.} \bibnamefont{Hebert}},
  \bibinfo{author}{\bibfnamefont{K.~M.} \bibnamefont{Liewer}},
  \bibinfo{author}{\bibfnamefont{C.~J.} \bibnamefont{Naudet}},
  \bibinfo{author}{\bibfnamefont{D.}~\bibnamefont{Saltzberg}}, and
  \bibinfo{author}{\bibfnamefont{D.}~\bibnamefont{Williams}},
  \bibinfo{year}{2004}, \bibinfo{journal}{Phys. Rev. Lett.}
  \textbf{\bibinfo{volume}{93}}(\bibinfo{number}{4}), \bibinfo{pages}{041101}.

\bibitem[{\citenamefont{Gottschalk}(1981)}]{Gottschalk:1980rv}
\bibinfo{author}{\bibnamefont{Gottschalk}, \bibfnamefont{T.}},
  \bibinfo{year}{1981}, \bibinfo{journal}{Phys.Rev.}
  \textbf{\bibinfo{volume}{D23}}, \bibinfo{pages}{56}.

\bibitem[{Grabosch \emph{et~al.}(1989)\citenamefont{Grabosch}
  \emph{et~al.}}]{Grabosch:1988gw}
\bibinfo{author}{\bibnamefont{Grabosch}, \bibfnamefont{H.}}, \emph{et~al.}
  (\bibinfo{collaboration}{SKAT Collaboration}), \bibinfo{year}{1989},
  \bibinfo{journal}{Z.Phys.} \textbf{\bibinfo{volume}{C41}},
  \bibinfo{pages}{527}.

\bibitem[{Gran \emph{et~al.}(2006)\citenamefont{Gran}
  \emph{et~al.}}]{modern-ma}
\bibinfo{author}{\bibnamefont{Gran}, \bibfnamefont{R.}}, \emph{et~al.},
  \bibinfo{year}{2006}, \bibinfo{journal}{Phys. Rev.}
  \textbf{\bibinfo{volume}{D74}}, \bibinfo{pages}{052002}.

\bibitem[{Grassler \emph{et~al.}(1982)\citenamefont{Grassler}
  \emph{et~al.}}]{kaon-bc-9}
\bibinfo{author}{\bibnamefont{Grassler}, \bibfnamefont{H.}}, \emph{et~al.},
  \bibinfo{year}{1982}, \bibinfo{journal}{Nucl. Phys.}
  \textbf{\bibinfo{volume}{B194}}, \bibinfo{pages}{1}.

\bibitem[{\citenamefont{Haatuft} \emph{et~al.}(1983)\citenamefont{Haatuft,
  Myklebost, Olsen, Willutzky, and Petitjean}}]{Haatuft:1983it}
\bibinfo{author}{\bibnamefont{Haatuft}, \bibfnamefont{A.}},
  \bibinfo{author}{\bibfnamefont{K.}~\bibnamefont{Myklebost}},
  \bibinfo{author}{\bibfnamefont{J.}~\bibnamefont{Olsen}},
  \bibinfo{author}{\bibfnamefont{M.}~\bibnamefont{Willutzky}}, and
  \bibinfo{author}{\bibfnamefont{P.}~\bibnamefont{Petitjean}}
  (\bibinfo{collaboration}{Bergen-CERN-Strasbourg Collaboration}),
  \bibinfo{year}{1983}, \bibinfo{journal}{Nucl.Phys.}
  \textbf{\bibinfo{volume}{B222}}, \bibinfo{pages}{365}.

\bibitem[{\citenamefont{Hampel} \emph{et~al.}(1998)\citenamefont{Hampel,
  Heusser, Kiko, Kirsten, Laubenstein, Pernicka, Rau, Ršnn, Schlosser, W—jcik,
  v.~Ammon, Ebert} \emph{et~al.}}]{Hampel1998114}
\bibinfo{author}{\bibnamefont{Hampel}, \bibfnamefont{W.}},
  \bibinfo{author}{\bibfnamefont{G.}~\bibnamefont{Heusser}},
  \bibinfo{author}{\bibfnamefont{J.}~\bibnamefont{Kiko}},
  \bibinfo{author}{\bibfnamefont{T.}~\bibnamefont{Kirsten}},
  \bibinfo{author}{\bibfnamefont{M.}~\bibnamefont{Laubenstein}},
  \bibinfo{author}{\bibfnamefont{E.}~\bibnamefont{Pernicka}},
  \bibinfo{author}{\bibfnamefont{W.}~\bibnamefont{Rau}},
  \bibinfo{author}{\bibfnamefont{U.}~\bibnamefont{Ršnn}},
  \bibinfo{author}{\bibfnamefont{C.}~\bibnamefont{Schlosser}},
  \bibinfo{author}{\bibfnamefont{M.}~\bibnamefont{W—jcik}},
  \bibinfo{author}{\bibfnamefont{R.}~\bibnamefont{v.~Ammon}},
  \bibinfo{author}{\bibfnamefont{K.}~\bibnamefont{Ebert}}, \emph{et~al.},
  \bibinfo{year}{1998}, \bibinfo{journal}{Physics Letters B}
  \textbf{\bibinfo{volume}{420}}(\bibinfo{number}{1-2}), \bibinfo{pages}{114 },
  ISSN \bibinfo{issn}{0370-2693},
  \urlprefix\url{http://www.sciencedirect.com/science/article/B6TVN-3VN3VHD-N/%
2/c95067b3390abcb166427346266b34c9}.

\bibitem[{\citenamefont{Hardy and Towner}(1999)}]{hardy:129}
\bibinfo{author}{\bibnamefont{Hardy}, \bibfnamefont{J.~C.}}, and
  \bibinfo{author}{\bibfnamefont{I.~S.} \bibnamefont{Towner}},
  \bibinfo{year}{1999}, \bibinfo{journal}{AIP Conference Proceedings}
  \textbf{\bibinfo{volume}{481}}(\bibinfo{number}{1}), \bibinfo{pages}{129},
  \urlprefix\url{http://link.aip.org/link/?APC/481/129/1}.

\bibitem[{\citenamefont{Harvey} \emph{et~al.}(2007)\citenamefont{Harvey, Hill,
  and Hill}}]{Harvey07}
\bibinfo{author}{\bibnamefont{Harvey}, \bibfnamefont{J.~A.}},
  \bibinfo{author}{\bibfnamefont{C.~T.} \bibnamefont{Hill}}, and
  \bibinfo{author}{\bibfnamefont{R.~J.} \bibnamefont{Hill}},
  \bibinfo{year}{2007}, \bibinfo{journal}{Phys.Rev.Lett.}
  \textbf{\bibinfo{volume}{99}}, \bibinfo{pages}{261601}.

\bibitem[{Hasegawa \emph{et~al.}(2005)\citenamefont{Hasegawa}
  \emph{et~al.}}]{k2k-coh}
\bibinfo{author}{\bibnamefont{Hasegawa}, \bibfnamefont{M.}}, \emph{et~al.},
  \bibinfo{year}{2005}, \bibinfo{journal}{Phys. Rev. Lett.}
  \textbf{\bibinfo{volume}{95}}, \bibinfo{pages}{252301}.

\bibitem[{\citenamefont{Hasert} \emph{et~al.}(1973)\citenamefont{Hasert,
  Faissner, Krenz, Von~Krogh, Lanske} \emph{et~al.}}]{Hasert:1973cr}
\bibinfo{author}{\bibnamefont{Hasert}, \bibfnamefont{F.}},
  \bibinfo{author}{\bibfnamefont{H.}~\bibnamefont{Faissner}},
  \bibinfo{author}{\bibfnamefont{W.}~\bibnamefont{Krenz}},
  \bibinfo{author}{\bibfnamefont{J.}~\bibnamefont{Von~Krogh}},
  \bibinfo{author}{\bibfnamefont{D.}~\bibnamefont{Lanske}}, \emph{et~al.},
  \bibinfo{year}{1973}, \bibinfo{journal}{Phys.Lett.}
  \textbf{\bibinfo{volume}{B46}}, \bibinfo{pages}{121}.

\bibitem[{Hasert \emph{et~al.}(1973)\citenamefont{Hasert}
  \emph{et~al.}}]{Hasert:1973ff}
\bibinfo{author}{\bibnamefont{Hasert}, \bibfnamefont{F.}}, \emph{et~al.}
  (\bibinfo{collaboration}{Gargamelle Neutrino Collaboration}),
  \bibinfo{year}{1973}, \bibinfo{journal}{Phys.Lett.}
  \textbf{\bibinfo{volume}{B46}}, \bibinfo{pages}{138}.

\bibitem[{Hasert \emph{et~al.}(1978)\citenamefont{Hasert}
  \emph{et~al.}}]{kaon-bc-5}
\bibinfo{author}{\bibnamefont{Hasert}, \bibfnamefont{F.~J.}}, \emph{et~al.},
  \bibinfo{year}{1978}, \bibinfo{journal}{Phys. Lett.}
  \textbf{\bibinfo{volume}{B73}}, \bibinfo{pages}{487}.

\bibitem[{\citenamefont{Hawker}(2002)}]{hawker}
\bibinfo{author}{\bibnamefont{Hawker}, \bibfnamefont{E.}},
  \bibinfo{year}{2002}.

\bibitem[{\citenamefont{Haxton}(1998)}]{Haxton1998110}
\bibinfo{author}{\bibnamefont{Haxton}, \bibfnamefont{W.~C.}},
  \bibinfo{year}{1998}, \bibinfo{journal}{Physics Letters B}
  \textbf{\bibinfo{volume}{431}}(\bibinfo{number}{1-2}), \bibinfo{pages}{110 },
  ISSN \bibinfo{issn}{0370-2693},
  \urlprefix\url{http://www.sciencedirect.com/science/article/B6TVN-3VXYC04-K/%
2/9c070d51b43cf47f81bb5bb58940cc32}.

\bibitem[{\citenamefont{Hayes and S}(2000)}]{bib:Towner}
\bibinfo{author}{\bibnamefont{Hayes}, \bibfnamefont{A.~C.}}, and
  \bibinfo{author}{\bibfnamefont{T.~I.} \bibnamefont{S}}, \bibinfo{year}{2000},
  \bibinfo{journal}{Phys. Rev. C} \textbf{\bibinfo{volume}{61}},
  \bibinfo{pages}{044603}.

\bibitem[{\citenamefont{Heger} \emph{et~al.}(2005)\citenamefont{Heger, Kolbe,
  Haxton, Langanke, MARTINEZPINEDO, and WOOSLEY}}]{Heger:2005dj}
\bibinfo{author}{\bibnamefont{Heger}, \bibfnamefont{A.}},
  \bibinfo{author}{\bibfnamefont{E.}~\bibnamefont{Kolbe}},
  \bibinfo{author}{\bibfnamefont{W.}~\bibnamefont{Haxton}},
  \bibinfo{author}{\bibfnamefont{K.}~\bibnamefont{Langanke}},
  \bibinfo{author}{\bibfnamefont{G.}~\bibnamefont{MARTINEZPINEDO}}, and
  \bibinfo{author}{\bibfnamefont{S.}~\bibnamefont{WOOSLEY}},
  \bibinfo{year}{2005}, \bibinfo{journal}{Physics Letters B}
  \textbf{\bibinfo{volume}{606}}(\bibinfo{number}{3-4}), \bibinfo{pages}{258}.

\bibitem[{Hernandez \emph{et~al.}(2009)\citenamefont{Hernandez}
  \emph{et~al.}}]{pcac-coh-models-5}
\bibinfo{author}{\bibnamefont{Hernandez}, \bibfnamefont{E.}}, \emph{et~al.},
  \bibinfo{year}{2009}, \bibinfo{journal}{Phys. Rev.}
  \textbf{\bibinfo{volume}{D80}}, \bibinfo{pages}{013003}.

\bibitem[{Hernandez \emph{et~al.}(2010)\citenamefont{Hernandez}
  \emph{et~al.}}]{delta-coh-models-6}
\bibinfo{author}{\bibnamefont{Hernandez}, \bibfnamefont{E.}}, \emph{et~al.},
  \bibinfo{year}{2010}, \bibinfo{journal}{Phys. Rev.}
  \textbf{\bibinfo{volume}{D82}}, \bibinfo{pages}{077303}.

\bibitem[{\citenamefont{de~los Heros}(2011)}]{bib:IceCube}
\bibinfo{author}{\bibnamefont{de~los Heros}, \bibfnamefont{C.~P.}}
  (\bibinfo{collaboration}{IceCube}), \bibinfo{year}{2011},
  \bibinfo{journal}{Nucl. Instrum. Meth.} \textbf{\bibinfo{volume}{A630}},
  \bibinfo{pages}{119}.

\bibitem[{\citenamefont{Hill}(2011)}]{Hill11}
\bibinfo{author}{\bibnamefont{Hill}, \bibfnamefont{R.~J.}},
  \bibinfo{year}{2011}, \bibinfo{journal}{Phys.Rev.}
  \textbf{\bibinfo{volume}{D84}}, \bibinfo{pages}{017501}.

\bibitem[{\citenamefont{Hiraide}(2009)}]{sb-new-coh}
\bibinfo{author}{\bibnamefont{Hiraide}, \bibfnamefont{K.}},
  \bibinfo{year}{2009}, \bibinfo{journal}{AIP Conf. Proc.}
  \textbf{\bibinfo{volume}{1189}}, \bibinfo{pages}{249}.

\bibitem[{Hiraide \emph{et~al.}(2008)\citenamefont{Hiraide}
  \emph{et~al.}}]{sb-coh-ccpip}
\bibinfo{author}{\bibnamefont{Hiraide}, \bibfnamefont{K.}}, \emph{et~al.},
  \bibinfo{year}{2008}, \bibinfo{journal}{Phys. Rev.}
  \textbf{\bibinfo{volume}{D78}}, \bibinfo{pages}{112004}.

\bibitem[{\citenamefont{'t~Hooft}(1971)}]{'tHooft:1971ht}
\bibinfo{author}{\bibnamefont{'t~Hooft}, \bibfnamefont{G.}},
  \bibinfo{year}{1971}, \bibinfo{journal}{Phys. Lett.}
  \textbf{\bibinfo{volume}{B37}}, \bibinfo{pages}{195}.

\bibitem[{\citenamefont{Hoummada} \emph{et~al.}(1995)\citenamefont{Hoummada,
  Mikou, Avenier, Bagieu, Cavaignac, and Koang}}]{bib:ILL95}
\bibinfo{author}{\bibnamefont{Hoummada}, \bibfnamefont{A.}},
  \bibinfo{author}{\bibfnamefont{S.~L.} \bibnamefont{Mikou}},
  \bibinfo{author}{\bibfnamefont{M.}~\bibnamefont{Avenier}},
  \bibinfo{author}{\bibfnamefont{G.}~\bibnamefont{Bagieu}},
  \bibinfo{author}{\bibfnamefont{J.~F.} \bibnamefont{Cavaignac}}, and
  \bibinfo{author}{\bibfnamefont{D.~H.} \bibnamefont{Koang}},
  \bibinfo{year}{1995}, \bibinfo{journal}{Applied Radiation and Isotopes}
  \textbf{\bibinfo{volume}{46}}(\bibinfo{number}{6-7}), \bibinfo{pages}{449 },
  ISSN \bibinfo{issn}{0969-8043},
  \urlprefix\url{http://www.sciencedirect.com/science/article/B6TJ0-40T9KPN-R/%
2/f69e68516dbe55913487b052752d8998}.

\bibitem[{\citenamefont{Iancu and Venugopalan}(2003)}]{Iancu:2003xm}
\bibinfo{author}{\bibnamefont{Iancu}, \bibfnamefont{E.}}, and
  \bibinfo{author}{\bibfnamefont{R.}~\bibnamefont{Venugopalan}},
  \bibinfo{year}{2003}, \eprint{hep-ph/0303204}.

\bibitem[{\citenamefont{Ikeda}(1964)}]{bib:IkedaSumRule}
\bibinfo{author}{\bibnamefont{Ikeda}, \bibfnamefont{K.}}, \bibinfo{year}{1964},
  \bibinfo{journal}{Progress of Theoretical Physics}
  \textbf{\bibinfo{volume}{31}}(\bibinfo{number}{3}), \bibinfo{pages}{434},
  \urlprefix\url{http://ptp.ipap.jp/link?PTP/31/434/}.

\bibitem[{\citenamefont{Jalilian-Marian and
  Kovchegov}(2006)}]{JalilianMarian:2005jf}
\bibinfo{author}{\bibnamefont{Jalilian-Marian}, \bibfnamefont{J.}}, and
  \bibinfo{author}{\bibfnamefont{Y.~V.} \bibnamefont{Kovchegov}},
  \bibinfo{year}{2006}, \bibinfo{journal}{Prog. Part. Nucl. Phys.}
  \textbf{\bibinfo{volume}{56}}, \bibinfo{pages}{104}.

\bibitem[{\citenamefont{Jenkins and Goldman}(2009)}]{Jenkins09}
\bibinfo{author}{\bibnamefont{Jenkins}, \bibfnamefont{J.}}, and
  \bibinfo{author}{\bibfnamefont{T.}~\bibnamefont{Goldman}},
  \bibinfo{year}{2009}, \bibinfo{journal}{Phys.Rev.}
  \textbf{\bibinfo{volume}{D80}}, \bibinfo{pages}{053005}.

\bibitem[{Jones \emph{et~al.}(1989)\citenamefont{Jones}
  \emph{et~al.}}]{Jones:1989vt}
\bibinfo{author}{\bibnamefont{Jones}, \bibfnamefont{G.}}, \emph{et~al.}
  (\bibinfo{collaboration}{WA21 Collaboration,
  Birmingham-CERN-Imperial-Coll-Munich-Oxford-University Coll Collaboration}),
  \bibinfo{year}{1989}, \bibinfo{journal}{Z.Phys.}
  \textbf{\bibinfo{volume}{C43}}, \bibinfo{pages}{527}.

\bibitem[{Jones \emph{et~al.}(1993)\citenamefont{Jones}
  \emph{et~al.}}]{kaon-bc-14}
\bibinfo{author}{\bibnamefont{Jones}, \bibfnamefont{G.~T.}}, \emph{et~al.},
  \bibinfo{year}{1993}, \bibinfo{journal}{Z. Phys.}
  \textbf{\bibinfo{volume}{C57}}, \bibinfo{pages}{197}.

\bibitem[{\citenamefont{deForest Jr. and Walecka}(1966)}]{bib:deForest1966}
\bibinfo{author}{\bibnamefont{deForest Jr.}, \bibfnamefont{T.}}, and
  \bibinfo{author}{\bibfnamefont{D.}~\bibnamefont{Walecka}},
  \bibinfo{year}{1966}, \bibinfo{journal}{Adv. in Phys.}
  \textbf{\bibinfo{volume}{15}}, \bibinfo{pages}{1}.

\bibitem[{\citenamefont{Juszczak} \emph{et~al.}(2010)\citenamefont{Juszczak,
  Sobczyk, and Zmuda}}]{Juszczak10}
\bibinfo{author}{\bibnamefont{Juszczak}, \bibfnamefont{C.}},
  \bibinfo{author}{\bibfnamefont{J.~T.} \bibnamefont{Sobczyk}}, and
  \bibinfo{author}{\bibfnamefont{J.}~\bibnamefont{Zmuda}},
  \bibinfo{year}{2010}, \bibinfo{journal}{Phys.Rev.}
  \textbf{\bibinfo{volume}{C82}}, \bibinfo{pages}{045502}.

\bibitem[{\citenamefont{Kayis-Topaksu}
  \emph{et~al.}(2011)\citenamefont{Kayis-Topaksu, Onengut, van Dantzig,
  de~Jong, Oldeman} \emph{et~al.}}]{KayisTopaksu:2011mx}
\bibinfo{author}{\bibnamefont{Kayis-Topaksu}, \bibfnamefont{A.}},
  \bibinfo{author}{\bibfnamefont{G.}~\bibnamefont{Onengut}},
  \bibinfo{author}{\bibfnamefont{R.}~\bibnamefont{van Dantzig}},
  \bibinfo{author}{\bibfnamefont{M.}~\bibnamefont{de~Jong}},
  \bibinfo{author}{\bibfnamefont{R.}~\bibnamefont{Oldeman}}, \emph{et~al.},
  \bibinfo{year}{2011}, \bibinfo{journal}{New J.Phys.}
  \textbf{\bibinfo{volume}{13}}, \bibinfo{pages}{093002}.

\bibitem[{Kayis-Topaksu \emph{et~al.}(2005)\citenamefont{Kayis-Topaksu}
  \emph{et~al.}}]{KayisTopaksu:2005je}
\bibinfo{author}{\bibnamefont{Kayis-Topaksu}, \bibfnamefont{A.}}, \emph{et~al.}
  (\bibinfo{collaboration}{CHORUS Collaboration}), \bibinfo{year}{2005},
  \bibinfo{journal}{Phys.Lett.} \textbf{\bibinfo{volume}{B626}},
  \bibinfo{pages}{24}.

\bibitem[{Kayis-Topaksu
  \emph{et~al.}(2008{\natexlab{a}})\citenamefont{Kayis-Topaksu}
  \emph{et~al.}}]{chorus-dimuon}
\bibinfo{author}{\bibnamefont{Kayis-Topaksu}, \bibfnamefont{A.}},
  \emph{et~al.}, \bibinfo{year}{2008}{\natexlab{a}}, \bibinfo{journal}{Nucl.
  Phys.} \textbf{\bibinfo{volume}{B798}}, \bibinfo{pages}{1}.

\bibitem[{Kayis-Topaksu
  \emph{et~al.}(2008{\natexlab{b}})\citenamefont{Kayis-Topaksu}
  \emph{et~al.}}]{Topaksu:2008xp}
\bibinfo{author}{\bibnamefont{Kayis-Topaksu}, \bibfnamefont{A.}}, \emph{et~al.}
  (\bibinfo{collaboration}{CHORUS Collaboration}),
  \bibinfo{year}{2008}{\natexlab{b}}, \bibinfo{journal}{Nucl.Phys.}
  \textbf{\bibinfo{volume}{B798}}, \bibinfo{pages}{1}.

\bibitem[{Kelkar \emph{et~al.}(1997)\citenamefont{Kelkar}
  \emph{et~al.}}]{delta-coh-models}
\bibinfo{author}{\bibnamefont{Kelkar}, \bibfnamefont{N.~G.}}, \emph{et~al.},
  \bibinfo{year}{1997}, \bibinfo{journal}{Phys. Rev.}
  \textbf{\bibinfo{volume}{C55}}, \bibinfo{pages}{1964}.

\bibitem[{\citenamefont{Kim and Primakoff}(1965)}]{Kim:1965zzc}
\bibinfo{author}{\bibnamefont{Kim}, \bibfnamefont{C.~W.}}, and
  \bibinfo{author}{\bibfnamefont{H.}~\bibnamefont{Primakoff}},
  \bibinfo{year}{1965}, \bibinfo{journal}{Phys. Rev.}
  \textbf{\bibinfo{volume}{140}}, \bibinfo{pages}{B566}.

\bibitem[{\citenamefont{Kolbe}
  \emph{et~al.}(1999{\natexlab{a}})\citenamefont{Kolbe, Langanke, and
  Mart\'\i{}nez-Pinedo}}]{bib:KolbeFe}
\bibinfo{author}{\bibnamefont{Kolbe}, \bibfnamefont{E.}},
  \bibinfo{author}{\bibfnamefont{K.}~\bibnamefont{Langanke}}, and
  \bibinfo{author}{\bibfnamefont{G.}~\bibnamefont{Mart\'\i{}nez-Pinedo}},
  \bibinfo{year}{1999}{\natexlab{a}}, \bibinfo{journal}{Phys. Rev. C}
  \textbf{\bibinfo{volume}{60}}(\bibinfo{number}{5}), \bibinfo{pages}{052801}.

\bibitem[{\citenamefont{Kolbe}
  \emph{et~al.}(1999{\natexlab{b}})\citenamefont{Kolbe, Langanke, and
  Vogel}}]{bib:Kolbe99}
\bibinfo{author}{\bibnamefont{Kolbe}, \bibfnamefont{E.}},
  \bibinfo{author}{\bibfnamefont{K.}~\bibnamefont{Langanke}}, and
  \bibinfo{author}{\bibfnamefont{P.}~\bibnamefont{Vogel}},
  \bibinfo{year}{1999}{\natexlab{b}}, \bibinfo{journal}{Nucl. Phys.}
  \textbf{\bibinfo{volume}{A652}}, \bibinfo{pages}{91}.

\bibitem[{\citenamefont{Kopeliovich}(2005)}]{pcac-coh-models-2}
\bibinfo{author}{\bibnamefont{Kopeliovich}, \bibfnamefont{B.~Z.}},
  \bibinfo{year}{2005}, \bibinfo{journal}{Nucl. Proc. Suppl.}
  \textbf{\bibinfo{volume}{139}}, \bibinfo{pages}{219}.

\bibitem[{Kozlov \emph{et~al.}(2000)\citenamefont{Kozlov}
  \emph{et~al.}}]{bib:KrasnoyarskDeut}
\bibinfo{author}{\bibnamefont{Kozlov}, \bibfnamefont{Y.}}, \emph{et~al.},
  \bibinfo{year}{2000}, \bibinfo{journal}{Phys. At. Nucl.}
  \textbf{\bibinfo{volume}{63}}, \bibinfo{pages}{1016}.

\bibitem[{\citenamefont{Krakauer} \emph{et~al.}(1992)\citenamefont{Krakauer,
  Talaga, Allen, Chen, Hausammann, Lee, Mahler, Lu, Wang, Bowles, Burman,
  Carlini} \emph{et~al.}}]{Krakauer1992}
\bibinfo{author}{\bibnamefont{Krakauer}, \bibfnamefont{D.~A.}},
  \bibinfo{author}{\bibfnamefont{R.~L.} \bibnamefont{Talaga}},
  \bibinfo{author}{\bibfnamefont{R.~C.} \bibnamefont{Allen}},
  \bibinfo{author}{\bibfnamefont{H.~H.} \bibnamefont{Chen}},
  \bibinfo{author}{\bibfnamefont{R.}~\bibnamefont{Hausammann}},
  \bibinfo{author}{\bibfnamefont{W.~P.} \bibnamefont{Lee}},
  \bibinfo{author}{\bibfnamefont{H.~J.} \bibnamefont{Mahler}},
  \bibinfo{author}{\bibfnamefont{X.~Q.} \bibnamefont{Lu}},
  \bibinfo{author}{\bibfnamefont{K.~C.} \bibnamefont{Wang}},
  \bibinfo{author}{\bibfnamefont{T.~J.} \bibnamefont{Bowles}},
  \bibinfo{author}{\bibfnamefont{R.~L.} \bibnamefont{Burman}},
  \bibinfo{author}{\bibfnamefont{R.~D.} \bibnamefont{Carlini}}, \emph{et~al.},
  \bibinfo{year}{1992}, \bibinfo{journal}{Phys. Rev. C}
  \textbf{\bibinfo{volume}{45}}(\bibinfo{number}{5}), \bibinfo{pages}{2450}.

\bibitem[{Kravchenko \emph{et~al.}(2003)\citenamefont{Kravchenko}
  \emph{et~al.}}]{bib:RICE}
\bibinfo{author}{\bibnamefont{Kravchenko}, \bibfnamefont{I.}}, \emph{et~al.},
  \bibinfo{year}{2003}, \bibinfo{journal}{Astropart. Phys.}
  \textbf{\bibinfo{volume}{20}}, \bibinfo{pages}{195}.

\bibitem[{Krenz \emph{et~al.}(1978{\natexlab{a}})\citenamefont{Krenz}
  \emph{et~al.}}]{nc-cc-single-pi-4}
\bibinfo{author}{\bibnamefont{Krenz}, \bibfnamefont{W.}}, \emph{et~al.},
  \bibinfo{year}{1978}{\natexlab{a}}, \bibinfo{journal}{Nucl. Phys.}
  \textbf{\bibinfo{volume}{B135}}, \bibinfo{pages}{45}.

\bibitem[{Krenz \emph{et~al.}(1978{\natexlab{b}})\citenamefont{Krenz}
  \emph{et~al.}}]{kaon-bc}
\bibinfo{author}{\bibnamefont{Krenz}, \bibfnamefont{W.}}, \emph{et~al.},
  \bibinfo{year}{1978}{\natexlab{b}}, \bibinfo{journal}{Phys. Lett.}
  \textbf{\bibinfo{volume}{B73}}, \bibinfo{pages}{493}.

\bibitem[{\citenamefont{Kretzer and Reno}(2002)}]{Kretzer:2002fr}
\bibinfo{author}{\bibnamefont{Kretzer}, \bibfnamefont{S.}}, and
  \bibinfo{author}{\bibfnamefont{M.}~\bibnamefont{Reno}}, \bibinfo{year}{2002},
  \bibinfo{journal}{Phys.Rev.} \textbf{\bibinfo{volume}{D66}},
  \bibinfo{pages}{113007}.

\bibitem[{Kullenberg \emph{et~al.}(2012)\citenamefont{Kullenberg}
  \emph{et~al.}}]{Kullenberg12}
\bibinfo{author}{\bibnamefont{Kullenberg}, \bibfnamefont{C.}}, \emph{et~al.}
  (\bibinfo{collaboration}{NOMAD Collaboration}), \bibinfo{year}{2012},
  \bibinfo{journal}{Phys.Lett.} \textbf{\bibinfo{volume}{B706}},
  \bibinfo{pages}{268}.

\bibitem[{Kullenberg \emph{et~al.}(2009)\citenamefont{Kullenberg}
  \emph{et~al.}}]{nomad-coh}
\bibinfo{author}{\bibnamefont{Kullenberg}, \bibfnamefont{C.~T.}},
  \emph{et~al.}, \bibinfo{year}{2009}, \bibinfo{journal}{Phys. Lett.}
  \textbf{\bibinfo{volume}{B682}}, \bibinfo{pages}{177}.

\bibitem[{\citenamefont{Kuraev} \emph{et~al.}(1977)\citenamefont{Kuraev,
  Lipatov, and Fadin}}]{Kuraev:1977fs}
\bibinfo{author}{\bibnamefont{Kuraev}, \bibfnamefont{E.~A.}},
  \bibinfo{author}{\bibfnamefont{L.~N.} \bibnamefont{Lipatov}}, and
  \bibinfo{author}{\bibfnamefont{V.~S.} \bibnamefont{Fadin}},
  \bibinfo{year}{1977}, \bibinfo{journal}{Sov. Phys. JETP}
  \textbf{\bibinfo{volume}{45}}, \bibinfo{pages}{199}.

\bibitem[{\citenamefont{Kuramoto} \emph{et~al.}(1990)\citenamefont{Kuramoto,
  Fukugita, Kohyama, and Kubodera}}]{Kuramoto1990711}
\bibinfo{author}{\bibnamefont{Kuramoto}, \bibfnamefont{T.}},
  \bibinfo{author}{\bibfnamefont{M.}~\bibnamefont{Fukugita}},
  \bibinfo{author}{\bibfnamefont{Y.}~\bibnamefont{Kohyama}}, and
  \bibinfo{author}{\bibfnamefont{K.}~\bibnamefont{Kubodera}},
  \bibinfo{year}{1990}, \bibinfo{journal}{Nuclear Physics A}
  \textbf{\bibinfo{volume}{512}}(\bibinfo{number}{4}), \bibinfo{pages}{711 },
  ISSN \bibinfo{issn}{0375-9474},
  \urlprefix\url{http://www.sciencedirect.com/science/article/B6TVB-473NHXP-25%
/2/cee6237ef1b96fb1007084c5b37e8141}.

\bibitem[{Kurimoto \emph{et~al.}(2010{\natexlab{a}})\citenamefont{Kurimoto}
  \emph{et~al.}}]{sb-ncpi0}
\bibinfo{author}{\bibnamefont{Kurimoto}, \bibfnamefont{Y.}}, \emph{et~al.},
  \bibinfo{year}{2010}{\natexlab{a}}, \bibinfo{journal}{Phys. Rev.}
  \textbf{\bibinfo{volume}{D81}}, \bibinfo{pages}{033004}.

\bibitem[{Kurimoto \emph{et~al.}(2010{\natexlab{b}})\citenamefont{Kurimoto}
  \emph{et~al.}}]{sb-coh-ncpi0}
\bibinfo{author}{\bibnamefont{Kurimoto}, \bibfnamefont{Y.}}, \emph{et~al.},
  \bibinfo{year}{2010}{\natexlab{b}}, \bibinfo{journal}{Phys. Rev.}
  \textbf{\bibinfo{volume}{D81}}, \bibinfo{pages}{111102(R)}.

\bibitem[{\citenamefont{Kurylov} \emph{et~al.}(2002)\citenamefont{Kurylov,
  M.J., and Vogel}}]{bib:Kurylov2002}
\bibinfo{author}{\bibnamefont{Kurylov}, \bibfnamefont{A.}},
  \bibinfo{author}{\bibfnamefont{R.-M.} \bibnamefont{M.J.}}, and
  \bibinfo{author}{\bibfnamefont{P.}~\bibnamefont{Vogel}},
  \bibinfo{year}{2002}, \bibinfo{journal}{Physical Review C}
  \textbf{\bibinfo{volume}{65}}, \bibinfo{pages}{055501}.

\bibitem[{Kuvshinnikov \emph{et~al.}(1991)\citenamefont{Kuvshinnikov}
  \emph{et~al.}}]{bib:ROVNO91}
\bibinfo{author}{\bibnamefont{Kuvshinnikov}, \bibfnamefont{V.}}, \emph{et~al.},
  \bibinfo{year}{1991}, \bibinfo{journal}{JETP}
  \textbf{\bibinfo{volume}{54}}(\bibinfo{number}{N5}), \bibinfo{pages}{259}.

\bibitem[{\citenamefont{Kuzmin and Naumov}(2009)}]{Kuzmin:2008zz}
\bibinfo{author}{\bibnamefont{Kuzmin}, \bibfnamefont{K.}}, and
  \bibinfo{author}{\bibfnamefont{V.}~\bibnamefont{Naumov}},
  \bibinfo{year}{2009}, \bibinfo{journal}{Phys.Atom.Nucl.}
  \textbf{\bibinfo{volume}{72}}, \bibinfo{pages}{1501}.

\bibitem[{Kuzmin \emph{et~al.}(2008)\citenamefont{Kuzmin}
  \emph{et~al.}}]{kuzmin-ma}
\bibinfo{author}{\bibnamefont{Kuzmin}, \bibfnamefont{K.~S.}}, \emph{et~al.},
  \bibinfo{year}{2008}, \bibinfo{journal}{Eur. Phys. J.}
  \textbf{\bibinfo{volume}{C54}}, \bibinfo{pages}{517}.

\bibitem[{\citenamefont{Kwon} \emph{et~al.}(1981)\citenamefont{Kwon, Boehm,
  Hahn, Henrikson, Vuilleumier, Cavaignac, Koang, Vignon, Feilitzsch, and
  M\"ossbauer}}]{bib:ILL81}
\bibinfo{author}{\bibnamefont{Kwon}, \bibfnamefont{H.}},
  \bibinfo{author}{\bibfnamefont{F.}~\bibnamefont{Boehm}},
  \bibinfo{author}{\bibfnamefont{A.~A.} \bibnamefont{Hahn}},
  \bibinfo{author}{\bibfnamefont{H.~E.} \bibnamefont{Henrikson}},
  \bibinfo{author}{\bibfnamefont{J.~L.} \bibnamefont{Vuilleumier}},
  \bibinfo{author}{\bibfnamefont{J.~F.} \bibnamefont{Cavaignac}},
  \bibinfo{author}{\bibfnamefont{D.~H.} \bibnamefont{Koang}},
  \bibinfo{author}{\bibfnamefont{B.}~\bibnamefont{Vignon}},
  \bibinfo{author}{\bibfnamefont{F.~v.} \bibnamefont{Feilitzsch}}, and
  \bibinfo{author}{\bibfnamefont{R.~L.} \bibnamefont{M\"ossbauer}},
  \bibinfo{year}{1981}, \bibinfo{journal}{Phys. Rev. D}
  \textbf{\bibinfo{volume}{24}}(\bibinfo{number}{5}), \bibinfo{pages}{1097}.

\bibitem[{\citenamefont{Lang} \emph{et~al.}(1987)\citenamefont{Lang, Bodek,
  Borcherding, Giokaris, Stockdale} \emph{et~al.}}]{Lang:1986at}
\bibinfo{author}{\bibnamefont{Lang}, \bibfnamefont{K.}},
  \bibinfo{author}{\bibfnamefont{A.}~\bibnamefont{Bodek}},
  \bibinfo{author}{\bibfnamefont{F.}~\bibnamefont{Borcherding}},
  \bibinfo{author}{\bibfnamefont{N.}~\bibnamefont{Giokaris}},
  \bibinfo{author}{\bibfnamefont{I.}~\bibnamefont{Stockdale}}, \emph{et~al.},
  \bibinfo{year}{1987}, \bibinfo{journal}{Z.Phys.}
  \textbf{\bibinfo{volume}{C33}}, \bibinfo{pages}{483}.

\bibitem[{\citenamefont{Langanke} \emph{et~al.}(2004)\citenamefont{Langanke,
  Mart{\'\i}nez-Pinedo, von Neumann-Cosel, and Richter}}]{Langanke:2004ek}
\bibinfo{author}{\bibnamefont{Langanke}, \bibfnamefont{K.}},
  \bibinfo{author}{\bibfnamefont{G.}~\bibnamefont{Mart{\'\i}nez-Pinedo}},
  \bibinfo{author}{\bibfnamefont{P.}~\bibnamefont{von Neumann-Cosel}}, and
  \bibinfo{author}{\bibfnamefont{A.}~\bibnamefont{Richter}},
  \bibinfo{year}{2004}, \bibinfo{journal}{Physical Review Letters}
  \textbf{\bibinfo{volume}{93}}(\bibinfo{number}{20}).

\bibitem[{Lee \emph{et~al.}(1977)\citenamefont{Lee}
  \emph{et~al.}}]{nc-cc-single-pi-5}
\bibinfo{author}{\bibnamefont{Lee}, \bibfnamefont{W.}}, \emph{et~al.},
  \bibinfo{year}{1977}, \bibinfo{journal}{Phys. Rev. Lett.}
  \textbf{\bibinfo{volume}{38}}, \bibinfo{pages}{202}.

\bibitem[{\citenamefont{Lehtinen} \emph{et~al.}(2004)\citenamefont{Lehtinen,
  Gorham, Jacobson, and Roussel-Dupr\'e}}]{bib:FORTE}
\bibinfo{author}{\bibnamefont{Lehtinen}, \bibfnamefont{N.~G.}},
  \bibinfo{author}{\bibfnamefont{P.~W.} \bibnamefont{Gorham}},
  \bibinfo{author}{\bibfnamefont{A.~R.} \bibnamefont{Jacobson}}, and
  \bibinfo{author}{\bibfnamefont{R.~A.} \bibnamefont{Roussel-Dupr\'e}},
  \bibinfo{year}{2004}, \bibinfo{journal}{Phys. Rev. D}
  \textbf{\bibinfo{volume}{69}}(\bibinfo{number}{1}), \bibinfo{pages}{013008}.

\bibitem[{\citenamefont{Leitner} \emph{et~al.}(2009)\citenamefont{Leitner,
  Buss, Alvarez-Ruso, and Mosel}}]{Alvarez-Ruso07}
\bibinfo{author}{\bibnamefont{Leitner}, \bibfnamefont{T.}},
  \bibinfo{author}{\bibfnamefont{O.}~\bibnamefont{Buss}},
  \bibinfo{author}{\bibfnamefont{L.}~\bibnamefont{Alvarez-Ruso}}, and
  \bibinfo{author}{\bibfnamefont{U.}~\bibnamefont{Mosel}},
  \bibinfo{year}{2009}, \bibinfo{journal}{Phys.Rev.}
  \textbf{\bibinfo{volume}{C79}}, \bibinfo{pages}{034601}.

\bibitem[{\citenamefont{Leitner and Mosel}(2009)}]{1pi-fsi-models-3}
\bibinfo{author}{\bibnamefont{Leitner}, \bibfnamefont{T.}}, and
  \bibinfo{author}{\bibfnamefont{U.}~\bibnamefont{Mosel}},
  \bibinfo{year}{2009}, \bibinfo{journal}{Phys. Rev.}
  \textbf{\bibinfo{volume}{C79}}, \bibinfo{pages}{038501}.

\bibitem[{\citenamefont{Leitner and
  Mosel}(2010{\natexlab{a}})}]{1pi-fsi-models-4}
\bibinfo{author}{\bibnamefont{Leitner}, \bibfnamefont{T.}}, and
  \bibinfo{author}{\bibfnamefont{U.}~\bibnamefont{Mosel}},
  \bibinfo{year}{2010}{\natexlab{a}}, \bibinfo{journal}{Phys. Rev.}
  \textbf{\bibinfo{volume}{C82}}, \bibinfo{pages}{035503}.

\bibitem[{\citenamefont{Leitner and
  Mosel}(2010{\natexlab{b}})}]{1pi-fsi-models-5}
\bibinfo{author}{\bibnamefont{Leitner}, \bibfnamefont{T.}}, and
  \bibinfo{author}{\bibfnamefont{U.}~\bibnamefont{Mosel}},
  \bibinfo{year}{2010}{\natexlab{b}}, \bibinfo{journal}{Phys. Rev.}
  \textbf{\bibinfo{volume}{C81}}, \bibinfo{pages}{064614}.

\bibitem[{Leitner \emph{et~al.}(2006)\citenamefont{Leitner}
  \emph{et~al.}}]{modern-qe-theory-3}
\bibinfo{author}{\bibnamefont{Leitner}, \bibfnamefont{T.}}, \emph{et~al.},
  \bibinfo{year}{2006}, \bibinfo{journal}{Phys. Rev.}
  \textbf{\bibinfo{volume}{C73}}, \bibinfo{pages}{065502}.

\bibitem[{Leitner \emph{et~al.}(2009)\citenamefont{Leitner}
  \emph{et~al.}}]{delta-coh-models-4}
\bibinfo{author}{\bibnamefont{Leitner}, \bibfnamefont{T.}}, \emph{et~al.},
  \bibinfo{year}{2009}, \bibinfo{journal}{Phys. Rev.}
  \textbf{\bibinfo{volume}{C79}}, \bibinfo{pages}{057601}.

\bibitem[{\citenamefont{Llewellyn-Smith}(1972)}]{ll-smith}
\bibinfo{author}{\bibnamefont{Llewellyn-Smith}, \bibfnamefont{C.~H.}},
  \bibinfo{year}{1972}, \bibinfo{journal}{Phys. Rep.}
  \textbf{\bibinfo{volume}{3C}}, \bibinfo{pages}{261}.

\bibitem[{\citenamefont{Luyten} \emph{et~al.}(1963)\citenamefont{Luyten, Rood,
  and Tolhoek}}]{Luyten1963236}
\bibinfo{author}{\bibnamefont{Luyten}, \bibfnamefont{J.~R.}},
  \bibinfo{author}{\bibfnamefont{H.~P.~C.} \bibnamefont{Rood}}, and
  \bibinfo{author}{\bibfnamefont{H.~A.} \bibnamefont{Tolhoek}},
  \bibinfo{year}{1963}, \bibinfo{journal}{Nuclear Physics}
  \textbf{\bibinfo{volume}{41}}, \bibinfo{pages}{236 }, ISSN
  \bibinfo{issn}{0029-5582},
  \urlprefix\url{http://www.sciencedirect.com/science/article/B73DR-470WMGG-VP%
/2/04384c35da3b2b726ea997545059460b}.

\bibitem[{Lyubushkin \emph{et~al.}(2009)\citenamefont{Lyubushkin}
  \emph{et~al.}}]{nomad-qe}
\bibinfo{author}{\bibnamefont{Lyubushkin}, \bibfnamefont{V.}}, \emph{et~al.},
  \bibinfo{year}{2009}, \bibinfo{journal}{Eur. Phys. J.}
  \textbf{\bibinfo{volume}{C63}}, \bibinfo{pages}{355}.

\bibitem[{Maieron \emph{et~al.}(2003)\citenamefont{Maieron}
  \emph{et~al.}}]{modern-qe-theory-1}
\bibinfo{author}{\bibnamefont{Maieron}, \bibfnamefont{C.}}, \emph{et~al.},
  \bibinfo{year}{2003}, \bibinfo{journal}{Phys. Rev.}
  \textbf{\bibinfo{volume}{C68}}, \bibinfo{pages}{048501}.

\bibitem[{\citenamefont{Marciano and Parsa}(2003)}]{bib:Marciano03}
\bibinfo{author}{\bibnamefont{Marciano}, \bibfnamefont{W.~J.}}, and
  \bibinfo{author}{\bibfnamefont{Z.}~\bibnamefont{Parsa}},
  \bibinfo{year}{2003}, \bibinfo{journal}{Journal of Physics G: Nuclear and
  Particle Physics} \textbf{\bibinfo{volume}{29}}(\bibinfo{number}{11}),
  \bibinfo{pages}{2629},
  \urlprefix\url{http://stacks.iop.org/0954-3899/29/i=11/a=013}.

\bibitem[{\citenamefont{Marciano and Sirlin}(1981)}]{bib:MS}
\bibinfo{author}{\bibnamefont{Marciano}, \bibfnamefont{W.~J.}}, and
  \bibinfo{author}{\bibfnamefont{A.}~\bibnamefont{Sirlin}},
  \bibinfo{year}{1981}, \bibinfo{journal}{Phys. Rev. Lett.}
  \textbf{\bibinfo{volume}{46}}(\bibinfo{number}{3}), \bibinfo{pages}{163}.

\bibitem[{Mariani \emph{et~al.}(2011)\citenamefont{Mariani}
  \emph{et~al.}}]{k2k-ccpi0}
\bibinfo{author}{\bibnamefont{Mariani}, \bibfnamefont{C.}}, \emph{et~al.},
  \bibinfo{year}{2011}, \bibinfo{journal}{Phys. Rev.}
  \textbf{\bibinfo{volume}{D83}}, \bibinfo{pages}{054023}.

\bibitem[{Martinez \emph{et~al.}(2006)\citenamefont{Martinez}
  \emph{et~al.}}]{modern-qe-theory-4}
\bibinfo{author}{\bibnamefont{Martinez}, \bibfnamefont{M.~C.}}, \emph{et~al.},
  \bibinfo{year}{2006}, \bibinfo{journal}{Phys. Rev.}
  \textbf{\bibinfo{volume}{C73}}, \bibinfo{pages}{024607}.

\bibitem[{\citenamefont{Martini}(2011)}]{Martini:2011ui}
\bibinfo{author}{\bibnamefont{Martini}, \bibfnamefont{M.}},
  \bibinfo{year}{2011}, \eprint{1110.5895}.

\bibitem[{\citenamefont{Martini} \emph{et~al.}(2011)\citenamefont{Martini,
  Ericson, and Chanfray}}]{Martini:2011wp}
\bibinfo{author}{\bibnamefont{Martini}, \bibfnamefont{M.}},
  \bibinfo{author}{\bibfnamefont{M.}~\bibnamefont{Ericson}}, and
  \bibinfo{author}{\bibfnamefont{G.}~\bibnamefont{Chanfray}},
  \bibinfo{year}{2011}, \bibinfo{journal}{Phys.Rev.}
  \textbf{\bibinfo{volume}{C84}}, \bibinfo{pages}{055502}.

\bibitem[{\citenamefont{Maschuw}(1998)}]{Maschuw1998183}
\bibinfo{author}{\bibnamefont{Maschuw}, \bibfnamefont{R.}},
  \bibinfo{year}{1998}, \bibinfo{journal}{Progress in Particle and Nuclear
  Physics} \textbf{\bibinfo{volume}{40}}, \bibinfo{pages}{183 }, ISSN
  \bibinfo{issn}{0146-6410}, \bibinfo{note}{neutrinos in Astro, Particle and
  Nuclear Physics},
  \urlprefix\url{http://www.sciencedirect.com/science/article/B6TJC-3TB671T-T/%
2/5c6338b954079498544e055f49564be3}.

\bibitem[{Mason \emph{et~al.}(2007{\natexlab{a}})\citenamefont{Mason}
  \emph{et~al.}}]{nutev-dimuon}
\bibinfo{author}{\bibnamefont{Mason}, \bibfnamefont{D.}}, \emph{et~al.},
  \bibinfo{year}{2007}{\natexlab{a}}, \bibinfo{journal}{Phys. Rev. Lett.}
  \textbf{\bibinfo{volume}{99}}, \bibinfo{pages}{192001}.

\bibitem[{Mason \emph{et~al.}(2007{\natexlab{b}})\citenamefont{Mason}
  \emph{et~al.}}]{Mason:2007zz}
\bibinfo{author}{\bibnamefont{Mason}, \bibfnamefont{D.}}, \emph{et~al.}
  (\bibinfo{collaboration}{NuTeV Collaboration}),
  \bibinfo{year}{2007}{\natexlab{b}}, \bibinfo{journal}{Phys.Rev.Lett.}
  \textbf{\bibinfo{volume}{99}}, \bibinfo{pages}{192001}.

\bibitem[{\citenamefont{McFarland and Moch}(2003)}]{McFarland:2003jw}
\bibinfo{author}{\bibnamefont{McFarland}, \bibfnamefont{K.~S.}}, and
  \bibinfo{author}{\bibfnamefont{S.-O.} \bibnamefont{Moch}},
  \bibinfo{year}{2003}, \bibinfo{pages}{61}, \eprint{hep-ph/0306052}.

\bibitem[{Mention \emph{et~al.}(2011)\citenamefont{Mention}
  \emph{et~al.}}]{Mention:2011rk}
\bibinfo{author}{\bibnamefont{Mention}, \bibfnamefont{G.}}, \emph{et~al.},
  \bibinfo{year}{2011}, \bibinfo{journal}{Phys. Rev.}
  \textbf{\bibinfo{volume}{D83}}, \bibinfo{pages}{073006}.

\bibitem[{\citenamefont{Meucci} \emph{et~al.}(2004)\citenamefont{Meucci,
  Giusti, and Pacati}}]{Meucci2004277}
\bibinfo{author}{\bibnamefont{Meucci}, \bibfnamefont{A.}},
  \bibinfo{author}{\bibfnamefont{C.}~\bibnamefont{Giusti}}, and
  \bibinfo{author}{\bibfnamefont{F.~D.} \bibnamefont{Pacati}},
  \bibinfo{year}{2004}, \bibinfo{journal}{Nuclear Physics A}
  \textbf{\bibinfo{volume}{739}}(\bibinfo{number}{3-4}), \bibinfo{pages}{277 },
  ISSN \bibinfo{issn}{0375-9474},
  \urlprefix\url{http://www.sciencedirect.com/science/article/B6TVB-4CBVMBN-2/%
2/af7e53c4833c32f782d0ecf17f011f9a}.

\bibitem[{\citenamefont{Meucci} \emph{et~al.}(2011)\citenamefont{Meucci,
  Giusti, and Pacati}}]{Meucci11}
\bibinfo{author}{\bibnamefont{Meucci}, \bibfnamefont{A.}},
  \bibinfo{author}{\bibfnamefont{C.}~\bibnamefont{Giusti}}, and
  \bibinfo{author}{\bibfnamefont{F.~D.} \bibnamefont{Pacati}},
  \bibinfo{year}{2011}, \bibinfo{journal}{Phys.Rev.}
  \textbf{\bibinfo{volume}{D84}}, \bibinfo{pages}{113003}.

\bibitem[{\citenamefont{Mikheyev and Smirnov}(1985)}]{bib:MSW2}
\bibinfo{author}{\bibnamefont{Mikheyev}, \bibfnamefont{S.}}, and
  \bibinfo{author}{\bibfnamefont{A.}~\bibnamefont{Smirnov}},
  \bibinfo{year}{1985}, \bibinfo{journal}{Sov. J. Nucl. Phys.}
  \textbf{\bibinfo{volume}{42}}, \bibinfo{pages}{913}.

\bibitem[{\citenamefont{Mintz and Wen}(2007)}]{Mintz:2007zz}
\bibinfo{author}{\bibnamefont{Mintz}, \bibfnamefont{S.}}, and
  \bibinfo{author}{\bibfnamefont{L.}~\bibnamefont{Wen}}, \bibinfo{year}{2007},
  \bibinfo{journal}{Eur.Phys.J.} \textbf{\bibinfo{volume}{A33}},
  \bibinfo{pages}{299}.

\bibitem[{\citenamefont{Mishra} \emph{et~al.}(1990)\citenamefont{Mishra,
  Bachmann, Blair, Foudas, King} \emph{et~al.}}]{Mishra:1990yf}
\bibinfo{author}{\bibnamefont{Mishra}, \bibfnamefont{S.}},
  \bibinfo{author}{\bibfnamefont{K.}~\bibnamefont{Bachmann}},
  \bibinfo{author}{\bibfnamefont{R.}~\bibnamefont{Blair}},
  \bibinfo{author}{\bibfnamefont{C.}~\bibnamefont{Foudas}},
  \bibinfo{author}{\bibfnamefont{B.}~\bibnamefont{King}}, \emph{et~al.},
  \bibinfo{year}{1990}, \bibinfo{journal}{Phys.Lett.}
  \textbf{\bibinfo{volume}{B252}}, \bibinfo{pages}{170}.

\bibitem[{\citenamefont{Moch and Vermaseren}(2000)}]{Moch:1999eb}
\bibinfo{author}{\bibnamefont{Moch}, \bibfnamefont{S.}}, and
  \bibinfo{author}{\bibfnamefont{J.}~\bibnamefont{Vermaseren}},
  \bibinfo{year}{2000}, \bibinfo{journal}{Nucl.Phys.}
  \textbf{\bibinfo{volume}{B573}}, \bibinfo{pages}{853}.

\bibitem[{\citenamefont{Mosconi} \emph{et~al.}(2006)\citenamefont{Mosconi,
  Ricci, and Truhlik}}]{MOSCONI:2006jn}
\bibinfo{author}{\bibnamefont{Mosconi}, \bibfnamefont{B.}},
  \bibinfo{author}{\bibfnamefont{P.}~\bibnamefont{Ricci}}, and
  \bibinfo{author}{\bibfnamefont{E.}~\bibnamefont{Truhlik}},
  \bibinfo{year}{2006}, \bibinfo{journal}{Nuclear Physics A}
  \textbf{\bibinfo{volume}{772}}(\bibinfo{number}{1-2}), \bibinfo{pages}{81}.

\bibitem[{\citenamefont{Mosconi} \emph{et~al.}(2007)\citenamefont{Mosconi,
  Ricci, Truhlik, and Vogel}}]{bib:Mosconi2007ku}
\bibinfo{author}{\bibnamefont{Mosconi}, \bibfnamefont{B.}},
  \bibinfo{author}{\bibfnamefont{P.}~\bibnamefont{Ricci}},
  \bibinfo{author}{\bibfnamefont{E.}~\bibnamefont{Truhlik}}, and
  \bibinfo{author}{\bibfnamefont{P.}~\bibnamefont{Vogel}},
  \bibinfo{year}{2007}, \bibinfo{journal}{Physical Review C}
  \textbf{\bibinfo{volume}{75}}(\bibinfo{number}{4}).

\bibitem[{Mueller \emph{et~al.}(2011)\citenamefont{Mueller}
  \emph{et~al.}}]{Mueller:2011nm}
\bibinfo{author}{\bibnamefont{Mueller}, \bibfnamefont{T.~A.}}, \emph{et~al.},
  \bibinfo{year}{2011}, \eprint{1101.2663}.

\bibitem[{Nakajima \emph{et~al.}(2011)\citenamefont{Nakajima}
  \emph{et~al.}}]{Nakajima:2010fp}
\bibinfo{author}{\bibnamefont{Nakajima}, \bibfnamefont{Y.}}, \emph{et~al.}
  (\bibinfo{collaboration}{SciBooNE Collaboration}), \bibinfo{year}{2011},
  \bibinfo{journal}{Phys.Rev.} \textbf{\bibinfo{volume}{D83}},
  \bibinfo{pages}{012005}.

\bibitem[{\citenamefont{Nakamura and Seki}(2002)}]{Nakamura02}
\bibinfo{author}{\bibnamefont{Nakamura}, \bibfnamefont{H.}}, and
  \bibinfo{author}{\bibfnamefont{R.}~\bibnamefont{Seki}}, \bibinfo{year}{2002},
  \bibinfo{journal}{Nucl.Phys.Proc.Suppl.} \textbf{\bibinfo{volume}{112}},
  \bibinfo{pages}{197}.

\bibitem[{Nakamura \emph{et~al.}(2010{\natexlab{a}})\citenamefont{Nakamura}
  \emph{et~al.}}]{ga}
\bibinfo{author}{\bibnamefont{Nakamura}, \bibfnamefont{K.}}, \emph{et~al.},
  \bibinfo{year}{2010}{\natexlab{a}}, \bibinfo{journal}{J. Phys.}
  \textbf{\bibinfo{volume}{G37}}, \bibinfo{pages}{075021}.

\bibitem[{\citenamefont{Nakamura} \emph{et~al.}(2001)\citenamefont{Nakamura,
  Sato, Gudkov, and Kubodera}}]{Nakamura:2001fr}
\bibinfo{author}{\bibnamefont{Nakamura}, \bibfnamefont{S.}},
  \bibinfo{author}{\bibfnamefont{T.}~\bibnamefont{Sato}},
  \bibinfo{author}{\bibfnamefont{V.}~\bibnamefont{Gudkov}}, and
  \bibinfo{author}{\bibfnamefont{K.}~\bibnamefont{Kubodera}},
  \bibinfo{year}{2001}, \bibinfo{journal}{Physical Review C}
  \textbf{\bibinfo{volume}{63}}(\bibinfo{number}{3}).

\bibitem[{Nakamura \emph{et~al.}(2010{\natexlab{b}})\citenamefont{Nakamura}
  \emph{et~al.}}]{delta-coh-models-7}
\bibinfo{author}{\bibnamefont{Nakamura}, \bibfnamefont{S.~X.}}, \emph{et~al.},
  \bibinfo{year}{2010}{\natexlab{b}}, \bibinfo{journal}{Phys. Rev.}
  \textbf{\bibinfo{volume}{C81}}, \bibinfo{pages}{035502}.

\bibitem[{Nakayama \emph{et~al.}(2005)\citenamefont{Nakayama}
  \emph{et~al.}}]{k2k-ncpi0}
\bibinfo{author}{\bibnamefont{Nakayama}, \bibfnamefont{S.}}, \emph{et~al.},
  \bibinfo{year}{2005}, \bibinfo{journal}{Phys. Lett.}
  \textbf{\bibinfo{volume}{B619}}, \bibinfo{pages}{255}.

\bibitem[{Naumov \emph{et~al.}(2004)\citenamefont{Naumov}
  \emph{et~al.}}]{nomad-strange-1}
\bibinfo{author}{\bibnamefont{Naumov}, \bibfnamefont{D.}}, \emph{et~al.},
  \bibinfo{year}{2004}, \bibinfo{journal}{Nucl. Phys.}
  \textbf{\bibinfo{volume}{B700}}, \bibinfo{pages}{41}.

\bibitem[{\citenamefont{Navarro}(2006)}]{Navarro:2006ke}
\bibinfo{author}{\bibnamefont{Navarro}, \bibfnamefont{J.}},
  \bibinfo{year}{2006}, \bibinfo{journal}{Physics in Perspective}
  \textbf{\bibinfo{volume}{8}}(\bibinfo{number}{1}), \bibinfo{pages}{64}.

\bibitem[{\citenamefont{Nguyen}(1975)}]{NguyenTienNguyen1975485}
\bibinfo{author}{\bibnamefont{Nguyen}, \bibfnamefont{N.~T.}},
  \bibinfo{year}{1975}, \bibinfo{journal}{Nuclear Physics A}
  \textbf{\bibinfo{volume}{254}}(\bibinfo{number}{2}), \bibinfo{pages}{485 },
  ISSN \bibinfo{issn}{0375-9474},
  \urlprefix\url{http://www.sciencedirect.com/science/article/B6TVB-471YM00-4F%
/2/421cf0965e4bcbc8fde3ec4df908bc8c}.

\bibitem[{\citenamefont{Nico and Snow}(2005)}]{bib:SnowARNPS}
\bibinfo{author}{\bibnamefont{Nico}, \bibfnamefont{J.~S.}}, and
  \bibinfo{author}{\bibfnamefont{W.~M.} \bibnamefont{Snow}},
  \bibinfo{year}{2005}, \bibinfo{journal}{Annual Review of Nuclear and Particle
  Science} \textbf{\bibinfo{volume}{55}}(\bibinfo{number}{1}),
  \bibinfo{pages}{27},
  \urlprefix\url{http://www.annualreviews.org/doi/abs/10.1146/annurev.nucl.55.%
090704.151611}.

\bibitem[{\citenamefont{Nienaber}(1988)}]{nienaber}
\bibinfo{author}{\bibnamefont{Nienaber}, \bibfnamefont{P.}},
  \bibinfo{year}{1988}, Ph.D. thesis, \bibinfo{school}{Univerity of Illinois at
  Urbana-Champaign}.

\bibitem[{\citenamefont{Nieves} \emph{et~al.}(2004)\citenamefont{Nieves, Amaro,
  and Valverde}}]{Nieves04}
\bibinfo{author}{\bibnamefont{Nieves}, \bibfnamefont{J.}},
  \bibinfo{author}{\bibfnamefont{J.~E.} \bibnamefont{Amaro}}, and
  \bibinfo{author}{\bibfnamefont{M.}~\bibnamefont{Valverde}},
  \bibinfo{year}{2004}, \bibinfo{journal}{Phys.Rev.}
  \textbf{\bibinfo{volume}{C70}}, \bibinfo{pages}{055503}.

\bibitem[{\citenamefont{Nieves} \emph{et~al.}(2011)\citenamefont{Nieves,
  Ruiz-Simo, and Vicente-Vacas}}]{Nieves2011sc}
\bibinfo{author}{\bibnamefont{Nieves}, \bibfnamefont{J.}},
  \bibinfo{author}{\bibfnamefont{I.}~\bibnamefont{Ruiz-Simo}}, and
  \bibinfo{author}{\bibfnamefont{M.}~\bibnamefont{Vicente-Vacas}},
  \bibinfo{year}{2011}, \eprint{1110.1200}.

\bibitem[{\citenamefont{Nieves} \emph{et~al.}(2012)\citenamefont{Nieves, Simo,
  and Vacas}}]{Nieves:2011yp}
\bibinfo{author}{\bibnamefont{Nieves}, \bibfnamefont{J.}},
  \bibinfo{author}{\bibfnamefont{I.}~\bibnamefont{Simo}}, and
  \bibinfo{author}{\bibfnamefont{M.}~\bibnamefont{Vacas}},
  \bibinfo{year}{2012}, \bibinfo{journal}{Phys.Lett.}
  \textbf{\bibinfo{volume}{B707}}, \bibinfo{pages}{72}.

\bibitem[{\citenamefont{Nieves} \emph{et~al.}(2006)\citenamefont{Nieves,
  Valverde, and Vicente~Vacas}}]{Nieves06}
\bibinfo{author}{\bibnamefont{Nieves}, \bibfnamefont{J.}},
  \bibinfo{author}{\bibfnamefont{M.}~\bibnamefont{Valverde}}, and
  \bibinfo{author}{\bibfnamefont{M.}~\bibnamefont{Vicente~Vacas}},
  \bibinfo{year}{2006}, \bibinfo{journal}{Phys.Rev.}
  \textbf{\bibinfo{volume}{C73}}, \bibinfo{pages}{025504}.

\bibitem[{Nieves \emph{et~al.}(2006)\citenamefont{Nieves}
  \emph{et~al.}}]{modern-qe-theory-5}
\bibinfo{author}{\bibnamefont{Nieves}, \bibfnamefont{J.}}, \emph{et~al.},
  \bibinfo{year}{2006}, \bibinfo{journal}{Phys. Rev.}
  \textbf{\bibinfo{volume}{C73}}, \bibinfo{pages}{025501}.

\bibitem[{Onegut \emph{et~al.}(2006)\citenamefont{Onegut}
  \emph{et~al.}}]{chorus-dis}
\bibinfo{author}{\bibnamefont{Onegut}, \bibfnamefont{G.}}, \emph{et~al.},
  \bibinfo{year}{2006}, \bibinfo{journal}{Phys. Lett.}
  \textbf{\bibinfo{volume}{B632}}, \bibinfo{pages}{65}.

\bibitem[{Onengut \emph{et~al.}(2004)\citenamefont{Onengut}
  \emph{et~al.}}]{Onengut:2004vd}
\bibinfo{author}{\bibnamefont{Onengut}, \bibfnamefont{G.}}, \emph{et~al.}
  (\bibinfo{collaboration}{CHORUS Collaboration}), \bibinfo{year}{2004},
  \bibinfo{journal}{Phys.Lett.} \textbf{\bibinfo{volume}{B604}},
  \bibinfo{pages}{145}.

\bibitem[{Onengut \emph{et~al.}(2005)\citenamefont{Onengut}
  \emph{et~al.}}]{Onengut:2005sy}
\bibinfo{author}{\bibnamefont{Onengut}, \bibfnamefont{G.}}, \emph{et~al.}
  (\bibinfo{collaboration}{CHORUS Collaboration}), \bibinfo{year}{2005},
  \bibinfo{journal}{Phys.Lett.} \textbf{\bibinfo{volume}{B613}},
  \bibinfo{pages}{105}.

\bibitem[{\citenamefont{Paschos and Kartavtsev}(2003)}]{pcac-coh-models-1}
\bibinfo{author}{\bibnamefont{Paschos}, \bibfnamefont{E.~A.}}, and
  \bibinfo{author}{\bibfnamefont{A.~V.} \bibnamefont{Kartavtsev}},
  \bibinfo{year}{2003}, \eprint{hep-ph/0309148}.

\bibitem[{Paschos \emph{et~al.}(2006)\citenamefont{Paschos}
  \emph{et~al.}}]{pcac-coh-models-3}
\bibinfo{author}{\bibnamefont{Paschos}, \bibfnamefont{E.~A.}}, \emph{et~al.},
  \bibinfo{year}{2006}, \bibinfo{journal}{Phys. Rev.}
  \textbf{\bibinfo{volume}{D74}}, \bibinfo{pages}{054007}.

\bibitem[{Paschos \emph{et~al.}(2007)\citenamefont{Paschos}
  \emph{et~al.}}]{1pi-fsi-models}
\bibinfo{author}{\bibnamefont{Paschos}, \bibfnamefont{E.~A.}}, \emph{et~al.},
  \bibinfo{year}{2007}, \eprint{hep-ph/0704.1991}.

\bibitem[{Paschos \emph{et~al.}(2009)\citenamefont{Paschos}
  \emph{et~al.}}]{pcac-coh-models-6}
\bibinfo{author}{\bibnamefont{Paschos}, \bibfnamefont{E.~A.}}, \emph{et~al.},
  \bibinfo{year}{2009}, \bibinfo{journal}{Phys. Rev.}
  \textbf{\bibinfo{volume}{D80}}, \bibinfo{pages}{033005}.

\bibitem[{\citenamefont{Pasierb} \emph{et~al.}(1979)\citenamefont{Pasierb,
  Gurr, Lathrop, Reines, and Sobel}}]{bib:SavannahRiver}
\bibinfo{author}{\bibnamefont{Pasierb}, \bibfnamefont{E.}},
  \bibinfo{author}{\bibfnamefont{H.~S.} \bibnamefont{Gurr}},
  \bibinfo{author}{\bibfnamefont{J.}~\bibnamefont{Lathrop}},
  \bibinfo{author}{\bibfnamefont{F.}~\bibnamefont{Reines}}, and
  \bibinfo{author}{\bibfnamefont{H.~W.} \bibnamefont{Sobel}},
  \bibinfo{year}{1979}, \bibinfo{journal}{Phys. Rev. Lett.}
  \textbf{\bibinfo{volume}{43}}(\bibinfo{number}{2}), \bibinfo{pages}{96}.

\bibitem[{\citenamefont{Peccei and Donnelly}(1979)}]{bib:Peccei1979}
\bibinfo{author}{\bibnamefont{Peccei}, \bibfnamefont{R.~D.}}, and
  \bibinfo{author}{\bibfnamefont{T.}~\bibnamefont{Donnelly}},
  \bibinfo{year}{1979}, \bibinfo{journal}{Physics Reports}
  \textbf{\bibinfo{volume}{50}}(\bibinfo{number}{1}), \bibinfo{pages}{1}.

\bibitem[{Pohl \emph{et~al.}(1978)\citenamefont{Pohl}
  \emph{et~al.}}]{nc-elastic-5}
\bibinfo{author}{\bibnamefont{Pohl}, \bibfnamefont{M.}}, \emph{et~al.},
  \bibinfo{year}{1978}, \bibinfo{journal}{Phys. Lett.}
  \textbf{\bibinfo{volume}{72B}}, \bibinfo{pages}{489}.

\bibitem[{\citenamefont{R.~P.~Feynman}(1971)}]{rein-sehgal-1}
\bibinfo{author}{\bibnamefont{R.~P.~Feynman}, \bibfnamefont{F.~R.,
  M.~Kislinger}}, \bibinfo{year}{1971}, \bibinfo{journal}{Phys. Rev.}
  \textbf{\bibinfo{volume}{D3}}, \bibinfo{pages}{2706}.

\bibitem[{\citenamefont{Rabinowitz}
  \emph{et~al.}(1993)\citenamefont{Rabinowitz, Arroyo, Bachmann, Bazarko,
  Bolton} \emph{et~al.}}]{Rabinowitz:1993xx}
\bibinfo{author}{\bibnamefont{Rabinowitz}, \bibfnamefont{S.}},
  \bibinfo{author}{\bibfnamefont{C.}~\bibnamefont{Arroyo}},
  \bibinfo{author}{\bibfnamefont{K.}~\bibnamefont{Bachmann}},
  \bibinfo{author}{\bibfnamefont{A.}~\bibnamefont{Bazarko}},
  \bibinfo{author}{\bibfnamefont{T.}~\bibnamefont{Bolton}}, \emph{et~al.},
  \bibinfo{year}{1993}, \bibinfo{journal}{Phys.Rev.Lett.}
  \textbf{\bibinfo{volume}{70}}, \bibinfo{pages}{134}.

\bibitem[{\citenamefont{Rein}(1987)}]{rein-sehgal-2}
\bibinfo{author}{\bibnamefont{Rein}, \bibfnamefont{D.}}, \bibinfo{year}{1987},
  \bibinfo{journal}{Z. Phys.} \textbf{\bibinfo{volume}{C35}},
  \bibinfo{pages}{43}.

\bibitem[{\citenamefont{Rein and Sehgal}(1981)}]{rein-sehgal}
\bibinfo{author}{\bibnamefont{Rein}, \bibfnamefont{D.}}, and
  \bibinfo{author}{\bibfnamefont{L.~M.} \bibnamefont{Sehgal}},
  \bibinfo{year}{1981}, \bibinfo{journal}{Annals Phys}
  \textbf{\bibinfo{volume}{133}}, \bibinfo{pages}{79}.

\bibitem[{\citenamefont{Rein and Sehgal}(1983)}]{rein-sehgal-coherent}
\bibinfo{author}{\bibnamefont{Rein}, \bibfnamefont{D.}}, and
  \bibinfo{author}{\bibfnamefont{L.~M.} \bibnamefont{Sehgal}},
  \bibinfo{year}{1983}, \bibinfo{journal}{Nucl. Phys.}
  \textbf{\bibinfo{volume}{B223}}, \bibinfo{pages}{29}.

\bibitem[{\citenamefont{Reines} \emph{et~al.}(1976)\citenamefont{Reines, Gurr,
  and Sobel}}]{Reines:1976pv}
\bibinfo{author}{\bibnamefont{Reines}, \bibfnamefont{F.}},
  \bibinfo{author}{\bibfnamefont{H.~S.} \bibnamefont{Gurr}}, and
  \bibinfo{author}{\bibfnamefont{H.~W.} \bibnamefont{Sobel}},
  \bibinfo{year}{1976}, \bibinfo{journal}{Phys. Rev. Lett.}
  \textbf{\bibinfo{volume}{37}}, \bibinfo{pages}{315}.

\bibitem[{\citenamefont{Ricci and Truhlik}(2010)}]{Ricci:2010sk}
\bibinfo{author}{\bibnamefont{Ricci}, \bibfnamefont{P.}}, and
  \bibinfo{author}{\bibfnamefont{E.}~\bibnamefont{Truhlik}},
  \bibinfo{year}{2010}, \eprint{1012.2216}.

\bibitem[{\citenamefont{Riley} \emph{et~al.}(1999)\citenamefont{Riley,
  Greenwood, Kropp, Price, Reines} \emph{et~al.}}]{bib:BugeyDeut}
\bibinfo{author}{\bibnamefont{Riley}, \bibfnamefont{S.}},
  \bibinfo{author}{\bibfnamefont{Z.}~\bibnamefont{Greenwood}},
  \bibinfo{author}{\bibfnamefont{W.}~\bibnamefont{Kropp}},
  \bibinfo{author}{\bibfnamefont{L.}~\bibnamefont{Price}},
  \bibinfo{author}{\bibfnamefont{F.}~\bibnamefont{Reines}}, \emph{et~al.},
  \bibinfo{year}{1999}, \bibinfo{journal}{Phys.Rev.}
  \textbf{\bibinfo{volume}{C59}}, \bibinfo{pages}{1780}.

\bibitem[{Riley \emph{et~al.}(1998)\citenamefont{Riley}
  \emph{et~al.}}]{bib:KrasnoyarskDeut2}
\bibinfo{author}{\bibnamefont{Riley}, \bibfnamefont{S.~P.}}, \emph{et~al.},
  \bibinfo{year}{1998}, \bibinfo{journal}{Phys. Rev. C}
  \textbf{\bibinfo{volume}{59}}, \bibinfo{pages}{1780}.

\bibitem[{Rodriguez \emph{et~al.}(2008)\citenamefont{Rodriguez}
  \emph{et~al.}}]{k2k-ccpip}
\bibinfo{author}{\bibnamefont{Rodriguez}, \bibfnamefont{A.}}, \emph{et~al.},
  \bibinfo{year}{2008}, \bibinfo{journal}{Phys. Rev.}
  \textbf{\bibinfo{volume}{D78}}, \bibinfo{pages}{032003}.

\bibitem[{\citenamefont{Ruf}(2005)}]{bib:Ruf}
\bibinfo{author}{\bibnamefont{Ruf}, \bibfnamefont{G.}}, \bibinfo{year}{2005},
  Master's thesis, \bibinfo{school}{University of Bonn},
  \bibinfo{address}{Germany}.

\bibitem[{\citenamefont{Sajjad~Athar}
  \emph{et~al.}(2010)\citenamefont{Sajjad~Athar, Chauhan, and Singh}}]{Athar10}
\bibinfo{author}{\bibnamefont{Sajjad~Athar}, \bibfnamefont{M.}},
  \bibinfo{author}{\bibfnamefont{S.}~\bibnamefont{Chauhan}}, and
  \bibinfo{author}{\bibfnamefont{S.}~\bibnamefont{Singh}},
  \bibinfo{year}{2010}, \bibinfo{journal}{Eur.Phys.J.}
  \textbf{\bibinfo{volume}{A43}}, \bibinfo{pages}{209}.

\bibitem[{Schienbein \emph{et~al.}(2008)\citenamefont{Schienbein}
  \emph{et~al.}}]{tmass}
\bibinfo{author}{\bibnamefont{Schienbein}, \bibfnamefont{I.}}, \emph{et~al.},
  \bibinfo{year}{2008}, \bibinfo{journal}{J. Phys.}
  \textbf{\bibinfo{volume}{G35}}, \bibinfo{pages}{053101}.

\bibitem[{\citenamefont{Scholberg}(2006)}]{bib:CLEAR}
\bibinfo{author}{\bibnamefont{Scholberg}, \bibfnamefont{K.}},
  \bibinfo{year}{2006}, \bibinfo{journal}{Phys. Rev. D}
  \textbf{\bibinfo{volume}{73}}(\bibinfo{number}{3}), \bibinfo{pages}{033005}.

\bibitem[{\citenamefont{Schreckenbach}
  \emph{et~al.}(1985)\citenamefont{Schreckenbach, Colvin, Gelletly, and
  Von~Feilitzsch}}]{Schreckenbach:1985ep}
\bibinfo{author}{\bibnamefont{Schreckenbach}, \bibfnamefont{K.}},
  \bibinfo{author}{\bibfnamefont{G.}~\bibnamefont{Colvin}},
  \bibinfo{author}{\bibfnamefont{W.}~\bibnamefont{Gelletly}}, and
  \bibinfo{author}{\bibfnamefont{F.}~\bibnamefont{Von~Feilitzsch}},
  \bibinfo{year}{1985}, \bibinfo{journal}{Phys. Lett.}
  \textbf{\bibinfo{volume}{B160}}, \bibinfo{pages}{325}.

\bibitem[{Serebrov \emph{et~al.}(2005)\citenamefont{Serebrov}
  \emph{et~al.}}]{SEREBROV:2005je}
\bibinfo{author}{\bibnamefont{Serebrov}, \bibfnamefont{A.}}, \emph{et~al.},
  \bibinfo{year}{2005}, \bibinfo{journal}{Physics Letters B}
  \textbf{\bibinfo{volume}{605}}(\bibinfo{number}{1-2}), \bibinfo{pages}{72}.

\bibitem[{\citenamefont{Shrock}(1975)}]{single-k-models}
\bibinfo{author}{\bibnamefont{Shrock}, \bibfnamefont{R.}},
  \bibinfo{year}{1975}, \bibinfo{journal}{Phys. Rev.}
  \textbf{\bibinfo{volume}{D12}}, \bibinfo{pages}{2049}.

\bibitem[{\citenamefont{Singh and Oset}(1992)}]{singh-qe}
\bibinfo{author}{\bibnamefont{Singh}, \bibfnamefont{S.~K.}}, and
  \bibinfo{author}{\bibfnamefont{E.}~\bibnamefont{Oset}}, \bibinfo{year}{1992},
  \bibinfo{journal}{Nucl. Phys.} \textbf{\bibinfo{volume}{542A}},
  \bibinfo{pages}{587 .}

\bibitem[{\citenamefont{Singh and Vacas}(2006)}]{new-hyperon-models}
\bibinfo{author}{\bibnamefont{Singh}, \bibfnamefont{S.~K.}}, and
  \bibinfo{author}{\bibfnamefont{M.~J.~V.} \bibnamefont{Vacas}},
  \bibinfo{year}{2006}, \bibinfo{journal}{Phys. Rev.}
  \textbf{\bibinfo{volume}{D 74}}, \bibinfo{pages}{053309}.

\bibitem[{Singh \emph{et~al.}(2006)\citenamefont{Singh}
  \emph{et~al.}}]{delta-coh-models-1}
\bibinfo{author}{\bibnamefont{Singh}, \bibfnamefont{S.~K.}}, \emph{et~al.},
  \bibinfo{year}{2006}, \bibinfo{journal}{Phys. Rev. Lett.}
  \textbf{\bibinfo{volume}{96}}, \bibinfo{pages}{241801}.

\bibitem[{\citenamefont{Sirlin}(1980)}]{bib:Sirlin80}
\bibinfo{author}{\bibnamefont{Sirlin}, \bibfnamefont{A.}},
  \bibinfo{year}{1980}, \bibinfo{journal}{Phys. Rev. D}
  \textbf{\bibinfo{volume}{22}}(\bibinfo{number}{4}), \bibinfo{pages}{971}.

\bibitem[{\citenamefont{Sirlin and Marciano}(1981)}]{Sirlin:1981yz}
\bibinfo{author}{\bibnamefont{Sirlin}, \bibfnamefont{A.}}, and
  \bibinfo{author}{\bibfnamefont{W.}~\bibnamefont{Marciano}},
  \bibinfo{year}{1981}, \bibinfo{journal}{Nucl.Phys.}
  \textbf{\bibinfo{volume}{B189}}, \bibinfo{pages}{442}.

\bibitem[{\citenamefont{Smith and Moniz}(1972)}]{smith-moniz}
\bibinfo{author}{\bibnamefont{Smith}, \bibfnamefont{R.~A.}}, and
  \bibinfo{author}{\bibfnamefont{E.~J.} \bibnamefont{Moniz}},
  \bibinfo{year}{1972}, \bibinfo{journal}{Nucl. Phys.}
  \textbf{\bibinfo{volume}{B43}}, \bibinfo{pages}{605}.

\bibitem[{\citenamefont{Sobczyk}(2011)}]{Sobczyk11}
\bibinfo{author}{\bibnamefont{Sobczyk}, \bibfnamefont{J.~T.}},
  \bibinfo{year}{2011}, \bibinfo{journal}{AIP Conf.Proc.}
  \textbf{\bibinfo{volume}{1405}}, \bibinfo{pages}{59}.

\bibitem[{\citenamefont{Sobczyk}(2012)}]{Sobczyk:2012ah}
\bibinfo{author}{\bibnamefont{Sobczyk}, \bibfnamefont{J.~T.}},
  \bibinfo{year}{2012}, \eprint{1109.1081}.

\bibitem[{Son \emph{et~al.}(1983)\citenamefont{Son} \emph{et~al.}}]{kaon-bc-10}
\bibinfo{author}{\bibnamefont{Son}, \bibfnamefont{D.}}, \emph{et~al.},
  \bibinfo{year}{1983}, \bibinfo{journal}{Phys. Rev.}
  \textbf{\bibinfo{volume}{D28}}, \bibinfo{pages}{2129}.

\bibitem[{\citenamefont{Taddeucci} \emph{et~al.}(1987)\citenamefont{Taddeucci,
  Goulding, Carey, Byrd, Goodman, Gaarde, Larsen, Horen, Rapaport, and
  Sugarbaker}}]{Taddeucci1987125}
\bibinfo{author}{\bibnamefont{Taddeucci}, \bibfnamefont{T.~N.}},
  \bibinfo{author}{\bibfnamefont{C.~A.} \bibnamefont{Goulding}},
  \bibinfo{author}{\bibfnamefont{T.~A.} \bibnamefont{Carey}},
  \bibinfo{author}{\bibfnamefont{R.~C.} \bibnamefont{Byrd}},
  \bibinfo{author}{\bibfnamefont{C.~D.} \bibnamefont{Goodman}},
  \bibinfo{author}{\bibfnamefont{C.}~\bibnamefont{Gaarde}},
  \bibinfo{author}{\bibfnamefont{J.}~\bibnamefont{Larsen}},
  \bibinfo{author}{\bibfnamefont{D.}~\bibnamefont{Horen}},
  \bibinfo{author}{\bibfnamefont{J.}~\bibnamefont{Rapaport}}, and
  \bibinfo{author}{\bibfnamefont{E.}~\bibnamefont{Sugarbaker}},
  \bibinfo{year}{1987}, \bibinfo{journal}{Nuclear Physics A}
  \textbf{\bibinfo{volume}{469}}(\bibinfo{number}{1}), \bibinfo{pages}{125 },
  ISSN \bibinfo{issn}{0375-9474},
  \urlprefix\url{http://www.sciencedirect.com/science/article/B6TVB-4731354-15%
S/2/229fc497e78a91eb69dcc640ec9f484a}.

\bibitem[{\citenamefont{Tatara} \emph{et~al.}(1990)\citenamefont{Tatara,
  Kohyama, and Kubodera}}]{bib:Tatara90}
\bibinfo{author}{\bibnamefont{Tatara}, \bibfnamefont{N.}},
  \bibinfo{author}{\bibfnamefont{Y.}~\bibnamefont{Kohyama}}, and
  \bibinfo{author}{\bibfnamefont{K.}~\bibnamefont{Kubodera}},
  \bibinfo{year}{1990}, \bibinfo{journal}{Phys. Rev. C}
  \textbf{\bibinfo{volume}{42}}(\bibinfo{number}{4}), \bibinfo{pages}{1694}.

\bibitem[{\citenamefont{Towner}(1998)}]{bib:Towner1998}
\bibinfo{author}{\bibnamefont{Towner}, \bibfnamefont{I.~S.}},
  \bibinfo{year}{1998}, \bibinfo{journal}{Physical Review C}
  \textbf{\bibinfo{volume}{58}}, \bibinfo{pages}{1288}.

\bibitem[{Tzanov \emph{et~al.}(2006{\natexlab{a}})\citenamefont{Tzanov}
  \emph{et~al.}}]{nutev-dis}
\bibinfo{author}{\bibnamefont{Tzanov}, \bibfnamefont{M.}}, \emph{et~al.},
  \bibinfo{year}{2006}{\natexlab{a}}, \bibinfo{journal}{Phys. Rev.}
  \textbf{\bibinfo{volume}{D74}}, \bibinfo{pages}{012008}.

\bibitem[{Tzanov \emph{et~al.}(2006{\natexlab{b}})\citenamefont{Tzanov}
  \emph{et~al.}}]{Tzanov:2005kr}
\bibinfo{author}{\bibnamefont{Tzanov}, \bibfnamefont{M.}}, \emph{et~al.}
  (\bibinfo{collaboration}{NuTeV Collaboration}),
  \bibinfo{year}{2006}{\natexlab{b}}, \bibinfo{journal}{Phys.Rev.}
  \textbf{\bibinfo{volume}{D74}}, \bibinfo{pages}{012008}.

\bibitem[{Ushida \emph{et~al.}(1983)\citenamefont{Ushida}
  \emph{et~al.}}]{Ushida:1982ty}
\bibinfo{author}{\bibnamefont{Ushida}, \bibfnamefont{N.}}, \emph{et~al.}
  (\bibinfo{collaboration}{Canada-Japan-Korea-USA Hybrid Emulsion Spectrometer
  Collaboration}), \bibinfo{year}{1983}, \bibinfo{journal}{Phys.Lett.}
  \textbf{\bibinfo{volume}{B121}}, \bibinfo{pages}{292}.

\bibitem[{Vershinsky \emph{et~al.}(1991)\citenamefont{Vershinsky}
  \emph{et~al.}}]{bib:RovnoDeut}
\bibinfo{author}{\bibnamefont{Vershinsky}, \bibfnamefont{A.~G.}},
  \emph{et~al.}, \bibinfo{year}{1991}, \bibinfo{journal}{JETF Lett.}
  \textbf{\bibinfo{volume}{53}}, \bibinfo{pages}{513}.

\bibitem[{Vidyakin \emph{et~al.}(1987)\citenamefont{Vidyakin}
  \emph{et~al.}}]{bib:Krasnoyarsk}
\bibinfo{author}{\bibnamefont{Vidyakin}, \bibfnamefont{G.~S.}}, \emph{et~al.},
  \bibinfo{year}{1987}, \bibinfo{journal}{JETP} \textbf{\bibinfo{volume}{93}},
  \bibinfo{pages}{424}.

\bibitem[{Vilain \emph{et~al.}(1993)\citenamefont{Vilain}
  \emph{et~al.}}]{vilain}
\bibinfo{author}{\bibnamefont{Vilain}, \bibfnamefont{P.}}, \emph{et~al.},
  \bibinfo{year}{1993}, \bibinfo{journal}{Phys. Lett.}
  \textbf{\bibinfo{volume}{B313}}, \bibinfo{pages}{267}.

\bibitem[{Vilain \emph{et~al.}(1995{\natexlab{a}})\citenamefont{Vilain}
  \emph{et~al.}}]{Vilain:1996yf}
\bibinfo{author}{\bibnamefont{Vilain}, \bibfnamefont{P.}}, \emph{et~al.}
  (\bibinfo{collaboration}{CHARM-II Collaboration}),
  \bibinfo{year}{1995}{\natexlab{a}}, \bibinfo{journal}{Phys.Lett.}
  \textbf{\bibinfo{volume}{B364}}, \bibinfo{pages}{121}.

\bibitem[{Vilain \emph{et~al.}(1995{\natexlab{b}})\citenamefont{Vilain}
  \emph{et~al.}}]{Vilain:1994hm}
\bibinfo{author}{\bibnamefont{Vilain}, \bibfnamefont{P.}}, \emph{et~al.}
  (\bibinfo{collaboration}{CHARM-II Collaboration}),
  \bibinfo{year}{1995}{\natexlab{b}}, \bibinfo{journal}{Phys.Lett.}
  \textbf{\bibinfo{volume}{B345}}, \bibinfo{pages}{115}.

\bibitem[{Vilain \emph{et~al.}(1999)\citenamefont{Vilain}
  \emph{et~al.}}]{Vilain:1998uw}
\bibinfo{author}{\bibnamefont{Vilain}, \bibfnamefont{P.}}, \emph{et~al.}
  (\bibinfo{collaboration}{CHARM II Collaboration}), \bibinfo{year}{1999},
  \bibinfo{journal}{Eur.Phys.J.} \textbf{\bibinfo{volume}{C11}},
  \bibinfo{pages}{19}.

\bibitem[{\citenamefont{Volpe} \emph{et~al.}(2000)\citenamefont{Volpe,
  Auerbach, Col\`o, Suzuki, and Van~Giai}}]{bib:QRPA}
\bibinfo{author}{\bibnamefont{Volpe}, \bibfnamefont{C.}},
  \bibinfo{author}{\bibfnamefont{N.}~\bibnamefont{Auerbach}},
  \bibinfo{author}{\bibfnamefont{G.}~\bibnamefont{Col\`o}},
  \bibinfo{author}{\bibfnamefont{T.}~\bibnamefont{Suzuki}}, and
  \bibinfo{author}{\bibfnamefont{N.}~\bibnamefont{Van~Giai}},
  \bibinfo{year}{2000}, \bibinfo{journal}{Phys. Rev. C}
  \textbf{\bibinfo{volume}{62}}(\bibinfo{number}{1}), \bibinfo{pages}{015501}.

\bibitem[{\citenamefont{Walter}(2007)}]{ccpi-walter}
\bibinfo{author}{\bibnamefont{Walter}, \bibfnamefont{C.}},
  \bibinfo{year}{2007}, \bibinfo{journal}{AIP Conf. Proc.}
  \textbf{\bibinfo{volume}{967}}, \bibinfo{pages}{3}.

\bibitem[{\citenamefont{Watson} \emph{et~al.}(1985)\citenamefont{Watson,
  Pairsuwan, Anderson, Baldwin, Flanders, Madey, McCarthy, Brown, Wildenthal,
  and Foster}}]{bib:Watson1985}
\bibinfo{author}{\bibnamefont{Watson}, \bibfnamefont{J.~W.}},
  \bibinfo{author}{\bibfnamefont{W.}~\bibnamefont{Pairsuwan}},
  \bibinfo{author}{\bibfnamefont{B.~D.} \bibnamefont{Anderson}},
  \bibinfo{author}{\bibfnamefont{A.~R.} \bibnamefont{Baldwin}},
  \bibinfo{author}{\bibfnamefont{B.~S.} \bibnamefont{Flanders}},
  \bibinfo{author}{\bibfnamefont{R.}~\bibnamefont{Madey}},
  \bibinfo{author}{\bibfnamefont{R.~J.} \bibnamefont{McCarthy}},
  \bibinfo{author}{\bibfnamefont{B.~A.} \bibnamefont{Brown}},
  \bibinfo{author}{\bibfnamefont{B.~H.} \bibnamefont{Wildenthal}}, and
  \bibinfo{author}{\bibfnamefont{C.~C.} \bibnamefont{Foster}},
  \bibinfo{year}{1985}, \bibinfo{journal}{Phys. Rev. Lett.}
  \textbf{\bibinfo{volume}{55}}(\bibinfo{number}{13}), \bibinfo{pages}{1369}.

\bibitem[{Webber \emph{et~al.}(2011)\citenamefont{Webber}
  \emph{et~al.}}]{Webber:2010zf}
\bibinfo{author}{\bibnamefont{Webber}, \bibfnamefont{D.~M.}}, \emph{et~al.}
  (\bibinfo{collaboration}{MuLan}), \bibinfo{year}{2011},
  \bibinfo{journal}{Phys. Rev. Lett.} \textbf{\bibinfo{volume}{106}},
  \bibinfo{pages}{041803}.

\bibitem[{\citenamefont{Weinberg}(1962)}]{weinberg1962und}
\bibinfo{author}{\bibnamefont{Weinberg}, \bibfnamefont{S.}},
  \bibinfo{year}{1962}, \bibinfo{journal}{Phys. Rev.}
  \textbf{\bibinfo{volume}{128}}(\bibinfo{number}{3}), \bibinfo{pages}{1457}.

\bibitem[{\citenamefont{Weinberg}(1967)}]{Weinberg:1967tq}
\bibinfo{author}{\bibnamefont{Weinberg}, \bibfnamefont{S.}},
  \bibinfo{year}{1967}, \bibinfo{journal}{Phys. Rev. Lett.}
  \textbf{\bibinfo{volume}{19}}, \bibinfo{pages}{1264}.

\bibitem[{\citenamefont{Willis} \emph{et~al.}(1980)\citenamefont{Willis,
  Hughes, N\'emethy, Burman, Cochran, Frank, Redwine, Duclos, Kaspar, Hargrove,
  and Moser}}]{bib:LAMPFvd}
\bibinfo{author}{\bibnamefont{Willis}, \bibfnamefont{S.~E.}},
  \bibinfo{author}{\bibfnamefont{V.~W.} \bibnamefont{Hughes}},
  \bibinfo{author}{\bibfnamefont{P.}~\bibnamefont{N\'emethy}},
  \bibinfo{author}{\bibfnamefont{R.~L.} \bibnamefont{Burman}},
  \bibinfo{author}{\bibfnamefont{D.~R.~F.} \bibnamefont{Cochran}},
  \bibinfo{author}{\bibfnamefont{J.~S.} \bibnamefont{Frank}},
  \bibinfo{author}{\bibfnamefont{R.~P.} \bibnamefont{Redwine}},
  \bibinfo{author}{\bibfnamefont{J.}~\bibnamefont{Duclos}},
  \bibinfo{author}{\bibfnamefont{H.}~\bibnamefont{Kaspar}},
  \bibinfo{author}{\bibfnamefont{C.~K.} \bibnamefont{Hargrove}}, and
  \bibinfo{author}{\bibfnamefont{U.}~\bibnamefont{Moser}},
  \bibinfo{year}{1980}, \bibinfo{journal}{Phys. Rev. Lett.}
  \textbf{\bibinfo{volume}{44}}(\bibinfo{number}{8}), \bibinfo{pages}{522}.

\bibitem[{Willocq \emph{et~al.}(1992)\citenamefont{Willocq}
  \emph{et~al.}}]{kaon-bc-13}
\bibinfo{author}{\bibnamefont{Willocq}, \bibfnamefont{S.}}, \emph{et~al.},
  \bibinfo{year}{1992}, \bibinfo{journal}{Z. Phys.}
  \textbf{\bibinfo{volume}{C53}}, \bibinfo{pages}{207}.

\bibitem[{\citenamefont{Wolfenstein}(1978)}]{bib:MSW1}
\bibinfo{author}{\bibnamefont{Wolfenstein}, \bibfnamefont{L.}},
  \bibinfo{year}{1978}, \bibinfo{journal}{Phys. Rev. D}
  \textbf{\bibinfo{volume}{17}}, \bibinfo{pages}{2369}.

\bibitem[{Wu \emph{et~al.}(2008)\citenamefont{Wu} \emph{et~al.}}]{nomad-dis}
\bibinfo{author}{\bibnamefont{Wu}, \bibfnamefont{Q.}}, \emph{et~al.},
  \bibinfo{year}{2008}, \bibinfo{journal}{Phys. Lett.}
  \textbf{\bibinfo{volume}{B660}}, \bibinfo{pages}{19}.

\bibitem[{Zacek \emph{et~al.}(1986)\citenamefont{Zacek}
  \emph{et~al.}}]{bib:Gosgen86}
\bibinfo{author}{\bibnamefont{Zacek}, \bibfnamefont{G.}}, \emph{et~al.},
  \bibinfo{year}{1986}, \bibinfo{journal}{Phys. Rev. D}
  \textbf{\bibinfo{volume}{34}}(\bibinfo{number}{9}), \bibinfo{pages}{2621}.

\bibitem[{Zeitnitz \emph{et~al.}(1994)\citenamefont{Zeitnitz}
  \emph{et~al.}}]{bib:KARMENC12N12}
\bibinfo{author}{\bibnamefont{Zeitnitz}, \bibfnamefont{B.}}, \emph{et~al.},
  \bibinfo{year}{1994}, \bibinfo{journal}{Prog. Part. Nucl. Phys.}
  \textbf{\bibinfo{volume}{32}}, \bibinfo{pages}{351}.

\bibitem[{Zeller \emph{et~al.}(2002)\citenamefont{Zeller}
  \emph{et~al.}}]{nutev-stw}
\bibinfo{author}{\bibnamefont{Zeller}, \bibfnamefont{G.~P.}}, \emph{et~al.},
  \bibinfo{year}{2002}, \bibinfo{journal}{Phys. Rev. Lett.}
  \textbf{\bibinfo{volume}{88}}, \bibinfo{pages}{091802}.

\end{thebibliography}

\end{document}